# A novel, compact and portable 2-LTD-Brick $x$-pinch radiation source: its development and radiation performance

by

Roman Shapovalov

A dissertation

submitted in partial fulfilment

of the requirements for the degree of

Doctor of Philosophy in the Department of Physics

Idaho State University

Fall 2016

# TABLE OF CONTENTS

# LIST OF FIGURES

# LIST OF TABLES

# ABSTRACT

Almost all well-known *x*-pinch *x*-ray radiation machines are large, based on a conventional Marx generator, and lack portability. The literature suggests that a current rate of rise of 1 kA/ns or more is required for "good" *x*-pinch radiation performance, which, for reasonable current rise times, translates to a current requirement of 100 kA or more. Those requirements are difficult to achieve in a limited volume, if one wants to build a compact machine without the use of traditional Marx generators, pulse-forming lines, and transmission lines.

In this work we describe a new, compact and portable *x*-pinch driver based on two "slow" LTD bricks combined into one solid unit. The short-circuit tests demonstrated the required 1-kA/ns current rate-of-rise and *x*-pinch shots confirmed "good" *x*-pinch radiation performance and revealed the potential for many *x*-pinch applications.

# 1 INTRODUCTION

## 1.1 Overview

An *x*-pinch *x*-ray radiation source has many proven applications in plasma backlighting, phase-contrast imaging of soft biological objects, characterization of inertial-confinement-fusion capsules, and more. However, there are specific electrical requirements one must satisfy to achieve a "good" *x*-pinch radiation performance. Specifically, the current rise rate delivered to a low-inductance *x*-pinch load has to be at least 1 kA/ns. For reasonable current rise times that translates to at least a 100-kA peak-current requirement.

It is fairly simple to satisfy the aforementioned requirements and build an *x*-pinch radiation machine based on conventional Marx generators and pulse-forming lines, when there are no limitations in driver size and cost. Indeed, many such machines were built and are widely used in *x*-pinch experiments. However, in most cases such installations are bulky, expensive, contain insulating oil, require de-ionized water, and reside in large universities and laboratories. A few small pulsed-power drivers are known. Almost all of them are based on new low-inductance capacitor and switch technologies, but even more drivers are needed. The purpose of this work is to develop a new, compact and portable *x*-pinch radiation source generator. We investigate the possibilities behind the driver design, describe its fabrication, perform detailed driver testing, and characterize its *x*-pinch *x*-ray performance. Our 2-LTD-brick *x*-pinch driver is very compact and portable, contains no oil, is inexpensive, and can be easily relocated to where *x*-pinch radiation source is needed.

## 1.2 *X*-pinch *x*-ray radiation source and its applications

In its original configuration, the *x*-pinch *x*-ray radiation source (in which the crossing and just touching of two thin, 5 to 50 μm, wires forms the letter X, hence the name, *x* pinch) was first introduced in 1982 by Zakharov et al. [1]. Later, four or more wires were used, simple or nested configurations were tested, while the name "*x* pinch" has been retained for these generic configurations. The *x*-pinch load is placed in the anode-cathode gap of a high-current pulse generator and the initial current is divided among all *x*-pinch wires. The total current combines in the region where the wires cross, where, because a higher local magnetic field configuration, a few-hundred μm-long *z*-pinch plasma column is formed. After column implosion, one or more, so called, "hot spots" (also known as a "bright spot", or "micro-pinch") are formed in the middle of the wire intersection region. Such bright spots are reliably produced in predictable locations (within a few hundreds μm-long *z*-pinch plasma column) with predictable times (with ±2 ns time jitter). A single-wire explosion, in contrast, randomly generates hot-spots at different locations and times along the entire wire length. Figure 1.2-1 shows time-integrated pinhole images of a single wire *z* pinch [2] (on the left) and a two-wire *x* pinch [3] (on the right). The random locations of bright-spots are seen along the entire *z*-pinch wire length, while only one or two bright-spots are observed in a small wire crossing point for *x*-pinch images. The #1, #2 and #3 *x*-pinch images use different *x*-ray filters placed in front of radiography film, and a) and b) represents different wire materials and diameters (two, 18-um W wires for a), and two, 25-um Mo wires for b). The *z*-pinch image presented on the left is performed using XP pulser (480 kA, 100 ns), and *x*-pinch images presented next done using the PIAF (250-kA peak current, 170-ns rise time in this configuration) LC generator.

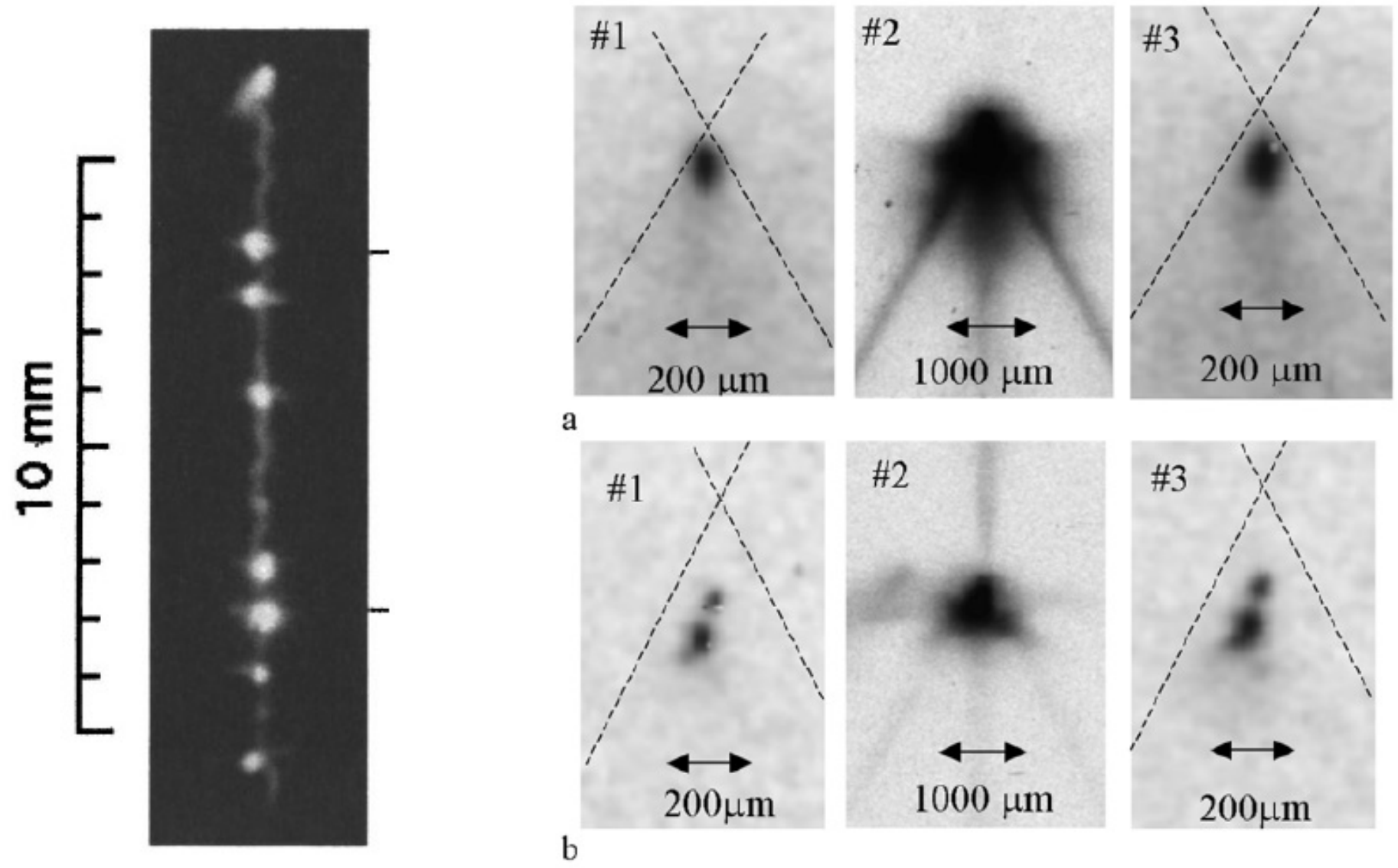

Figure 1.2-1 *Z*-pinch (left) and x-pinch (right) time-integrated pinhole images.

These *x*-pinch "hot spots" are believed to be among the brightest laboratory soft *x*-ray (1-10 keV) sources [4]. At the same time, the radiating size of a hot-spot is extremely small (a few μm), and the radiation pulse is extremely fast (ps time scales). Such radiation parameters are hard, if impossible, to achieve simultaneously in all other known *x*-ray radiation sources. The hot-spot formation can be theoretically understood in the approximation of radiative magnetohydrodynamics (RMHD). It can be shown, that if the pinch current exceeds some critical value, the radiation energy loss will dominate the Joule energy deposition, and *z*-pinch plasma column will undergo radiation collapse. It can also be shown, that such radiation collapse and hot-spot formation occurs under conditions close to Bennett equilibrium [4]. The Bennett equilibrium simply describes the balance of magnetic- and plasma-kinetic pressure, and it can be expressed as:

$$2Nk(T_e + T_i) = \left(\frac{\mu_0}{4\pi}\right) I^2 \qquad (1.2\text{-}1)$$

where N is the electron or ion density per unit length, k is the Boltzmann constant, $T_e$ and $T_i$ are electron and ion temperature, $\mu_0$ is a vacuum magnetic permeability, and I is the total pinch current. The derivation of the Bennett equation for z-pinch geometry is presented in Appendix A.

When the pinch current exceeds a critical value, radiation losses will dominate over Ohmic heating, the plasma temperature will drop, and radiation collapse will take place. Because of MHD instabilities, the collapsing plasma column will be disrupted, and one (or more) hot-spot will be formed. For the case of a completely ionized plasma, the critical current is called the Braginsky-Pease current given by:

$$I_{BP} = 0.22\sqrt{\lambda}(Z_n + 1)/Z_n \qquad (1.2\text{-}2)$$

Here, $Z_n$ is the nucleus charge and $\lambda$ is the Coulomb logarithm (~10). For high-Z materials (such as *x* pinches), the ratio (Z+1)/Z is approximately equal to one, and $I_{BP}$ approximately equals to 700 kA, which is independent on the *x*-pinch wire materials and mass. Obviously, that such a high current can only be achievable in a large pulse-power generators. In the case of partially ionized plasma, the critical current strongly depends on the linear mass of the wire and can be on the order of several tens of kA [4]. It is important to emphasize, that such a relatively small current can be produced with even a very small pulsed-power generator, and hot-spot can be formed at this case.

A "hot spot" is thought to be a high temperature (> 500 eV), small sized (~ 1-2 μm), near solid-density plasma object with a ps life time. Figure 1.2-2 presents MHD simulations [5] of $x$-pinch explosion using the 3D GORGON code. The left picture shows 3D surface of constant density (red), mass density contours (color) and synthetic radiographs (grey) of the $x$-pinch evolution just before (28 ns), during (30 ns) and just after (32 ns) the radiation collapse. The formation of plasma jets, micro-Z pinch column, and two dislike plasma electrodes can be clearly observed from simulations. The right picture plots the plasma parameters from different models [5] with a time relative to radiation collapse. The development of the extreme plasma conditions with radius of about 2 µm, with density of near solid density, and with temperatures of about 2 keV are predicted from these set of simulations.

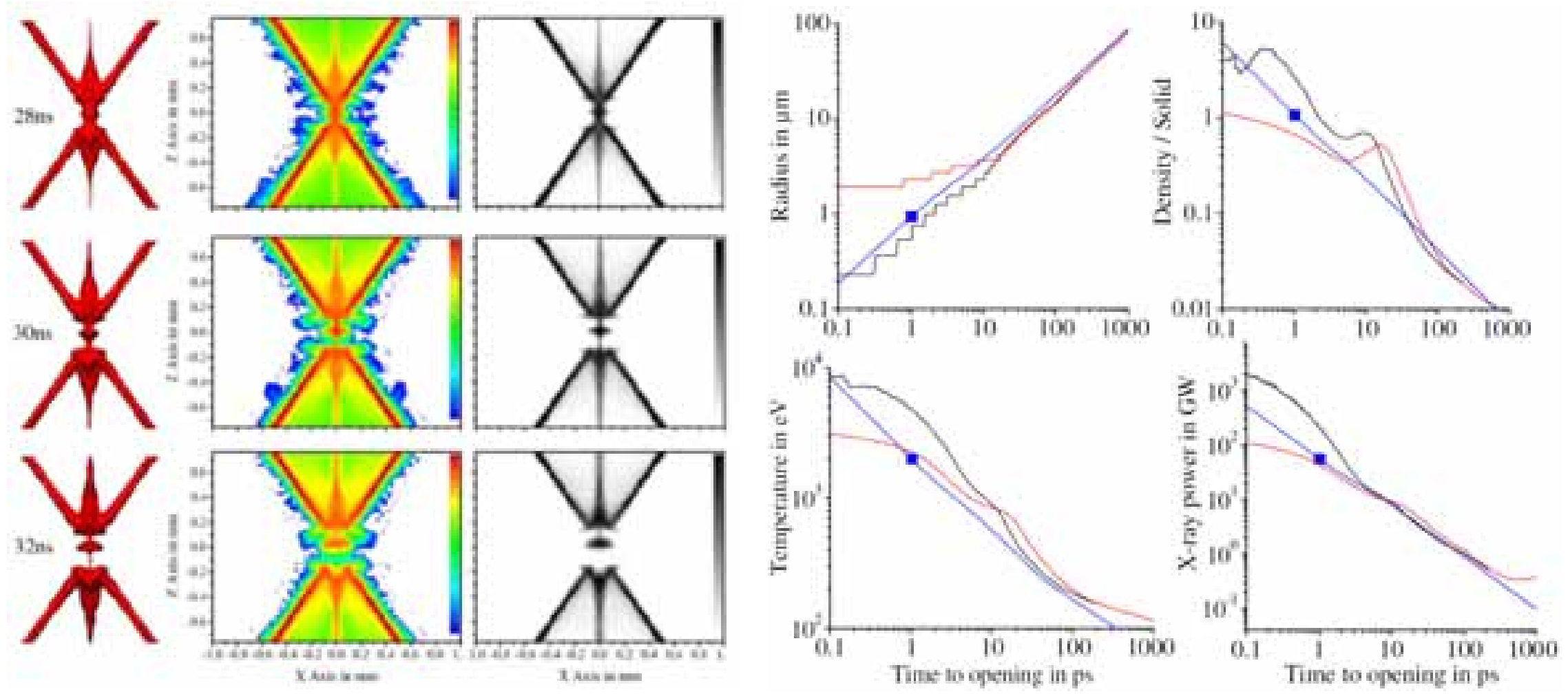

Figure 1.2-2 MHD simulations of "hot spot" formation [5]. Left is 3D surface of constant density (red), mass density contours (color), and synthetic radiographs (gray). Right is radius, peak density, temperature, and $x$-ray power versus time before radiation collapse.

Some detailed studies [6] [7] [8] [9] of $x$-pinch dynamics were performed at Cornell University using the XP facility (470 kA, 100 ns). Five radiographs of 17-µm Mo-wire $x$ pinches [8], just before and after the $x$-ray burst, are presented in Figure 1.2-3 and Figure 1.2-4, correspondingly. Two $x$ pinches were mounted in parallel between two electrodes of the pulser, so each $x$ pinch served as backlighter $x$-ray source for point-projection $x$-ray imaging of the other. The formation of a 200-300-μm-long, 100-μm-diameter *z*-pinch plasma column between two, dense-plasma electrodes ("mini-diode") can be clearly seen at the intersections of *x*-pinch wires (Figure 1.2-3), with times given relative to the moment of the *x*-ray burst. Starting with *t* = -1.9 ns image, the m=0 ("sausage") instability starts to develop, leading to formation of a very small diameter neck, less than 1 ns before the intense *x*-ray burst.

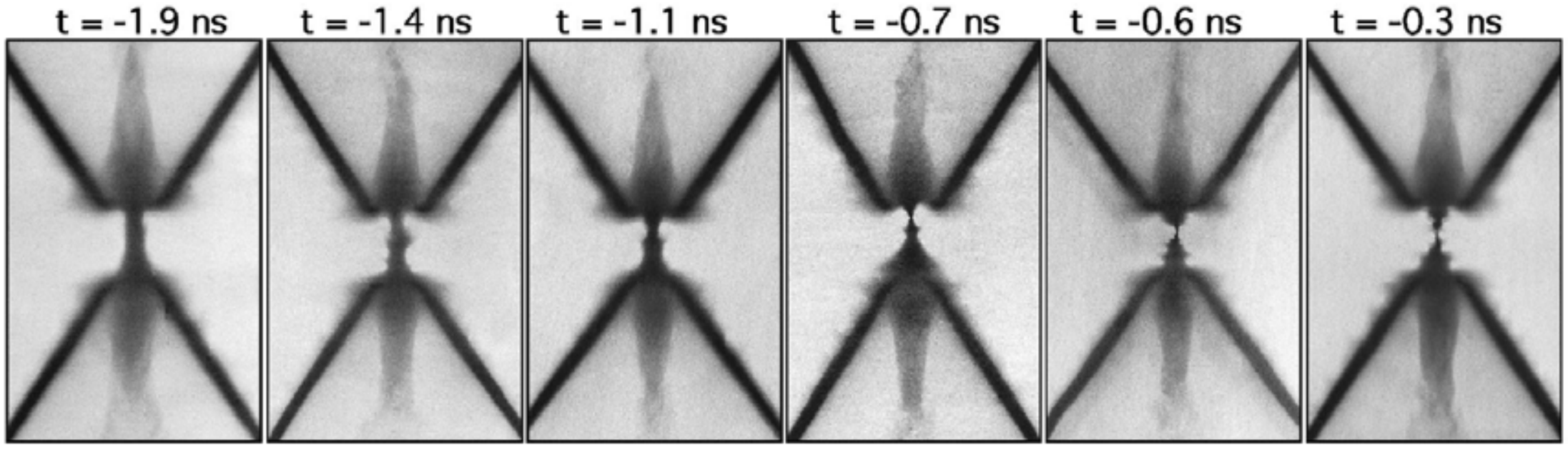

Figure 1.2-3 Radiographs of an *x*-pinch implosion just before the *x*-ray burst.

The radiographs of the final stage of *x*-pinch dynamics just after the *x*-ray burst are presented in Figure 1.2-4. In this stage, the whole column is disappearing within about 2 ns after the *x*-ray burst moment. It can be noted, that *x*-pinch implosion involves only the *x*-pinch crossing area, while the outside region remains almost unchanged.

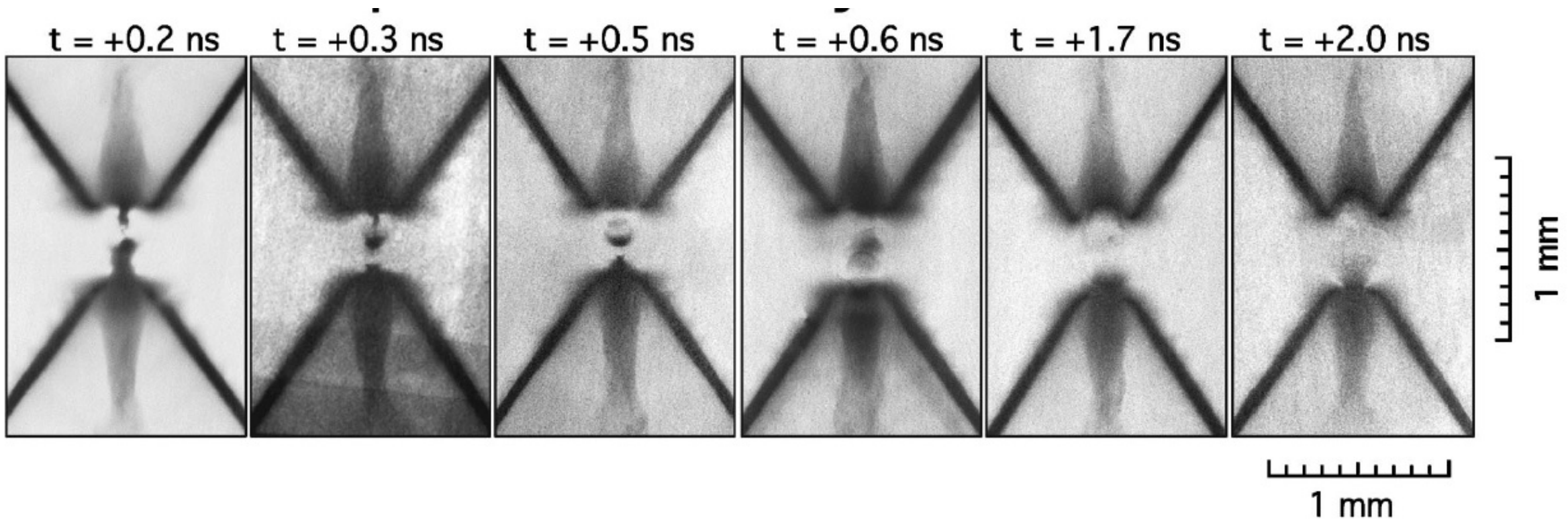

Figure 1.2-4 Radiographs of an *x*-pinch implosion just after the *x*-ray burst.

Typical *x*-ray diode (XRD) signal [3] from *x*-pinch burst is presented in Figure 1.2-5. The signal was produced in a 25-um-diameter Mo-wire *x*-pinch shot, and one can see a short-duration, single-peak radiation pulse generated in this shot. On the right we have two *x*-ray signals for the same shot, but with different filters placed in front of each XRD. The full-width at half-maximum (FWHM) are equal to 1.5 ns for XRD1 signal (Al 1.6 μm + Mylar 0.9 μm), and 2.5 ns for XRD2 signal (Mylar 3 μm).

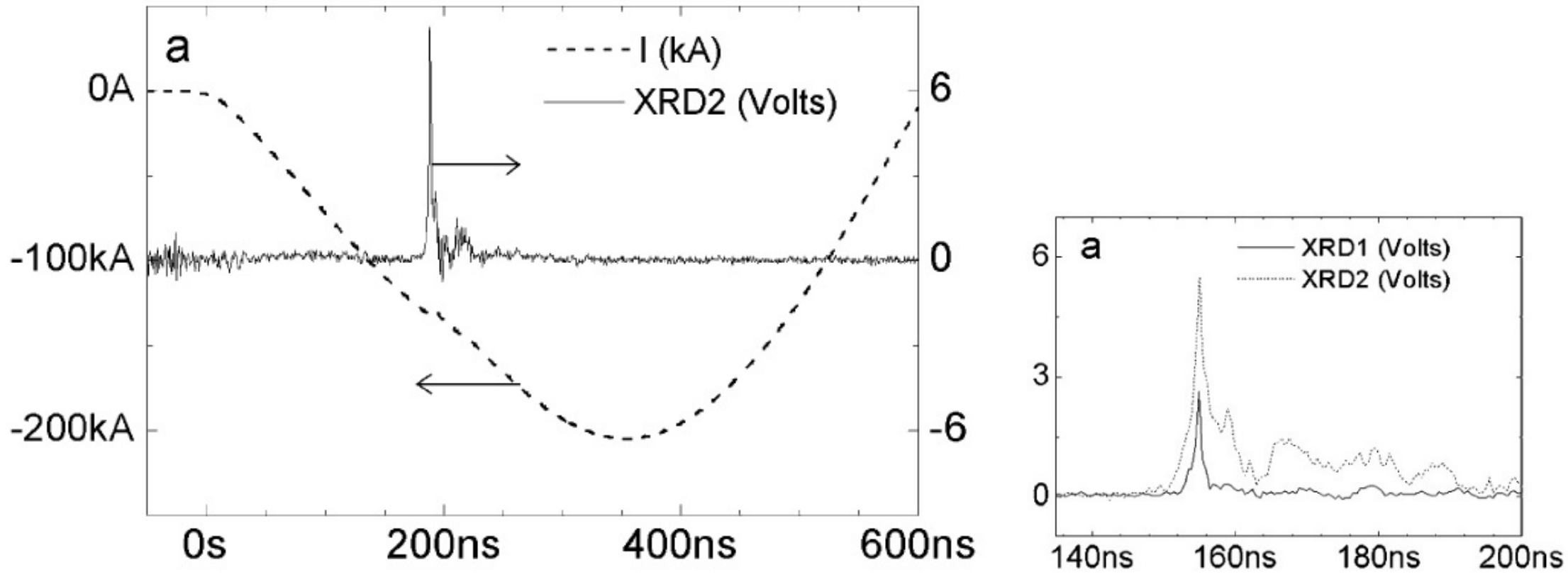

Figure 1.2-5 A typical *x*-ray XRD signal together with *x*-pinch current (left), XRD1 vs. XRD2 pulse duration (right)..

*X*-pinch *x*-ray pulse duration measurements are often limited by the time resolution of the available instrument, cable bandwidth, and the digitizer system. For example, the 1.5-ns value observed with XRD1 signal in Figure 1.2-5 was already limited by the XRD time resolution. (Higher bandwidth XRDs are possible.) Diamond photoconducting detectors (PCD) can increase the time resolution of measured signal to about 250 ps [10], and a Kentech fast streak camera up to 5-10 ps [11].Table 1.2-1 summarizes some *x*-pinch *x*-ray pulse-duration measurements. It can be seen that the duration of the measured signal depends on the *x*-pinch composition (wire masses and diameters) and the photon range.

But, in many cases (highlighted with red), the measurements were limited by the detector time resolution.

Table 1.2-1. Experimental data on the duration of the *x*-pinch *x*-ray burst.

| Driver | *X* pinch | Detector | Filter | Pulse duration | Ref |
|---|---|---|---|---|---|
| **PIAF** | Mo 2x25 µm | XRD | Mylar 3 µm (230-290 + > 800 eV) | 2.5 ns | **[3]** |
| | | | Al 1.6 + Mylar 0.9 µm (>700 eV) | 1.5 ns | |
| **XP** | Mo 2x13 µm + NiCr 2x25 µm | PCD (diamond) | Be 100 µm (> 1 keV) | 300 ps* | **[10]** |
| | | | Ti 12 µm (> 2.5 keV) | 250 ps* | |
| | 2xMo 2x17 µm | Kentech slow streak camera | Be 165 µm (> 2 keV) | 400 ps | **[11]** |
| | | | Be 165 µm + Al 40 µm (> 5 keV) | < 80 ps | |
| | 2xMo 2x17 µm | Kentech fast streak camera | Be 50 µm (>1.8 keV) | 50-100 ps | |
| | | | Be 50 µm + Al 80 µm (> 5 keV) | 5-10 ps | |
| | 2xW 2x13 µm | | Be 50 µm (>1.8 keV) | 15-40 ps | |
| | | | Be 50 µm + Al 80 µm (> 5 keV) | 3-15 ps | |

*PCD signals are from a 13-µm-diameter, Mo *x* pinch.

Yet another unique feature of *x*-pinch *x*-ray radiation sources, along with their ps pulse durations, is the µm radiation spot sizes. (These two numbers are clearly related.) Time-integrated radiographs of radiation spots obtained using 5-µm pinhole camera from 25-µm Nb *x* pinch [12] are presented in Figure 1.2-6 (left). Spot sizes from 20 µm to 100 µm were observed for radiation above 1.5 keV, however, only 5-µm source size (pinhole-limited) was observed for radiation above 6 keV. Figure 1.2-6 (right) shows a high-resolution image of an exploding 25-µm W wire using 25-µm Al *x* pinch [12]. Spatial structures as small as 3 µm to 4 µm were observed in this experiment.

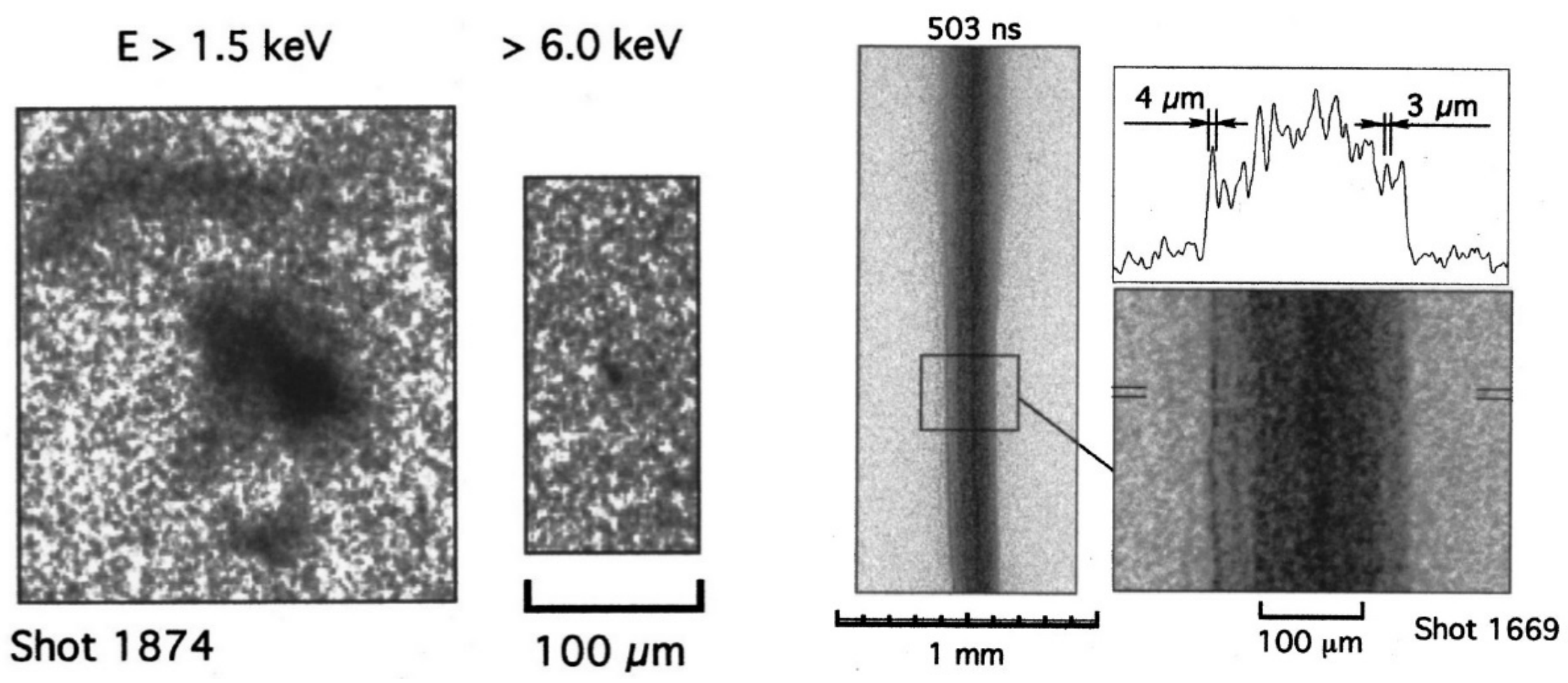

Figure 1.2-6. Radiographs of a 25-μm Nb *x* pinch obtained using 5-μm pinhole camera (left). Radiograph of exploding 25-um W wire from 25-um Al *x* pinch (right).

Table 1.2-2 presents more examples of *x*-pinch radiation source-size measurements. Spot sizes as small as a few μm were observed in many cases, which, indeed, were already limited by instrument resolution (red color highlighted).

Table 1.2-2. *X*-pinch radiation source size measurements.

| **Driver** | ***X* pinch** | **Diagnostics instrument** | **Energy range** | **Source size** | **Ref** |
|---|---|---|---|---|---|
| **PIAF** | W 2x18 μm | 10-μm pinhole 200-μm aperture penumbral images | > 2.4 keV | 12-14 μm | **[3]** |
| **XP** | Al 2x25 μm | x 4 magnification image of test object | 3-5 keV | 3-4 μm | **[12]** |
| | Nb 2x25 μm | x 90-120 magnification image of test object | 1.33-2 keV | 1.3-2 μm | |
| | varied | 5-μm aperture pinhole camera image | > 1.5 keV | 20-100 μm | |
| | | | > 6 keV | 5 μm | |
| | Mo 2x17 μm | Linear Bragg-Fresnel lens | > 5 keV | 3 μm | |
| | W 4x17 μm | Fresnel diffraction slit experiment | 2.7-4.9 keV | 1.5 μm | **[13]** |

The central event of the evolution of all *x* pinches is an almost instantaneous *x*-ray burst (Figure 1.2-3 and Figure 1.2-4) with pulse duration as small as a few ps (see Table 1.2-1), source size as small as a few μm (Table 1.2-2) and, continuum and line radiation [8] [12] [14] [15] in soft and hard *x*-ray region. The fact that one or more radiation spots are always formed in a predictable location (within the 100-200 μm *z*-pinch plasma column), in predictable times (with 2-ns jitter), along with its unique *x*-ray characteristics, make *x* pinches a better choice in many applications in plasma physics, biology and others, which we briefly referenced in Table 1.2-3.

Table 1.2-3. Applications of *x*-pinch *x*-ray radiation.

| **Application** | **Reference** |
|---|---|
| Radiography of other x pinches | [6] [7] [8] [9] [10] [11] |
| Radiography of single exploding wires | [2] [12] [16] [17] [18] [19] [20] |
| Radiography of wire array z pinches | [2] [21] [22] |
| Phase-contrast imaging of soft biological objects | [13] [23] [24] [25] |
| Study of CH foams inside wire array *z* pinches | [26] |
| Study of inertial-confinement-fusion capsule shells | [27] |

## 1.3 Pulsed-power high-current *x*-pinch drivers

For a long time, a conventional pulsed-power device for driving an *x*-pinch load was a high-voltage Marx generator coupled with one or more pulse-forming lines and transmission lines to compress an initially long (few μs) pulse from the Marx output into a short (a few hundred ns) pulse at the load. For example, the XP pulser operating at Cornell University, consists of a 10-stage Marx generator, 4 coaxial intermediate storage capacitors, 3 coaxial pulse forming lines, one self-breaking gas switch, and 8 self-breaking water switches. It was built about 20 years ago and can deliver about 450 kA with a 40-ns, 10-90% rise time to a low inductance load. Recent progress in the development of low-inductance, high-current capacitors [28] [29] [30] and switches [31] [32] [33] [34] opens up big opportunities in the design of new pulsed-power generators, which can directly supply a high-current pulse to a low-inductance *x*-pinch load. For example, PIAF, built at Ecole Polytechnique, France, is a LC generator, based on total of 6, low-inductance 180-nF, 50-kV capacitors connected in parallel, which can deliver a 250-kA, 180-ns high-current pulse to an *x*-pinch load.

We summarize some known pulsed-power, high-current *x*-pinch generators in the Table 1.3-1 below. The table is divided in three sections; section 1 combines all pulsed-power generators with current amplitude delivered to *x*-pinch load below 1 MA, section 2 – with current amplitude above 1 MA, and section 3 is a special section which contains all well-known small-sized *x*-pinch pulsed-power generators.

Table 1.3-1. Overview of some known *x*-pinch pulsed-power high-current generators.

| Name | Location | Technology | Parameters | Ref. |
|---|---|---|---|---|
| | ***x*-pinch drivers below 1 MA** | | | |
| Light-II-A | China Institute of Atomic Energy, China | Marx | 200 kA | **[35]** |
| PPG-1 | Tsinghua University, Beijing, China | Marx | 400 kA 100 ns | **[36]** |
| Llampüdkeñ | Pontifical Catholic University, Chile | Marx | 400 kA 260 ns | **[37]** |
| LION | Cornell University, NY | Marx | 470 kA 80 ns | **[38]** |
| XP pulser | Cornell University, NY | Marx | 470 kA 40 ns | **[39]** |
| | ***x*-pinch drivers above 1 MA** | | | |
| MAIZE | University of Michigan, US | LTD | 1 MA 100 ns | **[40]** |
| COBRA | Cornell University, NY | Marx | 1 MA 95 ns | **[41]** |
| ZEBRA | University of Nevada, Reno | Marx | 1.2 MA 90-ns | **[42]** |
| MAGPIE | Imperial College, London | Marx | 1.4 MA 240 ns | **[43]** |
| QiangGuang-1 | Institute of Northwest Nuclear Technology, China | LTD | 2 MA 80 ns | **[44]** |
| S-300 | Lebedev Physical Institute, Russia | Marx | 2.3 MA 150 ns | **[45]** |
| | **small-sized, table-top** | | | |
| x-pinch pulser | University of California, San-Diego | Marx | 80 kA 50 ns | **[27]** |
| table-top | Tsinghua University, China | Marx | 100 kA 60 ns | **[23]** |
| PIAF | Ecole Polytechnique, France | LC | 250 kA 180 ns | **[46]** |
| GenASIS | University of California, San Diego | LTD | 250 kA 150 ns | **[47]** |
| SPAS | Institute of High Current Electronics, Tomsk | LC | 300 kA 200 ns | **[48]** |
| Compact | Institute of High Current Electronics, Tomsk | LC | 650 kA 390 ns | **[49]** |

"Marx (technology)" in the table above assumes that pulsed-power generators are based on one or more oil-insulated Marx generators, pulse-forming and transmission lines to compress and deliver pulses to *x*-pinch load, and gas switches. All such pulsed-power systems are able to deliver a large, fast-rising current to a low-inductance, *x*-pinch load and have, for decades, generated a huge amount of *x*-pinch research around the world.

"LC (technology)" in the Table 1.3-1 assumes that generators are built based on new low-inductance capacitors and switches which can directly deliver a high-current pulse to a low-inductance *x*-pinch load without a need of transmission lines. Sometimes, two capacitors and a switch are placed into one solid unit, called a "brick", and then the whole pulsed-power driver is assembled based on linear transformer driver (LTD) architecture. We classify such a driver as "LTD (technology)".

When one wants to build a pulsed-power generator with a current amplitude in excess of 200 kA delivered to *x*-pinch load and when there are no limitations in available cost and space, it is a reasonable choice to build a pulsed-power system based on a conventional Marx generator and transmission line, but in general, all such installations are bulky and expensive to build, maintain and operate. Pulsed-power generators based on new low-inductance capacitors and switches offer many advantages compared to conventional Marx-based systems, which we briefly summarize in the Table 1.3-2. What is most important, such new technologies allows one to design and build compact, inexpensive *x*-pinch drivers which can in many cases replace pulsed-power systems based on conventional Marx generators.

Table 1.3-2. Comparison of pulsed-power *x*-pinch drivers' technologies.

| **Marx technology** | **LC technology** |
|---|---|
| need to compress initially long pulse and to deliver it to a low-inductance *x*-pinch load | directly deliver high-current pulse into a low-inductance *x*-pinch load |
| oil-filled Marx generators | no isolating oil |
| water transmission lines | no transmission line |
| gas switches | gas switches |
| low efficiency | high efficiency |
| bulky, expensive to build, maintained, and operate | small-sized and inexpensive |
| fixed to one location | can be designed to be portable |

## 1.4 "Good" *x*-pinch radiation performance requirements

By "good" *x*-pinch radiation performance we understand that only one bright radiation spot is formed, in predictable location and at a predictable time. These source characteristics are unique (not possible simultaneously in other known radiation sources). The radiation spot size is very small (a few μm), the radiation pulse is very fast (less than 1 ns), and the radiation power is huge (GW). The *x*-pinch driver overview (see Table 1.3-1) and the literature [41] [48] suggest that a minimum current rate-of-rise, dI/dt, of 1 kA/ns is required for such "good" *x*-pinch radiation performance. If current rate-of-rise is below this value, no such bright radiation spots can be detected in any experiments irrespective of the current amplitude [48].

For any reasonable current rise time, the 1-kA/ns requirement translates to a current amplitude of about 100 kA or more. Such a current amplitude correlates well with our earlier discussions (see equation (1.2-2) and text below) where only ~ 200 kA is required for "hot spot" formation in the case of incompletely ionized plasma.

One way to satisfy these current requirements (> 1 kA/ns, > 100 kA) is to build a large scale machine, where currents in excess of 500 kA are generated on 100 ns and less time scale. Such machines are based on standard Marx generators, with compression and pulse forming lines, are built worldwide (see Table 1.3-1), and the current rate of rise is usually much larger than the critical value of 1 kA/ns. However, it is hard to exceed the 1-kA/ns current requirement, if one wants to build a small-sized *x*-pinch pulsed-power machines with a "good" radiation performance. Only a few such small-sized pulsers have succeeded in overcoming this current-rise-rate requirement and almost all of them are based on new capacitor and switch technologies, as can be seen in Table 1.3-1.

## 1.5 Project goal and tasks

Several, small-scale *x*-pinch pulsers have been developed but more are needed. We propose to develop and demonstrate a new, compact and portable *x*-pinch radiation source based on two, slow LTD bricks combined into one single unit and to demonstrate "good" *x*-pinch radiation performance. The current rate-of-rise delivered to a low-inductance *x*-pinch load must be at least 1 kA/ns with a peak current amplitude of at least 100 kA. To achieve these requirements we propose to:

1. Develop an *x*-pinch driver electrical model based on two, slow LTD bricks and estimate its main electrical characteristics (peak current, rise time, current rate-of-rise and inductance).
2. Design and fabricate a driver based on two, LTD bricks. Develop and calibrate pulsed-power, high-current diagnostics.
3. Perform short-circuit tests on the 2-LTD-brick driver and show that it satisfies the 1 kA/ns current rate-of-rise requirement. Determine the total internal inductance of the driver.
4. Perform *x*-pinch driver tests and evaluate its *x*-pinch radiation performance.

Such an *x*-pinch driver configuration has never been tested, but, conceptually, it has the promise to be compact and to be portable, so the driver can be easily relocated to practically anywhere a bright, fast and small *x*-pinch radiation source is needed. As far as we know, no such compact and portable *x*-pinch radiation sources currently exist.

# 2 THEORY

## 2.1 Evolution of our driver design approaches

Initially, it was planned to non-destructively convert the Idaho State Induction System (ISIS) Marx generator into a low-impedance pulsed-power *x*-pinch driver. While it was technically possible to do so and simulations showed that such a system could deliver up to 300-kA peak-current with 70-ns, 10-90%, rise-time, a series of potential problems were discovered. Simulations showed that a negative reflective wave is formed inside the system, which can damage one of the pulse-forming-line (PFL) switches. To operate safety, damping resistors would have to be placed after the PFLs but, with a penalty, the maximum current at *x*-pinch load would be limited to about 200 kA. Taking into account this result and some other factors (time, effort and cost, needed to switch between the ISIS *x*-pinch operation mode and the normal operation mode of ISIS), the decision was made to design and build a new, stand-alone *x*-pinch driver. This initial idea was reported [50] [51] and briefly outlined in Appendix B.

Recent progress in the development of low-inductance, high-current capacitors and switches opens up opportunities in the design of compact, high-current pulse generators for driving low-inductance *x*-pinch loads. Such a design approach offers many advantages (see Table 1.3-2) when compared to traditional, Marx-based generators, and a few such drivers have been developed (see Table 1.3-1). To get a feeling what can be expected from these new capacitor and switch technologies, we designed a compact *x*-pinch driver, based on only four, fast high-current capacitors. Our simulation shows [52] that the driver can supply about 180-kA peak-current with a 150-ns time-to-peak into "matched" *x*-pinch load, which is briefly outlined in Appendix C.

Our final design is based on two, "slow" LTD bricks (a total of four capacitors and two switches) combined into a one, solid unit. This design was reported recently [53] [54]. It can be described by a simple RLC circuit with four, fast 140-nF, 100-kV capacitors that store up to 2.8 kJ initial energy. The short-circuit-load tests [55] [56] confirm the required 1-kA/ns current-rate of rise and shots with *x*-pinch loads [57] [58] [59] [60] [61] reveal its potential for *x*-pinch applications. The size of our driver is only 0.7×0.3×0.3 meters and it weighs about 90 kg. Our 2-LTD-brick *x*-pinch driver is compact and portable, contains no oil, and can be easily relocated to where *x*-pinch radiation source is needed. As far as we know, no such portable *x*-pinch drivers exist.

## 2.2 RLC circuit and matched load approach

For a small-sized generator without a long transmission line, the lumped-component approximation, $L_c \ll \lambda$, can be used, where $L_c$ denotes typical driver dimensions, and λ denotes typical wavelength of interest. Indeed, the typical time scale for small-sized generator is about

$$\mathrm{T} = \mathrm{L/c} = (3\ \mathrm{m})/\left(3 \times 10^{8}\ \frac{\mathrm{m}}{\mathrm{s}}\right) = 10\ \mathrm{ns}, \qquad (2.2\text{-}1)$$

which is much smaller than the typical current rise time of pulsed-power driver (see Table 1.3-1). Such lumped circuit approximation allows one to simplify the driver description, and the whole generator, including *x*-pinch load, can be described as a simple series RLC circuit. (Presented in Figure 2.2-1.)

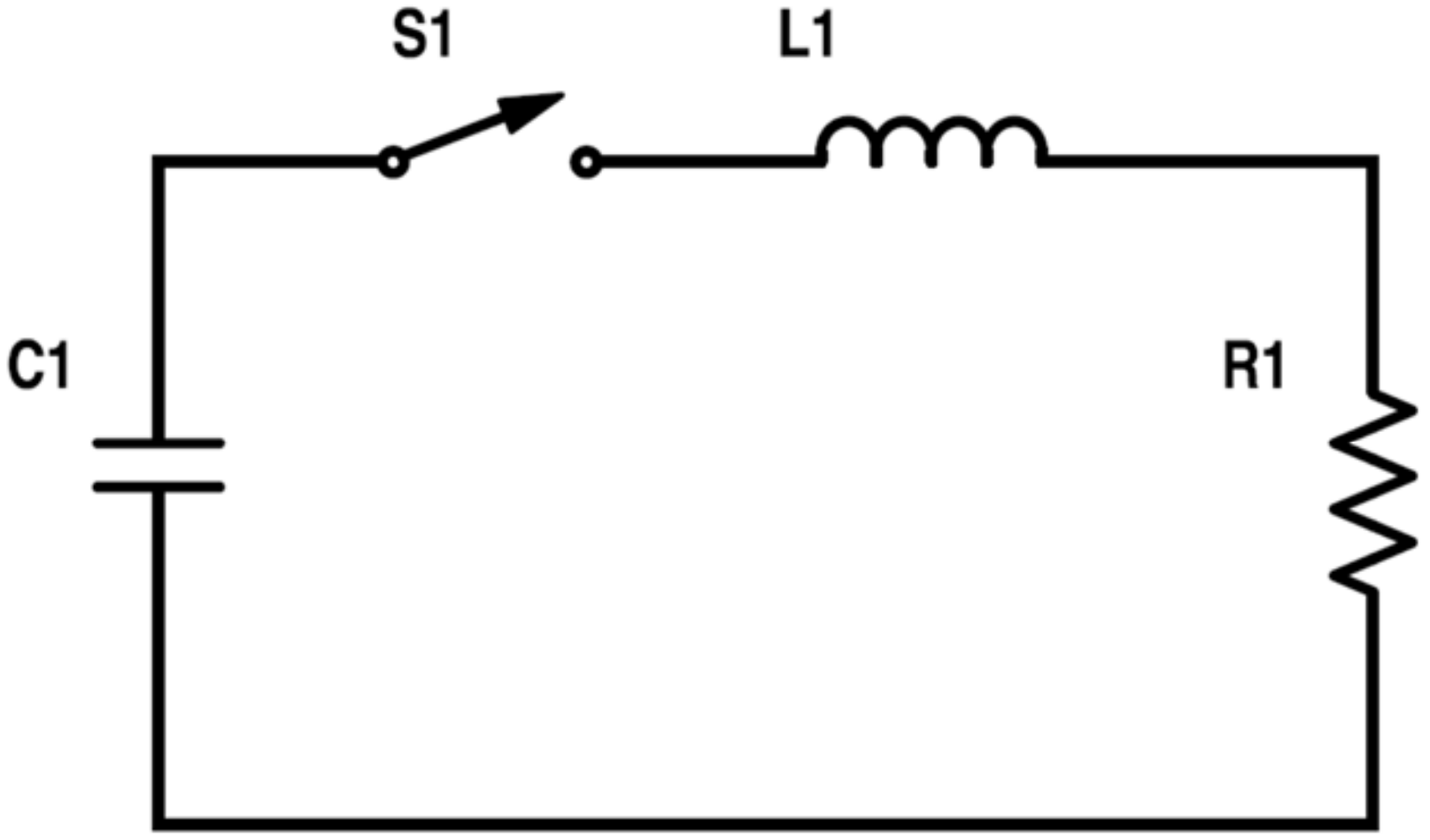

Figure 2.2-1. Simple series RLC circuit.

The equation governing the behavior of an RLC circuit has a form

$$L\frac{d^2i}{dt^2} + R\frac{di}{dt} + \frac{i}{C} = 0 \tag{2.2-2}$$

with initial conditions $i(t_0) = 0$ and $V_c(t_0) = V_0$. The general solution of this equation can easily be found (see derivation in Appendix D) as:

$$i(t) = \frac{V_0}{L\omega} e^{-\alpha t} \sin(\omega t) \tag{2.2-3}$$

where $\alpha = {}^{R}/_{2L}$ and $\omega = \sqrt{{}^{1}/_{LC} - {}^{R^2}/_{4L^2}}$ (2.2-4)

There are, usually, three general cases to be considered, depending on the value of oscillation constant ω:

Overdamped (ω < 0): $$R > 2\sqrt{L/C} \quad (2.2\text{-}5)$$

Underdamped (ω > 0): $$R < 2\sqrt{L/C} \quad (2.2\text{-}6)$$

Critically damped (ω = 0): $$R = 2\sqrt{L/C} \quad (2.2\text{-}7)$$

The solutions of Equation (2.2-2) for three different cases (2.2-5), (2.2-6) and (2.2-7) are plotted in Figure 2.2-2 in arbitrary units. For the underdamped case (a) the current has oscillatory behavior with a small decay constant, so it undergoes several oscillations before it reaches zero. In the overdamped case (b) the current reaches maximum value and then decays to zero. Here, the decay constant is large and the peak current is small. In the intermediate, critically damped case (c), the current also reaches some peak value but it decays faster compared to overdamped case (b).

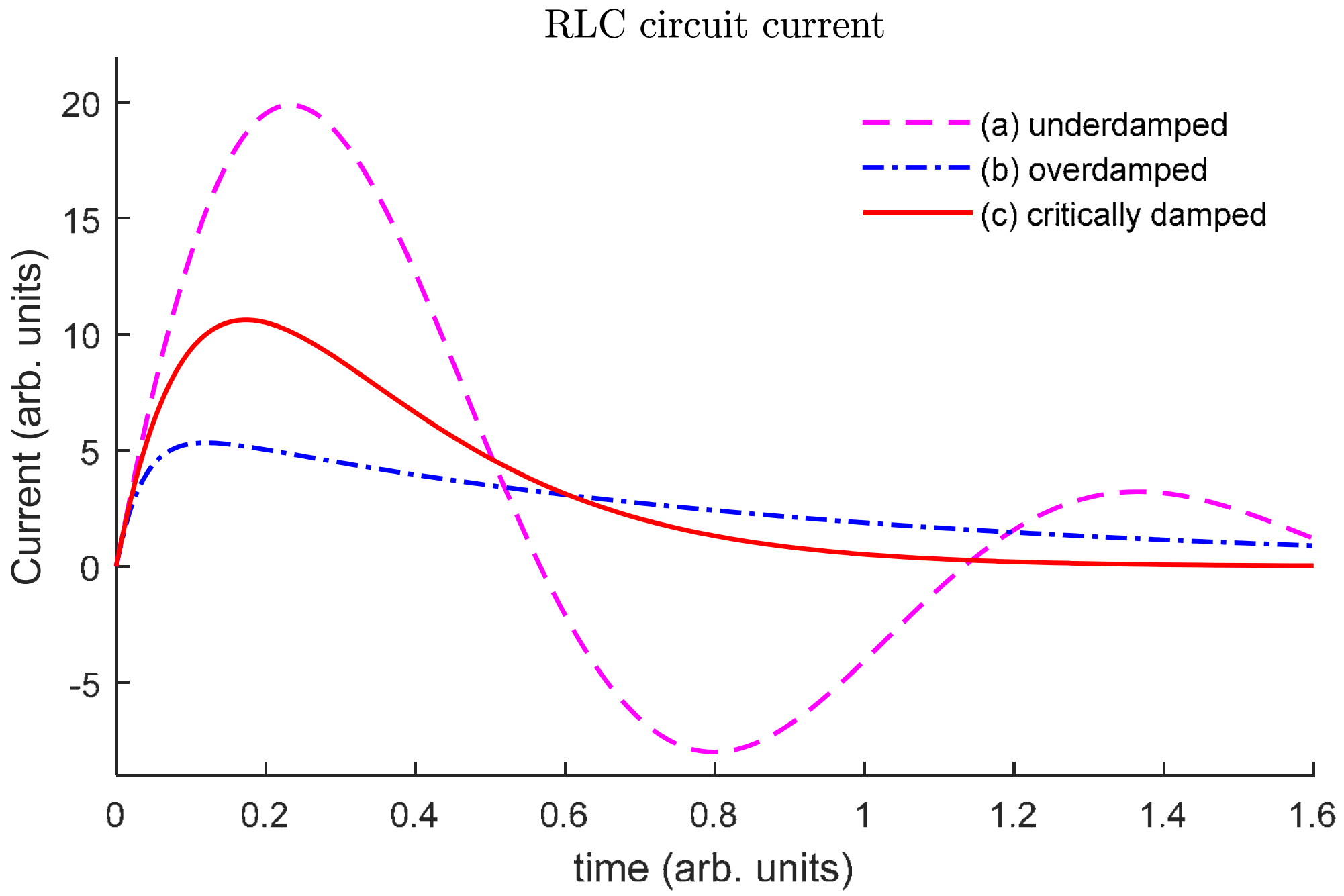

Figure 2.2-2. RLC circuit current for tree cases (a) underdamped and (b) overdamped and (c) critically damped;

The time to peak $t_p$ and peak current $i_p$ can be found by setting the first derivative of general current solution (2.2-3) to the first zero, so:

$$t_p = {}^{1}/_{\omega} \arctan({}^{\omega}/_{\alpha}) \tag{2.2-8}$$

$$i_p = \frac{V_o}{L\omega} e^{-{}^{\alpha}/_{\omega} \arctan({}^{\omega}/_{\alpha})} \sin[\arctan({}^{\omega}/_{\alpha})] \tag{2.2-9}$$

By setting the first derivative of (2.2-3) to the second zero, the time to peak current reversal $t_{min}$ and the corresponding minimum peak current $i_{min}$ can be found, which are:

$$t_{min} = {}^{1}/_{\omega} \arctan\left(\frac{2\alpha\omega}{\alpha^2 - \omega^2}\right) \tag{2.2-10}$$

$$i_{min} = \frac{V_o}{L\omega} e^{-{}^{\alpha}/_{\omega} \arctan\left(\frac{2\alpha\omega}{\alpha^2-\omega^2}\right)} \sin\left[\arctan\left(\frac{2\alpha\omega}{\alpha^2 - \omega^2}\right)\right] \tag{2.2-11}$$

The last equation (2.2-11) allows one to estimate the maximum reverse current which is important in selection of circuit capacitor.

The "critically damped" $R = 2\sqrt{L/C}$ (also known as "critically matched" load) case and, its close analog, $R = \sqrt{L/C}$ (called "matched" load) case are both important in design of small-sized pulsed-power drivers [62] [63], and we consider them in more details below. The solution for both cases is plotted in Figure 2.2-3 and some quantitative relations are derived then, but, as it can be seen from the plot, a pulse generator with a “matched” load is, in general, able to produce higher pulse current as compared to the “critically matched” case, and it is more suitable for design of *x*-pinch radiographic machine [62].

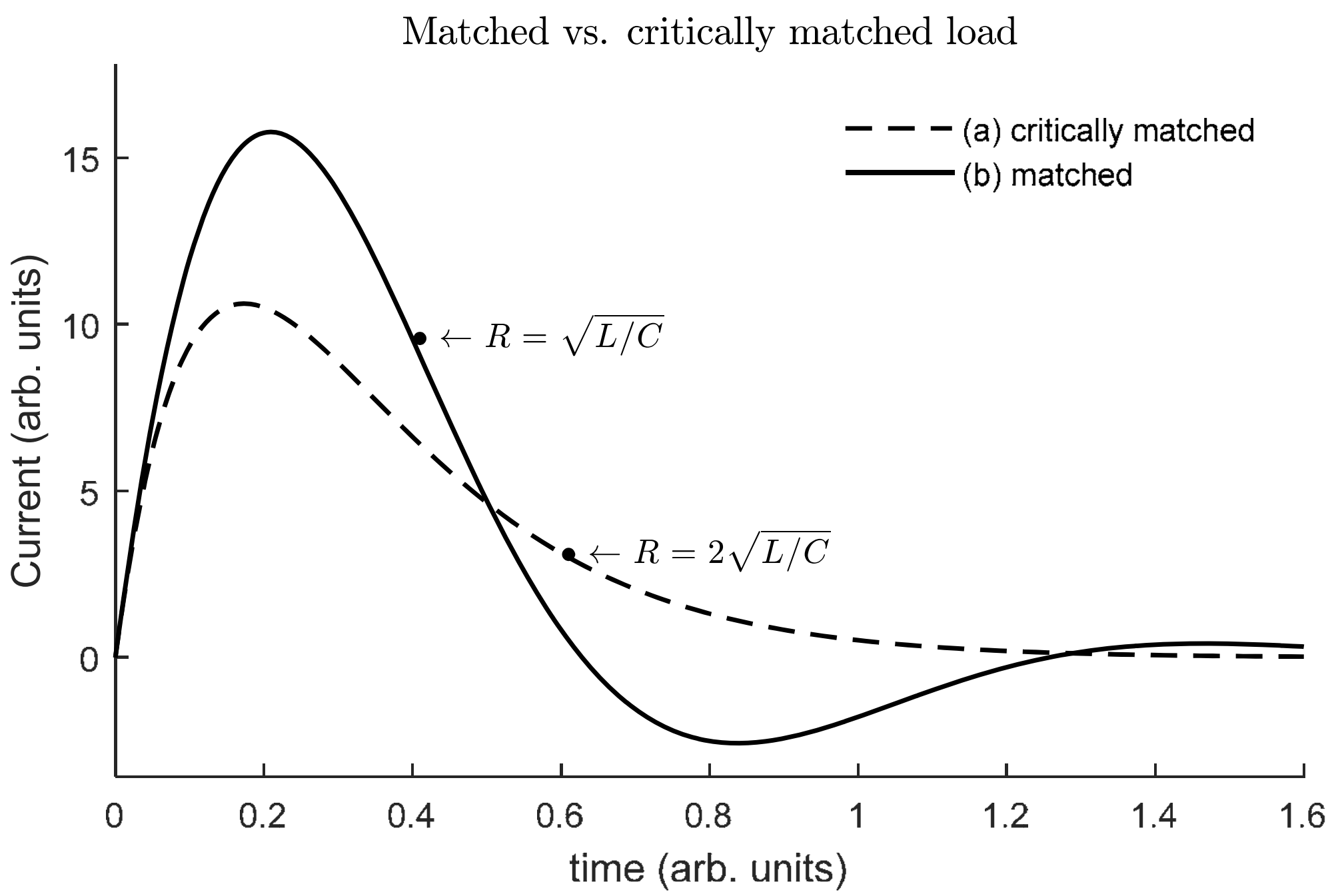

Figure 2.2-3. "Matched" load versus "critically matched" solutions.

In both cases, the decay and oscillation constants α and ω, correspondingly, are simplified to:

"matched" load $\omega = 0.86\sqrt{\frac{1}{LC}}$ and $\alpha = 0.5\sqrt{\frac{1}{LC}}$ (2.2-12)

"critically matched" $\omega = 0$ and $\alpha = \sqrt{\frac{1}{LC}}$ (2.2-13)

so, the ratio $\omega/\alpha$ becomes independent of L and C. The time to peak current (2.2-8), the peak current (2.2-9) and the peak voltage $v_{peak} = R \times i_{peak}$ for "matched" and "critically matched" load can be estimated as summarized in Table 2.2-1 below:

Table 2.2-1. "Matched" vs. "critically matched" load cases.

| "Matched" $R = \sqrt{L/C}$ | "Critically matched" $R = 2\sqrt{L/C}$ | |
|---|---|---|
| $t_{peak} = 1.21\sqrt{LC}$ | $t_{peak} = \sqrt{LC}$ | (2.2-14) |
| $i_{peak} = 0.55\dfrac{V_0}{\sqrt{L/C}}$ | $i_{peak} = 0.37\dfrac{V_0}{\sqrt{L/C}}$ | (2.2-15) |
| $v_{peak} = 0.55\,V_0$ | $v_{peak} = 0.74\,V_0$ | (2.2-16) |

As can be seen from (2.2-15), for given L and C, the "matched" load gives about 50% higher peak current and a slightly higher peak power compared to "critically matched" load, and so, it is more suitable for design of *x*-pinch drivers where a high current of 100 kA or more is desirable. The "critically matched" load gives, in general, higher load voltage (2.2-16) and is more suitable for design of compact high-voltage accelerators. The expression (2.2-14) and (2.2-15) for time to peak and peak current can also be used to estimate the driver total inductance L, if the total capacitance C is known (that is usually the case) and the experimental conditions are close to "matched" (or "critically matched") load cases.

Given (2.2-14) and (2.2-15), the current rate of rise (0-100%) can be estimated as:

$$\left(\frac{i_{peak}}{t_{peak}}\right)_{matched} = \frac{0.55V_0/\sqrt{L/C}}{1.21\sqrt{LC}} = 0.45\frac{V_0}{L} \qquad (2.2\text{-}17)$$

$$\left(\frac{i_{peak}}{t_{peak}}\right)_{cr.matched} = \frac{0.37V_0/\sqrt{L/C}}{\sqrt{LC}} = 0.37\frac{V_0}{L} \qquad (2.2\text{-}18)$$

The last two expressions give a feeling how to get at a higher current rate of rise, which is important in design of *x*-pinch radiation machine. As it was discussed earlier, the driver current rate of rise should be at least 1 kA/ns to have a "good" *x*-pinch radiation performance. As can be seen, the current rate of rise is proportional to initial charging voltage $V_0$, independent of the total capacitance C, and inversely proportional to the total inductance L. Obviously, to design a "good" radiation *x*-pinch machine, one wants to maximize $V_0$ and to minimize L, but, in general, the "matched" load gives about 22% higher current-rise-rate value compared to "critically matched" case.

The energy transferred to a load R by the time t can be estimated by integrating the instantaneous load power:

$$E_{load} = \int_0^t I^2 R dt \qquad (2.2\text{-}19)$$

and it is plotted for "matched" and "critically matched" loads in Figure 2.2-4.

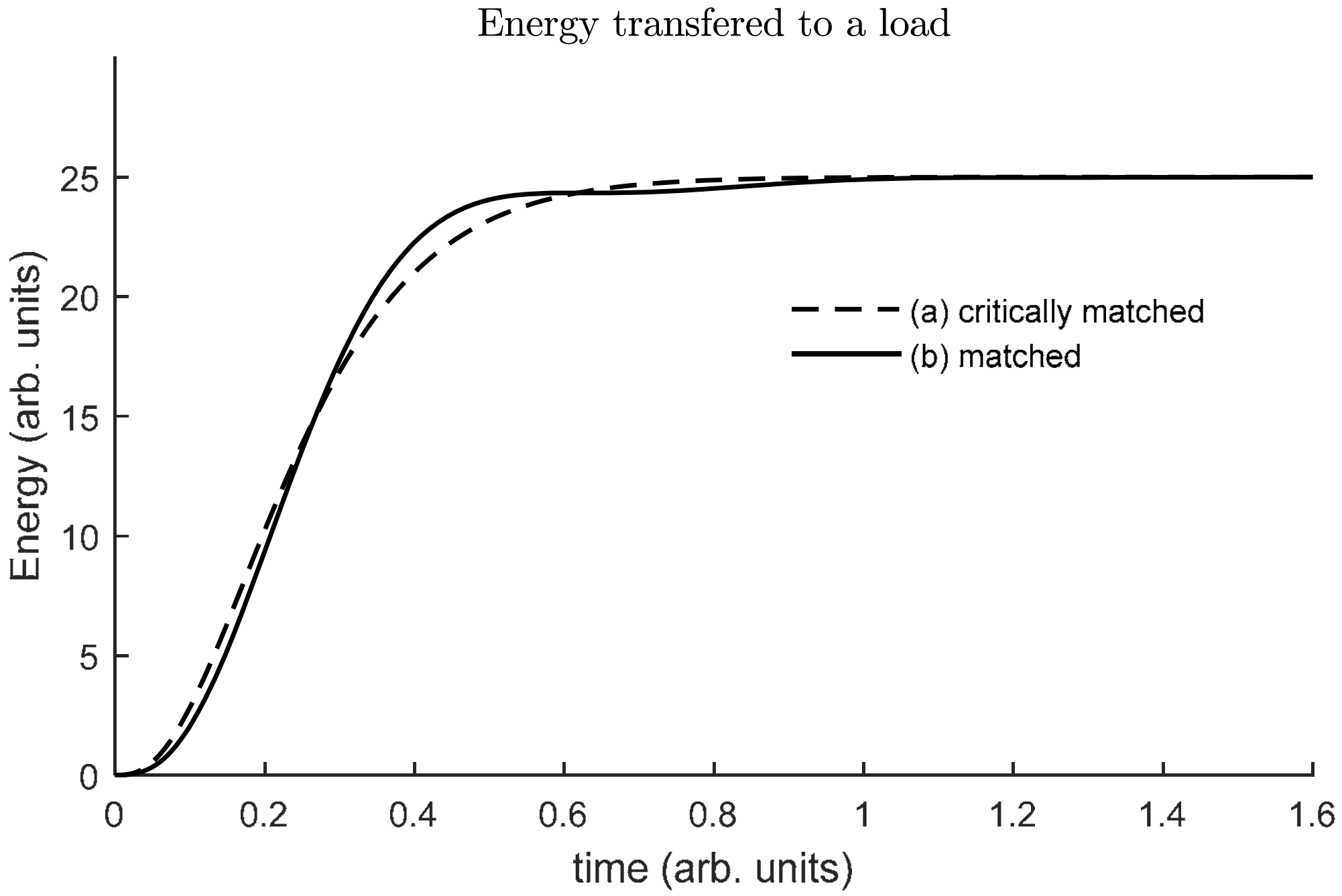

Figure 2.2-4. Energy transferred to a load in RLC circuit.

As can be seen, the energy transferred to a load by the time to peak is about 40% of the total energy initially stored in capacitors for the both cases. This value is important to keep in mind, as *x*-pinch radiation burst usually occurs at the time just before peak current, and, if a driver design is based on "matched" (or close to it) load, this value gives the best capacitor-to-load efficiency at the moment of *x*-ray burst.

At the end, we briefly discuss the short-circuit case as it is important for driver testing and characterization. The short-circuit condition is given by:

$$R \ll 2\sqrt{L/C}, \tag{2.2-20}$$

with the solution plotted in Figure 2.2-5.

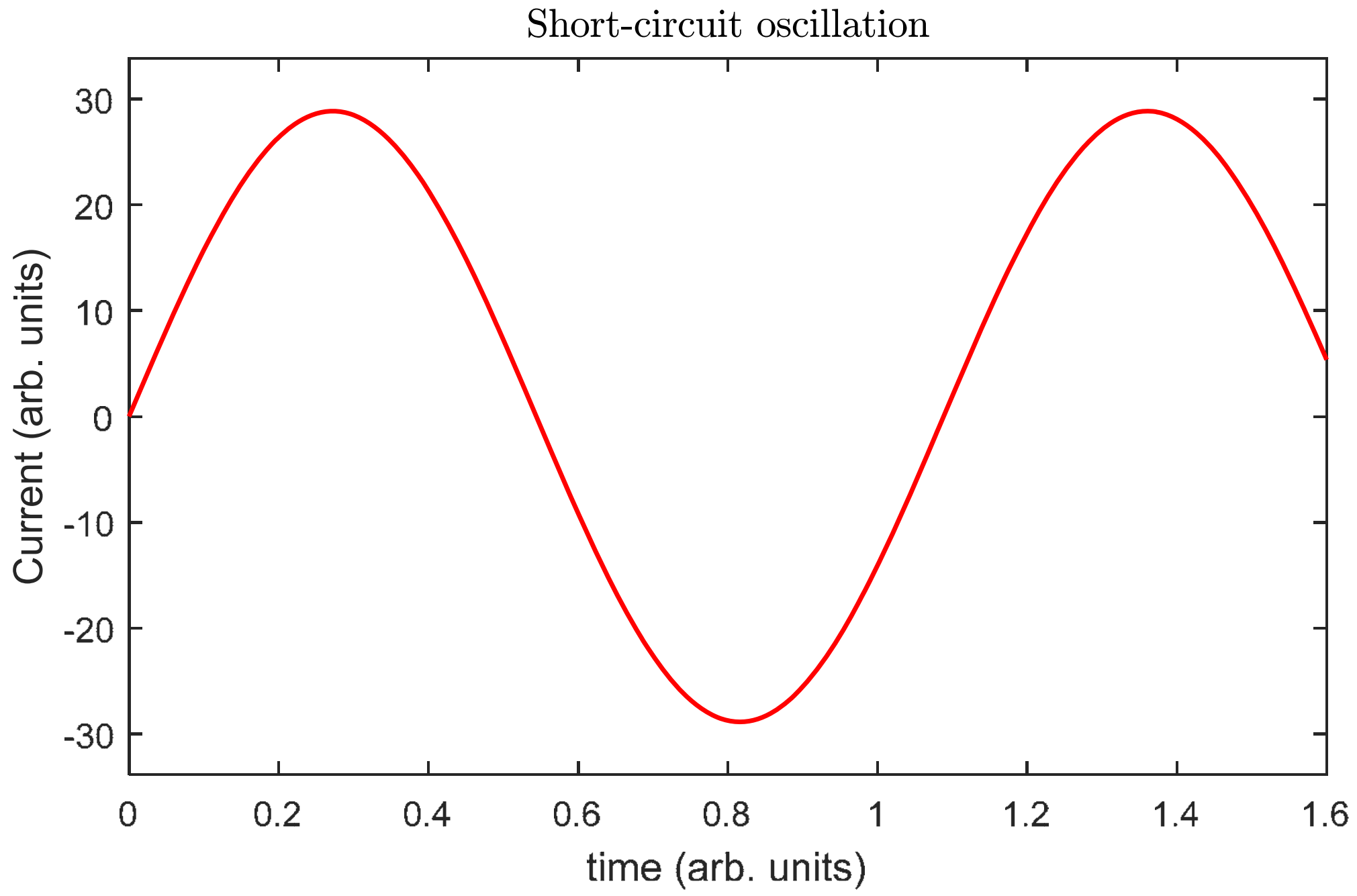

Figure 2.2-5. Short-circuit load current oscillations.

The time to first peak, time to first zero and peak current can be estimated as:

$$t_p = 1.57\sqrt{LC} \qquad (2.2\text{-}21)$$

$$t_0 = 3.14\sqrt{LC} \qquad (2.2\text{-}22)$$

$$i_p = V_0/\sqrt{L/C} \qquad (2.2\text{-}23)$$

Given the total driver capacitance C and initial charging voltage $V_0$, and using either of three expressions above, one can estimate the total driver inductance L from the experimental short-circuit test data. However, the expression for time to first zero (2.2-22) is more preferable, as a zero-crossing point can be estimated with a better accuracy from the experimental data.

We briefly discussed the lumped RLC circuit with emphasize on "critically matched" and "matched" load cases and derived some useful expressions, which can be used to estimate the driver time to peak (2.2-14), peak current (2.2-15), peak voltage (2.2-16) and current rise time (2.2-17) and (2.2-18). The last, "matched" load case is more suitable for *x*-pinch driver design, as it gives a higher peak current (see 2-15) with faster current rate of rise (see 2-17 vs. 2-18). We also derived a few expressions (2.2-21), (2.2-22) and (2.2-23) which can be usefull for driver characterization in short-circuit testing. All these formulas are for estimates only, and, in general, a driver circuit simulation has to be performed to predict the driver performance or to analyse its experimental data.

# 3 METHODS AND MATERIALS

## 3.1 Description of the "slow" LTD/NRL brick

The proposed final design of our compact and portable *x*-pinch driver is based on two, "slow" LTD bricks [64], which were generously loaned to us by the Naval Research Laboratory (NRL). These bricks were originally built for NRL by Ktech Corporation. Such (or similar) LTD bricks are usually arranged in circular-in-plane geometry around a magnetic core inside a LTD cavity to drive a low-inductance load. More LTD cavities can be later stackable so as to add voltage [40] [64].

The general view of one LTD/NRL brick is shown in Figure 3.1-1. It consists of two, low-inductance, 140-nF GA 35465 capacitors, connected in parallel, and one multichannel, multi-gap switch in series with each pair of capacitors. The capacitors are potted in 3M Scotchcast 8 epoxy, and gas, trigger, and charging feedthroughs are made through penetrations in the epoxy. The brick is about 25-cm wide, 40-cm long and 15-cm high.

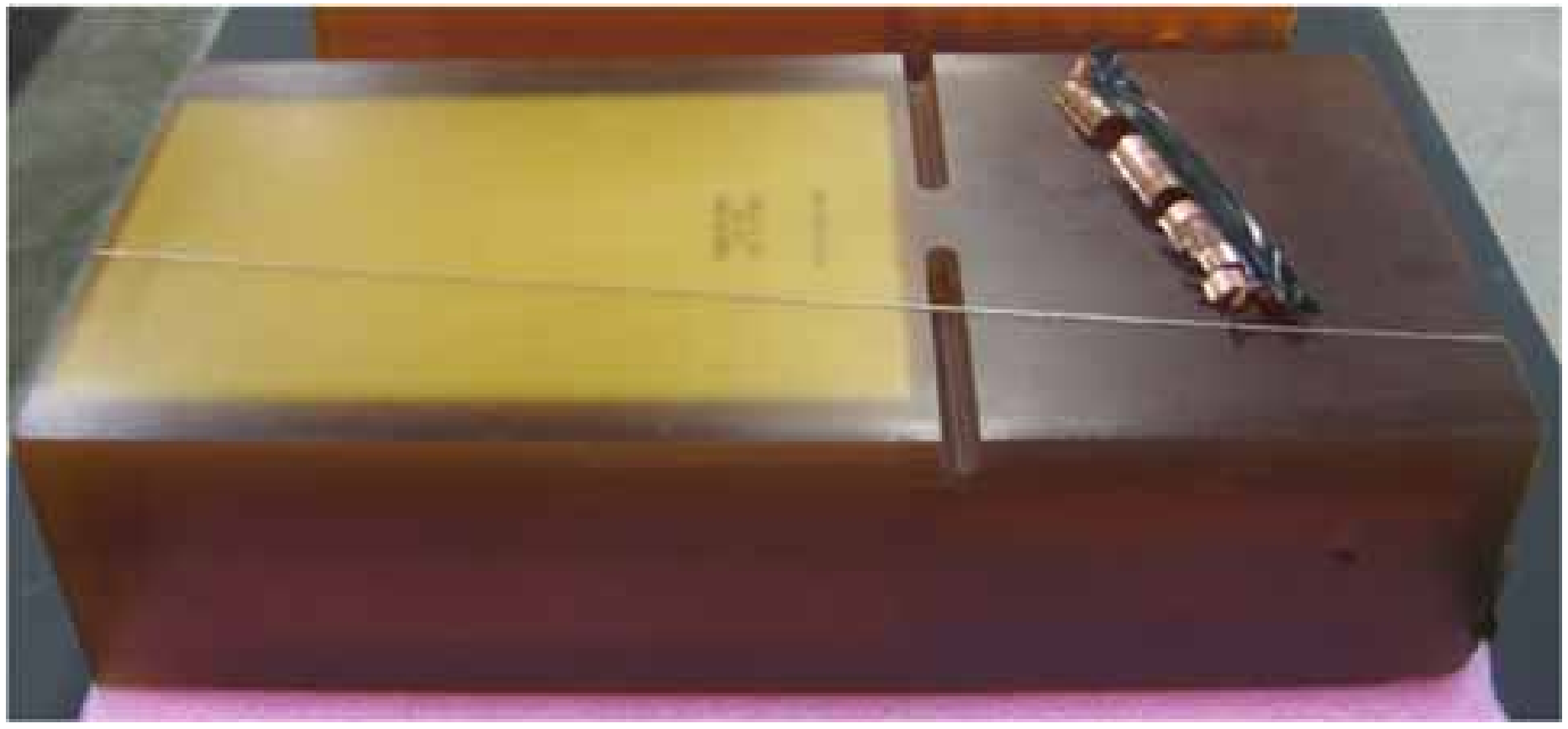

Figure 3.1-1. The LTD/NRL Brick.

A view of the LTD/NRL switch is presented in Figure 3.1-2. It has 5 channels, 7 gaps per channel, and has resistively enforced voltage grading from the HV electrode to ground. The individual gaps are 6 mm. The switch is designed to operate slightly above 1-bar absolute pressure with dry air. The switch has a very low, less than 10-nH, inductance value and it has a measured jitter of < 10 ns. This switch design gives the LTD/NRL brick its excellent rise time and allows the brick to directly drive a low-inductance load [64]. However, the best jitter time and inductance values are achieved when bricks are charged to greater than 80 kV. If the charging voltage is less, the switch inductance can be larger and the whole brick operation is less predictable and stable.

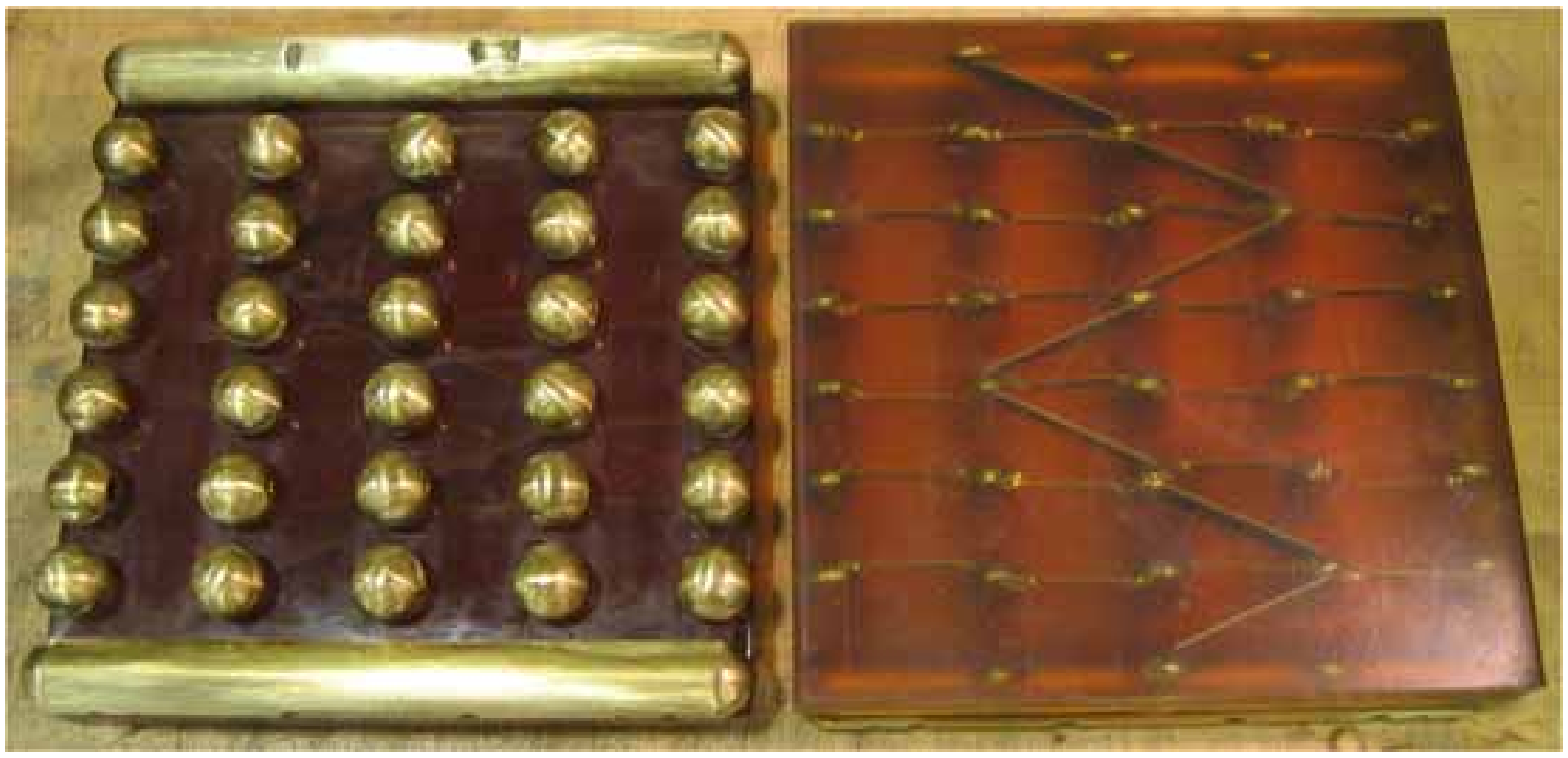

Figure 3.1-2. The LTD/NRL brick switch.

The electrical characteristics of the brick capacitor and switch are briefly summarized in Table 3.1-1. The rated capacitor peak current is about 50 kA/cap and the max peak current (fault) is 75 kA/cap. Because each brick is comprised of two capacitors connected in parallel, the total brick current normally should not exceed 100-110 kA/brick and, in any case, may never exceed more than 150 kA/brick. The rated/maximum voltage reversal of each capacitor is 10/80%, which is good enough for most pulsed-power applications. The capacitors/switch resistance is low and can be usually neglected. The "bare" (base) inductance of capacitor and switch is about 10 nH, but when placed in epoxy, the inductance is increased to about 16 nH and 20 nH for capacitor and switch, respectively.

Table 3.1-1. Electrical parameters of brick capacitor and switch.

| **Parameter** | **GA 35465 capacitor** | **LTD switch** |
|---|---|---|
| Capacitance | 140 nF | |
| Approx. base/in-brick inductance | 10/16 nH | 10/20 nH |
| Resistance | 8 mΩ | 20 mΩ |
| Rated voltage | 100 kV | |
| Rated/max voltage reversal | 10/80 % | |
| Rated/max(fault) peak current | 50/75 kA | |

The electrical circuit of LTD brick is presented in Figure 3.1-3. Two capacitors are connected in parallel with following switch in series. The total brick capacitance is 280 nF, and the total brick inductance is about 28 nH. Such a low brick inductance allows one to successfully drive many low-inductance loads, including an *x* pinch, for example.

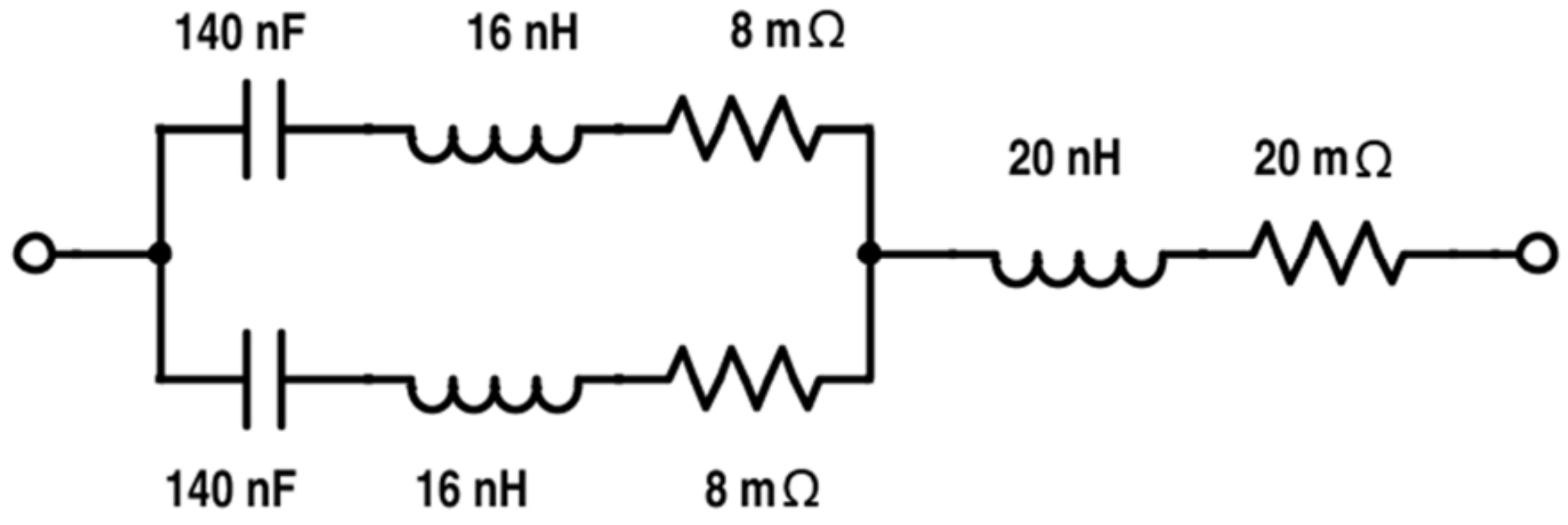

Figure 3.1-3. The LTD/NRL brick electrical circuit.

One LTD brick can supply about 100-110-kA maximum peak current into a low-inductance load with rise time depending on the load inductance. For *x*-pinch applications, usually more than 100-kA peak current is wanted, and to successfully drive an *x*-pinch load a minimum of two LTD bricks have to be used. Such a driver model based on two, "slow" LTD/NRL bricks will be discussed in the next sections.

## 3.2 Electrical circuit of the 2-LTD-Brick driver

As it was mentioned earlier, a minimum of two LTD bricks is needed to successfully drive a low-inductance *x*-pinch load. Our final design was reported elsewhere [54] [55] [56] and is based on two, LTD bricks placed in parallel inside a single housing. The mechanical design will be discussed in the next sections and here we concentrate on electrical model and simulations of our 2-LTD-Brick driver.

The electrical circuit of the proposed 2-LTD-brick *x*-pinch driver is presented in Figure 3.2-1. Two LTD bricks are simply connected in parallel with the switch sides facing to a common ground and capacitor sides connected to the load.

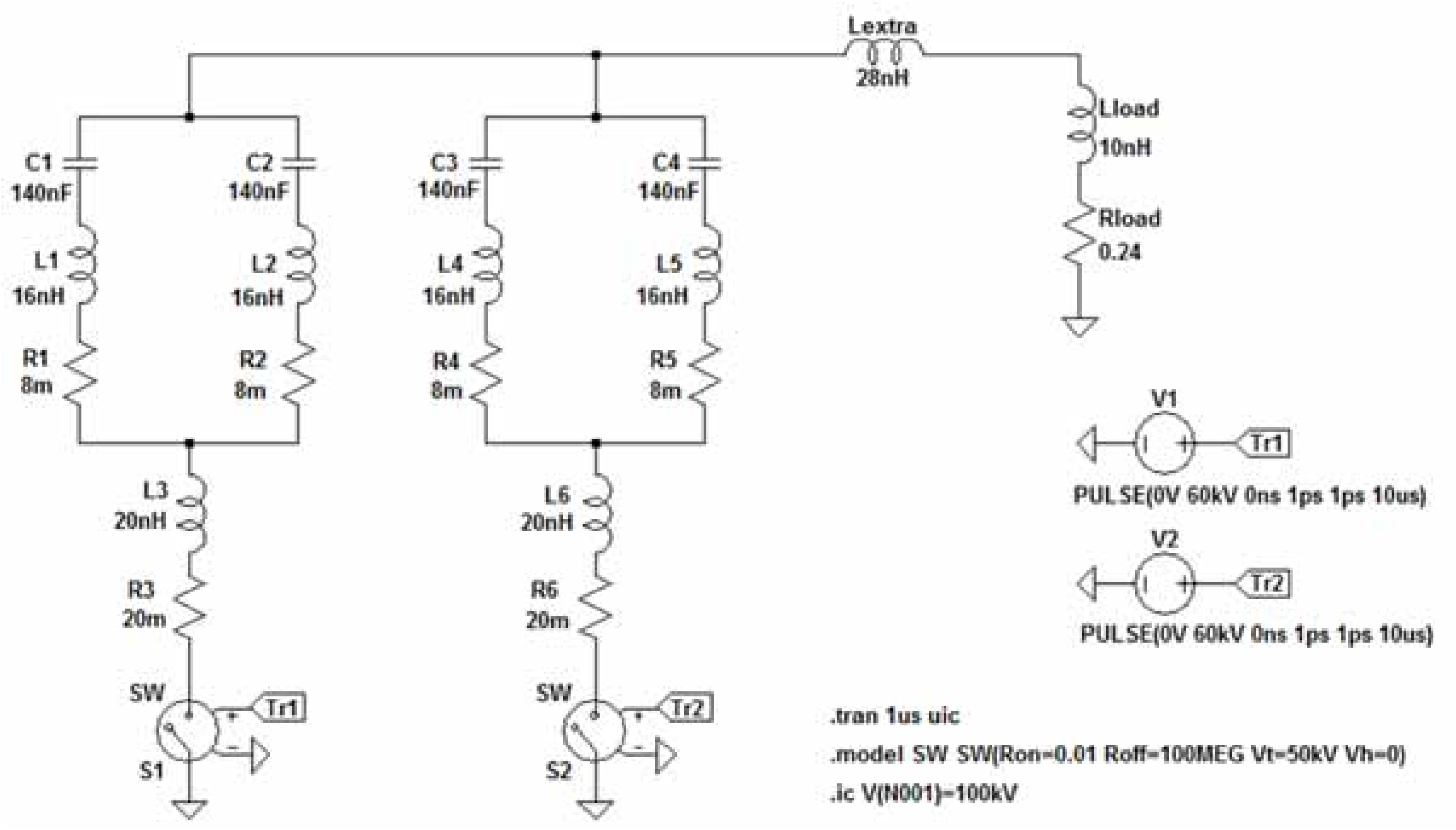

Figure 3.2-1. 2-LTD-brick *x*-pinch driver electrical diagram.

The total capacitance of our 2-LTD-Brick driver is $2 \times 280\text{nF} = 560\ \text{nF}$ so the driver can store a maximum total energy of,

$$E = \frac{1}{2}CV^2 = 2.8\ \text{kJ} \qquad (3.2\text{-}1)$$

when it is fully charged to 100 kV.

One brick inductance is about 28 nH, and inductance of two bricks connected in parallel is simply a half this value which is about 14 nH. The inductance of the *x*-pinch load is estimated to be about 10 nH with a resistance of 0.24 Ω. These numbers correspond to an $x$-pinch load comprised of two, 11-mm-long, 40-μm-diameter Mo wires in the initial "cold" state. The real values will be different depending on the electrical parameters of the real expanded, "ionized" wire state.

The tricky part is to estimate the extra inductance of the driver, $L_{extra}$, which is strongly dependent on driver geometry and can be only properly estimated when the driver dimensions are known. We will jump ahead and look at the driver drawings, which are discussed later in chapter 3.4, to estimate this driver extra inductance. The extra inductance is comprised of the inductance of bricks placed inside a finite metal volume (bricks-in-housing), inductance of header section, and inductance of vacuum section without $x$-pinch load (vacuum section). Each section is approximated by a coaxial transmission line for ease of calculation with the inductance in nH given by:

$$L_{load} = 2h \ln \frac{r_b}{r_a} \tag{3.2-2}$$

where, h is a section length in cm, $r_a$ is a section inner diameter, and $r_b$ is a section outer diameter. The dimensions of each driver section and the results of the inductance calculations are presented in Table 3.2-1.

Table 3.2-1. Extra inductance of driver for different sections.

| **Driver Section** | **$R_a$ (mm)** | **$R_b$ (mm)** | **h (cm)** | **L (nH)** |
|---|---|---|---|---|
| one brick-in-housing | 124 | 178 | 54.6 | 40 |
| Header part 1 section | 127 | 178 | 0.64 | 0.4 |
| Header part 2 section | 25 | 178 | 1.3 | 5.1 |
| Vacuum section without load | 25 | 32 | 4.1 | 2.0 |

As can be seen, the total inductance of header and vacuum sections is only about 7-8 nH, and, the extra inductance of a brick placed inside solid metal volume is about 40 nH (and about 20 nH for both bricks in parallel). So, the total extra inductance can be estimated to be about 28 nH. This value should be considered an estimate only and has to be measured. The driver electrical circuit model has to be updated, which will be discussed later in chapter 3.6.

Given two brick's inductance of 14 nH, extra driver inductance of 28 nH, and the *x*-pinch load inductance of 10 nH, the total driver internal inductance (without *x*-pinch load) and total driver inductance (including *x*-pinch load) can be estimated to be:

$$\mathrm{L_{internal}} = \mathrm{L_{brick}} \backslash 2 + \mathrm{L_{extra}} = 28 \backslash 2 + 28 = 42\ \mathrm{nH} \qquad (3.2\text{-}3)$$

$$\mathrm{L_{total}} = \mathrm{L_{internal}} + \mathrm{L_{load}} = 42 + 10 = 52\ \mathrm{nH} \qquad (3.2\text{-}4)$$

## 3.3 The Screamer code and simulations of 2-LTD-Brick driver

The Screamer computer code [65] was chosen to perform circuit simulations and to predict a behavior of proposed 2-LTD-brick *x*-pinch driver. Screamer was originally developed at Sandia National Laboratories in 1985 as a special purpose circuit code to simulate single module accelerators. It is written in Fortran 77 with a few Fortran 90 extensions, and it is highly optimized for speed and efficiency. Construction of a Screamer input file is very user friendly and many formats of output files are available for post-analysis of circuit parameters.

Screamer V.3.x (and earlier versions) solves electrical circuit problems with series and parallel elements with restricted in-line topology. The circuit is built by specifying one main branch and secondary branches directly attached to a main one. No branch-in-branch connections are allowed. Such a limited Screamer topology allows one to very efficiently simulate a large circuit with very large numbers of circuit elements and nodes with excellent time resolution. A more powerful version of Screamer V.4.0 [66] was recently developed, which allows for unlimited branch-in-branch topology, but still preserves most of the original efficiency and speed.

The Screamer circuit elements are organized in blocks (sections) connected in series to form a single branch. A π-section block and its subsets, mainly used in design of 2-LTD-brick driver, are presented in Figure 3.3-1. In addition to static elements, Screamer provides the user with capability to specify variable elements, including many switch models, diode models, gas-puff models and many more models that are specific for pulsed-power applications.

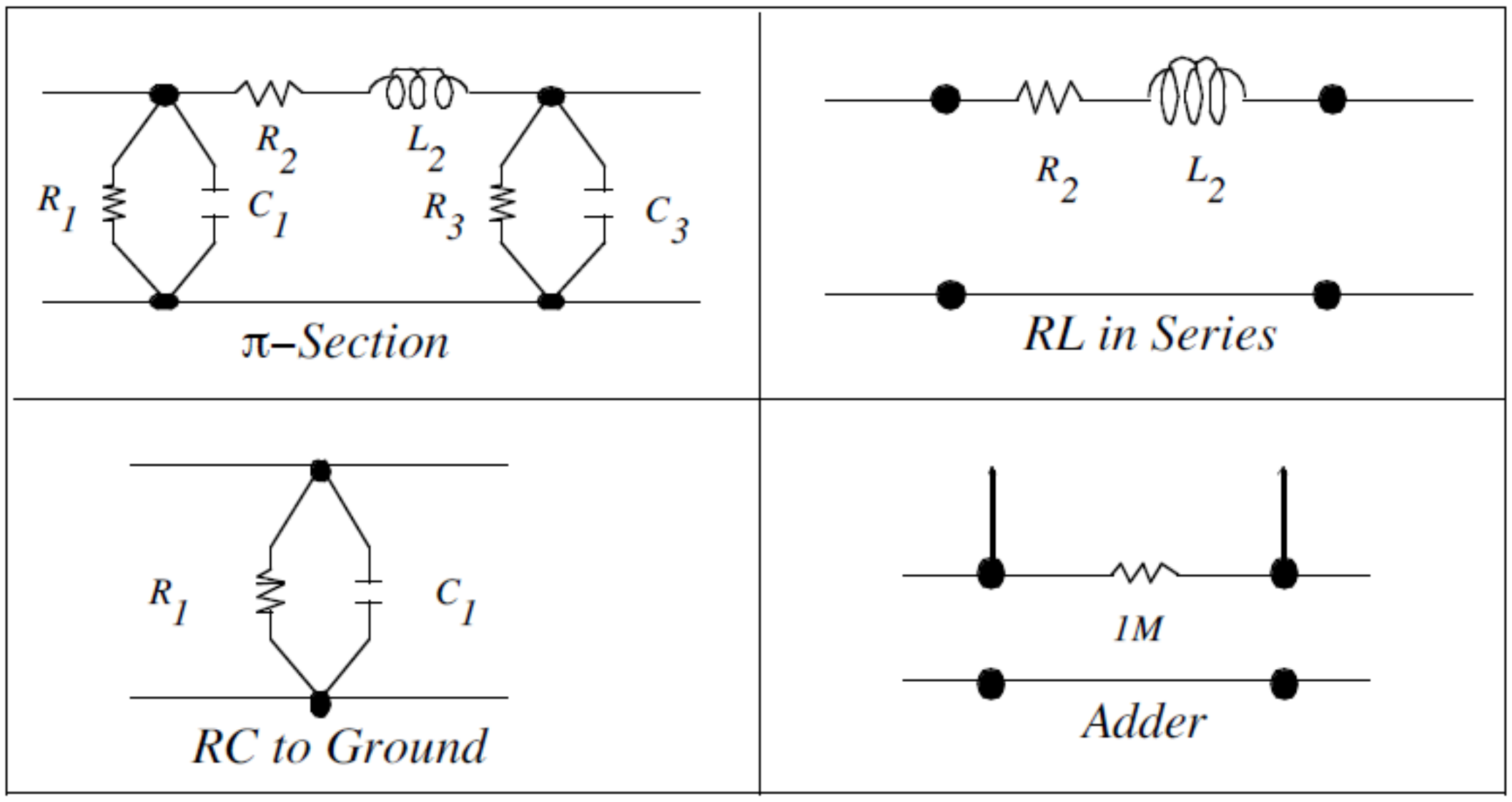

Figure 3.3-1. Screamer π-section block and subset.

A Screamer model of 2-LTD-brick driver circuit was developed and presented in Figure 3.3-2. The first RC-to-ground section represents the total driver capacitance. The following five RL-in-series sections break apart the total driver inductance for different driver sections (bricks, switch, bricks-in-housing, header and vacuum sections). The last two RL and RCG sections represent the *x*-pinch load inductance and resistance values. The screamer input file with all input parameters for 2-LTD-brick driver model is given in Appendix E.

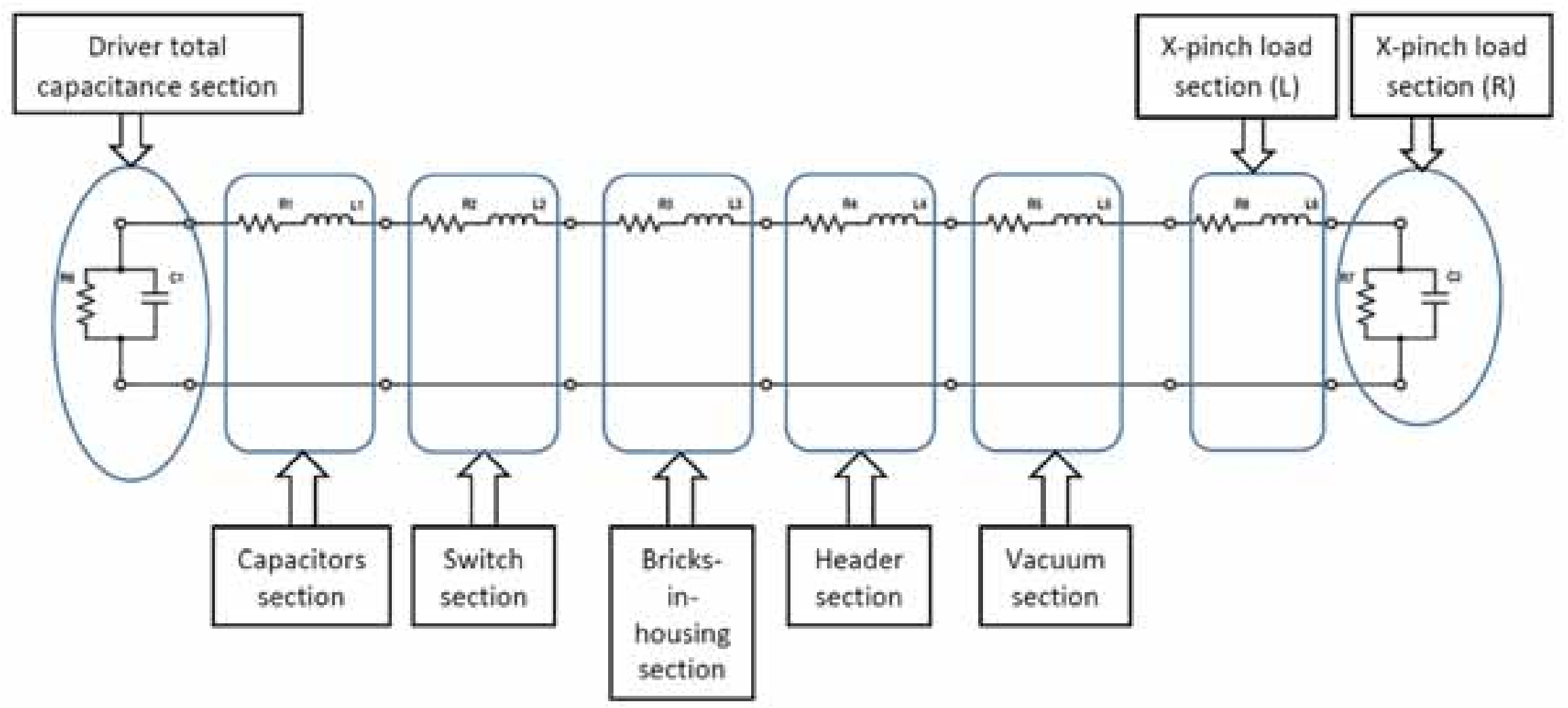

Figure 3.3-2. Screamer model of 2-LTD-brick *x*-pinch driver.

Screamer simulations of the *x*-pinch load current (red line) and the *x*-pinch load voltage (blue line) from a 2-LTD-brick driver are presented Figure 3.3-3. As can be seen, the peak current delivered to *x*-pinch load is about 195 kA with about 214-ns time-to-peak value when capacitors are fully charged to 100 kV. The maximum load voltage is about 47 kV with about -11-kV reversal at the load. The lower load voltage allows us to minimize the unwanted bremsstrahlung radiation from e-beam [25], which follows very shortly after *x*-ray burst from *x*-pinch "hot spot" is formed. Also, the lower load voltage makes it easy to design a load section, which allows one to use a smaller anode-cathode gap and, as result, allows one to minimize the load inductance value.

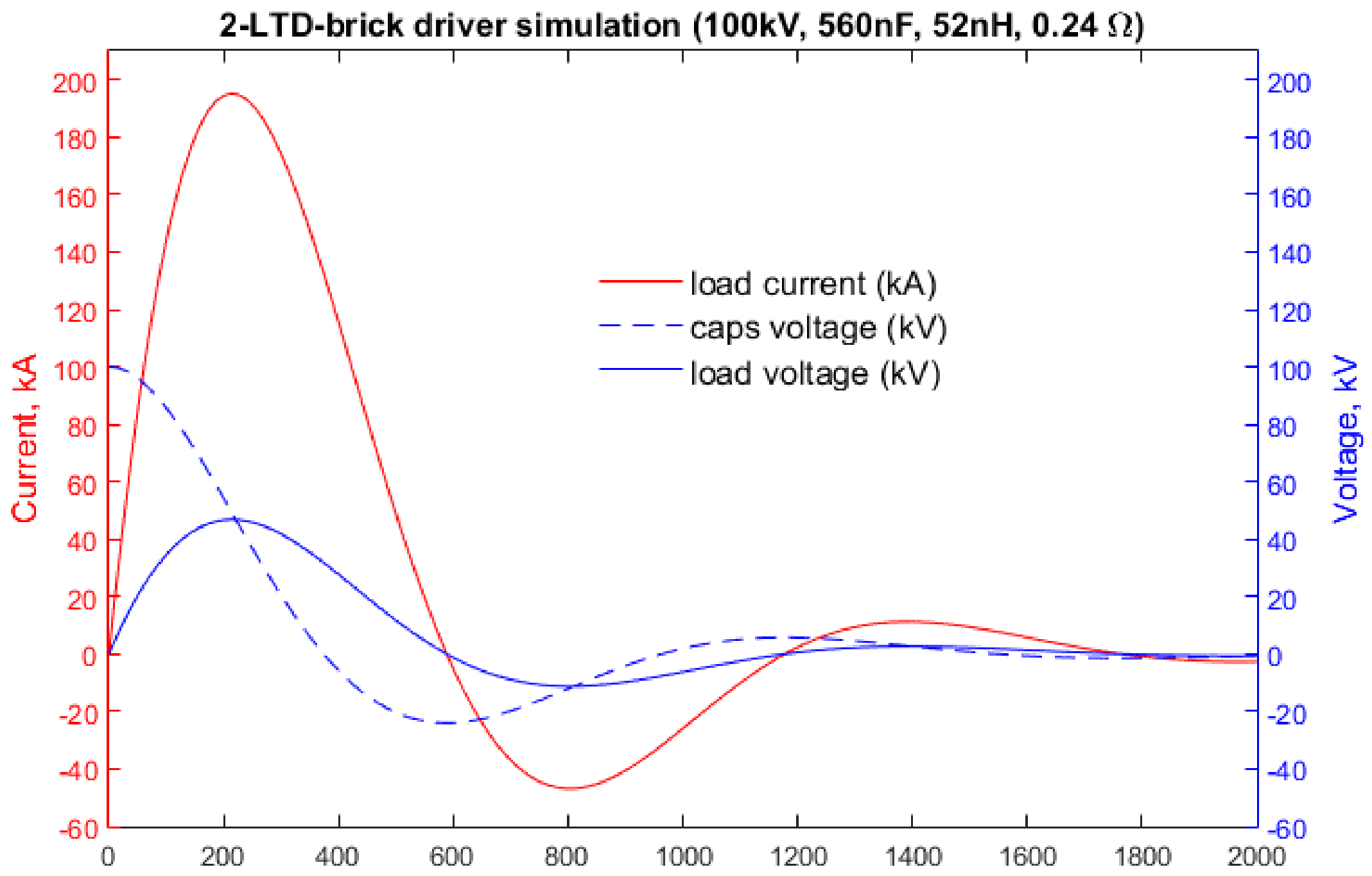

Figure 3.3-3. Output *x*-pinch load current and voltage.

The capacitor's voltage (blue dashed line) is also shown in Figure 3.3-3 to evaluate the capacitors voltage reversal, which is important parameter to be considered in the driver design. The maximum reversal voltage at bricks capacitors is about -24 kV, which is well inside the capacitor 10-80% safety margin (see Table 3.1-1).

The current rate of rise (10-90%) can be estimated from the load output current curve and is found to be about 1.5 kA/ns. To estimate the maximum absolute values of dI/dt, the numerical differentiation of the output load current was performed and presented in Figure 3.3-4. The absolute maximum value of dI/dt is about 1.9 kA/ns. Both values of current rate of rise (10-90% and maximum at point) are above the critical 1 kA/ns requirement, which has to be achieved in any high-current pulse generator in order to obtain a good *x*-pinch radiation performance as it was discussed earlier.

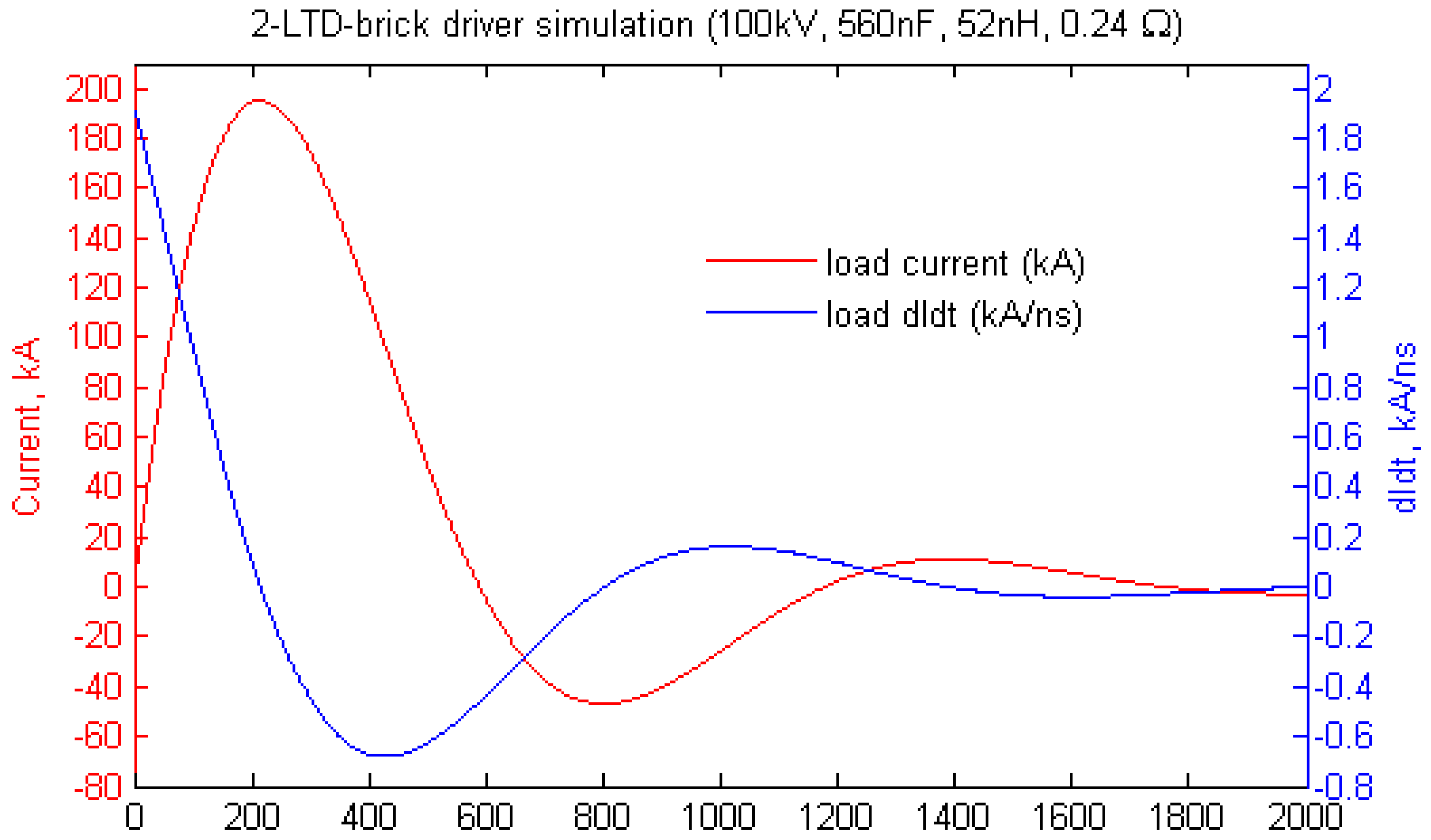

Figure 3.3-4. Output *x*-pinch load current rate of rise.

Figure 3.3-5 presents a Screamer simulation of the energy stored in all four capacitors (dashed black line) versus energy transferred to *x*-pinch load (solid black line). The *x*-pinch load current, for clarity, is also plotted (red line scaled by factor of 0.01). The total energy initially stored in four capacitors is 2.8 kJ and, as can be observed, almost all its energy is quickly transferred to *x*-pinch load. By the time of peak current, the energy transferred to *x*-pinch load reaches the value of about 1.1 kJ, which is almost 40% of the total energy initially stored inside the capacitors. As was discussed earlier in Chapter 2 and, as can be seen from these simulations, the proposed driver is naturally very efficient and almost half of the initial energy is transferred to the load by the time of peak current, when the *x*-pinch radiation burst usually happens.

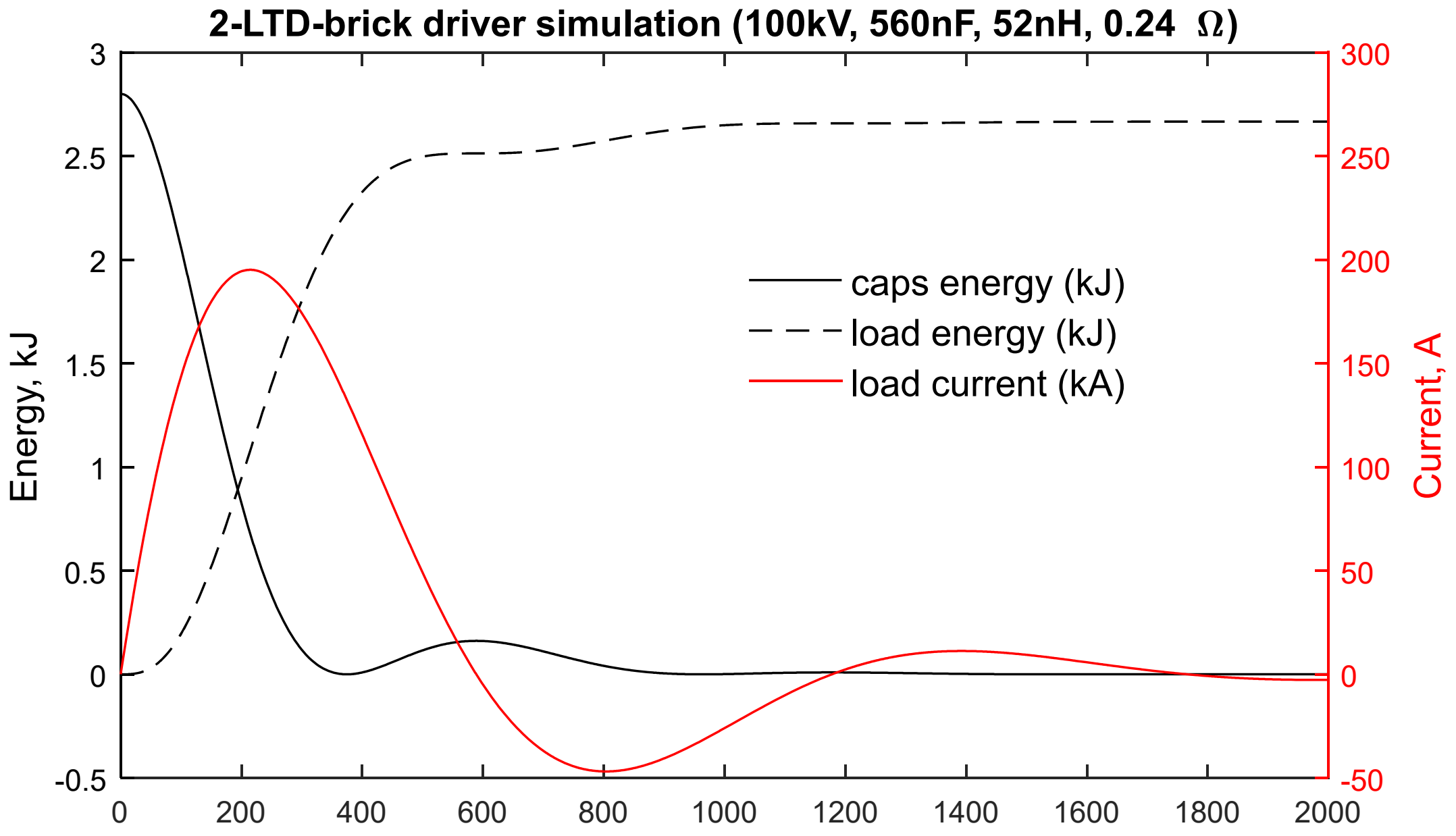

Figure 3.3-5. Energy transferred to *x*-pinch load.

It is useful to verify the results of Screamer simulations using a simple RLC model described in Section 2.2 earlier. The total resistance, inductance and capacitance of our 2-LTD-brick driver (see Figure 3.2-1 and Appendix E) equal to 0.24 Ω, 52 nH, and 560 nF, correspondingly. Such values are close to "matched" load case of RLC circuit, and, according to (2.2-14), (2.2-15) and (2.2-16), the peak current, peak voltage and time-to-peak can be estimated to be:

$$i_{peak} = 0.55 \times 100\ \text{kV} \,/\, \sqrt{52\ \text{nH} \,/\, 560\ \text{nF}} = 180\ \text{kA} \qquad (3.3\text{-}1)$$

$$V_{peak} = 0.55 \times 100\ \text{kV} = 55\ \text{kV} \qquad (3.3\text{-}2)$$

$$t_{peak} = 1.21 \times \sqrt{52\ \text{nH} \times 560\ \text{nF}} = 206\ \text{ns} \qquad (3.3\text{-}3)$$

which are in reasonably good agreement with Screamer simulations.

Presented in this section, the Screamer model of 2-LTD-Brick driver is very simple and allows for a fast calculation of many basic circuit parameters, needed for the driver design and development. However, the model is based on a static switch description, which is modeled with a simple RL-series block. A more sophisticated Screamer model of 2-LTD-Brick driver, which incorporates Tom Martin's variable-resistance switch model [65], is presented in Appendix F, and an input deck for the Screamer code is shown in Appendix G.

## 3.4 Mechanical design and fabrication

The mechanical design of our *x*-pinch *x*-ray generator is based on the general idea to make the driver compact and portable and to minimize the total driver inductance. The LTD bricks are usually arranged in a circular-in-plane geometry around a common load, as described, for example, in [64]. We choose to place two LTD/NRL bricks inside one single volume in side-by-side geometry. Such a configuration allows us to minimize the total driver inductance, and to make the whole driver assembly compact and portable.

A general artist's view of 2-LTD-brick *x*-pinch driver assembly is shown in Figure 3.4-1. Two LTD/NRL bricks are placed side-by-side inside a bricks housing with switches facing the ground plate and the capacitors facing the load section.

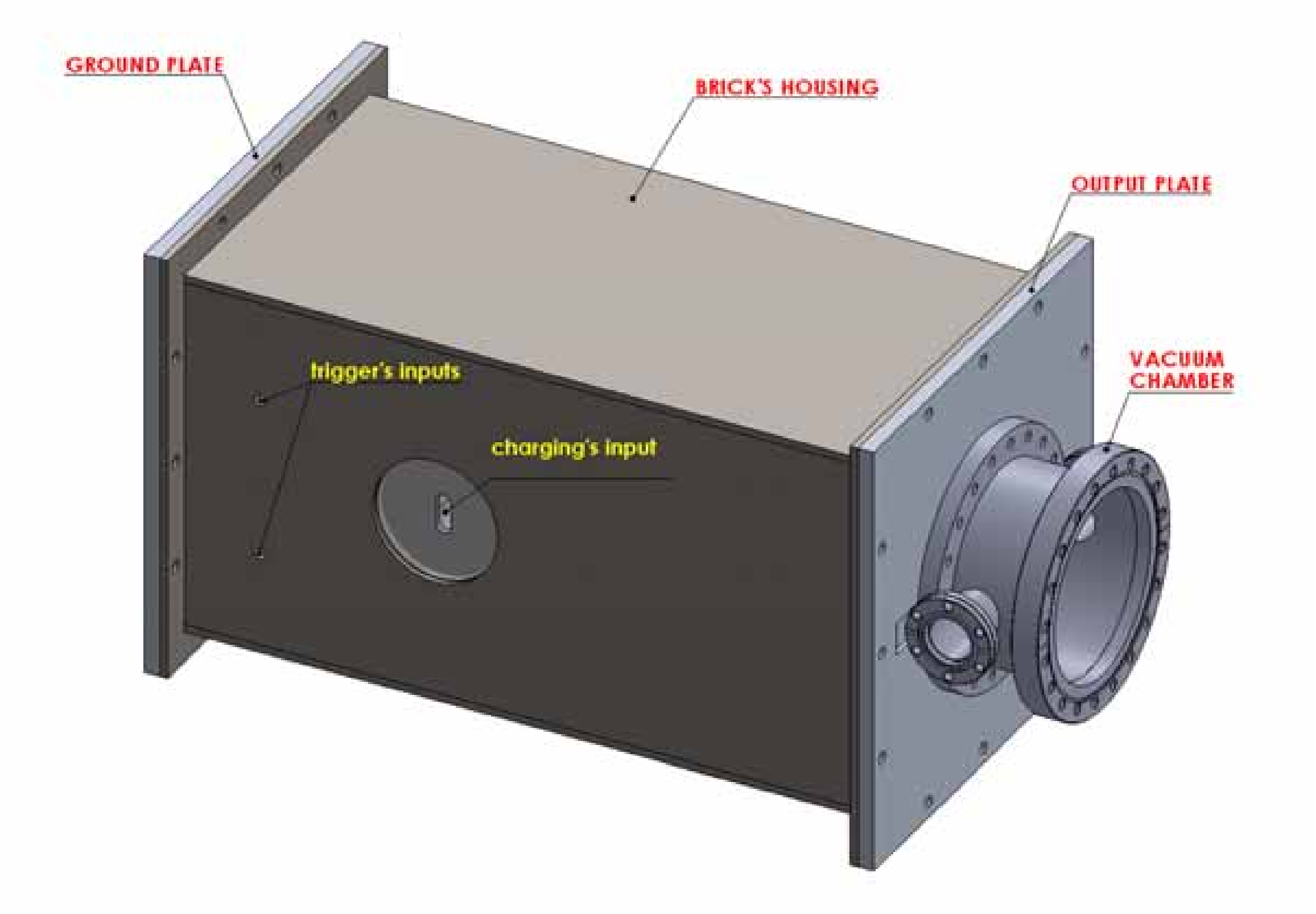

Figure 3.4-1. Artistic view of 2-LTD-brick *x*-pinch *x*-ray driver

The driver output plate drawing is presented in Figure 3.4-2. It is designed to press the bricks down inside the housing, to connect a vacuum chamber to a plate, and to serve as a return current pass to a ground. The Rogowski coil groove is made on the inner side of the plate to monitor a total current passing through the driver load. Other diagnostics, including B- or V-Dots, can be installed on this plate, if needed.

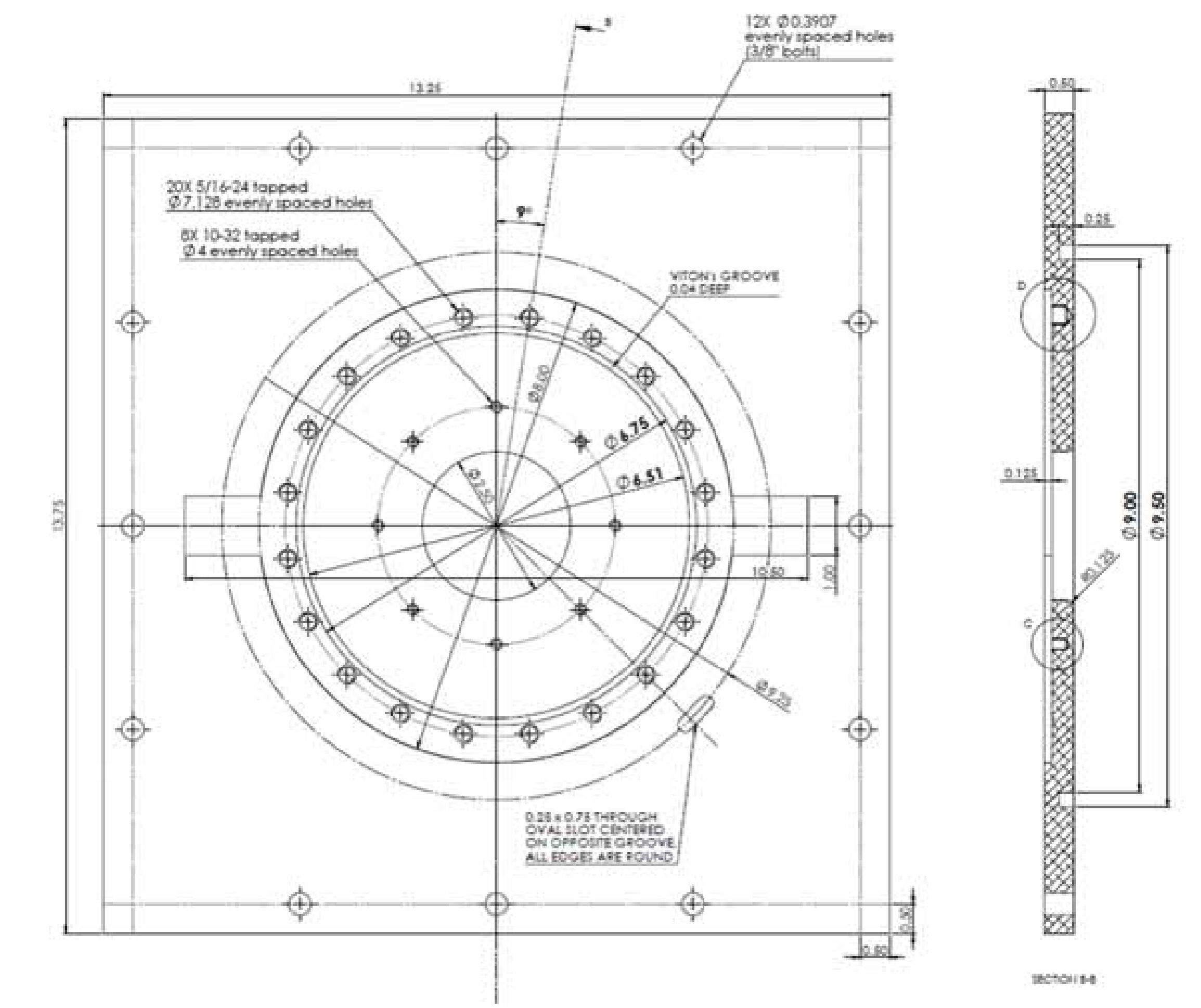

Figure 3.4-2. The drawing of driver output plate.

The cross-section view of vacuum chamber is presented in Figure 3.4-3. It is installed on top of the driver output plate and is designed to maintain the good vacuum needed for *x*-pinch *x*-ray generation. It is composed of a ConFlat™ Cross 6" (15.24 cm) in diameter, cylindrical body and two, 2-3/4" (6.99 cm) ConFlat™ output ports. One output port can be used for *x*-ray diagnostics and the other port for imaging.

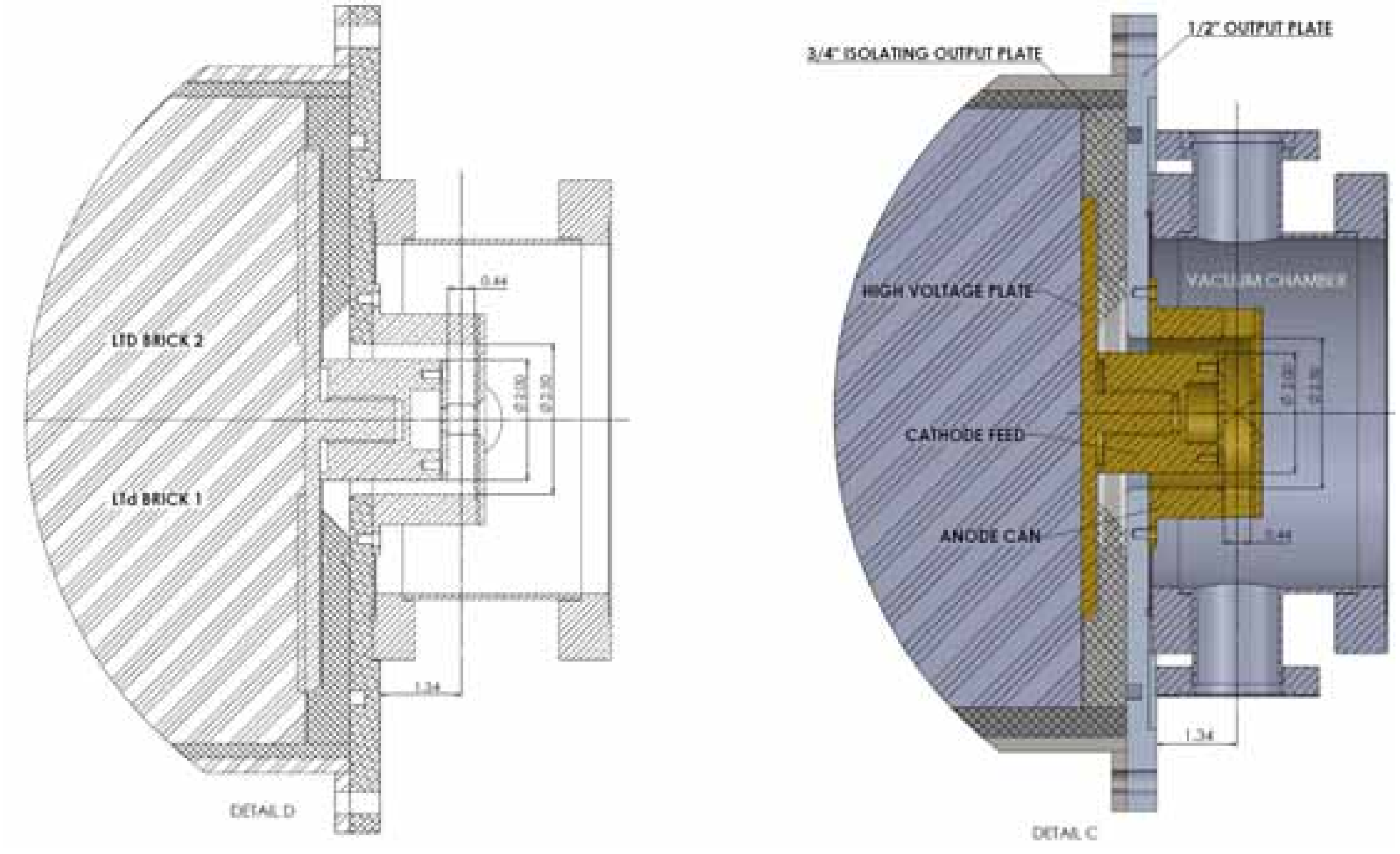

Figure 3.4-3. The output cross-section view of 2-LTD-brick *x*-pinch driver.

The high voltage plate combines two bricks and feeds the output current into the load section. A 3/4" (1.91 cm)-thick insulating acrylic plate is placed between the high voltage and output plates. A 2" (5.08 cm) long, 2" (5.08 cm) diameter cathode section is attached to the high voltage plate. The anode section is 1-3/4" (4.45 cm) long, has the 2-1/2" (6.35 cm) inner diameter and has two output *x*-ray windows aligned with vacuum chamber output ports. The insulator surface is at a -45º angle with respect to the cathode and is shielded from UV from the source by a minimum of two bounces (two boundary reflections).

As can be seen, the vacuum chamber is placed as close as possible to the LTD driver bricks, and the load section length is minimized. All these steps allowed to us to reduce the total inductance of header and vacuum flow sections. In addition, to further minimize the driver inductance, the anode-cathode gap was reduced to 6.3 mm. With such a small A-C gap, the power feed must operate in the self-magnetically-insulated mode. The $\boldsymbol{v} \times \boldsymbol{B}$ Lorentz force induced by a current pulse will turn back electrons emitted from the cathode and stop the voltage breakdown in the inter-electrode space. The minimum operating current, according to [46], can be estimated to be:

$$\mathrm{I_{min}} = 0.64\frac{\sqrt{\mathrm{V(kV)}}}{\ln(\mathrm{R_a}/\mathrm{R_b})} \tag{3.4-1}$$

where $R_a$ and $R_c$ are anode/cathode diameters. With the present geometry, the minimum operating current becomes:

$$\mathrm{I_{min}} = 2.87\sqrt{\mathrm{V(kV)}} \tag{3.4-2}$$

At a maximum anode-cathode voltage of 55 kV (see Figure 3.3-3), for the magnetic self-insulating principle to work, the minimum operating current becomes 20 kA. This is well below the simulated above 195-kA *x*-pinch peak load current value.

All *x*-pinch driver parts were carefully designed using SolidWorks 2013 [67], a 3D, solid-modeling engineering software package. All drawings (about 40 drawings) are stored on the IAC backup server for easy access, if needed. Some driver parts and assembly drawings are presented in Appendix F. The driver general view pictures are shown in Appendix G.

The whole driver is designed to be quickly assembled/disassembled for ease of inspection and modification. The brick switches, for example, can be inspected by removing the ground plate without taking apart a whole driver. All external driver supplies (HV charging line, HV trigger line, and switch air line) can be connected after driver is assembled through connections made in driver housing and ground plate. Such a design makes the driver easily to relocate to any place where the *x*-pinch radiation source is needed. The size of our driver is only 0.7x0.3x0.3 meters and it weighs about 90 kg.

## 3.5 High current diagnostics

### 3.5.1 Rogowski coil and calibration method

The calibrated Rogowski coil is extremely important for high-current pulsed measurements. We spent a significant time and effort, to build a well-calibrated Rogowski coil, which will be used in all future measurements. The basic principle of operation of Rogowski coil is described in Appendix J. Here, we describe an electrostatically shielded, non-integrating Rogowski coil with a very good signal-to-noise ratio, which was built as suggested by Dr. Spielman [68]. A 24-AWG insulated copper wire ("magnet wire" insulated with Kapton) was wrapped around the inner insulator of an RG223/U cable, stripped of the outer insulation and shielding braid. There are total of 29 turns with exactly 1-inch spacing. The view of a Rogowski coil, before and after it was electrostatically shielded with a Cu tape, is shown in Figure 3.5-1, on the left and on the right, respectively.

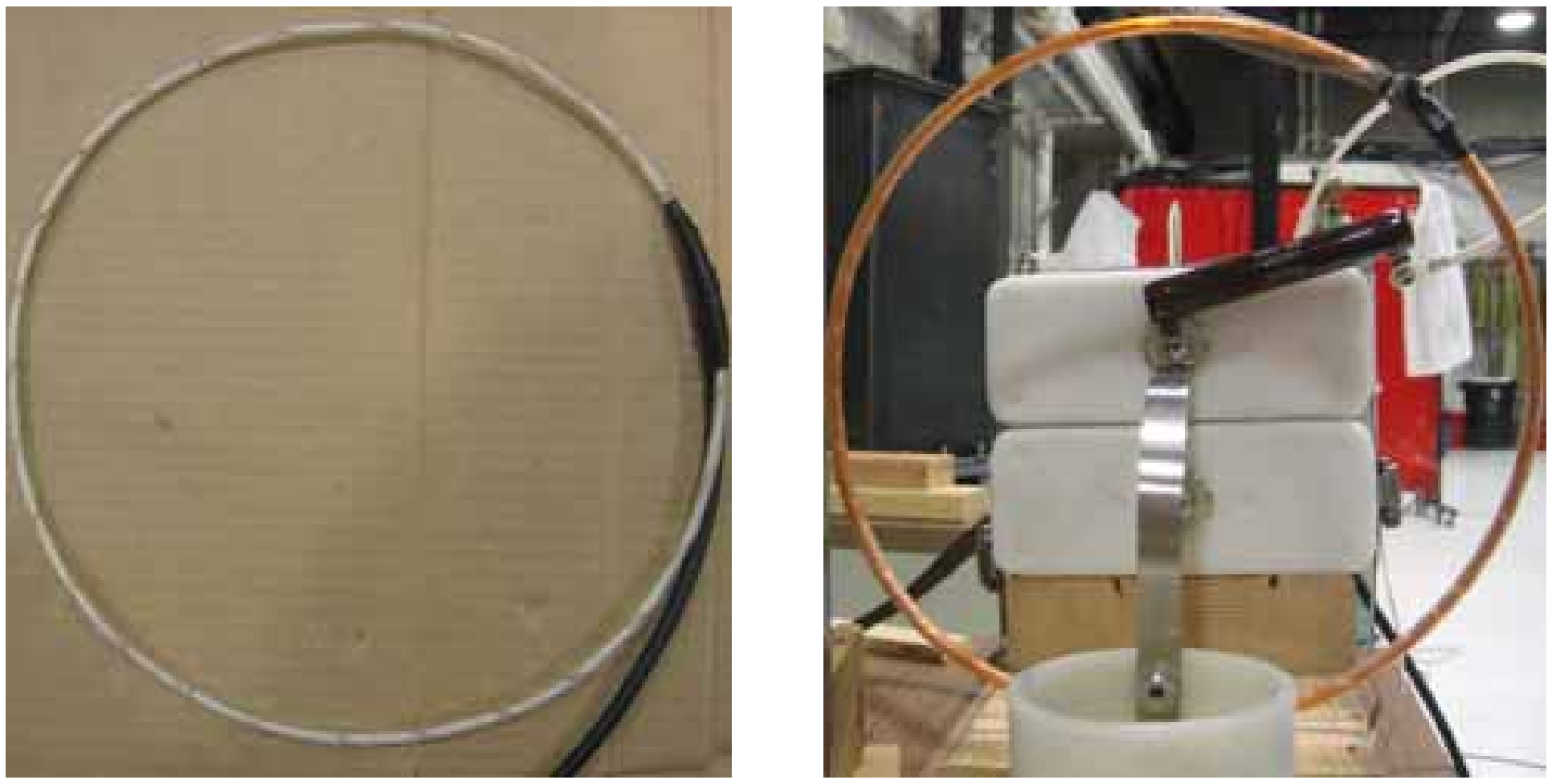

Figure 3.5-1. An unshielded Rogowski coil (left) and a shielded Rogowski coil placed around a calibration current pulse (right).

The inductance of the constructed Rogowski coil can be estimated as follow. The outer diameter of RG223 insulating core is 0.295 cm, the diameter of 24-AWG Cu-wire is 0.51 mm, and so, the area A of one small Rogowski loop is:

$$\mathrm{A} = \pi \mathrm{r}^2 = 3.14 \times ((0.295 + 0.051)/2)^2 = 9.4 \times 10^{-6}\ \mathrm{m}^2 \qquad (3.5\text{-}1)$$

and, by (J-2), the mutual inductance, M, and the coil inductance, L, equals to:

$$\mathrm{M} = -\left(1.26 \times 10^{-6}\ \frac{\mathrm{H}}{\mathrm{m}}\right) \times \left(\frac{1}{2.54 \times 10^{-2}}\frac{1}{\mathrm{m}}\right) \times (9.4 \times 10^{-6}\ \mathrm{m}^2) = 0.46\ \mathrm{H} \qquad (3.5\text{-}2)$$

$$\mathrm{L} = -\left(1.26 \times 10^{-6}\ \frac{\mathrm{H}}{\mathrm{m}}\right) \times \left(\frac{1}{2.54 \times 10^{-2}}\frac{1}{\mathrm{m}}\right)^2 \times (9.4 \times 10^{-6}\ \mathrm{m}^2) \times 0.74\ \mathrm{m} = 13.5\ \mathrm{nH} \qquad (3.5\text{-}3)$$

The general view of our calibration set-up is shown in Figure 3.5-2. It consists of a several main elements: 1) two General Atomics, fast, 20-nF, 35-kV capacitors connected in parallel to store the initial energy; 2) a 6-KV power supply with 50-kΩ charging resistor;

3) a high-voltage ball switch and 4) calibration current viewing resistor (CVR). The preliminary study of our Rogowski coil indicated that the calibration of the Rogowski coil depends only slightly on the relative position of the Rogowski coil with respect to the measured current. But to eliminate even this small calibration error, a metal conductor, carrying the calibration current, was placed at the center of our Rogowski coil, as can be seen Figure 3.5-1, on the right.

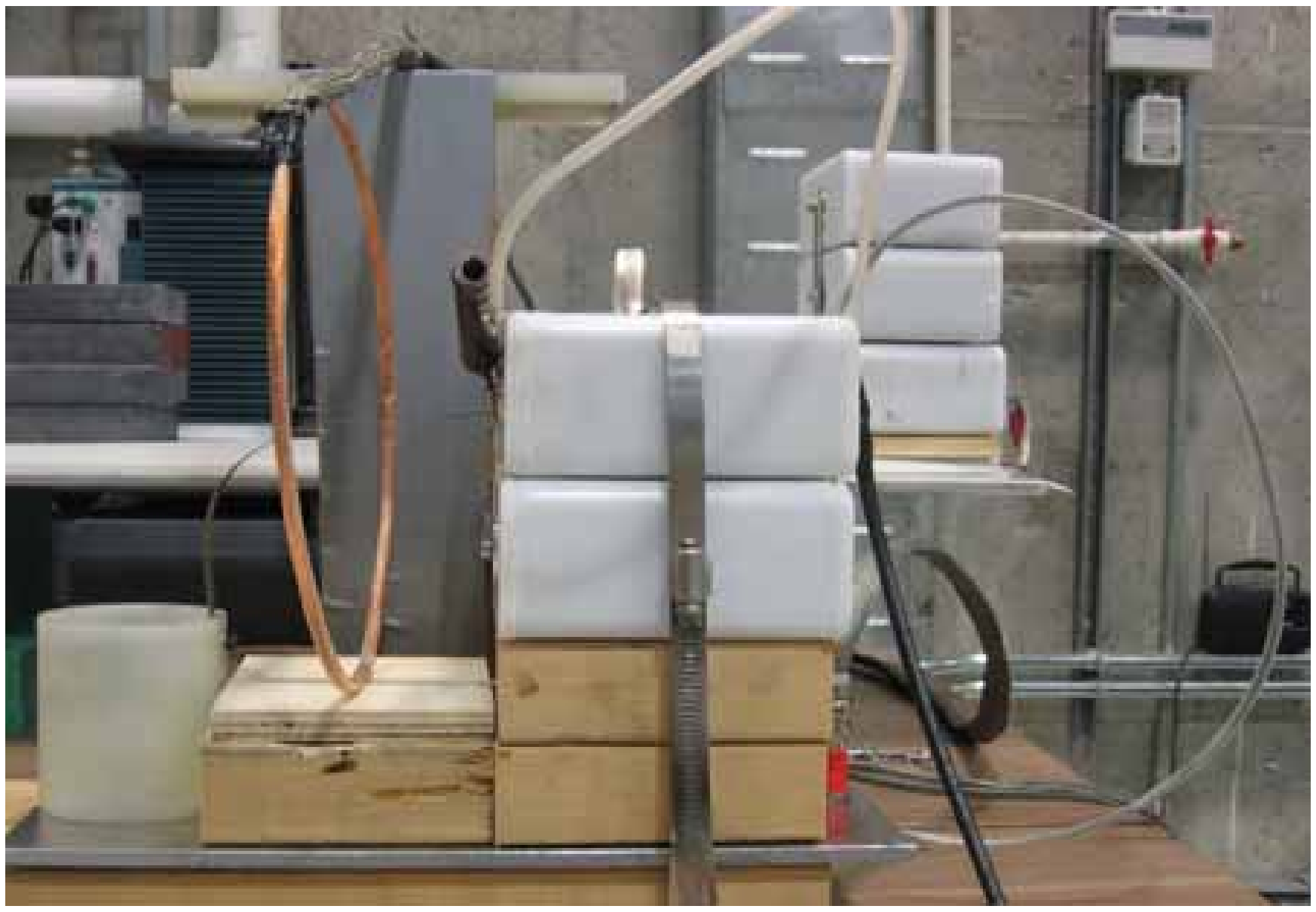

Figure 3.5-2. Rogowski coil calibration set-up.

A high-voltage ball switch was placed inside a specially designed oil-filled-cup, as can be seen in Figure 3.5-3, on the left, which allowed to us to reduce the switch arc length and to minimize the magnitude of the arc inductance and the arc resistance on the current waveform. We used a T&M Research, 0.1038-Ω CVR with NIST-traceable accuracy of 0.2% to calibrate the Rogowski coil. The CVR has been installed on the opposite side of the ground metal plate and was connected to capacitors with a flexible wire of varied length, as shown in Figure 3.5-3, on the right. By varying the length (inductance) of flexible wire, placed between CVR and capacitors, we were able to match the current oscillation time with the expected *x*-pinch current-rise-time of our 2-LTD-brick driver. In this way, the frequency content of the calibration current waveform matches that of the actual current pulse.

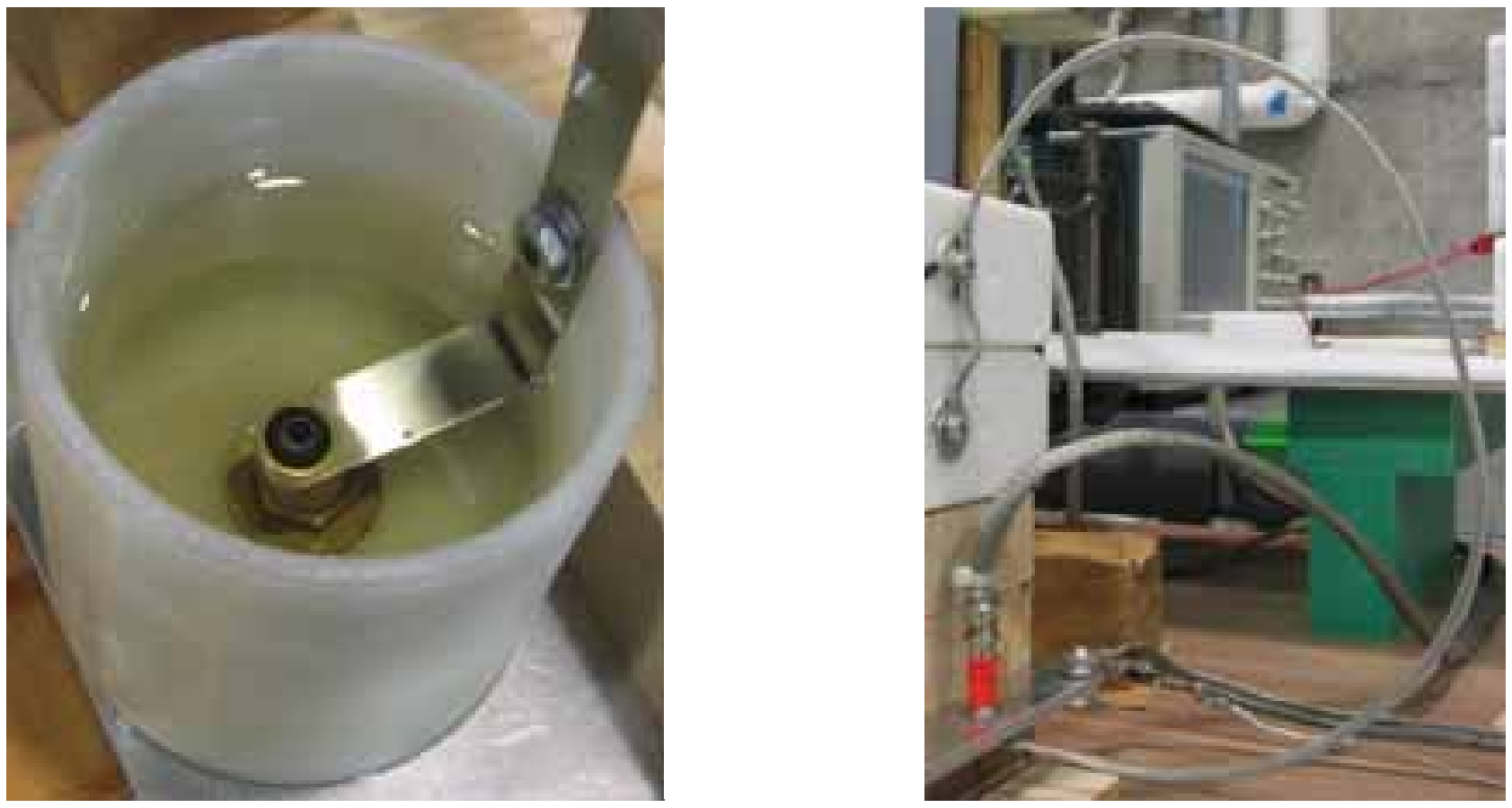

Figure 3.5-3. Oil filled cup (left) and CVR resistor (right).

The data was collected with an 8-bit Tektronix TDS 5104 digital oscilloscope, connected to the CVR and Rogowski coil by two, identical, 4-feet-long RG223/U cables. The signal cables were placed inside a shielding braid (tri-axial configuration) to eliminate any electrical noise and they were 50-Ω terminated into 1-MΩ scope inputs. It was found, that the most appropriate charging voltage for our calibration set-up was about 2 kV. First, it allowed to us to minimize a scope bit-error by maximizing the measured signal with respect to the screen size. In addition, at this charging voltage, we were still able to measure the CVR signal without using any attenuators, so the error associated with the attenuators was eliminated.

The "raw" CVR and Rogowski signals for one shot are presented in Figure 3.5-4. As can be seen, the amplitude of the CVR signal was about 44.8 V, and the amplitude of the Rogowski coil signal was about 1.3 V. The measured errors associated with these signals were calculated to be about 0.61% and 0.87% for the CVR and Rogowski-coil channels, respectively. These errors were evaluated as a ratio of standard deviation (SD) of pre-oscillation noise data, evaluated in the time interval from -4,000 ns to -50 ns, when oscillation is about to start, to the maximum amplitude of the corresponding signal. The SD of the noise data for this particular shot were 0.272 V for CVR channel and 0.011 V for the Rogowski coil signal.

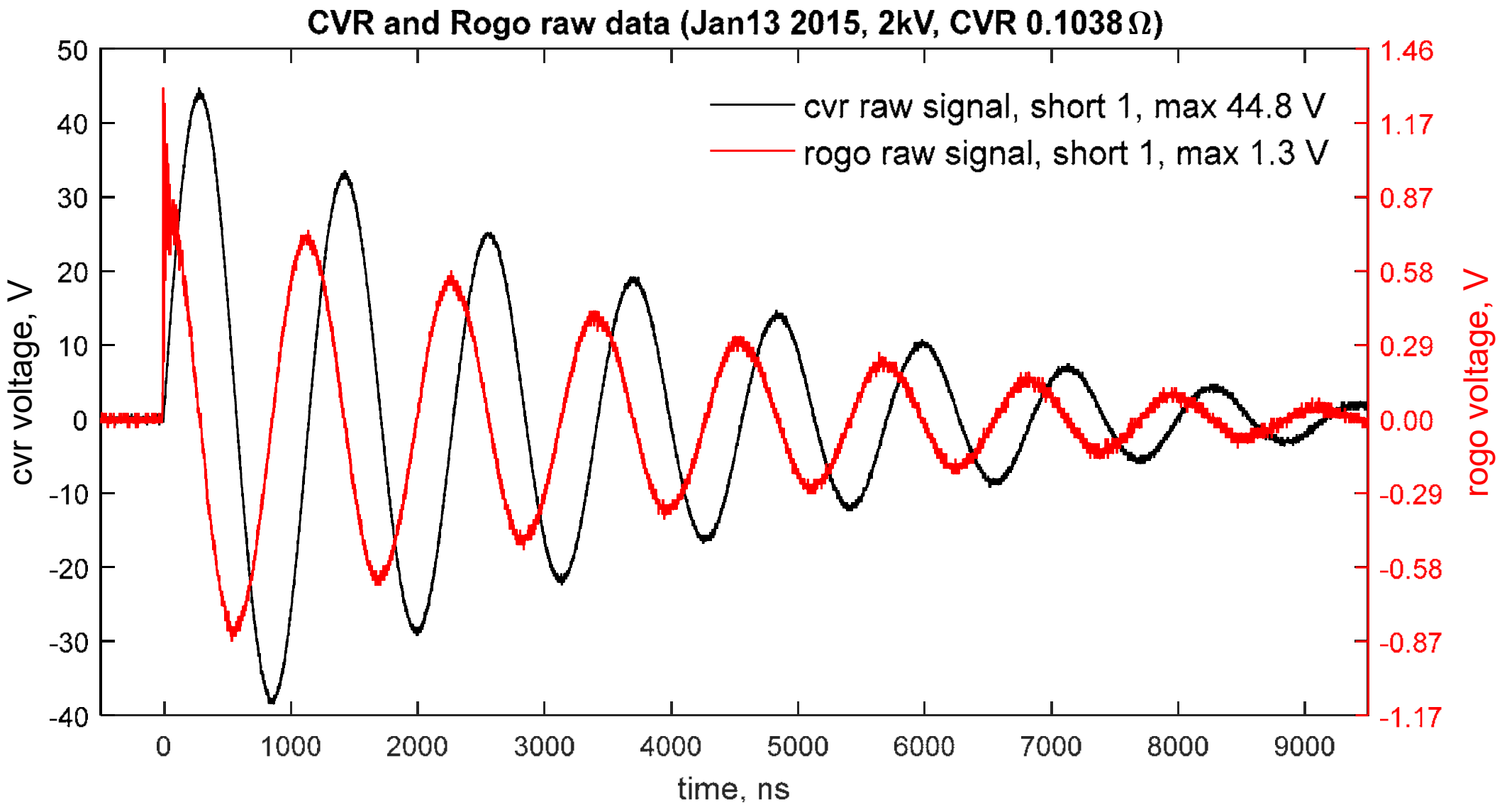

Figure 3.5-4 CVR and Rogowski-coil "raw" data for single shot.

To reduce the error associated with the single shot described above, the measurements were repeated, and then, the data was bin-by-bin averaged. The results of bin-by-bin averaging of the 25 measurements of CVR and Rogowski-coil data are presented in Figure 3.5-5. As can be seen, the quality (the noise amplitude) of the data was increased significantly. The error for the averaged shot was calculated to be about 0.12% for the CVR data, and about 0.18% for the Rogowski-coil data. That is about 5 times less than the errors associated with each single shot.

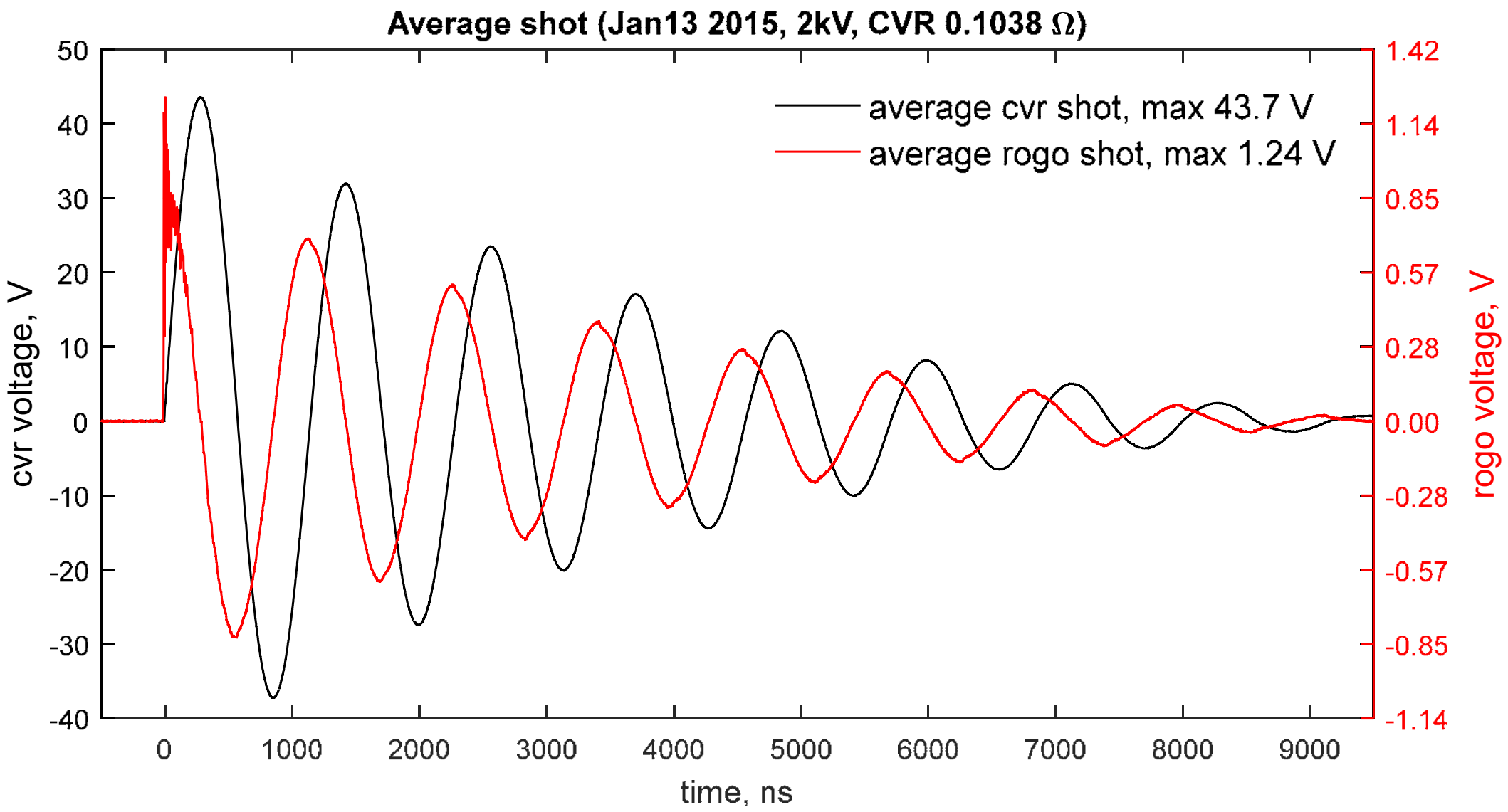

Figure 3.5-5 CVR and Rogowski-coil "raw" data for averaged shots.

To finally calibrate the Rogowski coil, we used the averaged shot, presented in Figure 3.5-5. The calibration procedure was as follows. First, the CVR current was calculated by dividing the averaged CVR "raw" data (black line in Figure 3.5-5) to the 0.1038-Ω CVR resistance value, and the Rogowski-coil current (uncalibrated) was calculated by integrating the averaged Rogowski "raw" data (red line in Figure 3.5-5). Next, the uncalibrated Rogowski current was least-squares fitted to the CVR current, and the calibration factor was found as a scaling parameter from a fitting procedure. Because the *x*-ray *x*-pinch burst usually happens in the first-half of the current pulse, the least-squares fit was done in this area of interest from 0 ns to about 1,000 ns.

The calibration results for the first two current cycles are presented in Figure 3.5-6. The top picture shows the CVR current versus the uncalibrated Rogowski-coil current, and the bottom compares the CVR current with the calibrated Rogowski-coil current shape. As can be seen, the agreement between the CVR and calibrated Rogowski current shapes is almost ideal.

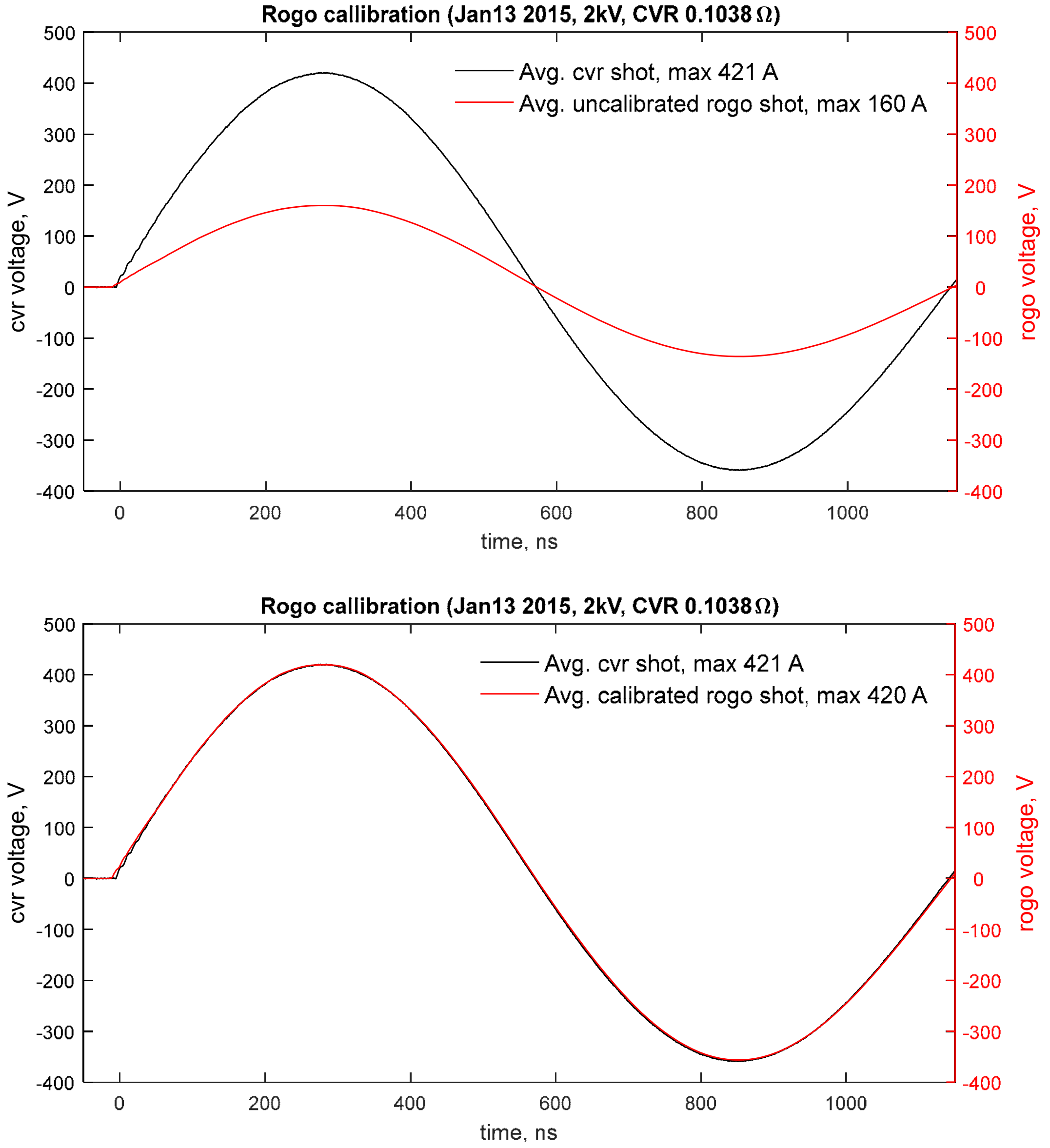

Figure 3.5-6 Rogowski coil calibration data for averaged shot. Top is CVR vs. Rogowski uncalibrated currents. Bottom is CVR vs. Rogowski calibrated currents.

The Rogowski coil calibration factor was found to be **2.62 A/V-ns**, and the error, associated with this value, is discussed below. The total calibration error can be divided in the following sub-errors:

1) **Data errors**. These type of errors were discussed and evaluated previously (see notes to Figure 3.5-4 and Figure 3.5-5). They were initially equal to about 0.61% and 0.87% for the single shot data, but were later reduced to about 0.12% and 0.18% for the averaged shot, for the CVR and Rogowski-coil channels, correspondingly. In addition, it was found, that after signal integration, the error associated with the Rogowski coil data was reduced by the factor of about 10 and was estimated to be 0.017%.

2) **CVR error**. The CVR error was associated with the CVR calibration accuracy and was equal to 0.2%.

3) **Rogowski coil calibration error associated with the fitting procedure**. This error was found from the least-square fitting of the Rogowski-coil current data to the CVR current data and was as low as 0.03%.

4) **Instrumentation errors**. These errors are usually hard to eliminate and hard to recognize, but we designed the calibration experiment in a way, that would minimize all possible errors of these kind. Below is a list off all instrumentation errors, which were recognized and properly eliminated:

- First, we used electrostatically shielded RG-223/U cables, all the same lengths, to eliminate any errors, associated with the cable length and attenuation.
- Second, we designed experiment in such a way that would eliminate the need for any attenuators, so we would not need to calibrate them.

- Also, the Rogowski coil was symmetrically positioned around the conducting metal strip, so the error associated with the Rogowski coil position (finite loop spacing) was also eliminated.
- In addition, the oil-filled cup was installed around the high-voltage spark switch to reduce errors associated with the spark behavior.
- Finally, the error associated with the current rise time was minimized by making the time-to-peak value in this calibration experiment about the same as expected in an *x*-pinch experiment.

The only known instrumentation error left is the error associated with the oscilloscope channel itself, which can be found in [69] and equals about 1.5%.

All these fractional percentage errors listed above (data, CVR, fitting and instrumentation errors) should be summed in quadrature, so the final calibration error can be estimated to be about:

$$\sqrt{(0.12^2 + 0.017^2) + 0.2^2 + 0.03^2 + 1.5^2} = 1.52\% \qquad (3.5\text{-}4)$$

It should be noted, that the main contribution to the total Rogowski coil calibration error is from the oscilloscope accuracy, and, while more measurements will reduce the data errors by averaging additional shots, to reduce the overall calibration error, a better oscilloscope accuracy will be needed. The accuracy of the CVR will limit the accuracy once the accuracy of the oscilloscope is improved.

The Rogowski coil calibration factor can also be independently calculated from the Rogowski coil design. The calibration factor between the Rogowski coil output signal, in

V, and the measured current, in A, is simply 1/M (where M is the mutual inductance of the coil), and is:

$$1/\mathrm{M} = -\frac{1}{0.46}\frac{\mathrm{A}}{\mathrm{V}\cdot\mathrm{ns}} = 2.17\,\frac{\mathrm{A}}{\mathrm{V}\cdot\mathrm{ns}} \tag{3.5-5}$$

which is about 17% less than the value found from calibration measurements. The discrepancy, probably, can be explained by the inaccurate calculated area, A, in the formula 3.1-4. Because of the small number of turns per unit length, and a circular bending of the active length of Rogowski coil, the effective area A of a one small loop will be smaller, and 1/M value will be larger. The calibration constant, 2.62 A/V-ns, found from measurements discussed above, is used in all following driver shots.

## 3.5.2 B-dot anode-cathode section current monitor

After some of *x*-pinch tests were performed it was established that it might be valuable to directly monitor the current flowing at the anode-cathode section in addition to the current monitor performed with Rogowski coil, which was described in the previous section. When driver operates with an *x*-pinch load, the acrylic insulating plate (see Figure 3.4-3) placed between the high-voltage and output driver's plates will eventually flash (breakdown) at the acrylic-vacuum interface preventing or limiting the total current flowing to the load section. It is crucial to know if the total driver energy is properly delivered to an *x*-pinch load at every shot.

To monitor the current flowing through the anode-cathode sections, a B-dot current monitor was built and calibrated as described below. We used 0.36-cm-diameter semi-rigid coaxial cable to fabricate a non-integrating, one-loop B-dot current monitor as it shown in

Figure 3.5-7. It was installed inside the anode cylinder just below the *x*-pinch section with a loop properly positioned a few mm below the inner surface of anode and filled with epoxy.

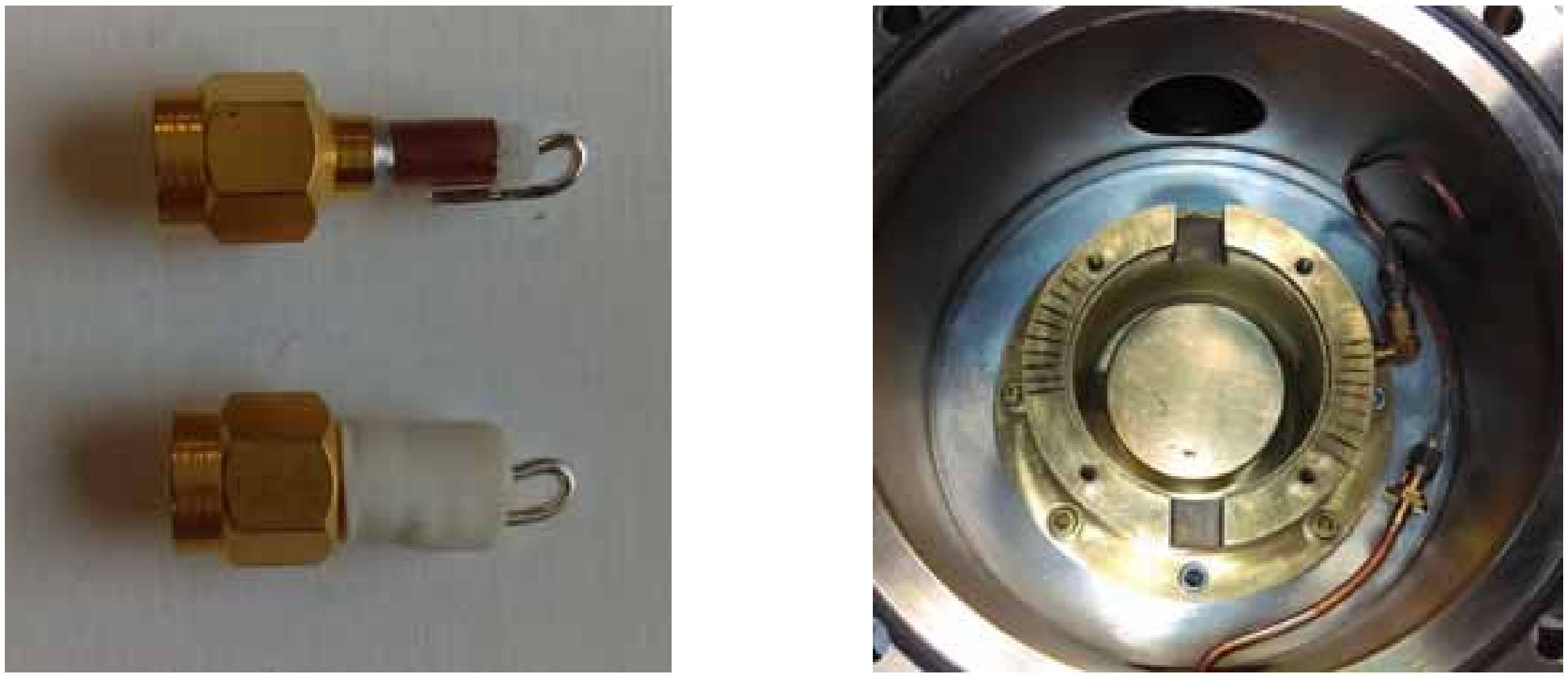

Figure 3.5-7. B-dot current diagnostic design (left) and set-up (right).

The fabricated as described above B-dot current monitor was cross-calibrated with Rogowski current monitor in a series of short-circuit shots performed with 2-inch-diameter, 1-inch-long brass cylinder load immersed in dielectric oil. As we do not need to precisely calibrate the B-dot as we did with the Rogowski coil, just one shot was selected and B-dot was cross-calibrated as shown below. The Rogowski and the B-dot signals, before cross-calibration, are presented in Figure 3.5-8. The B-dot signal is about 50 times less in amplitude then the Rogowski signal, and it appears, that it follows the shape of the Rogowski signal quite well, but, as can be expected, it is a little bit noisy and more sensitive to a higher oscillation frequency. If a higher quality B-dot signal will be needed, we could reduce this noise by increasing the signal amplitude of the B-dot by using a larger area loop and multiple turn loops.

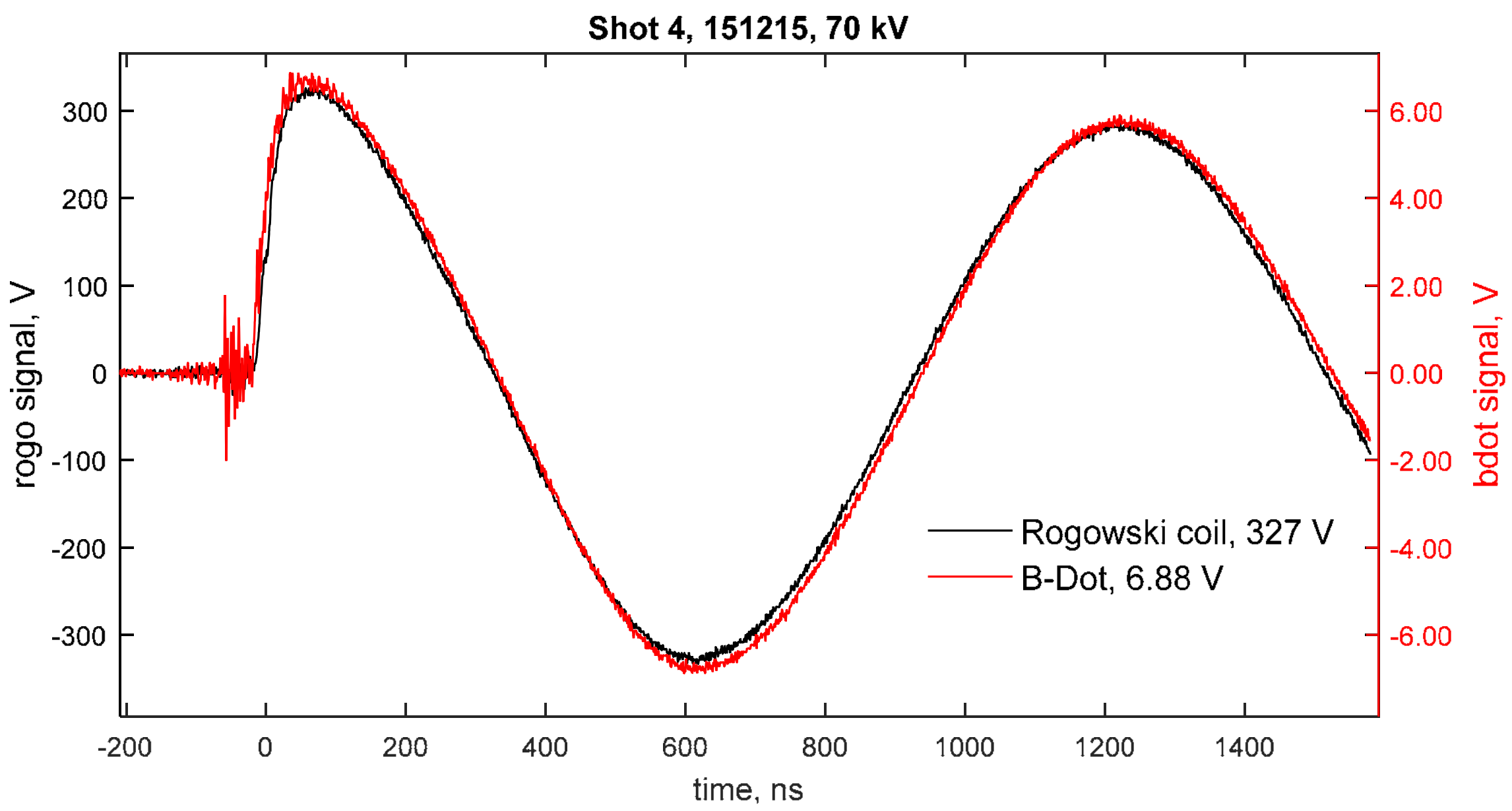

Figure 3.5-8 Rogowski and B-dot short-circuit shot data.

The results of cross-calibration are shown in Figure 3.5-9. The Rogowski and B-Dot signals were integrated and calibration was performed to match their first-peak current amplitudes. The calibration factor for the B-Dot current monitor was found to be 121.5 A/V-ns for this particular shot.

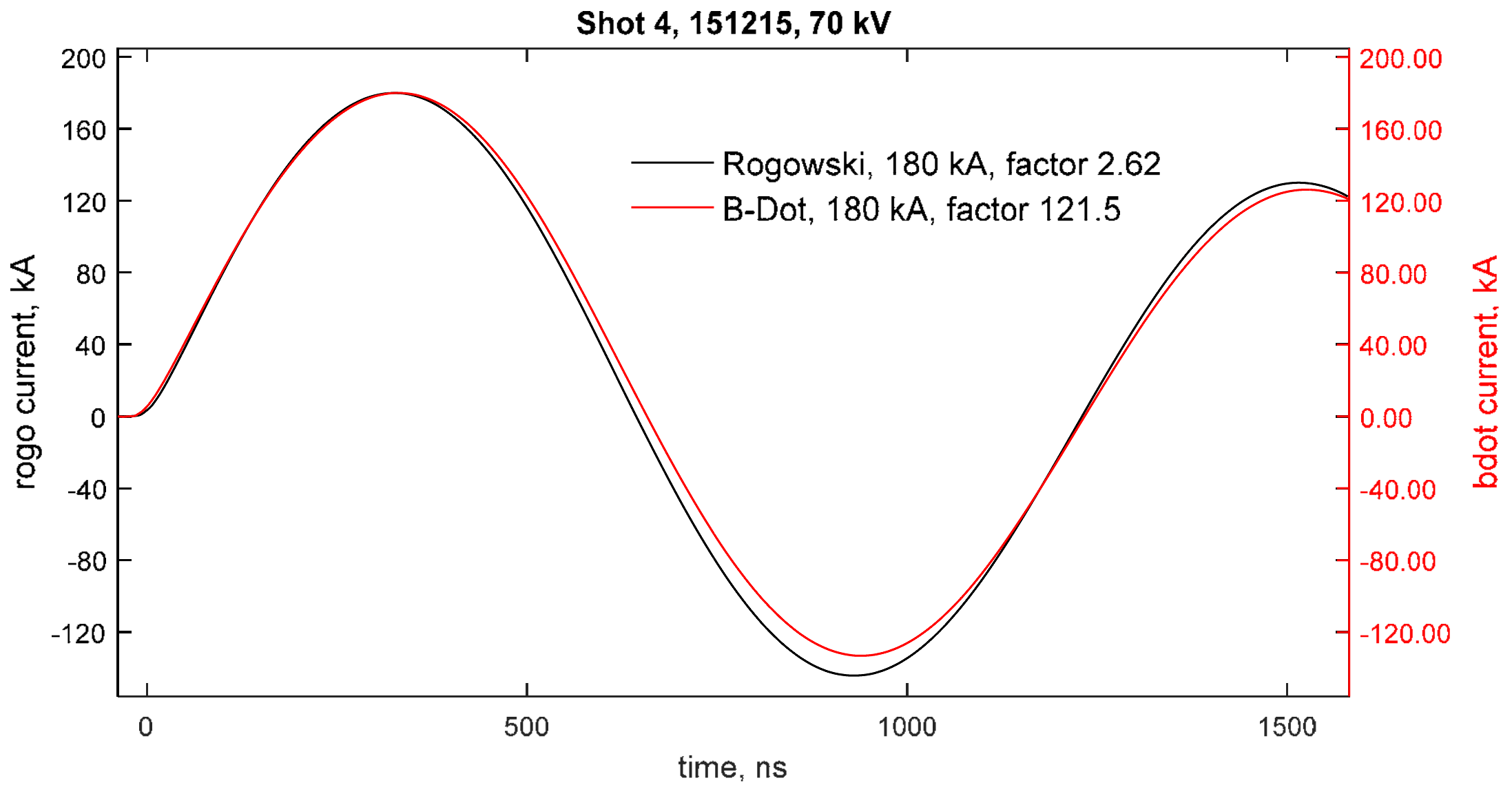

Figure 3.5-9 Rogowski and B-dot cross-calibrated currents.

There were total of 3 shots performed at 60 kV and 4 shots performed at 70 kV of initial driver charge and they are all in a good agreement with calibration factor found above.

## 3.6 Revised and improved driver models

In the last section of this chapter we revise the developed 2-LTD-Brick's driver electrical model based on the measured total, driver internal inductance, and discuss some reduction in a total driver inductance which can be made and how it may affect the performance of future drivers. These discussions are based on the experimental work performed and described later but these data are still more suitable to be shown in this chapter.

### 3.6.1 Revised driver model

As will be shown later (Chapter 4.2.1), the measured total driver internal inductance is estimated to be about 60 nH. This measured value represents the inductance of the whole generator without driver load. In this chapter, we present the revised 2-LTD-brick *x*-pinch driver model, based on measurements.

The revised driver circuit is presented in Figure 3.6-1. The extra driver inductance element (blue circled) was modified to reflect the measured total driver inductance value of 60 nH. As the brick's inductance was estimated earlier to be about 14 nH (two bricks in parallel), the extra driver inductance has to be about 46 nH. All other driver model elements are kept unchanged in these simulations.

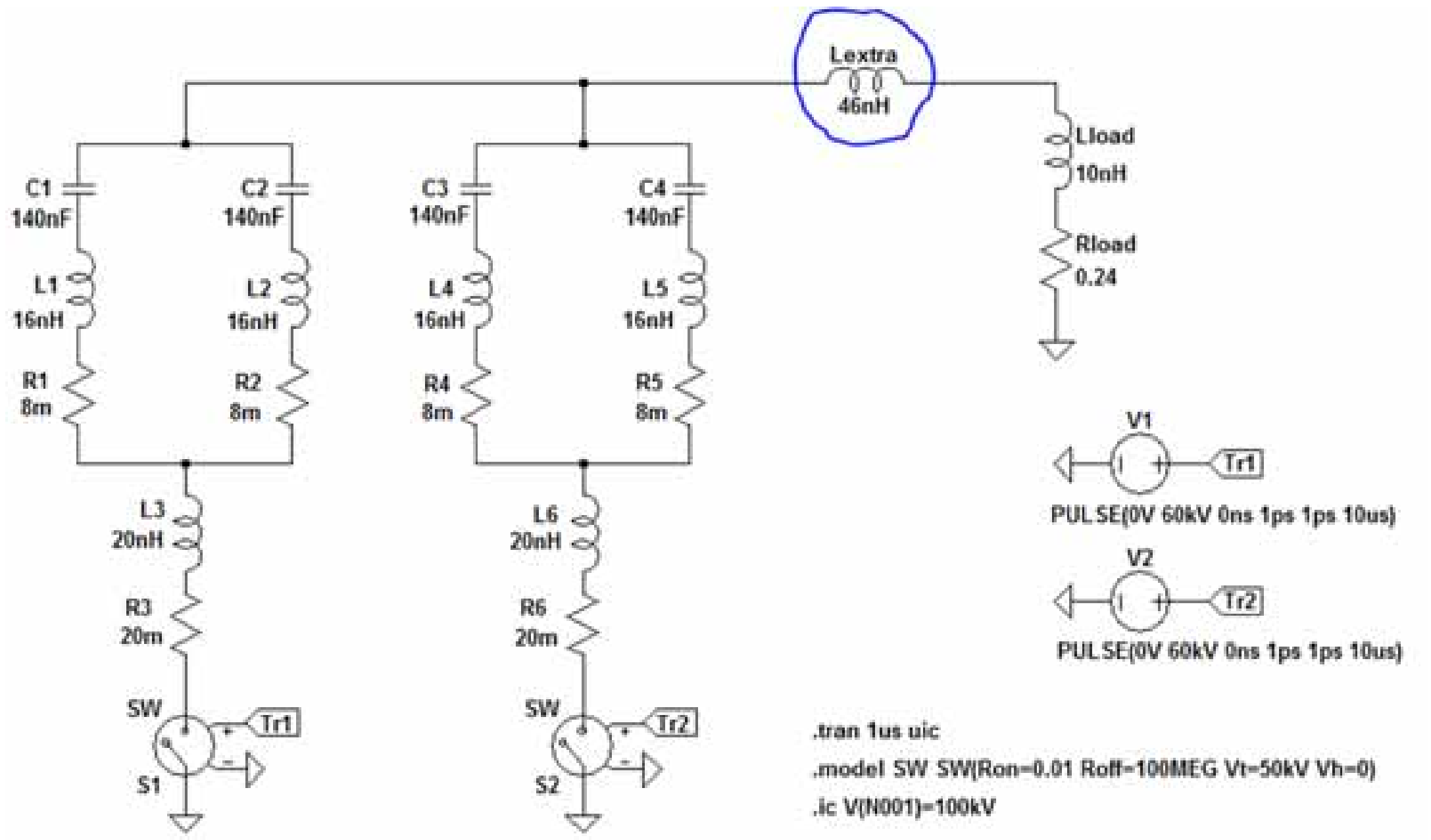

Figure 3.6-1. Revised 2-LTD-brick *x*-pinch driver electrical diagram.

Screamer simulations of revised 2-LTD-brick driver model are presented in Figure 3.6-2. The red line represents the output *x*-pinch load current, the blue line represents the output *x*-pinch load voltage, and the dashed blue line shows the voltage measured on brick's capacitors. As can be seen, the driver, when it is fully charged to 100 kV, can supply 178-kA peak-current with about a 158-ns, 10-90%, rise time. The corresponding current rate of rise (10-90), dI/dt, equals 1.1 kA/ns, and the maximum dI/dt equals to about 1.4 kA/ns. The maximum load voltage is about 43 kV with about -13-kV reversal. The maximum reversal voltage at the brick capacitors is about -30 kV, which is well inside the capacitor 10-80% safety margin.

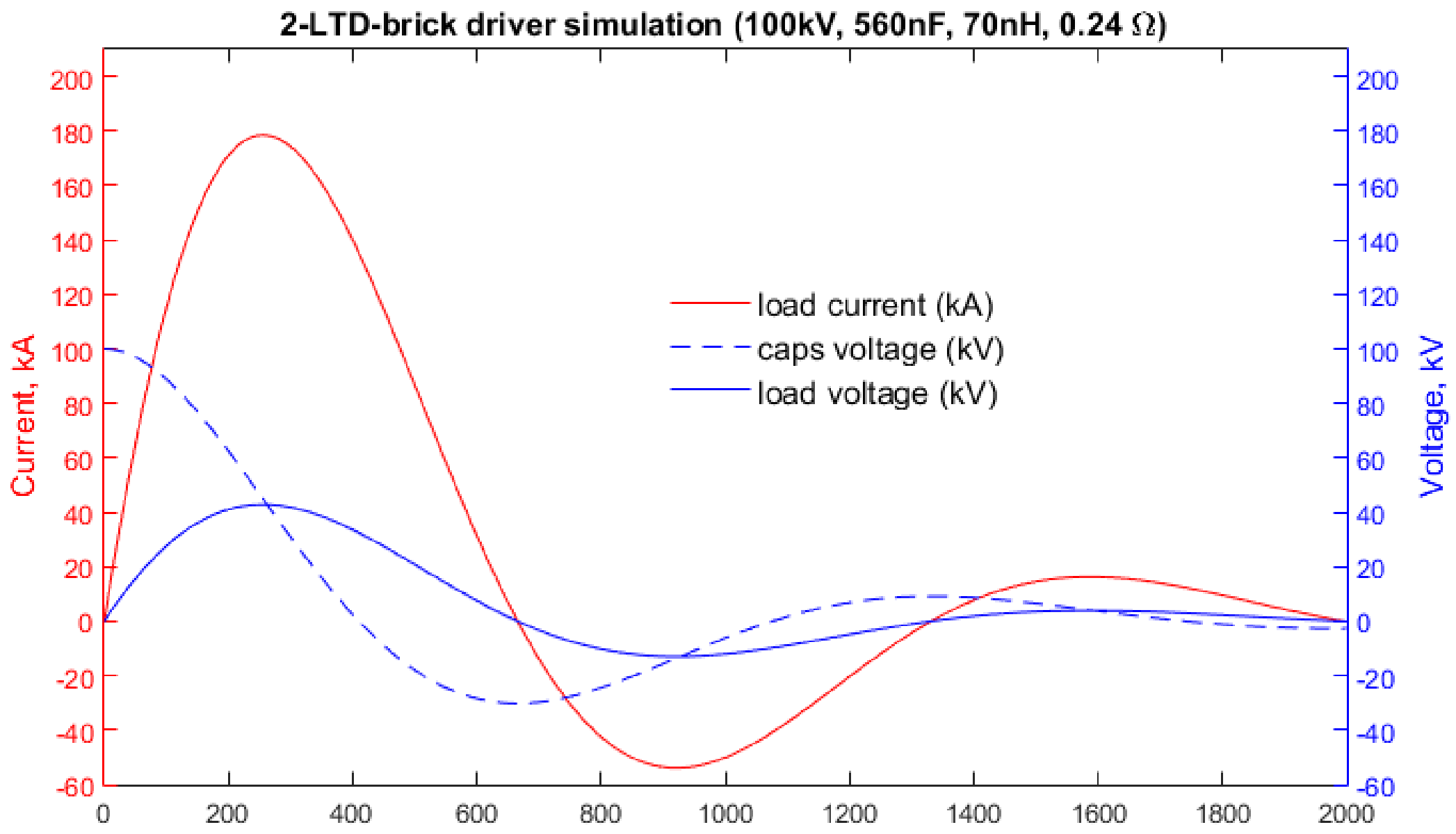

Figure 3.6-2. Simulations of 2-LTD-brick driver (revised model) output *x*-pinch load current and voltage.

## 3.6.2 Improved driver model

As it will be seen later from our experimental work (chapter 4 and 5), the developed 2-LTD-brick driver prototype works well and is amazingly stable. At the present time, the HV insulation acrylic plate has thickness of 1.3 cm between the HV and output driver's plate, and the anode-cathode gap is 7 mm. One question that can be asked here is what reduction in a total driver inductance can be made, and how it could affect the future driver's performance? Can we reduce the thickness of HV insulator and reduce the gap of magnetically-insulated transmission line (anode-cathode gap)? What other driver parts can be modified to reduce the total driver inductance?

According to [70], the acrylic has dielectric strength of about 400 V/mil. Assuming that a peak load voltage of 43 kV (see simulations Figure 3.6-2) is applied to an acrylic insulator, the thickness of acrylic, in principle, can be reduced to:

$$\frac{43\ \mathrm{kV}}{400\ \mathrm{V/mil}} = 107\ \mathrm{mil}\ \approx 0.3\ \mathrm{cm} \tag{3.6-1}$$

As it was already discussed earlier in a mechanical design section 3.4, the minimum operating current for a MITL equals:

$$\mathrm{I_{min}} = 0.64\frac{\sqrt{\mathrm{V(kV)}}}{\ln(\mathrm{R_a/R_b})} \tag{3.6-2}$$

This relation can be reversed, to estimate the smallest anode-cathode gap for given load voltage and the calculated load current:

$$\ln(\mathrm{R_a/R_b}) = \frac{0.64 \times \sqrt{43\ kV}}{178\ \mathrm{kA}} = 0.0236, \tag{3.6-3}$$

and, given the inner cathode diameter $R_b$ = 2.54 cm, the outer radius can be estimated to be $R_a$ = 2.60 cm, which can potentially reduce the anode-cathode gap to about 1 mm.

Such a decrease in the acrylic thickness and a reduction in anode-cathode gap will change the inductance of different parts of driver's power flow section as described in Table 3.6-1, where changes made compared to a previous calculations (see Table 3.2-1) are highlighted by gray colors.

Table 3.6-1. Improved driver inductance at different power flow sections.

| **Driver Section** | **$R_a$ (mm)** | **$R_b$ (mm)** | **h (cm)** | **L (nH)** |
|---|---|---|---|---|
| Header part 1 section | 127 | 178 | 0.64 | 0.4 |
| Header part 2 section | 25 | 178 | 1.0 | 3.9 |
| Vacuum section without load | 25 | 26 | 4.1 | 0.3 |

In addition, the *x*-pinch load section inductance was estimated earlier to be about 10 nH, but with some modifications (reducing the *x*-pinch height, or changing the load geometry) it can be reasonably reduced by factor of 2, resulting in the *x*-pinch inductance, in an initial "cold state", of about 3-5 nH.

All these improvements discussed above can effectively reduce the total driver inductance from about 70 nH, as it was assumed in a revised driver model, to about 60-62 nH. Screamer simulations of this improved 2-LTD-brick driver model are presented in Figure 3.6-3.

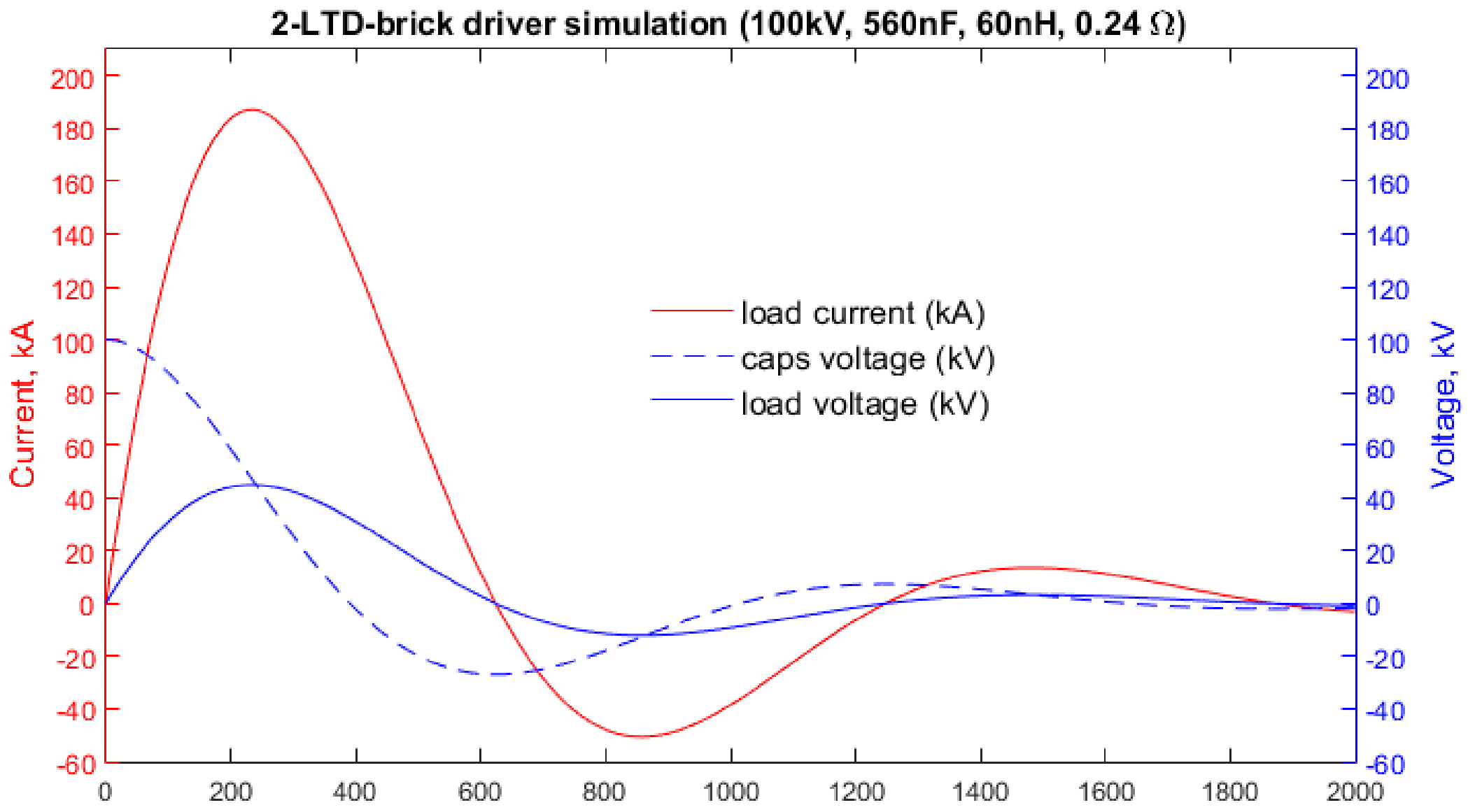

Figure 3.6-3. Simulations of 2-LTD-brick driver (improved model) output *x*-pinch load current and voltage.

As can be seen, the reduction in total driver inductance by 10 nH will increase the total driver current from 178-kA to about 187-kA peak-current and reduce the current 10-90% rise-time from about a 158-ns to about 144-ns. The corresponding current rate of rise (10-90), dI/dt, equals 1.4 kA/ns, and the maximum dI/dt equals about 1.6 kA/ns.

As can be seen, the improvements in the driver performance are not so significant compared to the revised driver model discussed in a previous chapter, as the main contribution to a total driver inductance comes from the bricks themselves and the extra brick inductance located in a surrounding housing. However, if desired, only the extensive driver testing and experimentations can validate the suggested future improvements of the 2-LTD-Brick discussed here. As experience shows, the driver failures can be unexpected and unpredictable, and a fine tuning of driver configurations is always needed.

### 3.6.3 Summary of 2-LTD-Driver model's circuit parameters

Table 3.6-2 bellow summarizes some electrical circuit parameters of the revised (model with a real, experimental measured driver parameters) and improved (possible reduction in a total driver inductance which can be made in a future) 2-LTD-brick's driver models and compares them with the initial driver model parameters discussed earlier in Chapter 3.3.

Table 3.6-2. Initial, revised and improved 2-LTD-Driver model's circuit parameters.

| **Driver parameters** | **Initial model (chapter 3.3)** | **Revised model (chapter 3.6.1)** | **Improved model (chapter 3.6.2)** |
|---|---|---|---|
| Initial charging voltage | 100 kV | 100 kV | 100 kV |
| Total capacitance | 560 nF | 560 nF | 560 nF |
| Extra driver inductance | 28 nH | 46 nH | 36 nH |
| Total driver inductance | 52 nH | 70 nH | 60 nH |
| Load peak/reversal current | 195/-47 kA | 178/-54 kA | 187/-51 kA |
| Load peak/reversal voltage | 47/-11 kV | 43/-13 kV | 45/-12 kV |
| Capacitors peak/reversal voltage | 100/-24 kV | 100/-30 | 100/-27 |
| Time-to-peak current | 214 ns | 255 ns | 233 ns |
| Rise time (10-90%) | 131 ns | 158 ns | 144 ns |
| Max dI/dt | 1.9 kA/ns | 1.4 kA/ns | 1.6 kA/ns |
| Energy (@ $T_{peak}$) | 38% | 38% | 38% |

**Revised vs. initial models:** The change in driver total inductance from 52 nH (initial model) to 70 nH (revised model) results in a drop of the load peak-current from 195 kA to about 178 kA and in an increase of the current rise time (10-90%) from 131 ns to about 158 ns. The load peak voltage is reduced from 47 kV to about 43 kV, and the capacitor's reversal voltage (max absolute value) increased from -24 kV to about -30 kV, still keeping the capacitors inside its safety margins. What is more important, the simulations show, that the driver is still able to deliver more than 1-kA/ns current rate-of-rise into the low inductance *x*-pinch load, when it is fully charged to 100 kV.

**Improved vs. revised models:** The reduction in total driver inductance from 70 nH (revised model) to 60 nH (improved model) results in an increase of the load peak-current from 178 kA to about 187 kA and in an decrease of the current rise time (10-90%) from 158 ns to about 144 ns. The peak load voltage and the capacitor's reversal voltage are almost unaffected, still keeping the capacitors inside its safety margins. The overall improvements in a total driver inductance can potentially increase the current rate of rise by about 14%, slightly improving the *x*-pinch radiation performance.

# 4 2-LTD-BRICK DRIVER SHORT-CIRCUIT TESTS AND RESULTS

## 4.1 Driver initial tests and turn-ups

The 2-LTD-Brick driver described here had never been built and tested before. It took an entire year to establish the proper driver configuration. A total of more than 210 short-circuit shots were performed, which are briefly summarized in Table 4.1-1 and discussed below. All shots are grouped in series of tests, by the time when some particular driver set-up was tested, or by some technical driver issue was resolved. For additional reference, all shots performed during this initial testing period and their main parameters are listed in Appendix L.

Table 4.1-1. Summary of 2-LTD-brick *x*-pinch driver initial test data.

<table>
<tr><th>Test №<br>(total shots)<br>Date</th><th>Shot ID*</th><th>Load</th><th>$V_{driver}$<br>kA</th><th>$dI/dt_{max}$<br>kA/ns</th><th>$I_{peak}$<br>kA</th><th>Driver set-up tested and performance notes</th></tr>
<tr><td>Test 1(6)<br>05/2014</td><td>101</td><td rowspan="3">1 Ω HVR**</td><td>70</td><td>0.23</td><td>39</td><td>Set-up: Common charging line;<br>Notes: Driver pre-fire and one brick firing;<br>Multiple sparking inside driver and load;</td></tr>
<tr><td>Test 2(4)<br>06/2014</td><td>202</td><td>80</td><td>0.35</td><td>102</td><td>Set-up: Separate charging line;<br>Notes: Pre-fire is eliminated, but bricks are fired not simultaneously;<br>Multiple sparking inside driver and at load;</td></tr>
<tr><td>Test 3(2)<br>07/2014</td><td>301</td><td>78</td><td>0.31</td><td>42</td><td>Set-up: Improved HV isolations;<br>Notes: HVR failure/explosion;</td></tr>
<tr><td colspan="7"></td></tr>
<tr><td rowspan="6">Test 4(44)<br>07/2014</td><td>412</td><td>SW*** R1</td><td>80</td><td>0.62</td><td>158</td><td rowspan="6">Set-up: Trigger line is pushed behind the last trigger’s ball;<br>Notes: Sparking at the end of trigger line is eliminated;<br>All bricks are fired simultaneously;</td></tr>
<tr><td>417</td><td>SW R2>R1</td><td>80</td><td>0.75</td><td>159</td></tr>
<tr><td>422</td><td>SW R3>R2</td><td>80</td><td>0.63</td><td>151</td></tr>
<tr><td>427</td><td>SW R4>R3</td><td>80</td><td>0.59</td><td>129</td></tr>
<tr><td>433</td><td>SW R5>R4</td><td>80</td><td>0.59</td><td>121</td></tr>
<tr><td>444</td><td>SW R6>R5</td><td>80</td><td>0.56</td><td>103</td></tr>
<tr><td>Test 5(5)<br>08/2014</td><td>505</td><td>SW***</td><td>80</td><td>0.87</td><td>174</td><td>Set-up: Cu sheet installed between two bricks;<br>Two PT-55 trigger system;</td></tr>
<tr><td colspan="7"></td></tr>
<tr><td>Test 6(6)</td><td>606</td><td>Ni-wire load</td><td>80</td><td>0.91</td><td>197</td><td>Set-up: same</td></tr>
</table>

| Test № (total shots) Date | Shot ID* | Load | $V_{driver}$ kA | $dI/dt_{max}$ kA/ns | $I_{peak}$ kA | Driver set-up tested and performance notes |
|---|---|---|---|---|---|---|
| 08/2014 | | | | | | |
| Test 7(92) 11/2014 | 728 | Brass Cylinder (1" long, 1" ⌀) | 80 | 0.97 | 212 | **Set-up:** Positive vs. negative power supply tested;<br>**Notes**: All shots are consistent; |
| | 729 | 40 AWG Cu | 80 | 1.05 | 219 | |
| | 738 | Brass Cylinder (1" long, 2" ⌀) | -50 | 0.45 | 114 | |
| | 747 | | -60 | 0.57 | 138 | |
| | 756 | | -70 | 0.67 | 164 | |
| | 757 | | 50 | 0.28 | 60 | |
| | 774 | | 60 | 0.76 | 160 | |
| | 783 | | 70 | 0.91 | 189 | |
| | 792 | | 80 | 1.07 | 222 | |
| | | | | | | |
| Test 8(4) Mar 2015 | 803 | Brass Cylinder (1" long, 2" ⌀) | 70 | 0.85 | 179 | **Set-up:** First driver's major repair (brick's cracks are removed and filed with a new epoxy);<br>Improved brick/Cu sheet/brick isolating;<br>Improved under-angle charging line design;<br>Improved through-bricks trigger line design;<br>**Notes**: 1st driver's major failure (multiple cracks in a brick's epoxy);<br>Some shots are problematic again (not-simultaneous, lower signal) |
| Test 9(39) April 2015 | 933 | | 60 | 0.60 | 144 | **Set-up:** Improved trigger system (HV1000 pre-pulser instead of PT-006, two positive PT-55 instead of two negative, through-bricks trigger line);<br>Tested + 125 kV PS up to 90 kV<br>**Notes**: Some shots are problematic (pre-fire);<br>Finally driver starts to pre-file even at lower, 30-40 kV, voltages |
| | 935 | | 70 | 0.80 | 182 | |
| | 937 | | 80 | 1.04 | 221 | |
| | 938 | | 85 | 1.2 | 240 | |
| | 939 | | 90 | 1.3 | 254 | |
| Test 10(14) June 2015 | 1010 | | 80 | 0.99 | 212 | **Set-up:** Second driver's major repair (cracks in brick's capacitor is sealed, brick's cracks are removed and filed with a new epoxy);<br>Improved charging line (same plus thicker tubing);<br>Improved trigger line (same plus wider bricks trigger line holes);<br>Driver mobility test at a new location;<br>**Notes**: 2nd driver's major failure (multiple cracks in a brick's epoxy, cracks and oil leakage in one of the bricks capacitors);<br>Charging line thin tubing failure;<br>Sparking between end of trigger line and bricks housing;<br>Finally all shots are "good" and driver operation is stable. |
| | 1012 | | 80 | 1.02 | 210 | |
| | 1014 | | 80 | 1.03 | 213 | |

*Shot ID- first digit is a test number, last two are the shot number;
**HVR - high voltage resistor load;
***SW - salted water load;

For safety considerations, the very first driver tests № 1-3 were performed with *1-Ω high-voltage resistor (HVR)* load installed between anode-cathode electrodes and immerged in dielectric oil. Tests № 4 and 5 were performed with a *salt-water load* that filled the anode-cathode gap and the lower part of the cathode feed. Test № 6 was done with a *Ni-wire load*, and tests № 7-10 was performed with a *brass cylindrical short-circuit load,* both immerged in a dielectric oil.

As can be seen from the Table 4.1-1, many driver improvements were made and many driver set-ups were tested, before a stable working driver configuration was established. Below, we briefly discuss some critical driver elements (trigger line, charging line, etc.), and how their different designs affected the driver performance.

## 4.1.1 Driver's charging line performance

Two different charging line designs, *common and separate*, were tested in a series of initial shots (tests № 1-2). The difference between the two charging designs are outlined in Figure 4.1-1. As can be seen, in a case of common charging line, a common metal rod was installed inside the brick housing to connect two HV brick electrodes on the switch side, so all 4 capacitors can be charged simultaneously. In a case of separate charging line, two bricks are charged independently by two separate charging lines.

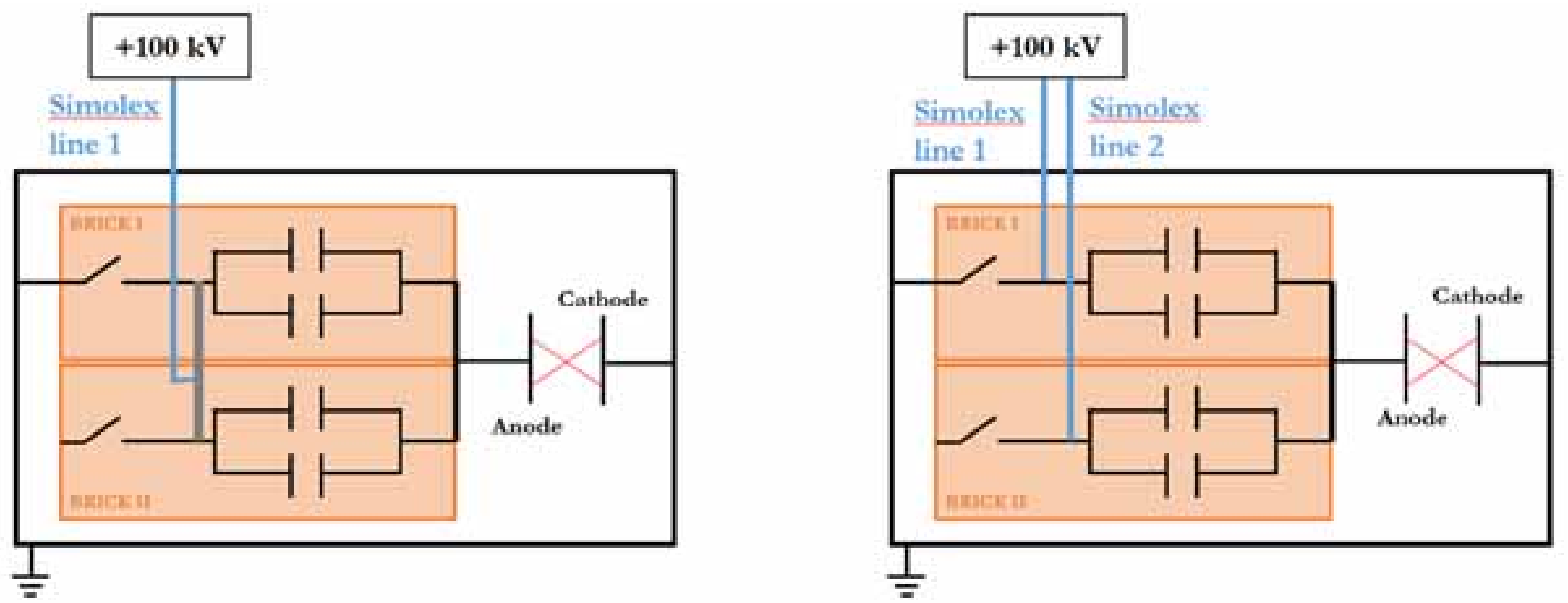

Figure 4.1-1. Common vs. separate charging line designs. Left shows the common charging line, right shows the separate charging line.

The *common charging line* was designed to limit the total current passing through each bricks in a case of a brick misfire, but the *separate charging* line is the most common and straightforward one. Designs and simulation results of these two charging lines are presented and discussed in Appendix K. Although, simulations show that a common charging line is safer, practically, it results in driver pre-fires. An example of two shots, one performed with a common charging line and other with separate charging lines, is presented in Figure 4.1-2.

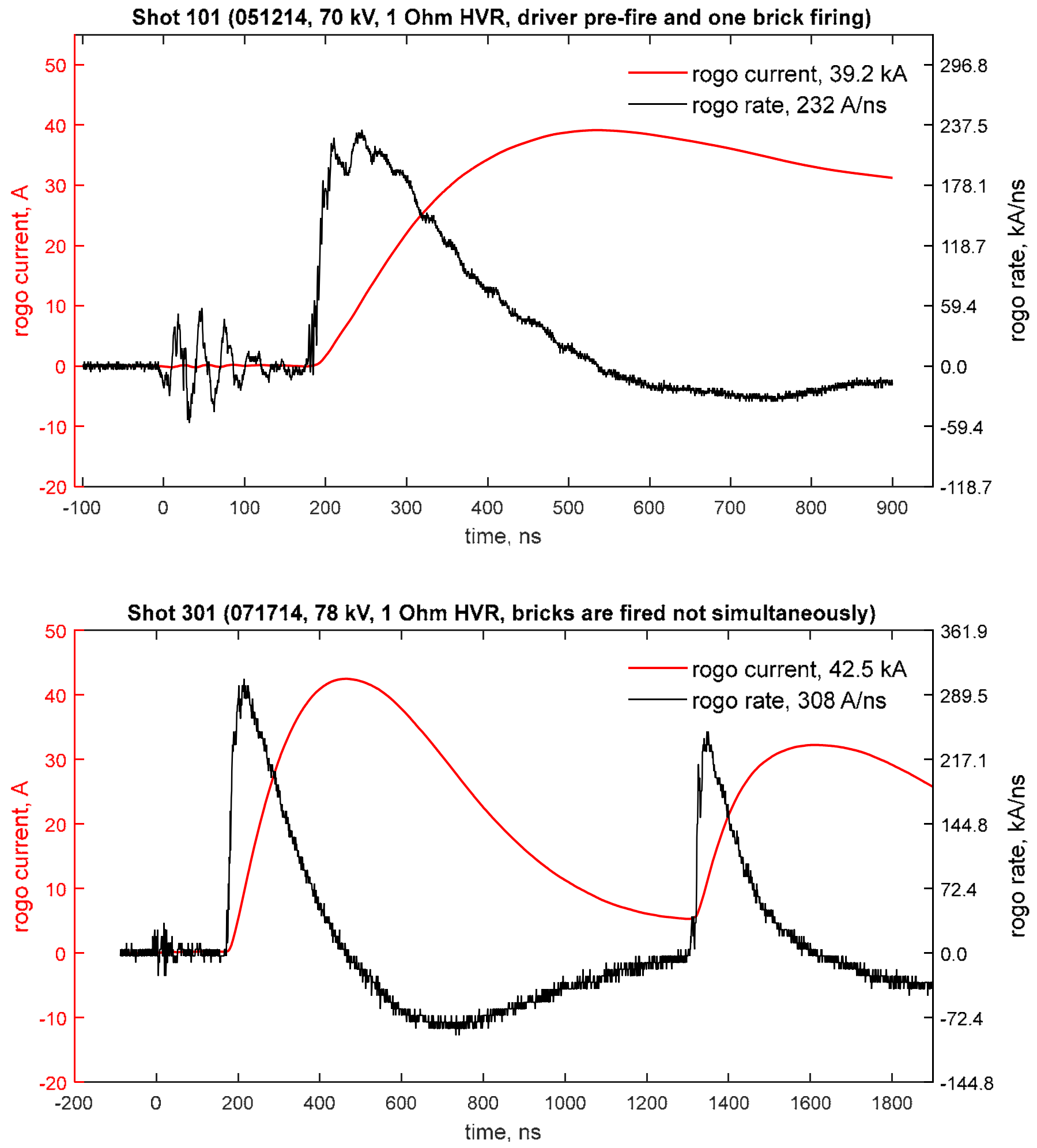

Figure 4.1-2. Shot № 101 (top) with a common charging line. Shot № 301 (bottom) with a separate charging line but with non-simultaneous brick firing.

For shots performed with a common charging line (Figure 4.1-2, top), the driver is self-triggered before the total charging voltage is reached. With a separate charging line (Figure 4.1-2, bottom), the driver is always manually fired at the proper time, although other triggering problems, not related to a charging line, were evident and will be discussed later.

Some additional charging line issues (arcing at the ends of the charging lines, charging line tubing failure, etc.) were found later, but they all were successfully eliminated as it is discussed in Table 4.1-1, test № 8-10 notes.

### 4.1.2 Driver's trigger line performance

The most common technical issues, observed in all our earlier tests, were non-simultaneous triggering of driver's bricks. The separation times between triggering of the two bricks could vary significantly, as shown, for example, Figure 4.1-3. The top event (shot № 202) shows a separation time of 200 ns between the brick's triggering. And the bottom event (shot № 302) shows a separation time of more than 1 μs between the brick's triggering.

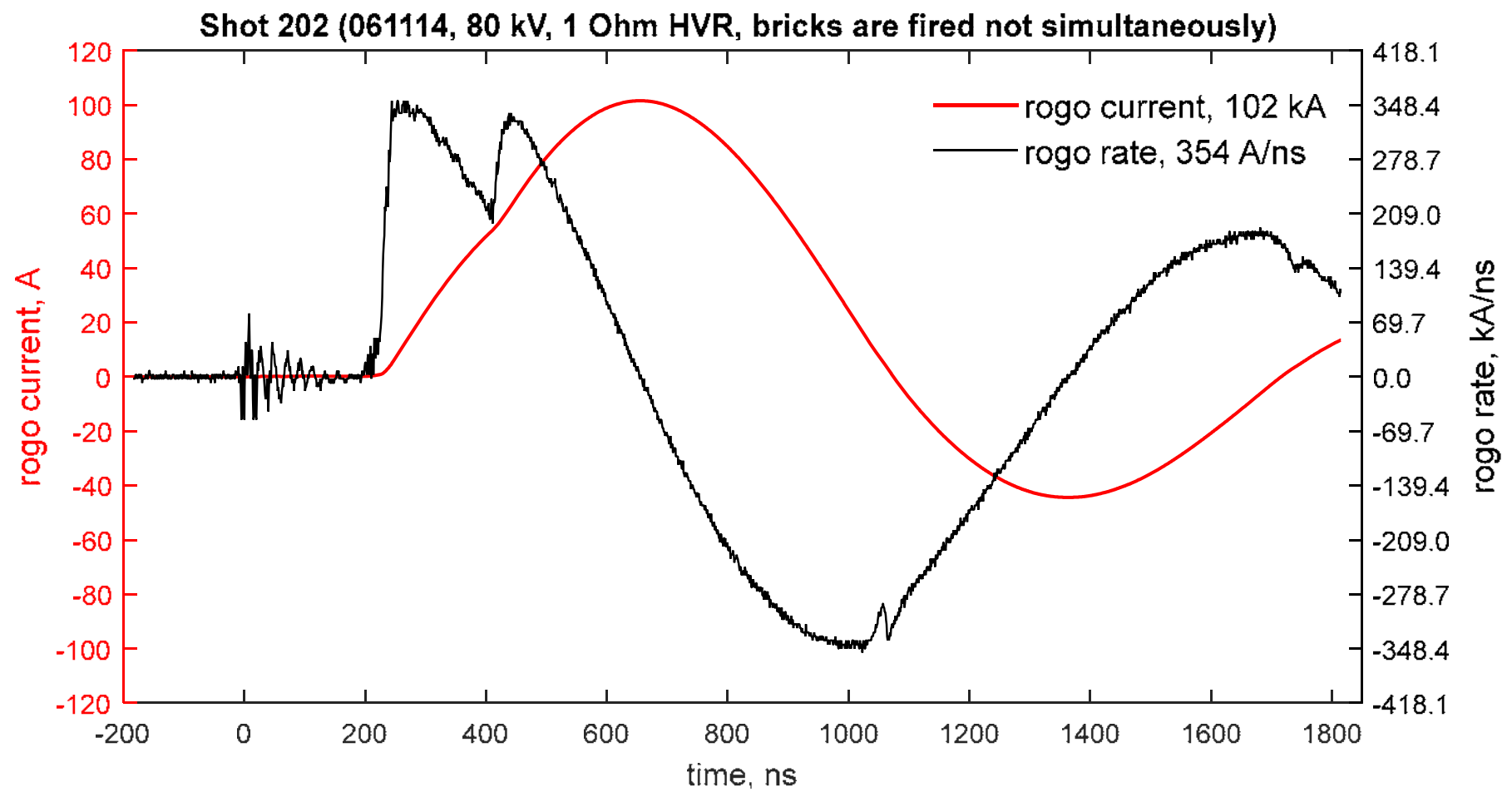

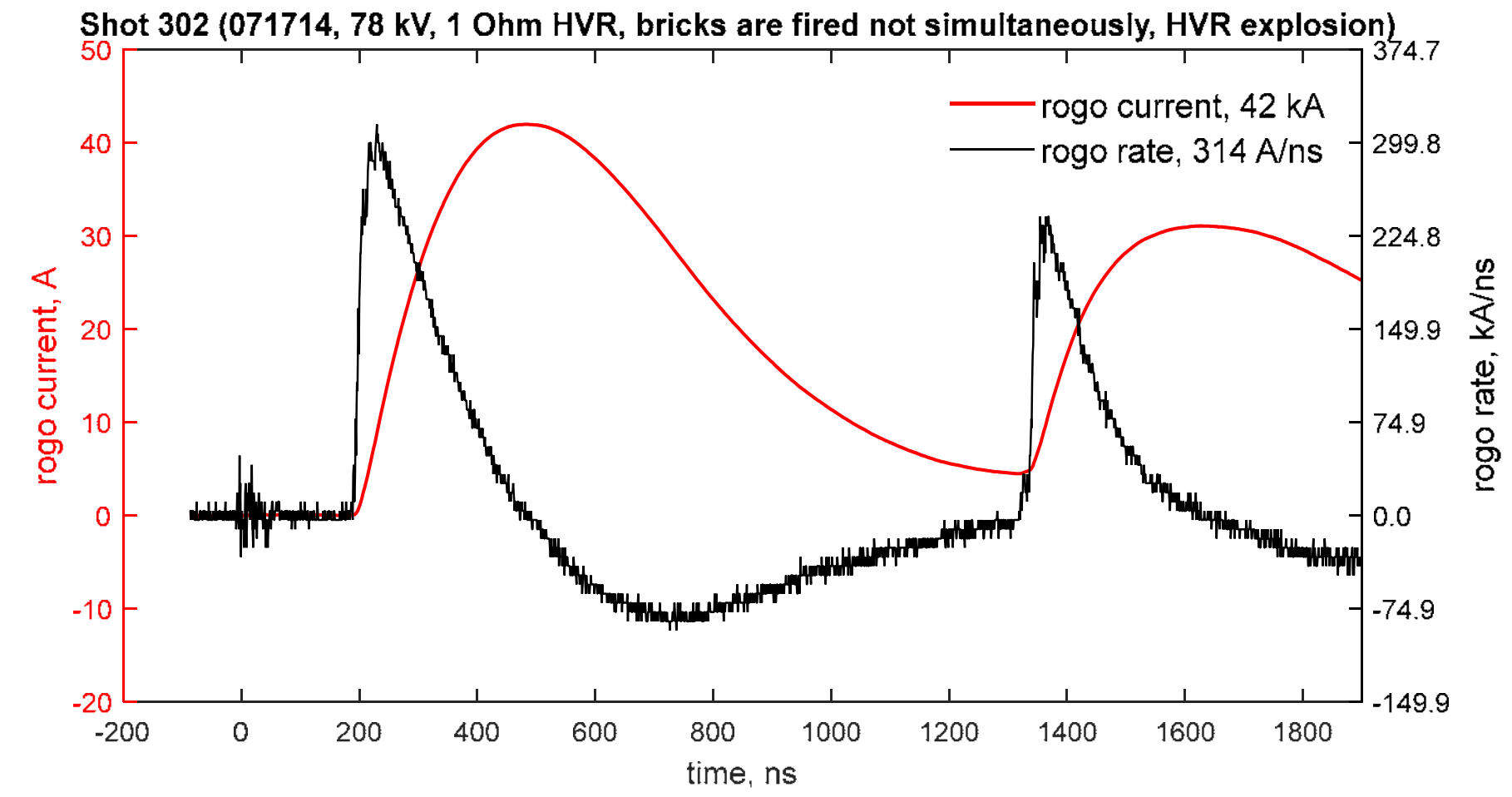

Figure 4.1-3. Shot № 202 (top) and shot № 302 (bottom) with different separation times between triggering of the two bricks.

All driver triggering issues were usually related to one or more sparks observed between the end of the trigger-line high-voltage cable and a vicinity of other driver elements, which resulted in earlier firing of the corresponding brick. The technical issues described above were usually resolved by pushing the trigger-line high-voltage cable farther and farther away from the brick's switch (as in test №4), or from the bricks housing (as in tests №8-9), or by increasing the trigger-line output housing holes (test №10).

In our earlier tests, one PT-55 is used to simultaneously trigger both driver's brick. Later, many other trigger set-ups were tested and compared: one PT-55 versus two PT-55's, positive versus negative PT-55, PT-003/006 versus HV-1000 pre-pulser, and more. Figure 4.1-4 shows our latest trigger box design with HV-100 pre-pulser and two negative PT-55 (left picture), and through-housing trigger-line design coming at the opposite side of driver's housing out (right picture).

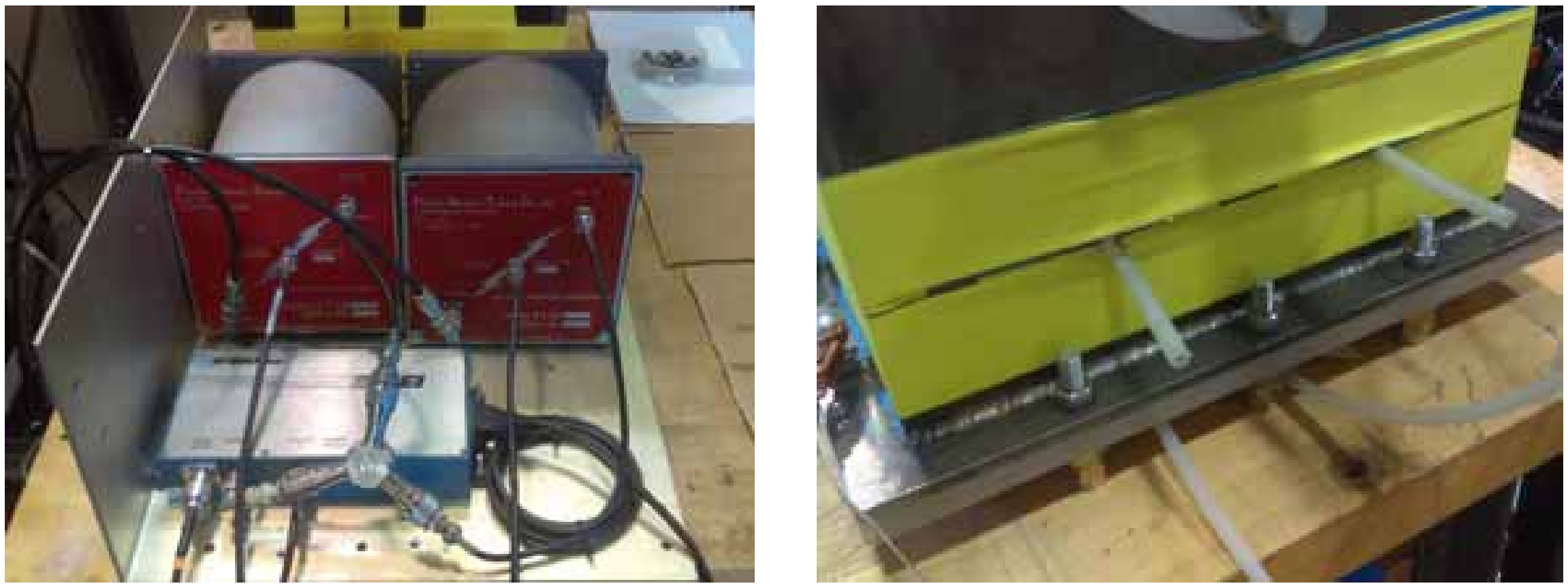

Figure 4.1-4. An *x*-pinch driver trigger system: trigger box (left); end of the through-housing trigger-line (right).

## 4.1.3 Driver brick failures and repairs

After extensive driver testing performed during tests № 1-7, a series of driver issues started to develop: bricks started to fire non-simultaneously and even started to self-fire at a low charging voltage again as it was observed in our earlier tests №1-3. The driver was taken apart, and some severe cracks were observed in the brick epoxy in the vicinity of the brick's-charging line-switch location. Figure 4.1-5 shows examples of such cracks observed outside and inside of the brick's epoxy.

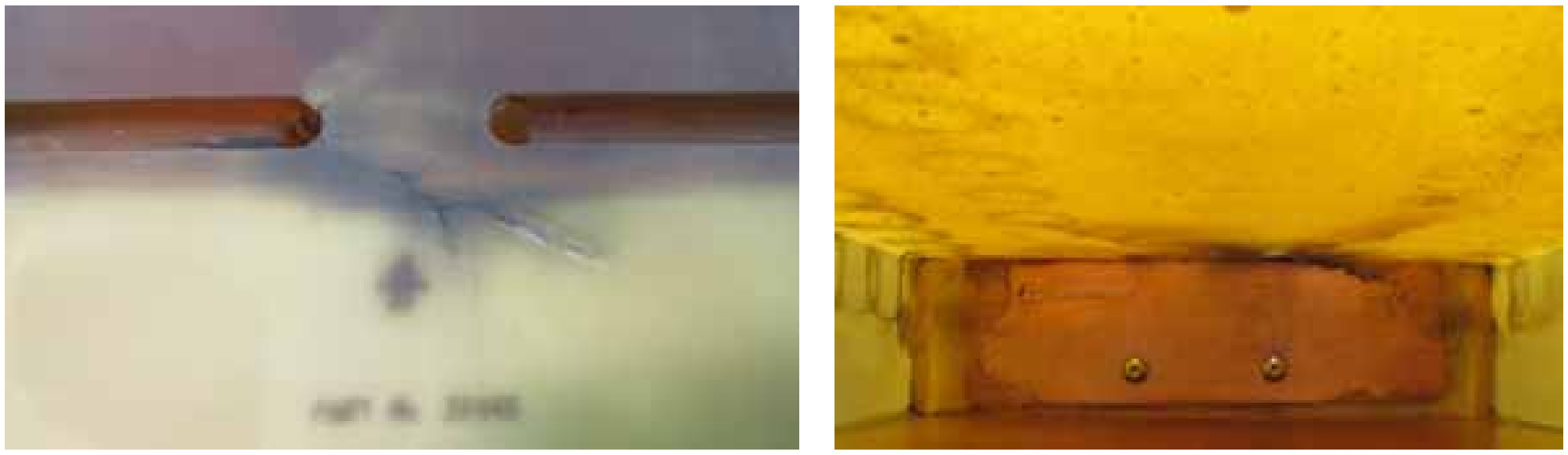

Figure 4.1-5. Cracks developed in a brick's epoxy: left is outside and right is inside.

All epoxy cracks were carefully cut out and refilling back with 3M Scotch-Cast electrical epoxy. Figure 4.1-6 shows brick before and after they were repaired.

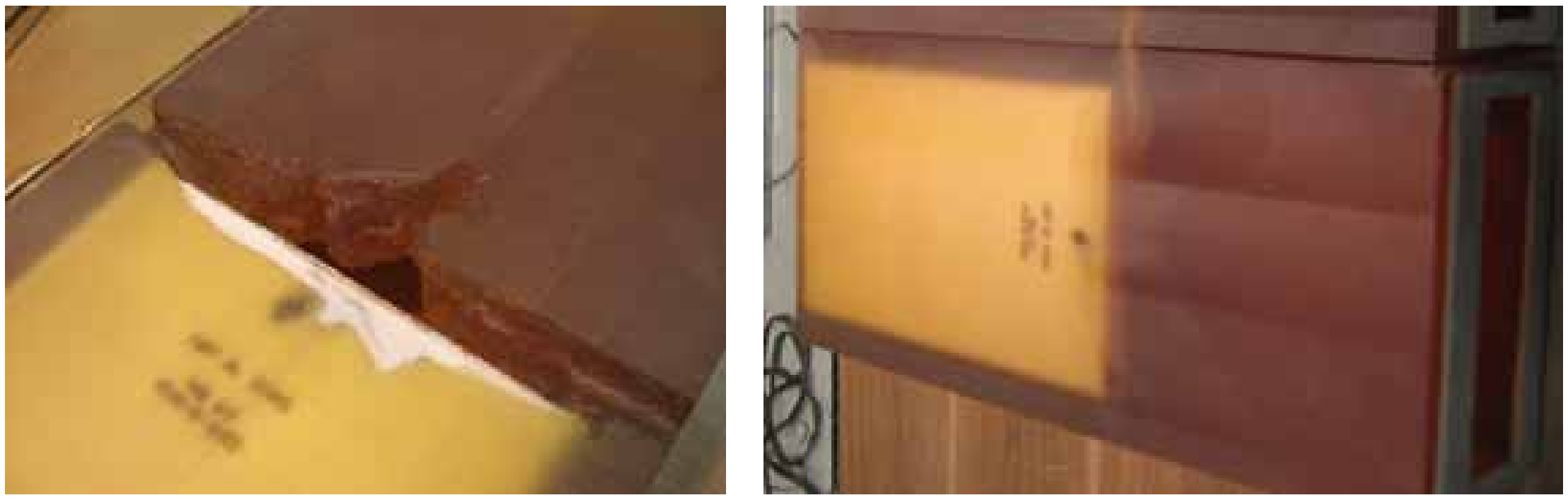

Figure 4.1-6. Bricks before (left) and after (right) repair.

After bricks were repaired, the driver was reassembled and tested again in a series of shots of test № 9. Unfortunately**, the driver failed the second time** during this testing period. It was taken apart and inspected again. After all epoxy cracks were removed, a further series of cracks was found in one of the brick capacitors as shown in Figure 4.1-7.

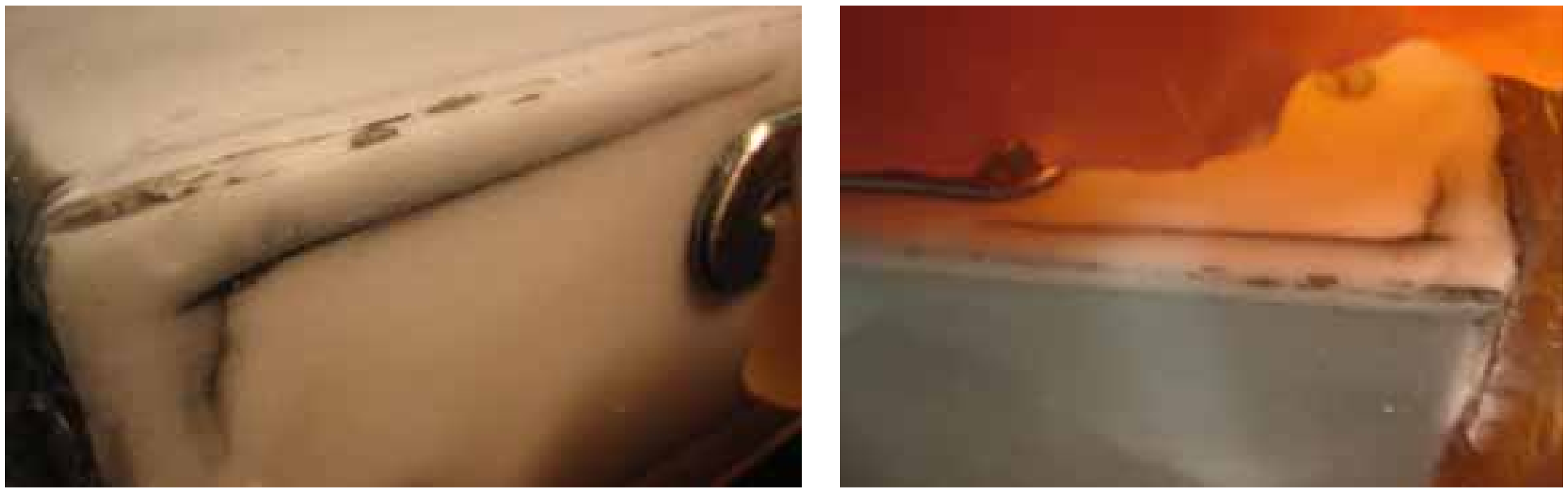

Figure 4.1-7. Cracks in a brick's capacitor: front crack (left), top crack (right).

The capacitor cracks were sealed under heat, the bricks were repaired back with epoxy, and the driver was tested again (test № 10). All shots were good; the driver operation was stable; and no more driver failures were observed after the second, brick repairs. Later, the whole driver was moved without being disassembled to a new location (see Figure 4.1-8). Afterwards, the driver performance was reproducible, reliable, and stable.

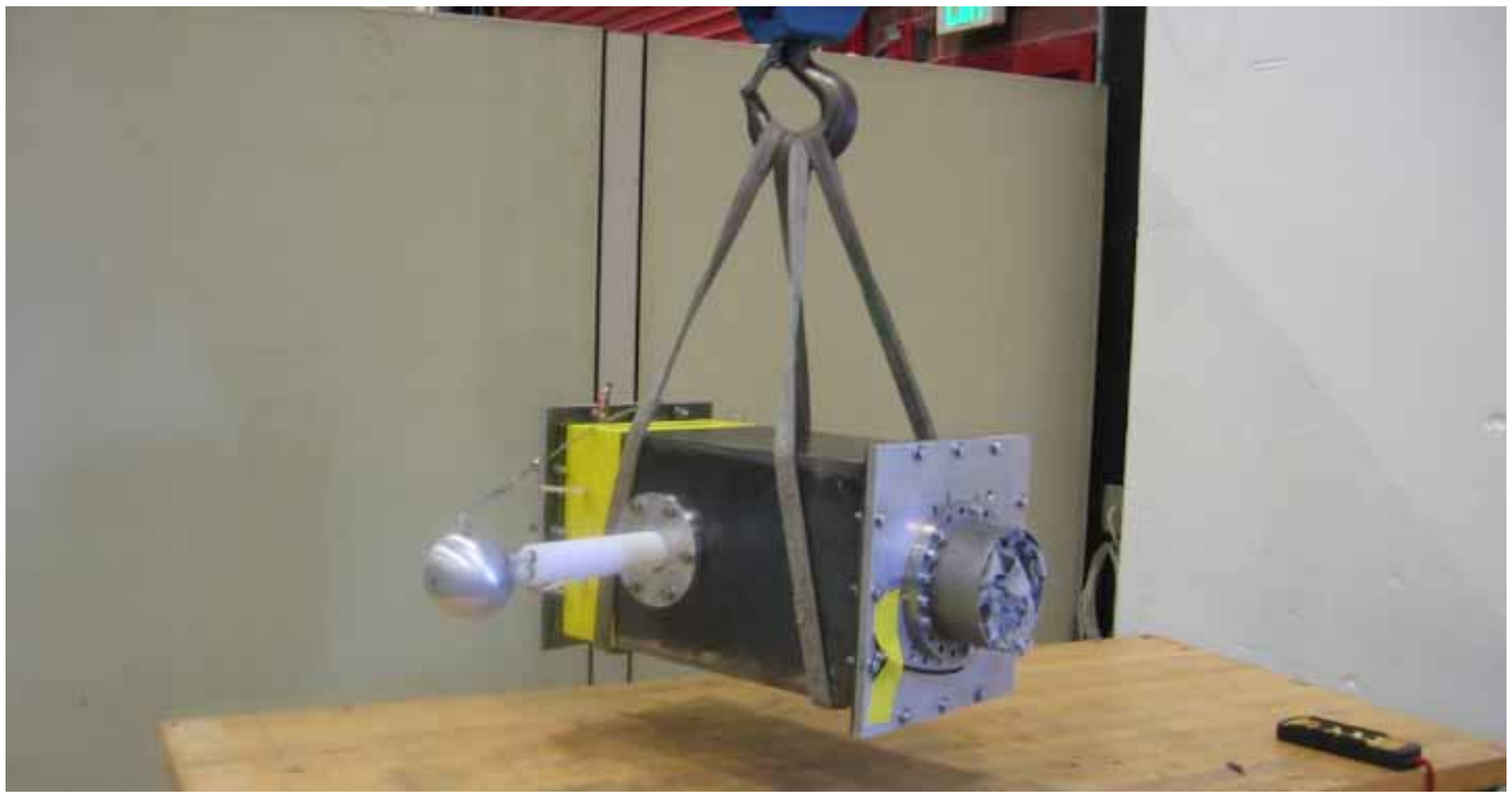

Figure 4.1-8. 2-LTD-Driver's movement to a new location.

## 4.2 2-LTD-Brick driver short-circuit test performance

Tests № 1-4 all were an initial, preliminary testing, designed to eliminate common problems associated with the fabricated 2-LTD *x*-pinch driver. Many issues, such as a multiple arcing inside the driver, pre-fires, and non-synchronized brick triggering, were addressed and resolved. Several configurations of charging and trigger lines were tested and the best for our driver were chosen. After a proper driver set-up was established, a series of tests were performed to estimate the main *x*-pinch driver electrical characteristics as discussed in the following sections. Test 6 was designed to measure the total, internal driver inductance and to estimate the driver resistance. Test 7 was performed to measure the driver behavior at different initial charge voltages.

### 4.2.1 Measurements of driver total internal inductance

Results of test № 6 are discussed below. A 1.48-cm-long, 2.9-mm-diameter Ni-wire load was installed in the anode-cathode gap, and immerged into a dielectric oil. The bricks were initially charged to 80 kV. The Rogowski coil "raw" signal and integrated current for short № 3 are presented in Figure 4.2-1. The maximum current rate-of-rise, achieved in this short, was about 0.84 kA/ns, the Rogowski peak current was about 185 kA, and the rise time, 10-90%, was about 220-ns.

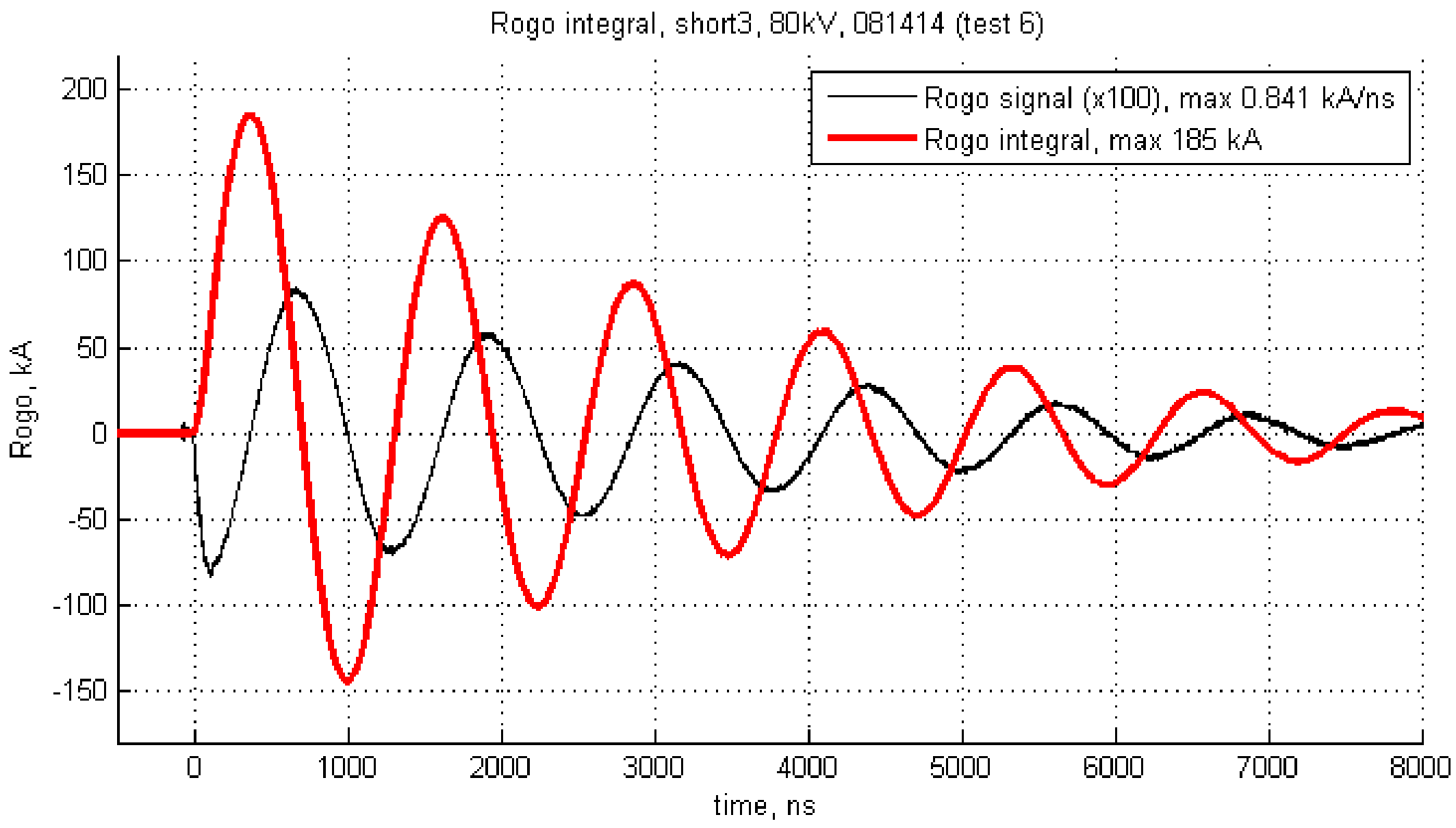

Figure 4.2-1. 2-LTD-brick driver test data with wire load, 08/14/2014, short 3, 80 kV; black line – Rogowski output current, kA/ns, red line – Rogowski current integral, kA.

The total current, measured in this shot, was compared with LTSpice [71] simulations, and results are presented in Figure 4.2-2. First, in simulations, the driver inductance was varied until a good agreement with measured oscillation time was achieved. The inductance found in simulation, which matches the observed oscillation time reasonably well, was about 69 nH. Next step was to vary the total driver resistance to match the observed time-decay constant; the resistance was found to be about 42 mΩ. In general, a good match between the measured currents and the simulated currents was achieved, however, simulations were not able to reproduce an initial, 80-kV capacitor charge voltage. Most likely, there were some losses inside the driver, which effectively reduced the initial voltage down to about 71 kV. After 5 μs, the experimental current starts to deviate from simulated one. This can be explained by changes in the switch resistance, which starts to increase at the end of the current pulse.

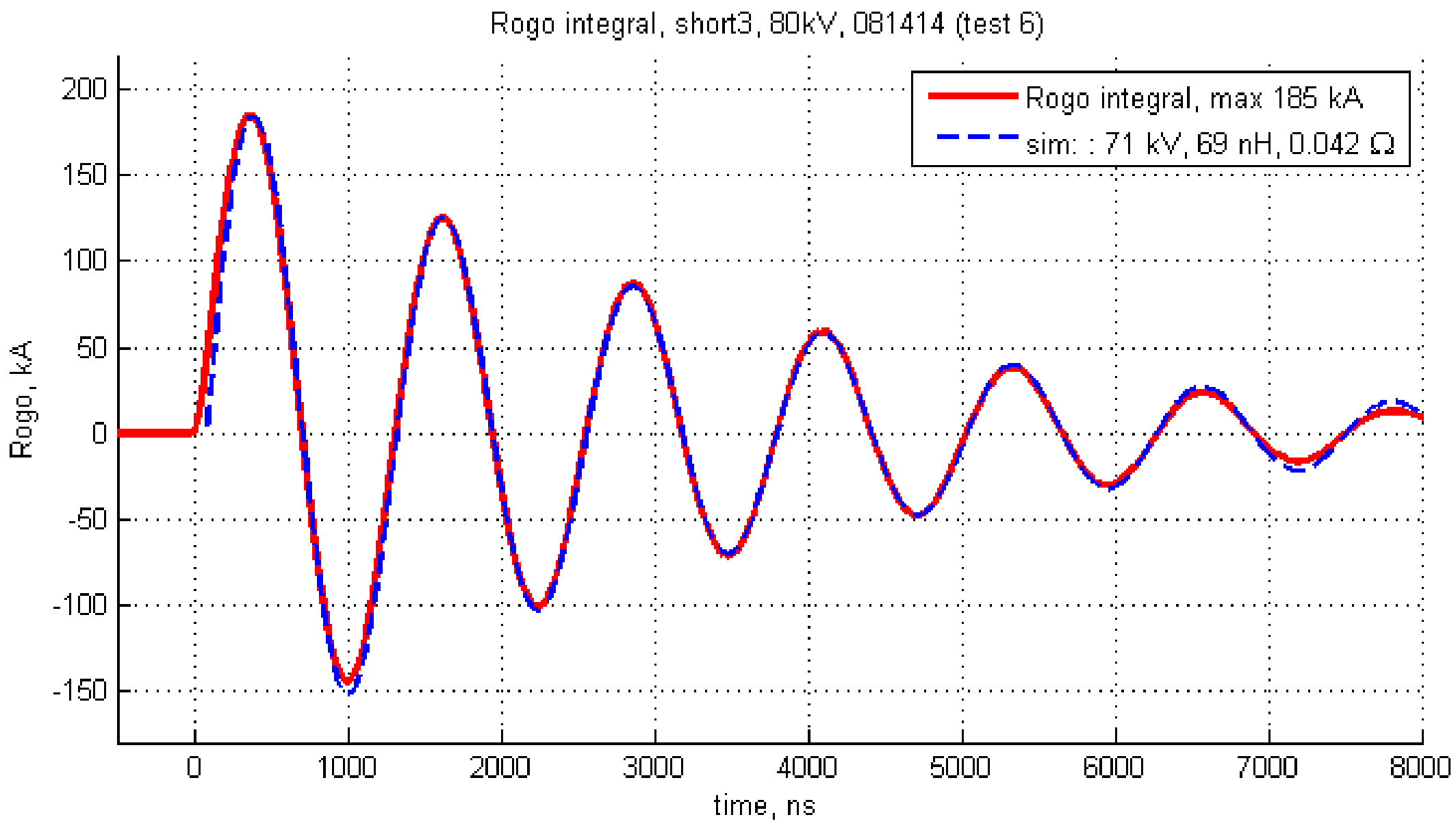

Figure 4.2-2. Experimental current with LTSpice simulation; red line – Rogowski current, 08/14/2014, short 3, 80 kV; blue (dotted) lines – LTSpice simulation.

The total internal inductance of our 2-LTD-brick *x*-pinch driver can be estimated as follows. The inductance value found above, 69 nH, is the total driver inductance, which includes the wire-load value. The wire load inductance in nH can be estimated as:

$$L_{load} = 2h \ln\frac{r_0}{r_1} = 9.12 \text{ nH} \qquad (4.2\text{-}1)$$

where, the wire length, **h**, is 1.48 cm, the wire radius, $\mathbf{r_1}$, is 1.45 mm, and the inner anode radius, $\mathbf{r_0}$, is 31.75 mm. So, the total internal driver inductance should be equal to:

$$L_{driver} = L_{measured} - L_{load} = (69 - 9.12)\text{nH} \approx 60 \text{ nH} \qquad (4.2\text{-}2)$$

## 4.2.2 Driver behavior at different charging voltages

A series of short-circuit shots, test № 7, were performed with a 2.54-cm-long, 5.08-cm-diameter brass cylinder placed between load-section electrodes and immersed in dielectric oil. The length of the cathode section was reduced to accommodate the load length. The non-integrated, electrostatically shielded Rogowski coil was placed inside the output plate and used to monitor the total load current. A total of 9 shots were performed for each 60/70/80 kV charging voltage and results are shown in Figure 4.2-3.

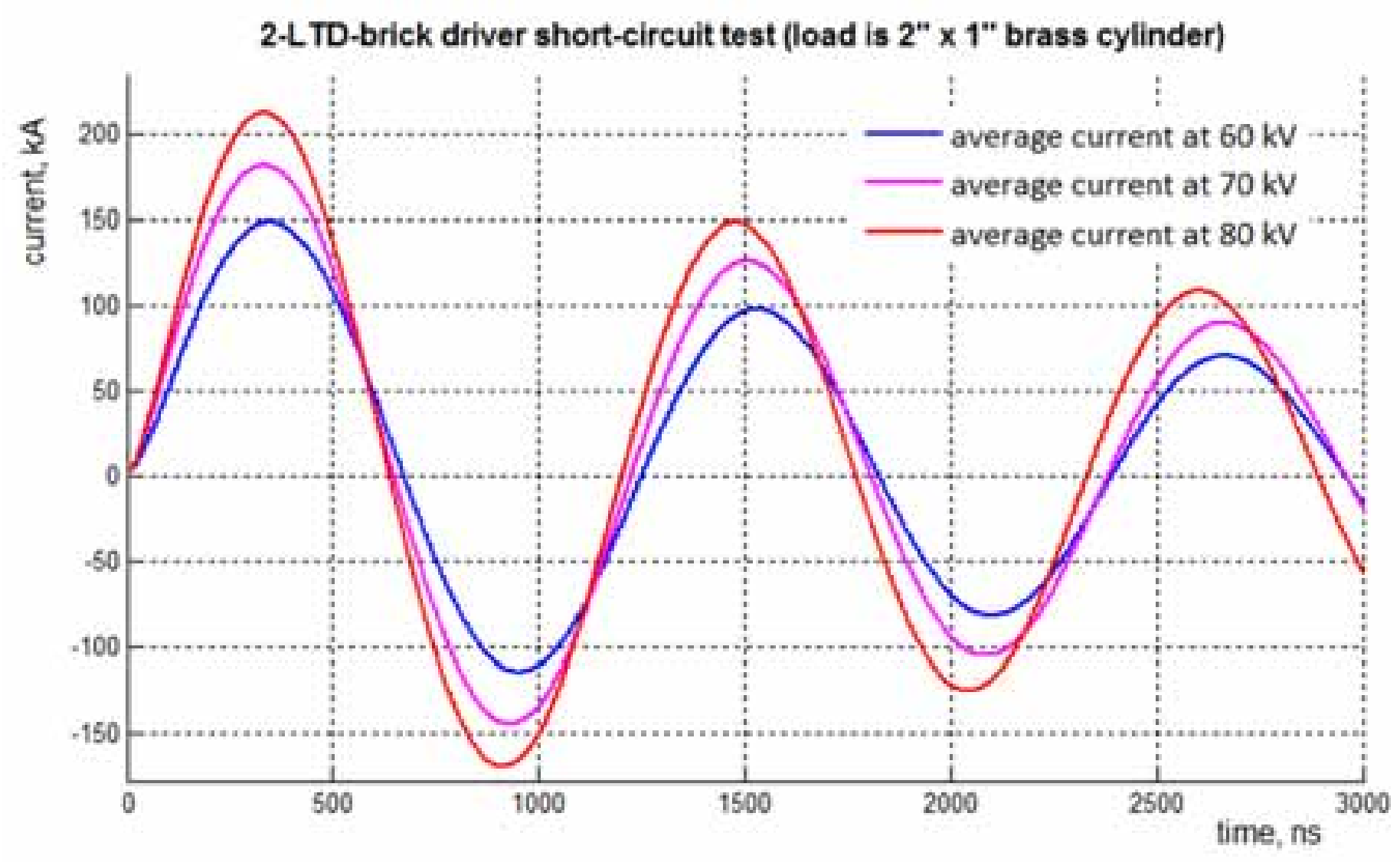

Figure 4.2-3. Short-circuit currents at different charging voltage. The load is a 2"-diameter, 1"-long brass cylinder in dielectric oil.

Table 4.2-1 summarizes the peak-current, rise-time and max current rate-of-rise values. As can be expected, the peak-current values increased, see (2.2-15), and the rise-time values decreased, see (2.2-14), as we increase the driver initial charge voltage. The maximum peak current is about 212 kA for 80-kV case with corresponding maximum dI/dt of 1.1 kA/ns.

Table 4.2-1. Short-circuit-test data with a 2"-diameter, 1"-long brass cylinder.

| **total number of shorts** | **date** | **charge voltage, kV** | **peak current, kA** | **rise time, 10-90%, ns** | **max dI/dt, kA/ns** |
|---|---|---|---|---|---|
| 9 | Nov 26 | 60 | 148 | 215 | 0.7 |
| 9 | Nov 26 | 70 | 181 | 209 | 0.9 |
| 9 | Nov 26 | 80 | 212 | 207 | 1.1 |

### 4.2.3 2-LTD-Brick 1-kA/ns current rate-of-rise requirement

In summary, the main electrical parameters of our 2-LTD-Brick *x*-pinch driver (the total driver internal inductance and resistance, test № 6) were evaluated and the driver performance at different initial charging voltages (peak-current, rise-time, and current rate-of-rise, test № 7) was characterized.

As was discussed in the introduction, the most important driver parameter, crucial for all future *x*-pinch driver performances, is the current rate-of-rise. The measured dI/dt, 1.1 kA/ns, presented in Table 4.2-1 satisfies the minimum 1-kA/ns requirement that was discussed in section 1.4. If the driver would operate at higher voltages, the current rate-of-rise would only increase, scaling proportionally to the initial charging voltage.

From the data presented in this section, we conclude that one of the main objectives of this work, which was to build and test a new compact and portable high-current pulse generator with 1 kA/ns current rate of rise requirement, was successfully met. The last section deals with testing a variety of *x*-pinch wire loads and evaluating their *x*-ray radiation performance.

# 5 2-LTD-BRICK DRIVER RADIATION PERFORMANCE

## 5.1 Experimental set-up and *x*-pinch radiation diagnostics

The typical diagnostics arrangement used in our *x*-pinch shots is schematically shown in Figure 5.1-1 and discussed below.

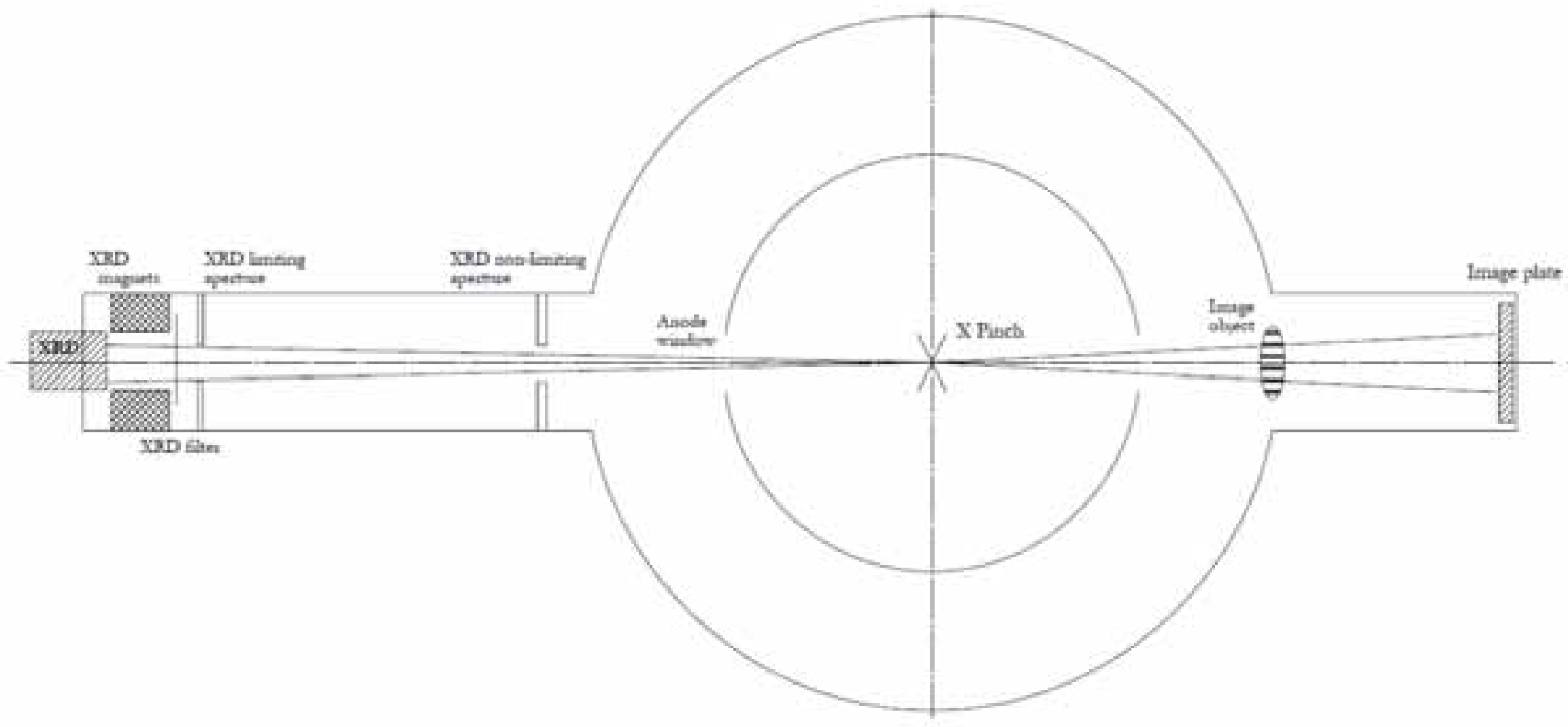

Figure 5.1-1. *X*-pinch diagnostics in a typical *x*-pinch shot.

**X Pinch:** The *x* pinch was installed at the center of the vacuum chamber between the anode and cathode gap having a separation distance of about 1.34 cm and a wire-crossing angle of about 70 º. The wires were first installed in parallel, and then were twisted 180º or slightly more to insure a proper "just-touching" *x*-pinch geometry. The 2-LTD-Brick driver was initially charged to 80 kV, the vacuum level in vacuum chamber was below $10^{-5}$ Torr and the brick-switch air pressure was just above 1 atm.

**X-Ray Diode:** The filtered vacuum *x*-ray diodes (XRD), described elsewhere [72] [73], were used as the primary diagnostics of *x*-pinch radiation and were mounted at the output flange at the end of the vacuum chamber's output port. The source-detector distance was varied from 20 cm up to 106 cm depending on the magnitude of the radiation signal. The XRD is composed of a polished vitreous-carbon photocathode and a highly transparent anode mesh in front of the photocathode to collect emitted photoelectrons. The photocathode-mesh separation was about 0.52 mm, and all parts were mounted inside an N-connector housing.

**X-Ray filter and aperture:** *X*-ray filters were placed in front of the XRD to protect the detector mesh and photocathode from *x*-pinch debris and to select different energy bands of measured *x*-ray radiation. The XRD filters were installed just before the XRD, for shots with ID 100-500, and about 8 cm away from XRD with a magnet installed between them, for shots with ID 500 and more. The XRD limiting aperture was always installed in-front of the XRD anode mesh. The inner diameter of the XRD aperture was 0.98 cm in all our *x*-pinch shots and used to calculate a solid angle for *x*-ray measurements.

**XRD special housing:** To accommodate all XRD, magnets, filter and limiting aperture inside the output line of vacuum chamber, a special housing was designed as presented in Appendix H and pictured in Figure 5.1-2 below. The filter was installed at one end of the housing and the XRD was placed at the opposite end. The Neodymium magnet was installed between them and served to cleanup any possible high-velocity electrons before they could reach the XRD mesh-photocathode volume.

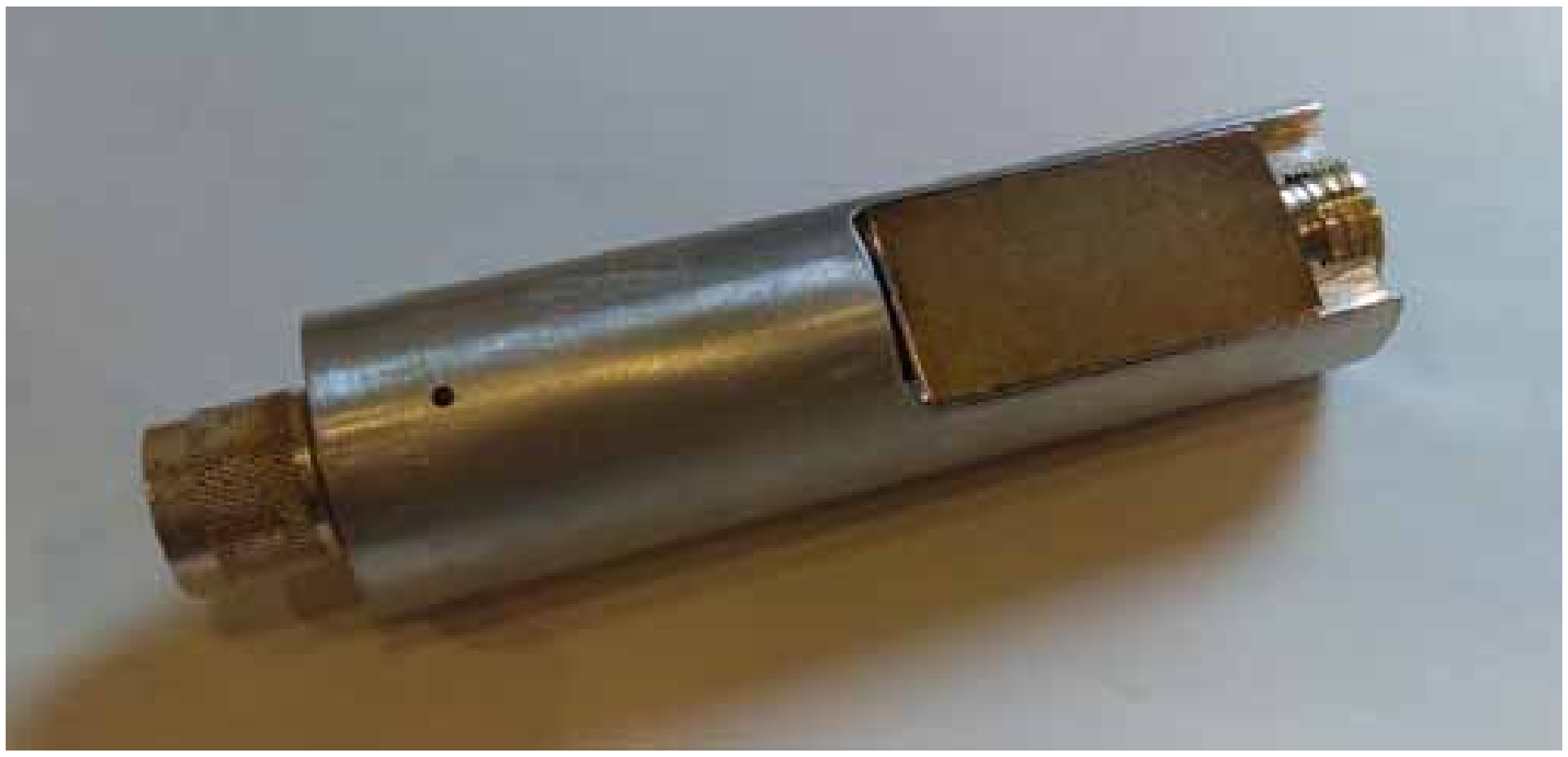

Figure 5.1-2. XRD special housing general view. .

**Screen box:** To protect our detection and recording system from electromagnetic noise, a small, 60-cm x 60-cm x 60-cm screen box was built as pictured in Figure 5.1-3. The screen box is double-shielded Faraday cage that completely isolates electromagnetic noise from our detection system during the *x*-pinch shots. The Tektronix 4-channel, 1-GHz, 5-GS/s oscilloscope, the XRD bias box, the XRD bias power supply, and the external battery were placed inside a screen box in the vicinity of the 2-LTD-Brick driver. All diagnostics (Rogowskii, B-Dot, and XRD) were properly connected to our screen box with a 10-feet-long wire-braid-shielded RG-223/U cables. The performance of our detection system before and after screen box implementation will be discussed later in this chapter.

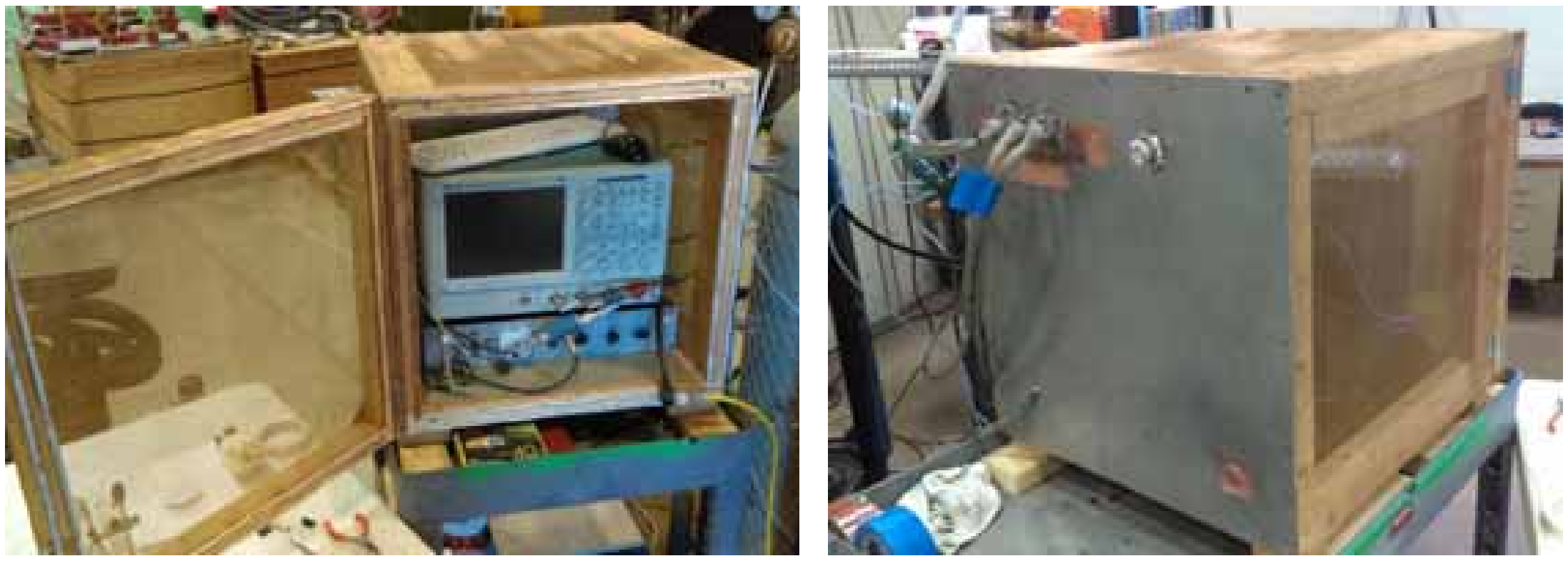

Figure 5.1-3. Double-shielded screen box. Front and back view.

**Data analysis code:** All experimental data (Rogowskii, B-Dot and XRD scope channel signals) were processed with a custom Matlab data analysis code, presented, for reference, in Appendix M. The code allows us to analyze scope data from different channels, subtract background noise, normalize, re-normalize and integrate signals, make document-ready figures and print some shot statistics on the screen.

**Current diagnostics:** In all *x*-pinch shots, the output driver current was monitored with the Rogowskii coil as described in Section 3.5.1. Starting from shot with ID 500, a B-Dot current monitor was installed in a load section and was used as a second current monitor to ensure that all driver energy was properly delivered to an *x*-pinch load in a reasonable time as was described in Section 3.5.2. If the B-Dot signal starts to deviate from the Rogowskii coil signal at some moment of time earlier than when the *x* pinch is formed, the driver's output acrylic insulator plate at vacuum-plate interface has to be cleaned. In general, the 2-LTD Driver does not require any maintenance, but after some number of shots a small deposit of wire material is seen on the acrylic vacuum insulator that can prevent the total driver energy from reaching the *x*-pinch load. The minimum required driver maintenance is described in Appendix N.

All *x*-pinch wire materials, geometries, and XRD filters, used in our *x*-pinch experiments, are listed in Table 5.1-1. The shot ID number is a unique *x*-pinch shot identifier composed from three digits: the first digit represents a set number, and the last two are a shot number for each particular set. The common features of some of our *x* pinches are discussed in a following section and a complete list of all shots performed with our 2-LTD-Brick driver is presented in Appendix O.

Table 5.1-1. Wire materials and XRD filters tested in *x*-pinch shots.

| ***x* pinch** | **XRD filters** | **Shots ID number*** |
|---|---|---|
| *80/120-µm Cu wire x pinch* | | |
| 2x80-µm Cu | 127-µm Mylar, 17-µm Al, 12/50-µm Kapton, 2-µm Aluminized Kimfol | 100-111, 131-139, 146-148, 150-152, 154, 159-161, 504, 509-511, 625-627 |
| 4x80-µm Cu | 17-µm Al, 50-µm Kapton | 112-115, 140 |
| 2x127-µm Cu | 12/50-µm Kapton | 142, 156 |
| 4x127-µm Cu | 12/50-µm Kapton, Al 1.5-µm | 116-129, 141, 149, 158 |
| *15-µm W-wire x pinch* | | |
| 2x15-µm W | 12/50-µm Kapton, 1.5/2.5/4/17- µm Al, 25/50/100-µm Fe | 201-258, 505-507, 512-513 |
| 4x15-µm W | 25-µm Fe, 12-µm Kapton | 259, 508 |
| *50-µm Mo wire x pinches* | | |
| 4x50-µm Mo | 25- µm Fe, 12-µm Kapton | 301-301, 501-503 |
| *Nilaco wire x pinches* | | |
| 2x30-µm W | 2-µm Aluminized Kimfol | 601-608, 628-652 |
| 2x20-µm W | | 609-613 |
| 2x30-µm Mo | | 614-618 |
| 2x30-µm Cu | | 619-624, 653-658 |

*Shot ID- first digit is a test number, last two are the shot number.

## 5.2 XRD filtered response

The XRD spectral sensitivity (also known as XRD filtered response) can be evaluated with Sandia *x*-ray radiation detector (XRD) code [74]. The XRD is an *x*-ray detector design code, contains library for most common filters and XRD photocathode materials, and allows for fast calculation of detector-filter package spectral sensitivity.

The XRD bare photocathode quantum efficiency (QE) for the polished carbon photocathode, used in all our experiments, is shown in Figure 5.2-1.

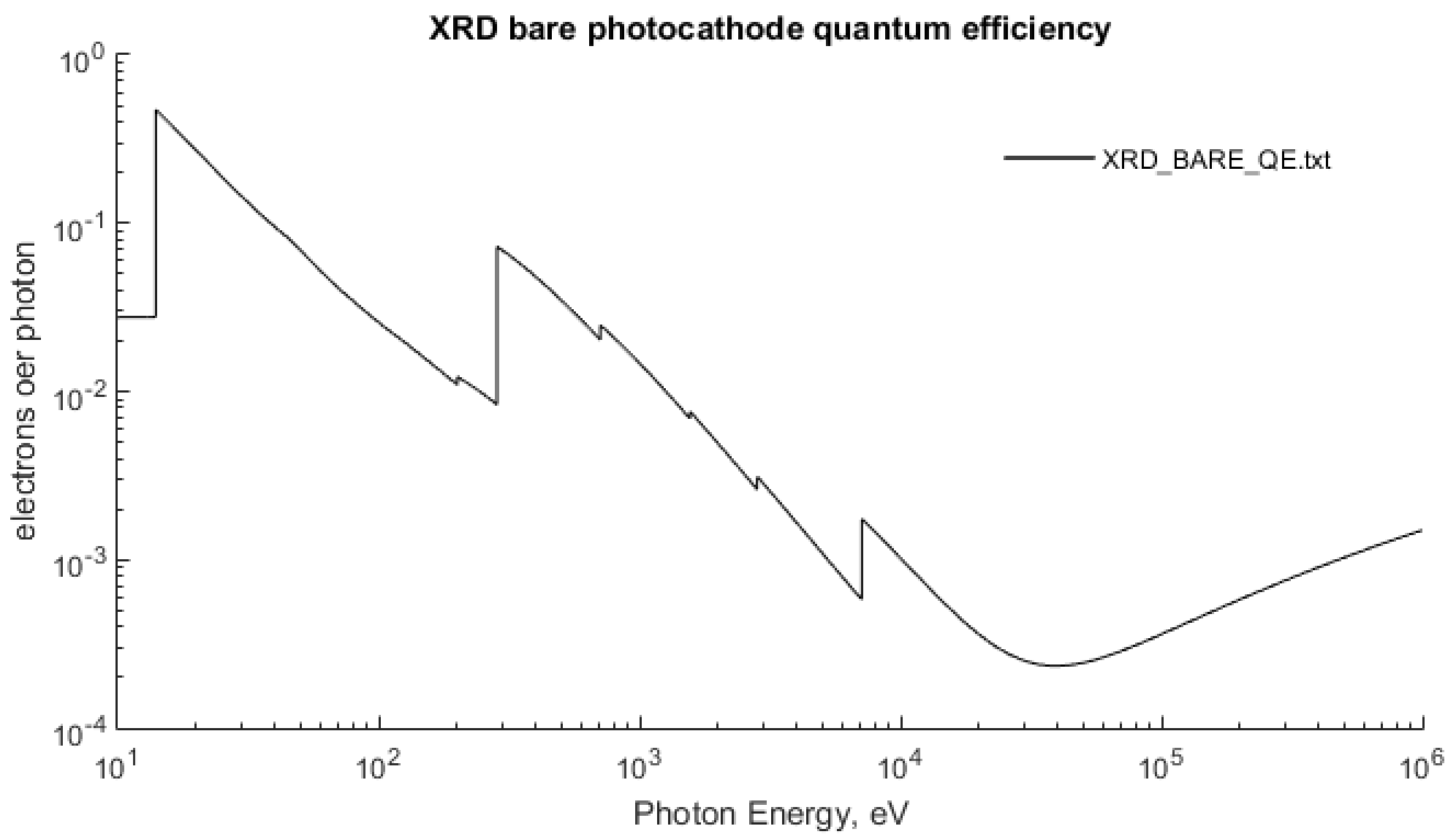

Figure 5.2-1. XRD bare quantum efficiency for polished vitreous-carbon photocathode.

The XRD bare photocathode efficiency is usually expressed in A/MW, and can be converted from quantum efficiency, QE, to detector response, R, as:

$$R\left[\frac{A}{MW}\right] = QE\left[\frac{\#e}{\gamma}\right] \times \frac{1}{E_\gamma[eV]} \times 10^6 \qquad (5.2\text{-}1)$$

The XRD filtered response is simply a product of XRD bare efficiency, R(E), times filter transmission, K(E) :

$$F(E) = R(E) \times K(E) \qquad (5.2\text{-}2)$$

Example of calculations of XRD filter efficiency for 2-μm Kimfol filter is presented in Figure 5.2-2. The red line shows the fractional filter transmission curve, and the black line represents XRD filtered response calculated according to (5.2-2).

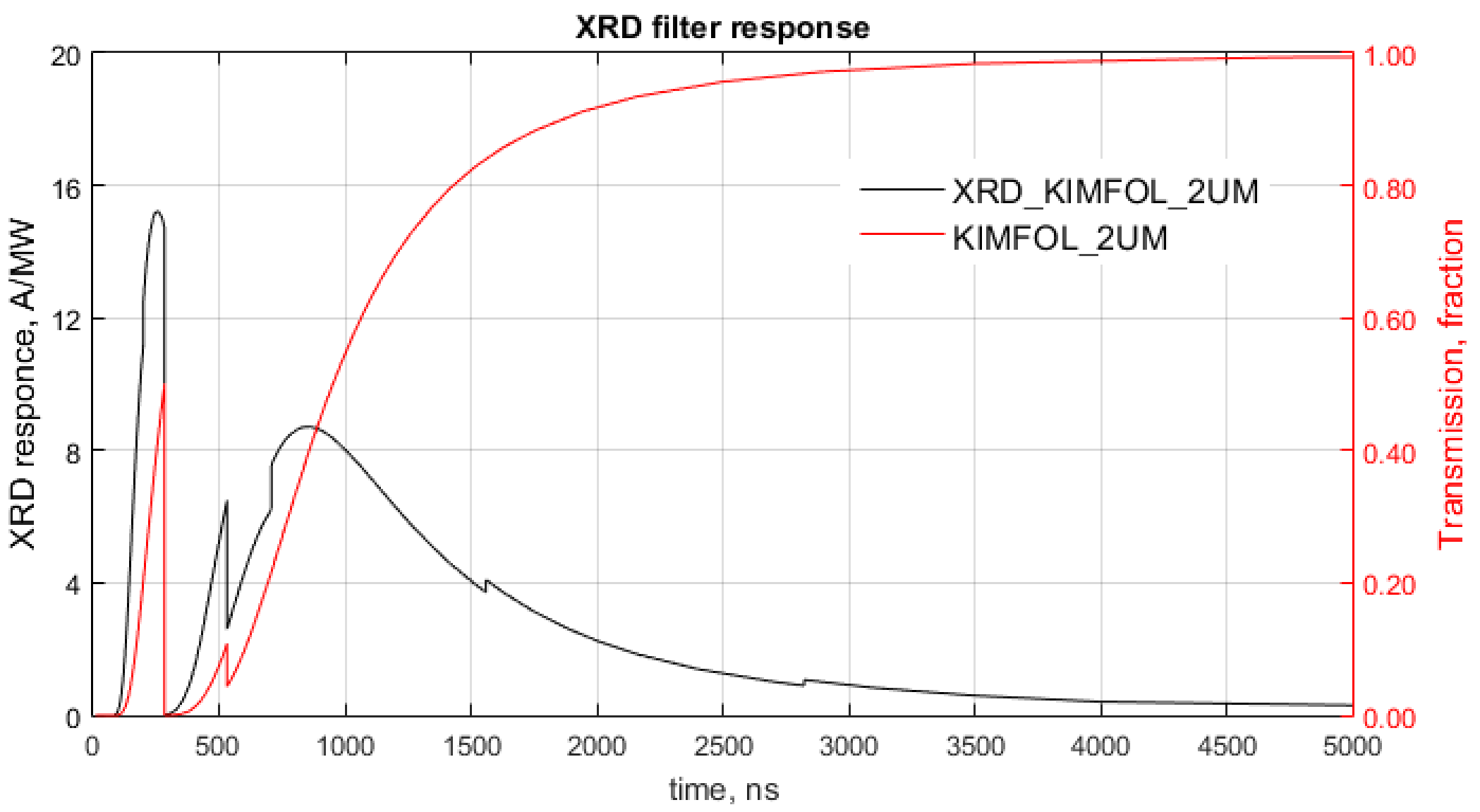

Figure 5.2-2. XRD polished carbon filtered response with 2-μm Kimfol window.

In this case, because the XRD bare efficiency is dropping logarithmically with increasing energy (See Figure 5.2-1), the total XRD filtered response is mostly sensitive to low energies photons, up to 3 keV. When a thicker filter is used, the total XRD response will be only the fractions of A/MW. Table 5.2-1 presents the maximum XRD filtered response for several filter materials used in our *x*-pinch testing experiments.

Table 5.2-1. Maximum XRD filtered response for different filter materials.

| **Filter** | **XRD photocathode** | **Maximum response $F_m$, A/MW** | **Energy at $F_m$, eV** |
|---|---|---|---|
| 2-μm Kimfol | Polished Carbon | 15.2 | 257 |
| 12-μm Kapton | Polished Carbon | 1.36 | 1720 |
| 50-μm Kapton | Polished Carbon | 0.41 | 2930 |
| 2.5-μm Al | Polished Carbon | 6.35 | 960 |
| 4-μm Al | Polished Carbon | 4.03 | 1150 |
| 17-μm Al | Polished Carbon | 0.92 | 1560 |
| 25-μm Fe | Polished Carbon | 0.03 | 7104 |
| 50-μm Fe | Polished Carbon | 0.01 | 7112 |

As can be seen, for example, when 50-μm Fe filter is used, the maximum XRD response is only 0.01 A/MW, which make it difficult to use and analyze XRD data. In general, the thinner filter is used with XRD, the better the results are.

## 5.3 Preliminary *x*-pinch testing

To primarily test our 2-LTD-Brick driver with *x*-pinch load, a series of initial shots were performed with 80/127-μm copper (Cu), and 15-μm tungsten (W) wire *x* pinches. Even if all of those shots did not produce any useable data, it was a good experience to have. First, many improvements in *x*-ray diagnostics were done, and second, it allowed us to understand why these initial wires did not generated any good *x*-pinch data.

**Screen box implementation:** It happened, that almost half of our initial shots (47 to be precise) were performed without a proper electrical shielding of our detector system. Figure 5.3-1 shows data for two typical 2x80-μm Cu *x*-pinch shots performed without (top) and with (bottom) a screen box. The Rogowskii signal (black line) was normalized to kA/ns, and XRD signal (red line) was observed using a 12-μm Kapton filter. As can be seen, the screen box implementation allowed us to significantly improve the XRD signal-to-noise ratio (up to 20 times), so XRD signal structure can be clearly observed now, especially at earlier, up to 300-ns times.

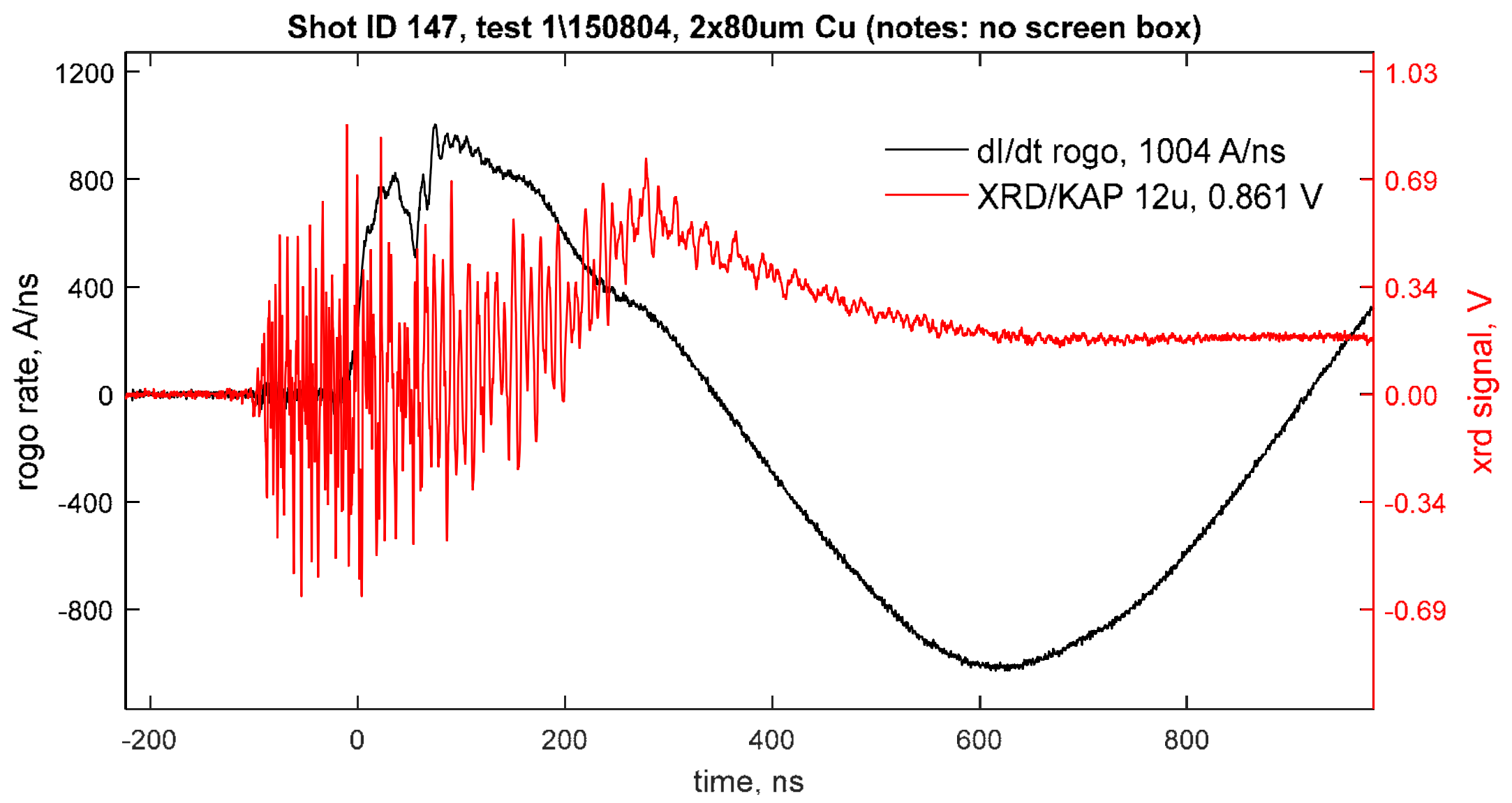

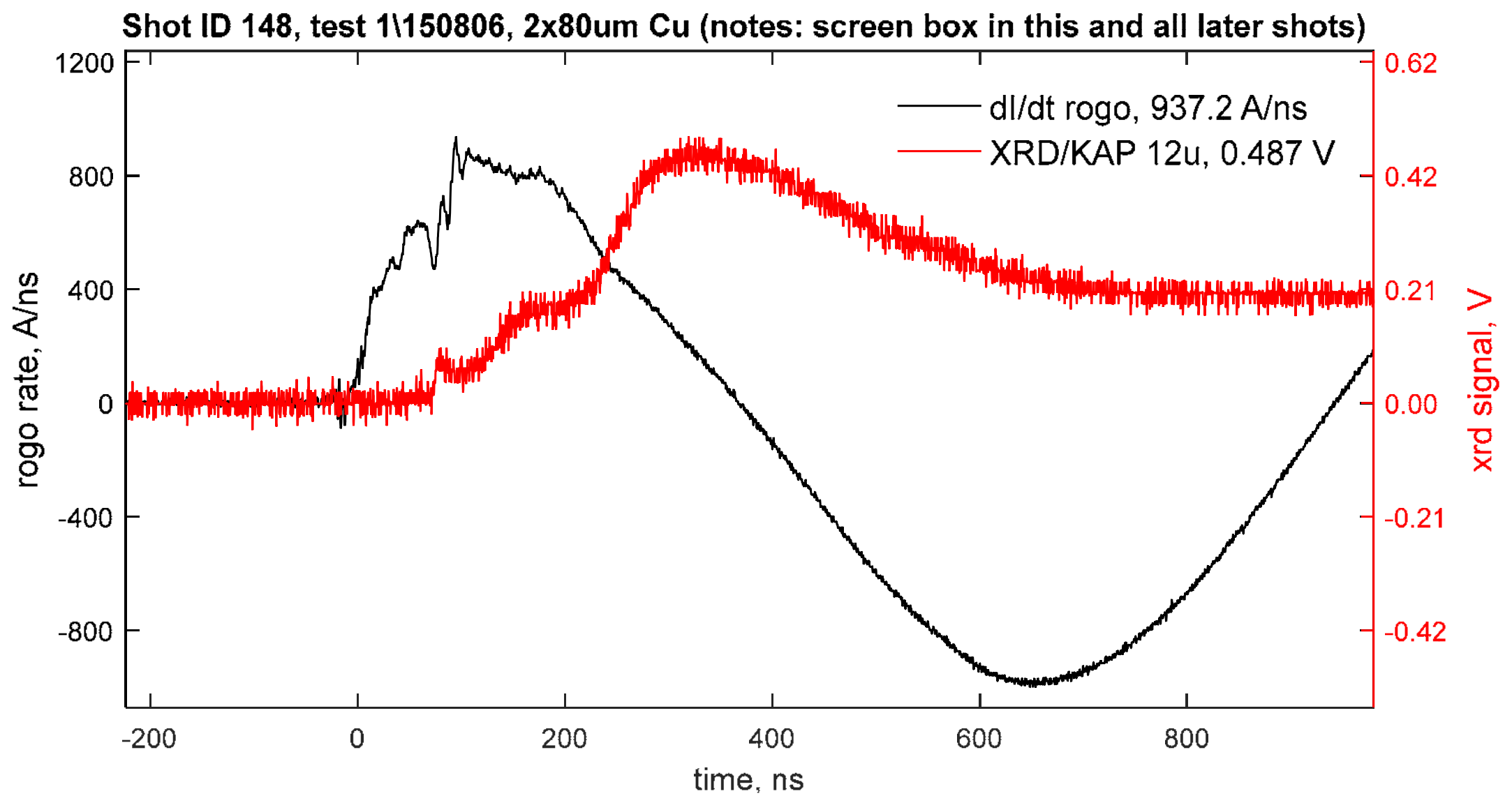

Figure 5.3-1. Screen box implementation: shot ID 147 no screen box (top), shot ID 148 with screen box (bottom). 2x80-µm Cu *x* pinch.

**Shots with 80/127-µm Cu *x*-pinches:** Figure 5.3-2 presents two typical shots performed with 80-µm Cu and 127-µm Cu wire *x* pinches. As can be seen, a characteristic dI/dt dip is formed at about 60-70-ns after the start of the current for 2x80-µm Cu *x* pinch, and at about 160-ns time moment for a shot performed with a 2x127-µm Cu *x* pinch. Those characteristic dI/dt dips are observed in all shots, shifted left or right in time, depending on

*x*-pinch configuration and wire masses, and, most likely, can be associated with a with wire plasma separation, when *x*-pinch diode starts forming, as it was discussed in the introduction section. Regardless, the XRD signals were always small and wide, but not high and narrow, as we expected.

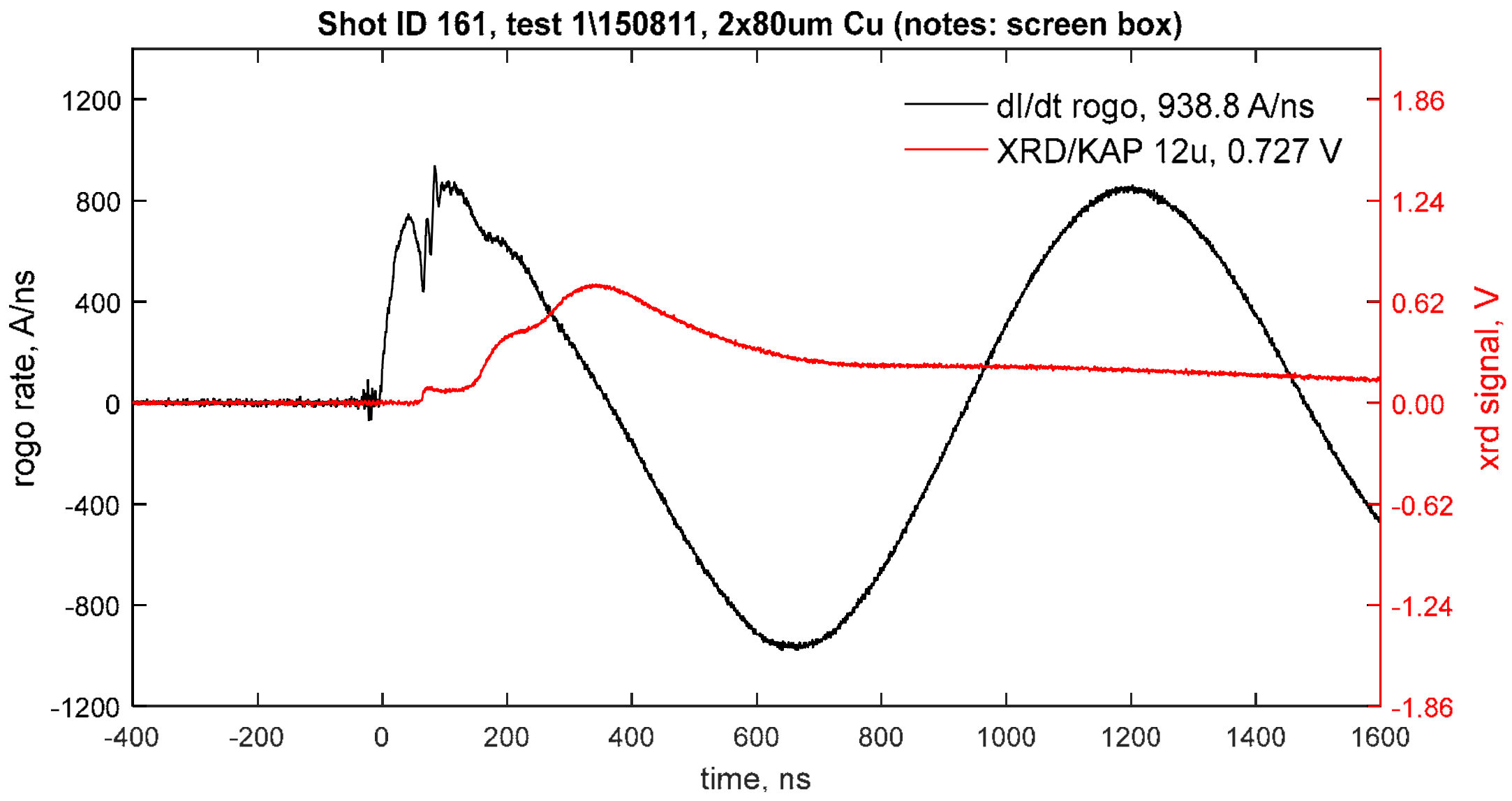

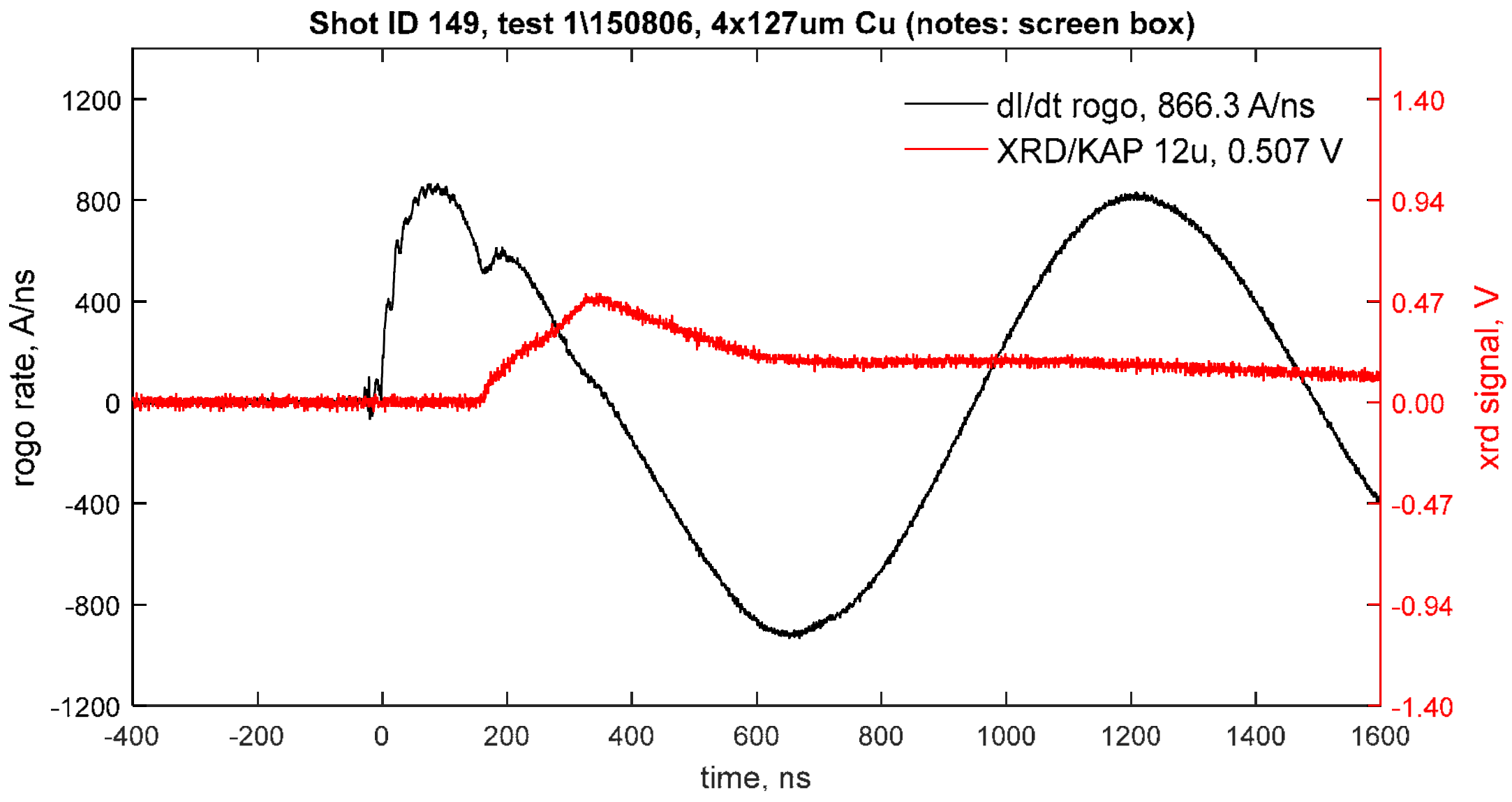

Figure 5.3-2. 2x80-µm (shot ID 161) vs. 2x127-µm (shot ID 149) Cu *x* pinches.

It appears, that 80/127-µm Cu wires are not "good" choices for our 2-LTD-Brick driver in order to generate a "good" *x*-pinch signal. More about how wire materials and its masses affect the *x*-pinch performance will be discussed in a section 5.4 later. **In just one case**, a sharp XRD signal was observed on shot ID 154 with an 80-μm Cu *x* pinch, as presented in Figure 5.3-3. The peak XRD signal was 0.24 V with FWHM of about 2 ns at the moment just before the dI/dt dip. We were not able to repeat this shot in any other trials, and it is not clear why this nice signal was observed.

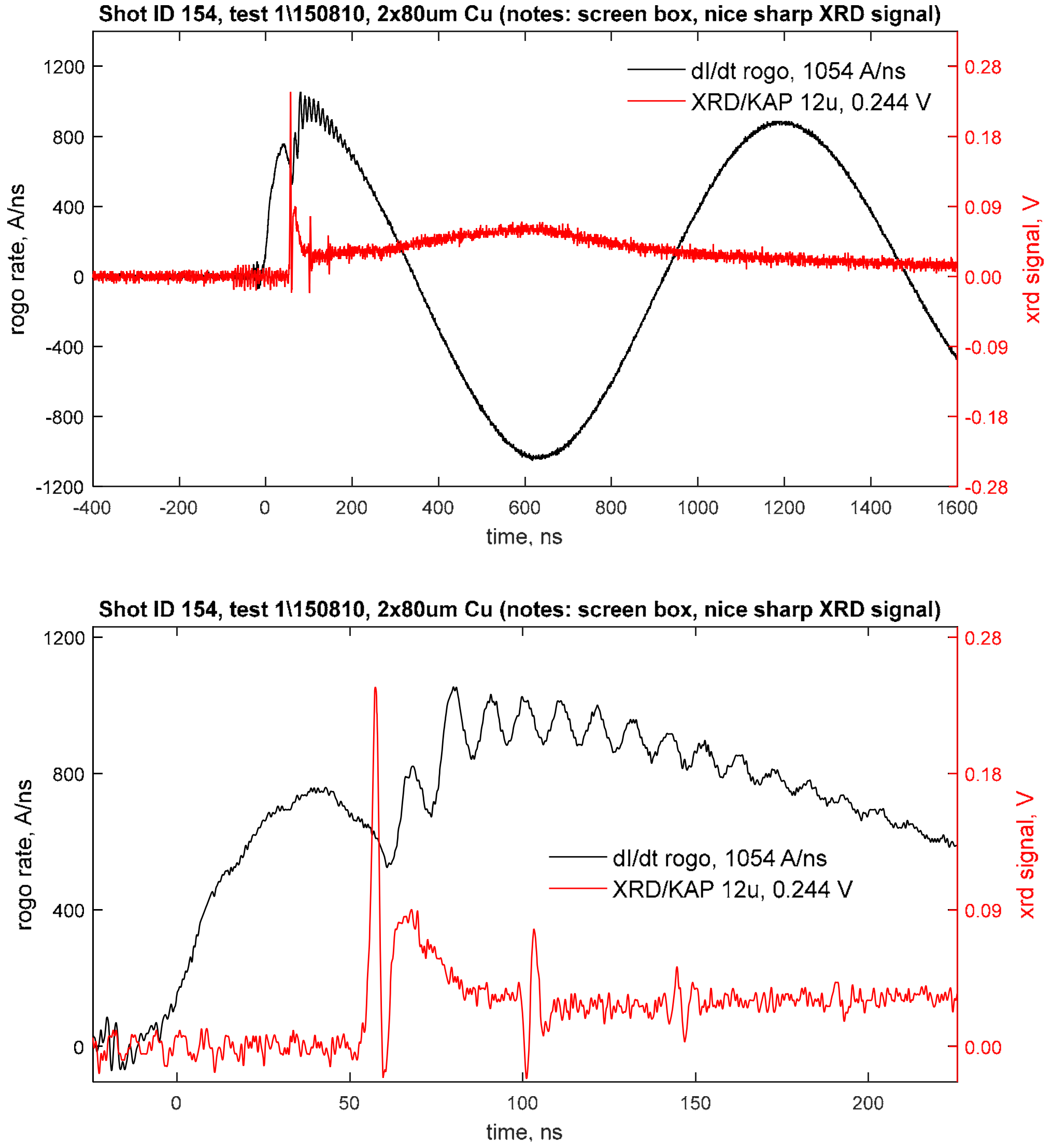

Figure 5.3-3. Shot ID 154 with 80-μm Cu *x* pinches and 2-ns wide XRD signal.

**Shots with 15-µm W *x* pinches:** To farther advance our understanding of *x*-pinches from different wire materials, a series of shots were performed with two and four 15-µm-diameter tungsten-wire *x* pinches, referred as 2x15-µm W or 4x15-µm W, correspondingly, as described below.

Figure 5.3-4 presents typical 2x15-µm W *x*-pinch shots performed with different filter materials. The Rogowskii signal was normalized to kA/ns (left plots) and integrated to a total current (right plots). The XRD signal was observed with 12-µm Kapton ($1^{st}$ row), 2.5-µm Al ($2^{nd}$ row), 4.0-µm Al ($3^{rd}$ row), and 25-µm Fe ($4^{th}$ row) filter windows with source-detector distance 90.2 cm, 105.9 cm, 49.5 cm, and 49.6 cm, correspondingly.

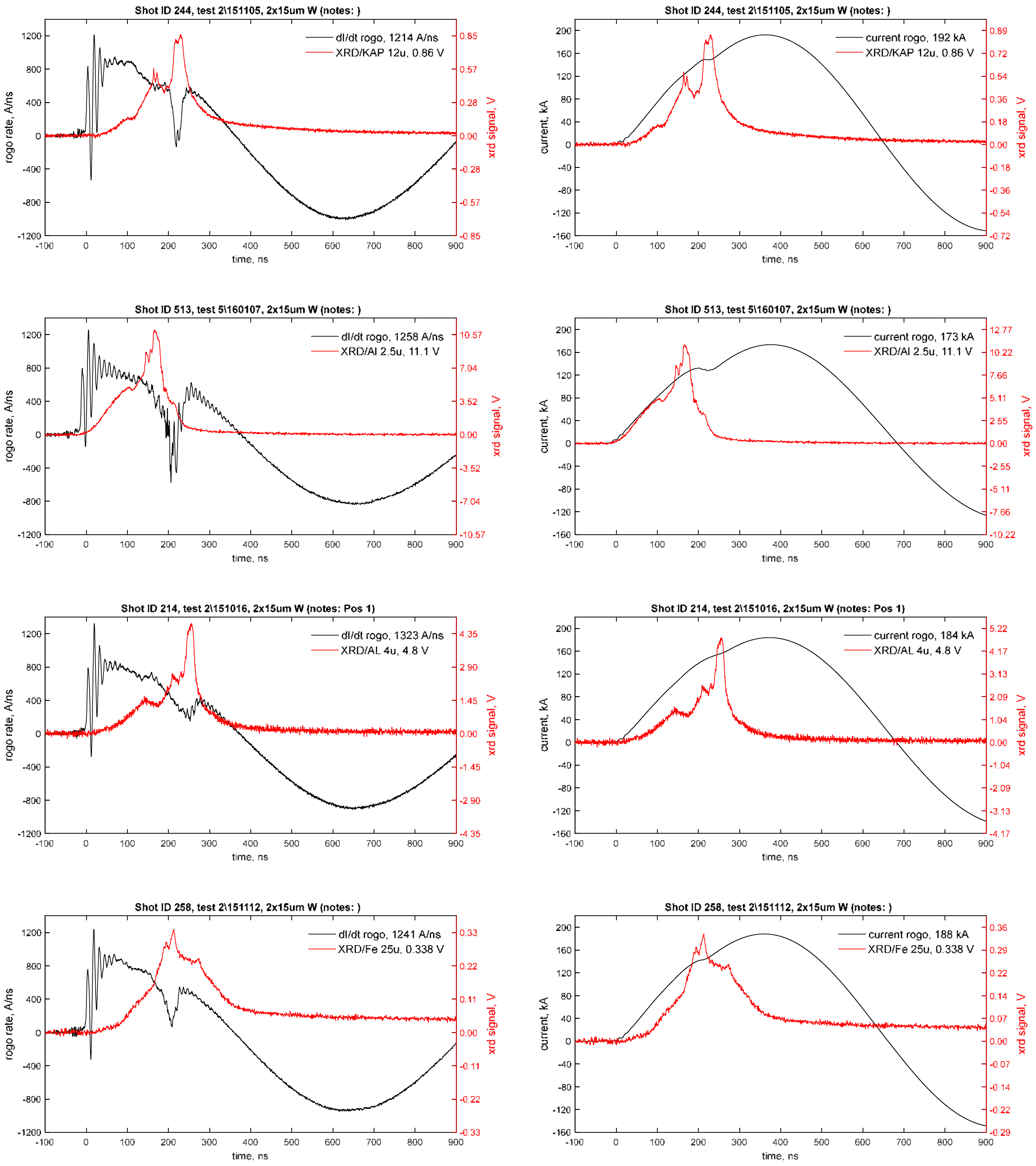

Figure 5.3-4. 2x15-µm W *x* pinches with 12-µm Kapton (1$^{st}$ row), 2.5-µm Al (2$^{nd}$ row), 4-µm Al (3$^{rd}$ row), and 25-µm Fe (4$^{th}$ row) filters.

The XRD signals, observed for all 2x15-µm W *x* pinches, in general, have complicated, not reproducible from shot-to-shot structures. Sometimes there was one, wide XRD peak, sometimes a few, wide peak structures were observed, or more complicated structures were developed at different moments of time.

**Just in a few cases**, narrow, well-separated XRD signals were observed with 2x15-μm W *x* pinches, as presented in Figure 5.3-5 for shot ID 201. The XRD was filtered with 12-μm Kapton window and source-detector distance was about 21 cm. The peak XRD signal was 3.6 V with FWHM of about 2 ns at the moment of about 230 ns just before the dI/dt dip was formed. The corresponding current value at the moment of current dip was about 160 kA. Even though it seems to be a "good" *x* pinch, it was never reproduced again.

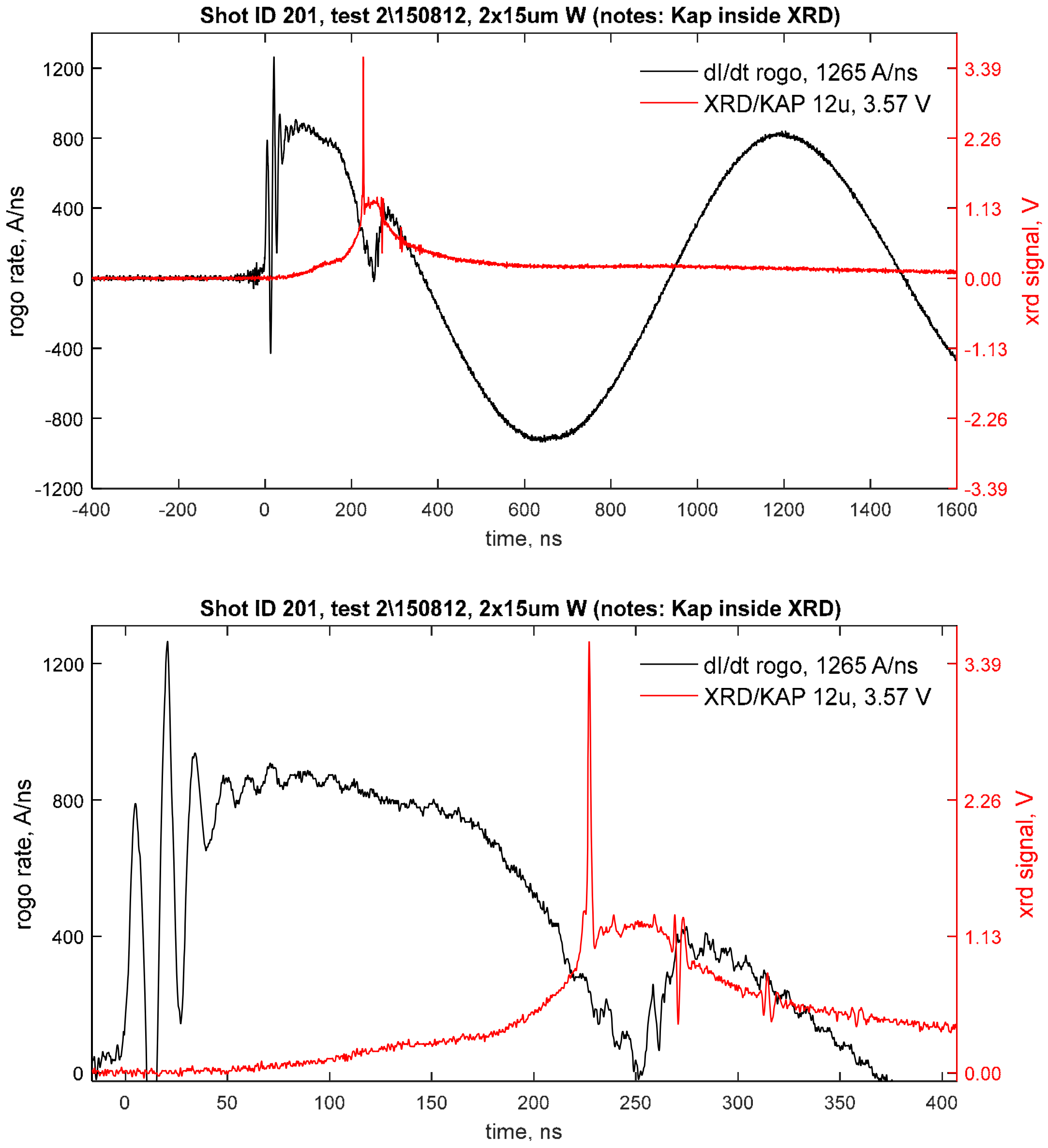

Figure 5.3-5. Shot ID 201 with 2x15-μm W *x* pinch and 2-ns wide XRD signal.

A few shots (ID 259 and ID 508) were also performed with a 4-wire *x*-pinch geometry (referred as 4x15-µm W *x* pinch) and both of them, for reference, are presented in Figure 5.3-6. For a first shot (ID 259), a clear dI/dt dip was observed, while for a second one (ID 508) no such a dip was visible, and XRD signal had multiple, complicated structures. The first shot was performed with 25-µm-thick Fe window in-front of XRD, and the second shot was with 12-µm Kapton window.

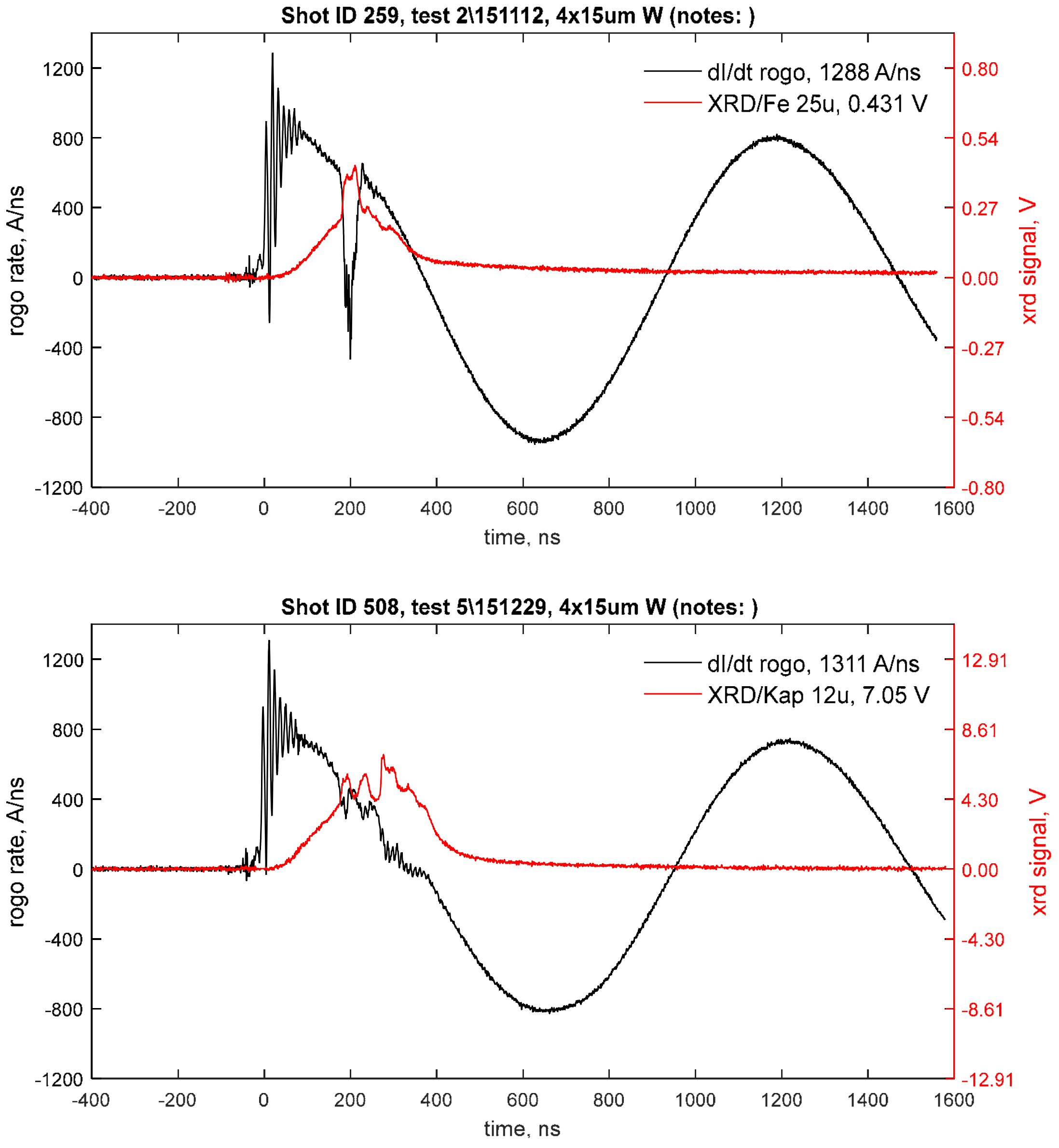

Figure 5.3-6. Shots with 4x15-µm W *x* pinches. ID 259 (left) and ID 508 (right).

As it was already with 80/127-μm Cu wire *x* pinches, the discussed here 15-μm W-wires *x* pinches are not a "good" choice for our 2-LTD-Brick driver as well. In a present case, the wires are too thin, and cannot be successfully pinched into a "hot spot" and to generate a sharp *x*-ray signal. Some shots (both with 80/127-μm W and 15-μm W wires) produced a sharp, about 2-ns wide, well-separate *x*-ray signal, which can be associated with a "good" *x*-pinch radiation performance. However such shots are likely exceptions, rather than a rule, and could not be repeated later. We will come back to this topic in a next section.

## 5.4 Selection of "good" *x* pinches for 2-LTD-Brick driver

It appears, that not all *x* pinches are able to generate a "good" *x*-pinch radiation signal. First, *x*-pinch radiation performance is strongly correlated with the choice of wire materials, wire geometries, and their masses [12] [75]. And, second, even if a proper *x*-pinch configuration has been established for a one particular radiation driver, it might not be a "good" choice for a different driver [41] [48].

Some authors suggest [48] that a simple scaling parameter, which can help to predict a proper *x* pinches for a different drivers, can be used:

$$X = \frac{I^2 t^2 \rho}{M^2} \approx const \tag{5.4-1}$$

Here, I is the driver's peak current, t is the time of *x*-pinch compression, M is linear mass and ρ is a wire density, usually, in the moment of *x*-pinch radiation burst. Because the actual state of the matter at the moment of *x*-pinch collapse is difficult, if not impossible, to evaluate, we will assume in all our next discussion, that the linear mass and density are in an initial "cold" wire state.

The scaling parameter (5.4-1) was analyzed in a way similar to [48] and results of these calculations are presented in Table 5.4-1. The data are grouped in a three different sections: the first part of table shows data taken from [48] with some extensions for other *x*-pinch drivers; the second part evaluates scaling parameters for shots performed with our 2-LTD-Brick driver discussed in previous sections; and a last, third part makes some predictions, which *x*-pinch configurations would be "good" choices for our next shots.

Table 5.4-1. *X*-pinch scaling parameters for different radiation machines.

| Material | Driver | *X* pinches | Linear mass M, μg/cm | Time t, ns | Current I, kA | It/M, kA*ns*cm/μg | X × $10^5$ | Ref/notes |
|---|---|---|---|---|---|---|---|---|
| *X pinches from different radiation machines* | | | | | | | | |
| Mo | BIN | 4 x 20 | 129 | 36 | 240 | 67 | 0.5 | [48] |
| | XP | 2 x 17.5 | 49 | 36 | 240 | 175 | 3.1 | [48] |
| | XP | 4 x 25 | 202 | 65 | 460 | 148 | 2.3 | [48] |
| | SPAS | 4 x 30 | 291 | 180 | 250 | 155 | 2.5 | [48] |
| | PIAF | 2 x 25 | 101 | 188 | 130 | 242 | 6.0 | [46] |
| | 1 LTD | 4 x 13 | 55 | 130 | 110 | 262 | 7.1 | [76] |
| W | Don | 4 x 8 | 39 | 42 | 120 | 130 | 3.3 | [48] |
| | BIN | 4 x 13 | 102 | 67 | 190 | 125 | 3.0 | [48] |
| | XP | 2 x 10 | 30 | 58 | 70 | 134 | 3.5 | [48] |
| | XP | 2 x 17.5 | 93 | 45 | 250 | 122 | 2.8 | [48] |
| | XP | 4 x 20 | 242 | 47 | 500 | 97 | 1.8 | [48] |
| | COBRA | 19 x 25 | 1794 | 100 | 1000 | 56 | 0.6 | [48] |
| | SPAS | 4 x 20 | 242 | 140 | 240 | 139 | 3.7 | [48] |
| *X pinches tested with 2-LTD-Brick driver in our initial shots* | | | | | | | | |
| Cu | 2-LTD | 2 x 80 | 900 | 60* | 37** | 2 | 0.001 | shot ID 151 |
| | | 4 x 80 | 1801 | 85* | 62** | 3 | 0.001 | shot ID 115 |
| | | 2 x 127 | 2026 | 100* | 73** | 4 | 0.001 | shot ID 142 |
| | | 4 x 127 | 4051 | 165* | 116** | 5 | 0.002 | shot ID 149 |
| W | 2-LTD | 2 x 15 | 68 | 220* | 160** | 518 | 51.6 | shot ID 203 |
| | | 4 x 15 | 136 | 200* | 131** | 193 | 7.1 | shot ID 259 |
| Mo | 2-LTD | 2 x 50 | 403 | | | | | dI/dt not observed |
| *X pinches suggested for a 2-LTD-Brick driver next testing* | | | | | | | | |
| Cu | 2-LTD | 2 x 25 | 88 | 200 | 130 | 296 | 7.8 | |
| | | 2 x 30 | 127 | 200 | 135 | 213 | 4.1 | |
| | | 2 x 35 | 172 | 200 | 140 | 162 | 2.4 | |
| Mo | 2-LTD | 2 x 25 | 101 | 220 | 140 | 305 | 9.6 | |
| | | 2 x 30 | 145 | 220 | 145 | 220 | 5.0 | |
| | | 2 x 35 | 198 | 220 | 150 | 167 | 2.9 | |
| W | 2-LTD | 2 x 20 | 121 | 240 | 160 | 318 | 19.4 | |
| | | 2 x 25 | 189 | 240 | 165 | 210 | 8.5 | |
| | | 2 x 30 | 272 | 240 | 170 | 150 | 4.3 | |

* time at the moment of dI/dt dip;
** current at the moment of dI/dt dip.

As can be seen, the scaling parameter (5.4-1) varies from just below 1 up to 7 for setups listed in first part of the table. For radiation machines with a higher dI/dt (BIN, XP, SPAS, etc.) the scaling parameters is usually low, and for radiation machines with dI/dt close to 1 (PIAF and similar) the scaling parameter is usually higher.

It appears, that almost all our *x* pinches, tested so far with our 2-LTD-Brick driver, were not "good" choices. For 80/127-μm Cu wires, the scaling parameters were too small, almost zero, and for 2x15-μm W *x* pinches, the scaling parameter was too high, above 50. The wires we tested were either too thick (80/127-μm Cu) or too thin (2x15-μm W) in order to be properly pinched by our 2-LTD-Brick driver. The shots performed with 4x15-μm W *x* pinches seems to have a right scaling parameter, about 7, but for some other reasons, no good *x*-pinch signals were "observed".

In a last section of table, we predict some *x* pinches, which could be "good" choices for our 2-LTD-Brick driver. The most difficult part of the prediction is the unknown values of time and current at the moment of "hot spot" formation, which are hard to predict until a real *x* pinch has been tested. Some reasonable values were assumed, and scaling parameters were calculated for a few Cu, Mo and W wires *x* pinches. Some of these *x* pinches will be tested and discussed in the following sections.

## 5.5 2-LTD-Brick Driver *x*-pinch timing performance

Based on anticipated "good" *x*-pinch radiation performance (see last section of Table 5.4-1), 30-μm Cu, 20/30-μm W, and 30-μm Mo wire materials were selected for our final *x*-pinch radiation performance testing experiment. The basic geometries and shots ID numbers for these *x* pinches are briefly summarized in Table 5.5-1 and a complete list of all shots performed is presented in Appendix N.

Table 5.5-1. List of shots with 20/30-μm W, 30-μm Mo and 30-μm Cu wires.

| ***X*-pinch config.** | **XRD** | | | **Shots ID *** |
|---|---|---|---|---|
| | **filter** | **distance, cm** | **aperture, cm** | |
| 2x30-μm Cu | 2-μm Kimfol | 105.9 | 0.98 | 619-624 |
| 2x20-μm W | 2-μm Kimfol | 105.9 | 0.98 | 609-613 |
| 2x30-μm W | 2-μm Kimfol | 105.9 | 0.98 | 601-608 |
| 2x30-μm Mo | 2-μm Kimfol | 105.9 | 0.98 | 614-618 |

* Shot ID- first digit is a test number, last two are the shot number.

A thin, 2-μm-thick aluminized (about 1-nm) Kimfol filter was installed in front of XRD with magnets placed between them as described in section 5.1. The filter transmission curve and XRD filtered response are presented in Figure 5.2-2. As can be seen, the XRD-filter is primarily sensitive to *x* rays from about 120 to 285 eV and from 400 up to 2400 eV for 10% sensitivity threshold with the maximum sensitivity of about 15.2 A/MW at about 260 eV.

## 5.5.1 Shots with 2x30-μm Cu *x* pinches

A series of 6 shots were performed with 2x30-μm Cu *x* pinches and some of them are in Figure 5.5-1. The Rogowskii signal is normalized to kA/ns (black line, left plot) and integrated to a total current (black line, right plot). The XRD signal (red line) is shown at both, left and right, plots. As can be seen, sharp, well-separated XRD signals are always formed about 200 ns after the start of the current. However, sometimes a second, more powerful peak is developed immediately after the first one (shot ID 620 in the second row), or more complicated, multi-peaks structures are developed at later times (shot ID 624 in the last row). Those later peaks do not seem to be reproducible and this load requires more study.

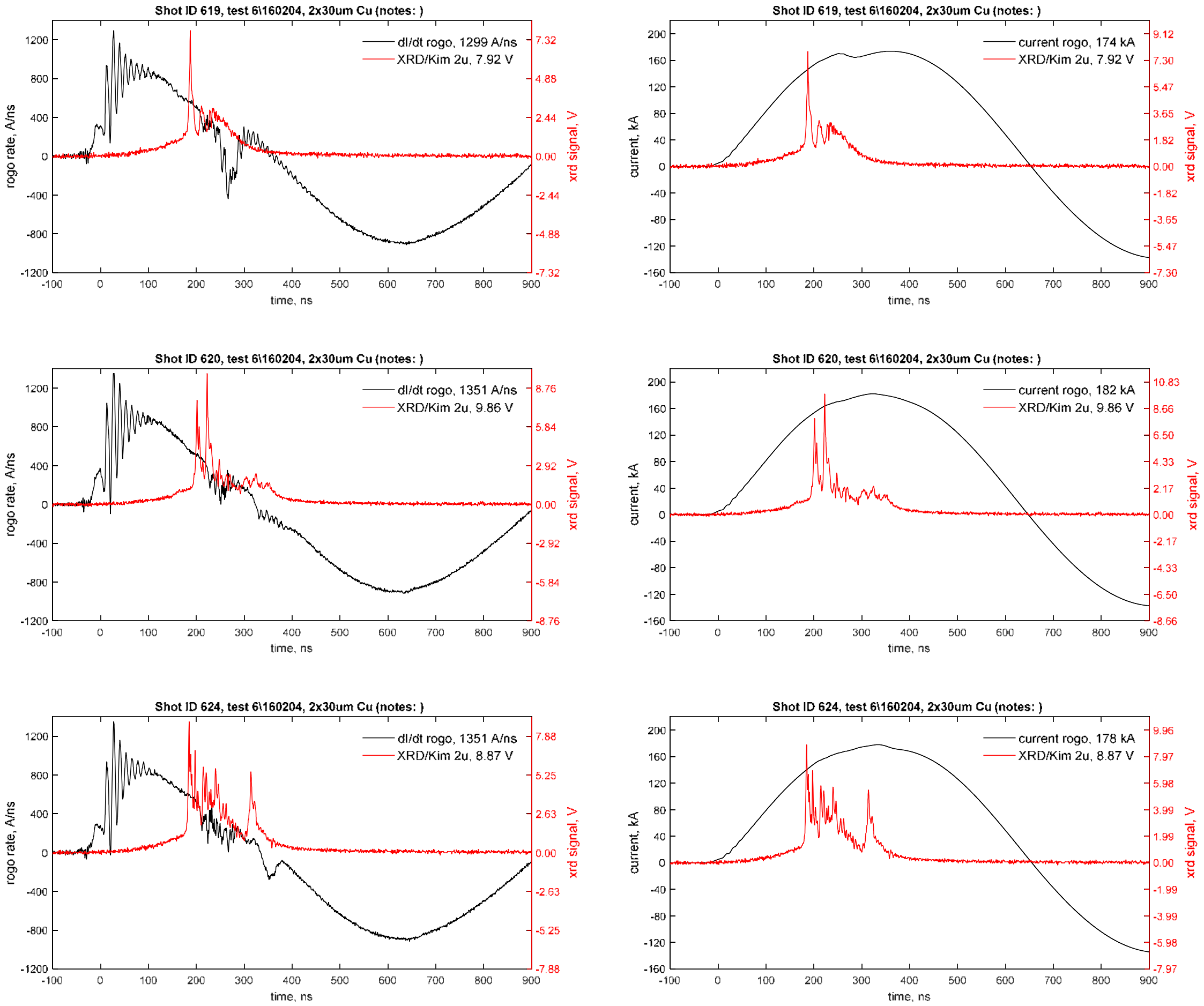

Figure 5.5-1. 2x30-µm Cu *x* pinches with single-peak (shot ID 619), two-peaks (shot ID 620), and multi-peaks (shot ID 624) structures.

In general, a sharp and fast XRD signals were always formed at 185-223-ns time window after the current start. The signal width varied from 3 to 6 ns and the peak amplitude varied from 6 to 9.9 V.

## 5.5.2 Shots with 2x20-μm W *x* pinches

A total of 5 shots were performed with 2x20-μm W *x* pinches, and three typical shots are shown in Figure 5.5-2. The situations with these *x* pinches are very similar to those with 2x20-μm Cu *x* pinches: a fast and bright XRD signal is initially formed at predictable time (first row of Figure 5.5-2), sometimes more complicated multi-peaks structures develop later in time (second row of Figure 5.5-2), and even first the peak develops at a later time when second peak structures are more pronounced (last row of Figure 5.5-2).

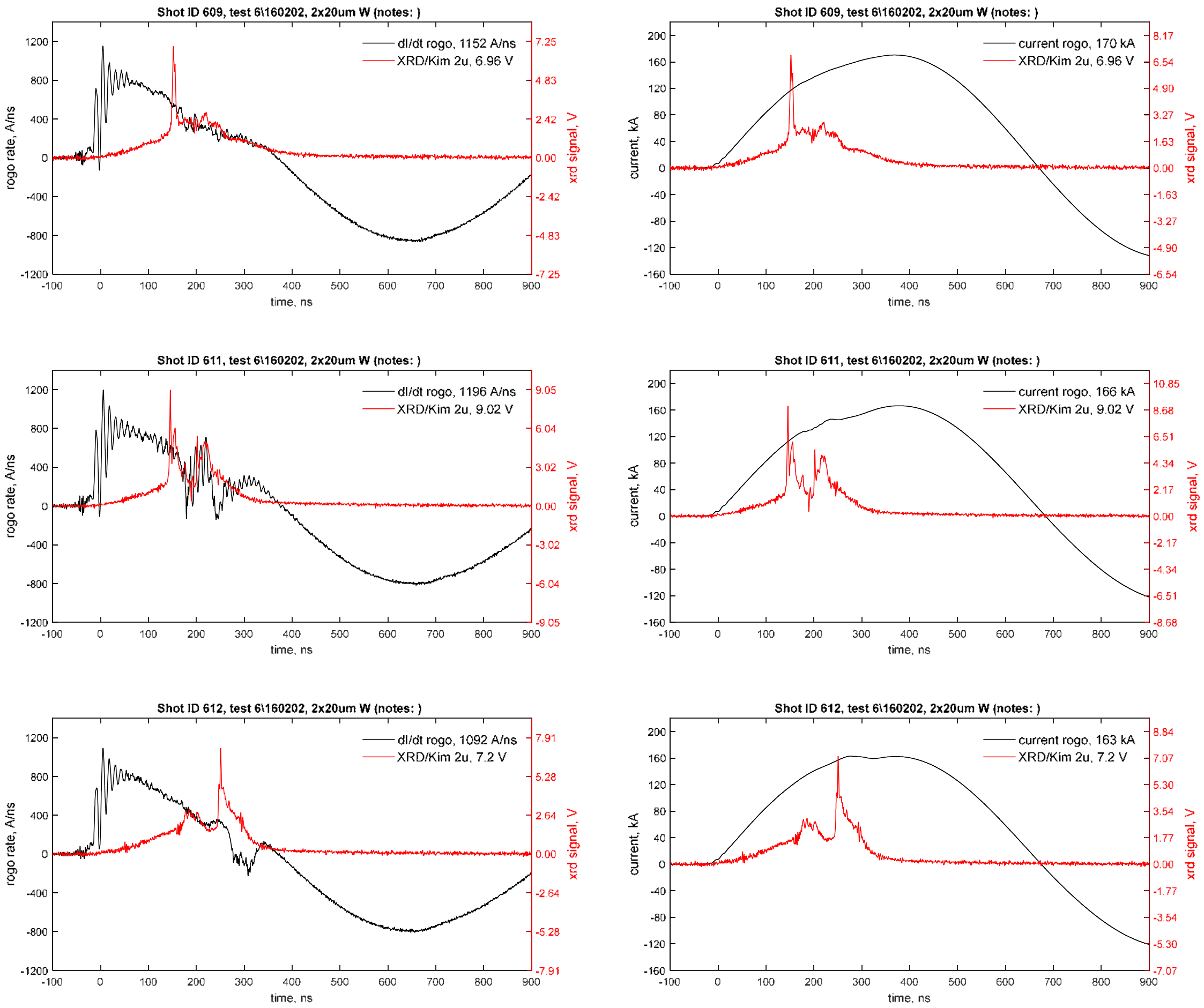

Figure 5.5-2. 2x20-μm W *x* pinches with single-peak (shot ID 609), multi-peaks (shot ID 611), and a later peak (shot ID 612) structures.

In general, out of 4 shots, except of shot with ID 612, when the peak of the XRD signal was formed at a later time, sharp fast XRD signals were always formed at 146-165-ns time window after the current start. The signal width varied from 3 to 8 ns and the peak amplitude varied from 5.6 to 9.0 V.

## 5.5.3 Shots with 2x30-μm W *x* pinches

A total of 33 shots were performed with 2x30-μm W *x* pinches, 8 initial shots to establish a timing performance, and 25 extra shots with an imaging plate that will be discussed at the next section. All shots, except several problematic ones when no clear XRD signal were observed, were included in timing performance analysis.

Figure 5.5-3 presents a few typical shots observed with 2x30-μm W *x* pinch. Shot ID 608 (first row of Figure 5.5-3) represents a "good" shot when a well-separated, about 3-ns wide XRD signal is formed just before the dI/dt dip at about 269 ns after the start of the current at the level of about 162 kA. Shot ID 632 (second row of Figure 5.5-3) is an example when secondary peaks develop immediately after the first peak. Shot ID 643 (last row of Figure 5.5-3) is an example when a not so narrow first peak is formed and when multiple XRD peak structures developed at a later time.

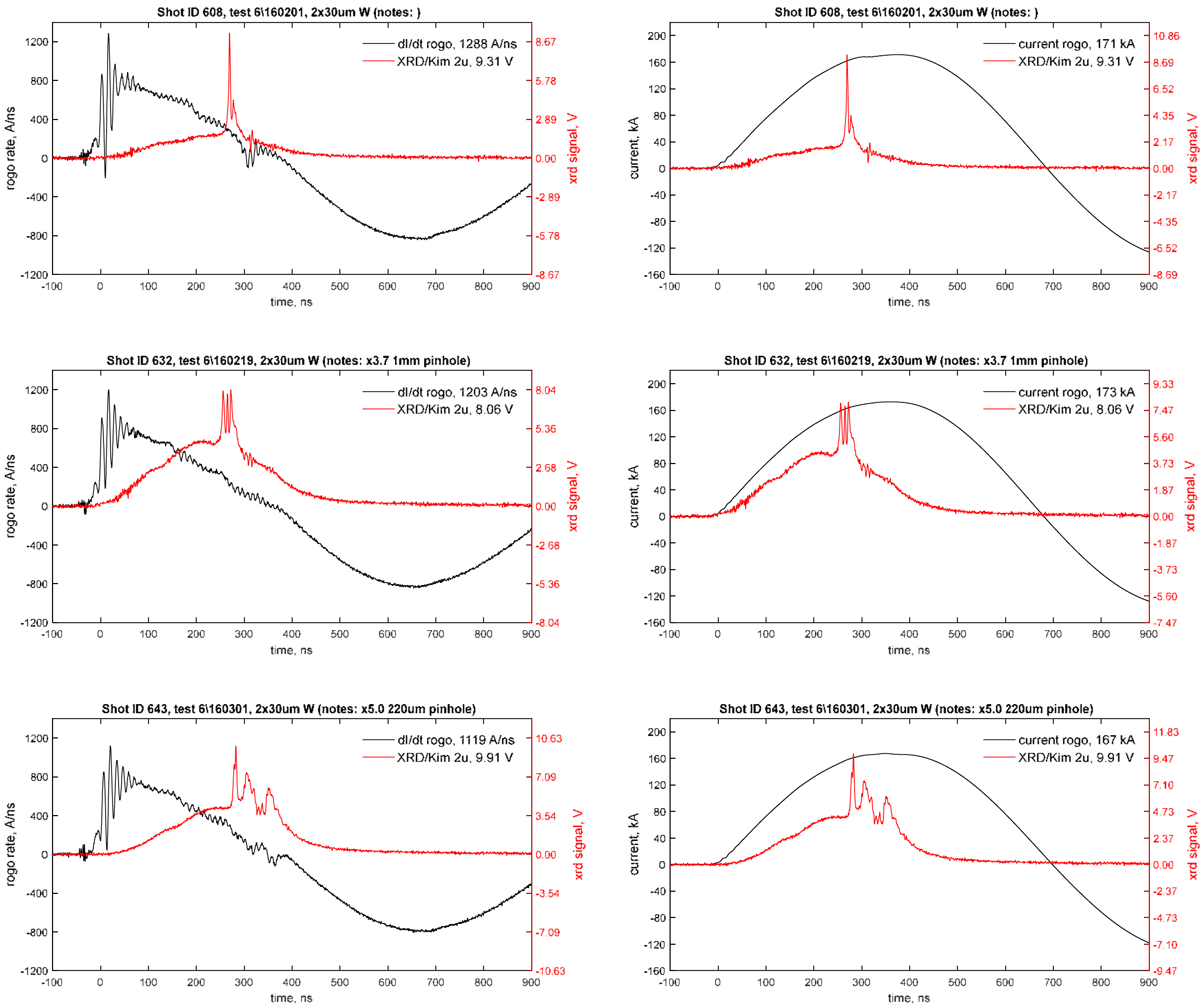

Figure 5.5-3. 2x30-µm W *x* pinches with single-peak (shot ID 609), three-peaks (shot ID 611), and multi-peaks (shot ID 612) structures.

In a series of 33 shots performed with 2x30-µm W *x* pinches, a total of 22 shots generated a fast and sharp XRD signal in the 219-359-ns time window after current start. The signal width varied from 2 to 16 ns and the peak amplitude from 3.6 to 17.4 V.

## 5.5.4 Shots with 2x30-µm Mo *x* pinches

The best and the most reproducible *x*-pinch signals were formed with 2x30-µm Mo *x* pinches. In a series of 11 shots, a total of 9 shots always produced bright and fast XRD peaks at 188-247-ns time window after current start. Typical *x*-pinch shots are presented in Figure 5.5-4. As can be seen, a very bright and fast XRD signal (first and second rows in Figure 5.5-4) is formed at about 200 ns after the start of the current. The width of the observed peaks is about 2 ns and the peak signals are about 10.6 V and 15.9 V for shot IDs 616 and 657, respectively. The shot ID 617 (third row in Figure 5.5-4) represents the case when the structure from two peaks overlaps and the total measured pulse width is about 6 ns.

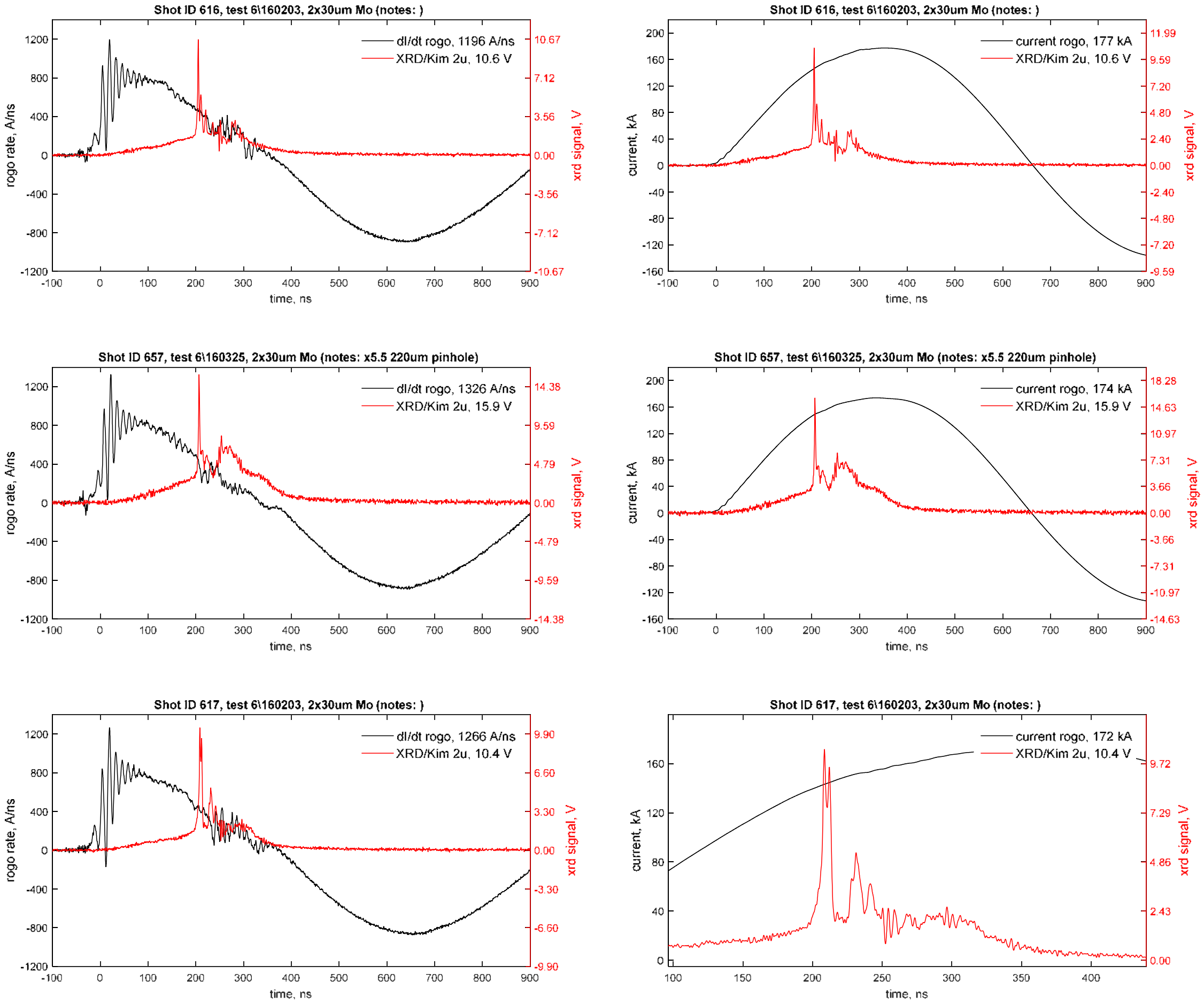

Figure 5.5-4. 2x30-µm Mo *x* pinches with single-peak (shot ID 616 and 627), and double-peaks (shot ID 617) structures.

In general, the measured width of the observed XRD signals varied from 2 to 6 ns and the peak amplitudes varied from 8.3 to 18.4 V. Sometimes secondary peaks were immediately following after the first one, but minimal or no secondary structures were developed in a later times as was observed with W and Cu *x* pinches.

## 5.5.5 Summary of timing performance of "good" *x* pinches

Table 5.5-2 summarizes the timing performance for shots with 20/30-µm W, 20-µm Mo and 30-µm Cu *x* pinches. Only shots, when a clear first XRD signal was observed, were analyzed, and for each shot, the XRD peak was selected when the maximum XRD signal was observed. It usually the first XRD peak for 20/30-µm W and 30-µm Mo *x* pinches, or can be a later peak for 30-µm Cu *x* pinches, when secondary peaks can be higher.

Table 5.5-2. *X* pinches timing performance.

| ***X*-pinch config.** | **XRD peak width** | **XRD peak time, ns** | | | **XRD peak current, kA** | | | **Comments** |
|---|---|---|---|---|---|---|---|---|
| | | **avg** | **std** | **range** | **avg** | **std** | **range** | |
| 2x30-µm Cu | 3-6 | 197 | 17 | 185-223 | 149 | 9 | 140-164 | 5 out of 6 shots |
| 2x20-µm W | 3-8 | 174 | 44 | 146-251 | 123 | 19 | 111-157 | 5 out of 5 shots |
| 2x30-µm W | 2-16 | 270 | 29 | 219-359 | 164 | 7 | 144-175 | 22 out of 33 shots |
| 2x30-µm Mo | 2-6 | 210 | 21 | 188-247 | 144 | 5 | 136-151 | 9 out of 11 shots |

Figure 5.5-5 represents the timing performance for the different *x* pinches discussed above, where for each *x*-pinch configuration, the average time when the maximum XRD peak is observed and its standard deviation is plotted. As can be seen, the time interval, when the first XRD peaks are observed, varies from about 150 ns up to 285 ns. That can be a useful tool, when a bright and fast *x*-pinch signal is needed to make images of different rapidly evolving systems at different time stamps.

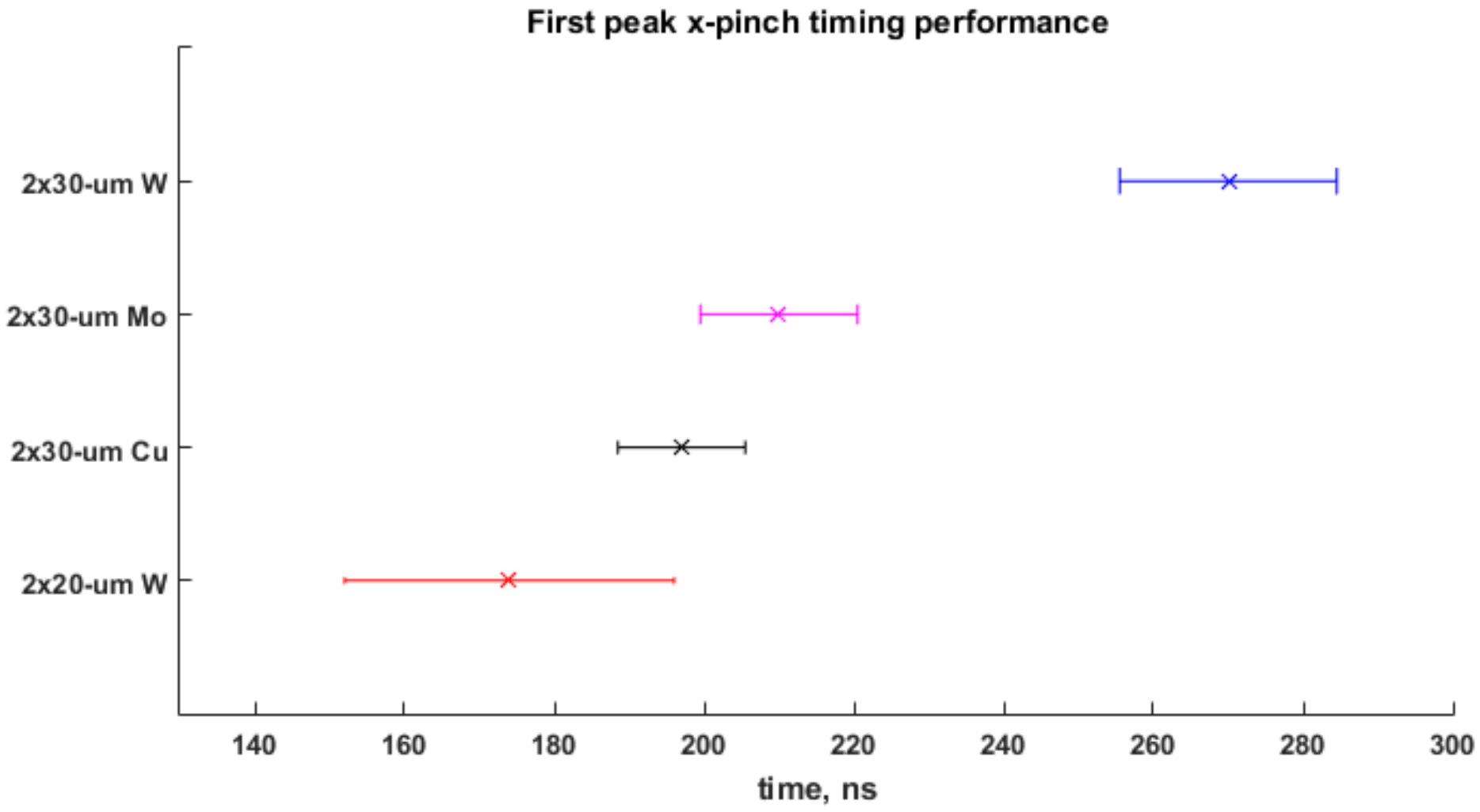

Figure 5.5-5. Timing performance for 2x20-μm W, 2x30-μm Cu, 2x30-μm Mo, and 2x30-μm W *x* pinches.

The best performance was achieved with 2x30-μm Mo *x* pinches, when sharp 2-6-ns wide XRD signals were always generated in the 190-250-ns time window after current start. Such XRD signals, most likely, can be associated with one "hot spot" formed at a predictable location and a predictable time, as was discussed in the introduction.

The worst timing performances were with the 2x30-μm Cu *x* pinches, when multi-peaks structures were often formed in a later times. Such multiple structure can be, most likely, associated with multiple "hot spots" formation at a later times, which is common for a low-Z wire materials.

The W *x* pinches seemed to perform as well as Mo *x* pinches, however, sometimes, no XRD sharp signals were observed at all, or secondary *x*-ray peaks were formed immediately after the first one. It looks like the wire diameter used for W *x*-pinch tests has to be further optimized in order to achieve the better *x*-pinch timing performance.

The fact that not all $x$-pinch shots always produced a bright and fast XRD signal can also be related to the way how the wires touch, what is the twist angle between two wires [77], what is the weight applied to wires, what is the wire material quality and purity, and more. We attempted to study some of those possible correlations, but we cannot make any sound conclusions at the present time.

## 5.6 X-pinch radiation power, energy and flux

To estimate the total radiation $x$-pinch power the following procedure was developed. The XRD signal at each moment $t$, assuming isotropic source, is a convolution of unknown $x$-ray spectrum S(E,t) with XRD filtered response F(E):

$$V(t) = 50\Omega \times \frac{a}{4\pi R^2} \int S(E,t)F(E)dE \qquad (5.6\text{-}1)$$

where V(t) is XRD signal in V;

$a$ is XRD aperture area;

R is a source-detector distance;

S(E,t) is unknown source power spectrum in MW/eV;

F(E) is XRD filtered response in A/MW.

It has to be noted, that the unknown spectrum S(E,$t$), as it written in (4.2-2), corresponds to some fraction of a total unknown source spectrum, visible with XRD detector. If one want to estimate the absolute total source $x$-ray spectrum, it has to be normalized to a total source area, which in many cases is unknown. For full description of the integral, see [73].

Replacing F(E) with its maximum values $F_{max}$, we can rewrite (5.6-2) with a following inequality:

$$V(t) \leq 50\Omega \times \frac{a}{4\pi R^2} \times F_{max} \int S(E,t)dE$$
$$= 50\Omega \times \frac{a}{4\pi R^2} \times F_{max} \times S(t) \quad (5.6\text{-}2)$$

Then the total *x*-ray radiation power at each moment *t* can be estimated as:

$$S(t) \geq V(t)/50\Omega/\left(\frac{a}{4\pi R^2}\right)/\mathrm{F_{max}} \quad (5.6\text{-}3)$$

where S(t) is unknown total power spectrum in MW;

The expression (5.6-3) gives a lower estimate of the total *x*-ray radiated power (in MW) measured by XRD detector at each moment *t*. The true *x*-ray source power value will always be higher.

The power calculation, performed according to (5.6-3) for shot ID 616 with 2x30-µm Mo *x* pinches, is presented in Figure 5.6-1. The source detector distance was about 105.9 cm, the XRD aperture diameter was about 0.98 cm and the corresponding solid angle was about $6.7 \times 10^{-5}$. The XRD filter was 2-µm-thick Aluminized Kimfol foil with the maximum sensitivity $F_{max}$ of about 15.2 A/MW at 260 eV as shown earlier in Figure 5.2-2.

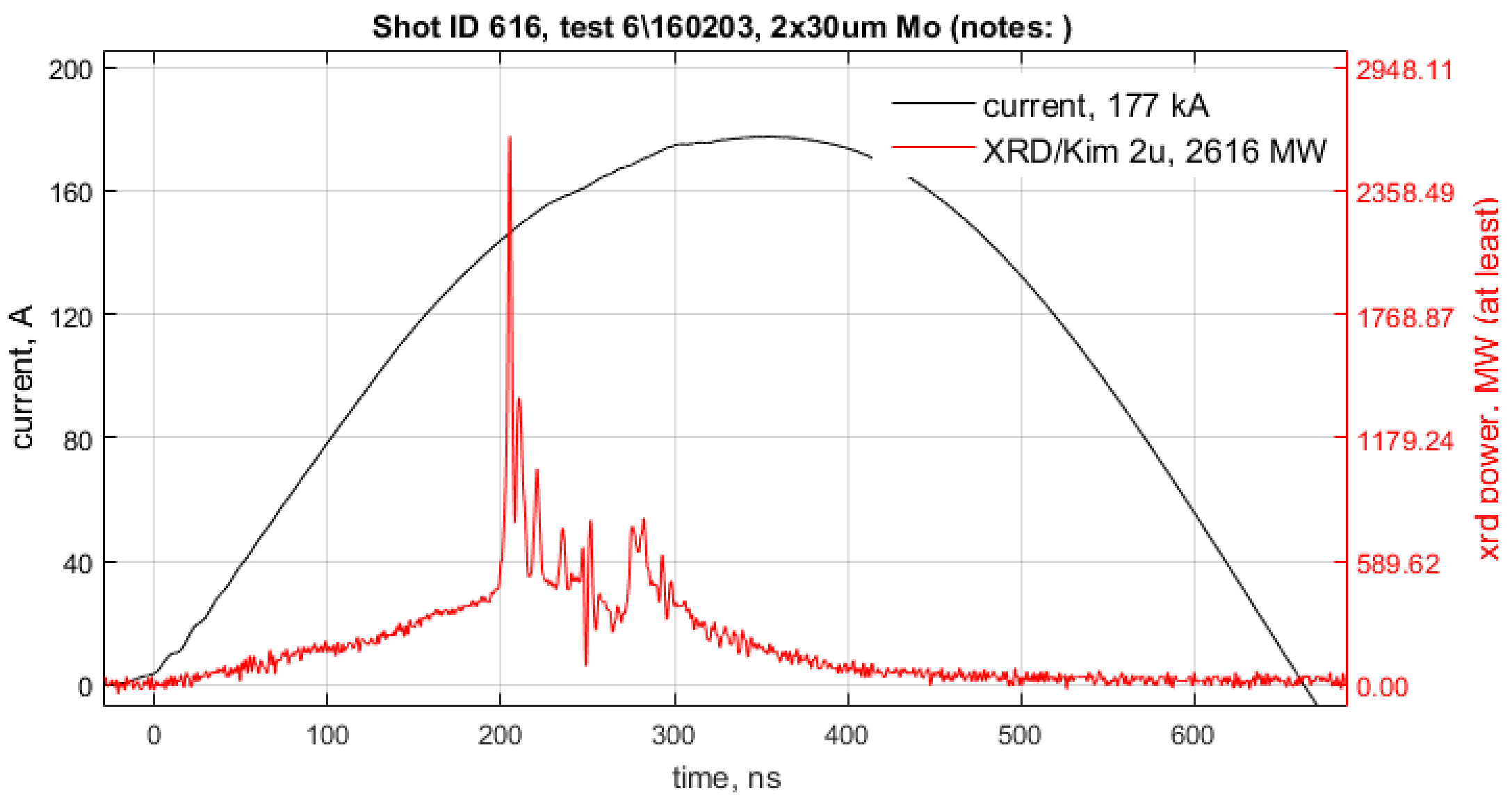

Figure 5.6-1. The minimum radiation power for shot ID 616 with 30-µm Mo *x* pinch and 2-µm-thick Kimfol XRD filter.

As can be seen, the measured power from the 30-µm Mo *x* pinch on this particular shot is about 2,616 MW for photon energies between 120 - 285 eV and 400 - 2400 eV. (See Figure 5.2-2 for the sensitivity range.) This measured power is a lower bound of the source power assuming all photons are emitted at 260 eV for a maximum XRD sensitivity. The actual *x*-pinch radiation power will be only larger, but cannot be estimated with a single XRD detector.

To estimate the total *x*-ray energy delivered near peak XRD power and during the entire shot, the XRD power signal was integrated from 0 to 205 ns and from 0 up to 1000 ns, correspondingly. In addition, to estimate the total *x*-ray energy radiated during the fast XRD pulse, the power curve at Figure 5.6-1 was also integrated from 202 to about 208 ns. The results of these integrations are summarized in Table 5.6-1. As can be seen, there are *at least* 9.2 J of photons between 120 - 285 eV and 400 - 2400 eV delivered at the moment of radiation burst, and *at least* 44 J and 135 J of photon energies delivered at the moment of XRD peak and during the entire shot, correspondingly.

Table 5.6-1. Total *x*-ray energy delivered at different time intervals for shot ID 616.

| *X* pinch | shot ID | filter | Time interval, ns | *at least* energy, J |
|---|---|---|---|---|
| 2x30 Mo | 616 | 2-µm Kimfol | 202-208 | 9.2 |
| 2x30 Mo | 616 | 2-µm Kimfol | 0-205 | 44 |
| 2x30 Mo | 616 | 2-µm Kimfol | 0-1000 | 135 |

It would be also interesting to estimate the corresponding photon flux. For simplicity, let say, all photons are at 260 eV at the maximum XRD sensitivity. Then, the total number of photons delivered during the XRD pulse is:

$$N_\gamma = 9.2\,\mathrm{J} \times (0.62 \times 10^{19}\,\mathrm{eV/J})\ /\ 260\,\mathrm{eV}\ = 2.2 \times 10^{17}\,\frac{\mathrm{photons}}{\mathrm{pulse}} \qquad (5.6\text{-}4)$$

As can be seen, our 30-µm Mo *x* pinch is capable of delivering an extremely bright radiation pulse with more than $10^{17}$ low-energy photons per pulse, that opens up possibilities for many applications when a high photon flux is desired.

## 5.7 "Hot spot" radiation size prediction

As was discussed in the introduction, "good" *x* pinches can be characterized by a few unique radiation parameters, which are extremely difficult to simultaneously achieve in other known radiation sources: the radiation pulse is fast (a few ns or less), the radiation source size is small (a few µm) and the radiation power is large (GW or more).

In a few previous sections, we showed that our 2-LTD-Brick driver is capable of generating a fast (2-4 ns) and bright (more than 2 GW) *x*-ray radiation pulse for shots performed with our best *x* pinches. We will discuss some direct measurements of the radiation source size in a next section, however, the radiation "hot spot" size can be also estimated using a simple scaling relation [27] as:

$$r \sim I^{-\frac{14}{9}} Z^{-\frac{10}{9}} \qquad (5.7\text{-}1)$$

where r is a "hot spot" size in μm, I is the load current in kA and Z is an *x*-pinch wire atomic number.

Table 5.7-1 presents the "hot spot" radius predictions for 20/30-μm W and 30-μm Mo *x* pinches. The 30-μm Cu *x* pinches often generate multiple radiation "hot spots", and are not included in this analysis. The current value corresponds to a moment in time when the XRD peak power was observed for each particular *x*-pinch shot.

Table 5.7-1. "Hot spot" size predictions.

| X-pinch config. | shot ID | Z | Current I, kA | hot spot radius, um |
|---|---|---|---|---|
| 2 x 30 W | 606 | 74 | 156 | 3 |
| 2 x 20 W | 611 | 74 | 112 | 5 |
| 2 x 30 Mo | 615 | 42 | 136 | 8 |

As can be seen, the predicted "hot spot" radii varies from 3-μm for 2 x 30-μm W *x*-pinches up to 8-μm for 2 x 30-μm Mo *x* pinches and are in a good agreement with some direct measurements [3] [12] [13].

# 6 X-PINCH IMAGING PERFORMANCE

After a proper set of *x*-pinch configurations was established and "good" radiation performance of the constructed 2-LTD-Brick driver was demonstrated, a series of imaging experiments were performed to further advance our understanding of *x*-pinch radiation source parameters and to show some possible applications. We have used a pinhole camera to measure the time-integrated *x*-pinch radiation "hot spot" size, and a step-wedge filter to evaluate a time-integrated spectrum in the energy range above 10 keV as described in the following few sections.

## 6.1 Experimental set-up

The experiment set up in our typical imaging experiment is schematically shown in Figure 6.1-1. The objects, which can be a pinhole camera, a step-wedge filter, or other small objects under irradiation, were placed at the distance R1 from the *x*-pinch radiation source. An imaging plate was usually installed at the end of the vacuum chamber output port at distance R2 from the object.

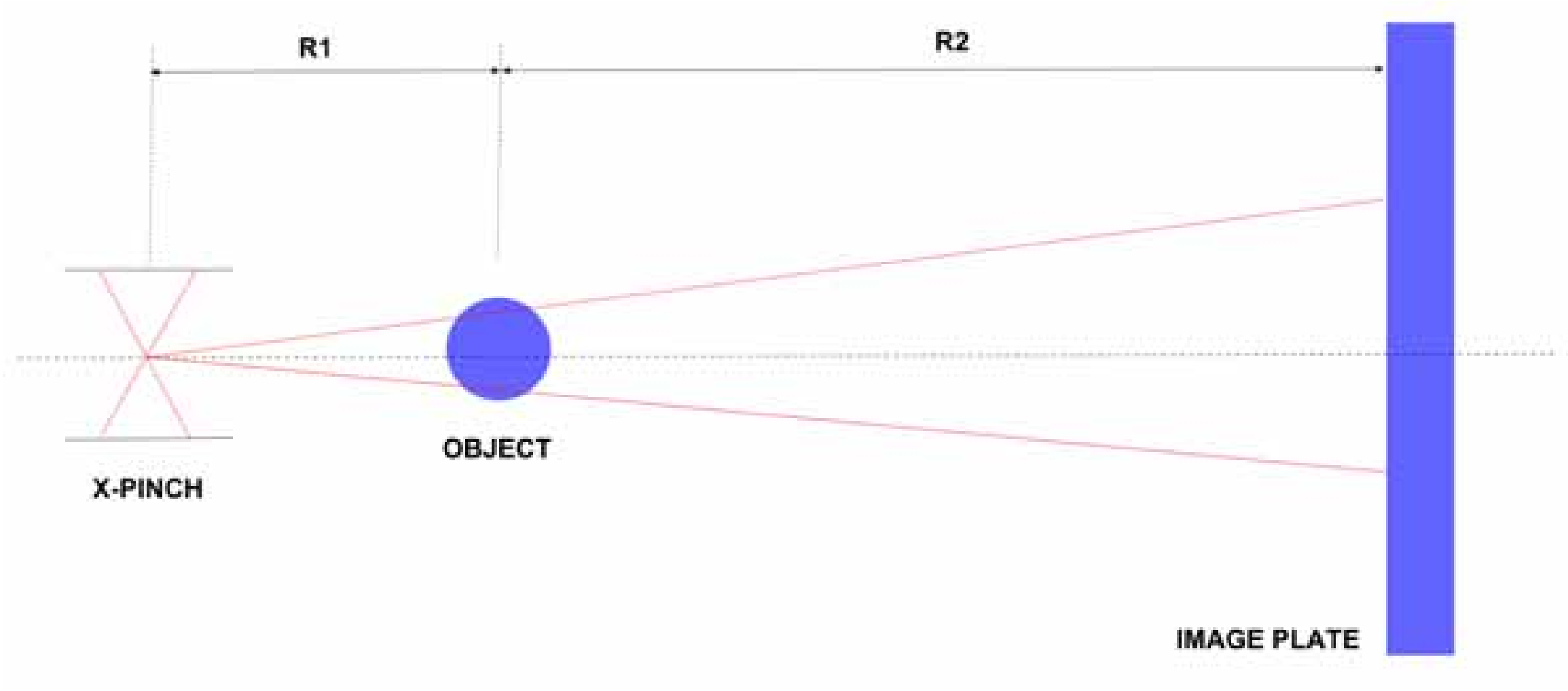

Figure 6.1-1. Source-object-plate arrangement in a typical imaging experiment.

The image was recorded with Gendex Phosphor Storage Plate (PSP) plate enclosed inside one or several layers of Al thin foil and scanned with Gendex DenOptix QSD system [78]. The spatial resolution of the Gendex PSP plate is reported [79] to be about 11 lp/mm (line pairs per millimeter) at DenOptix 600-dpi scan, but from [80] is usually limited to about 100 μm to 110 μm of the best resolved object size. The image was initially processed with ImageJ [81] and further analyzed with a custom Matlab code.

## 6.2 Pinhole camera image experiment

The pinhole camera can be used to measure the time-integrated radiation source size of a small emitting object as described below. If the radiation source can be approximated by the point-like object, the pinhole will form a sharp image as shown in Figure 6.2-1.

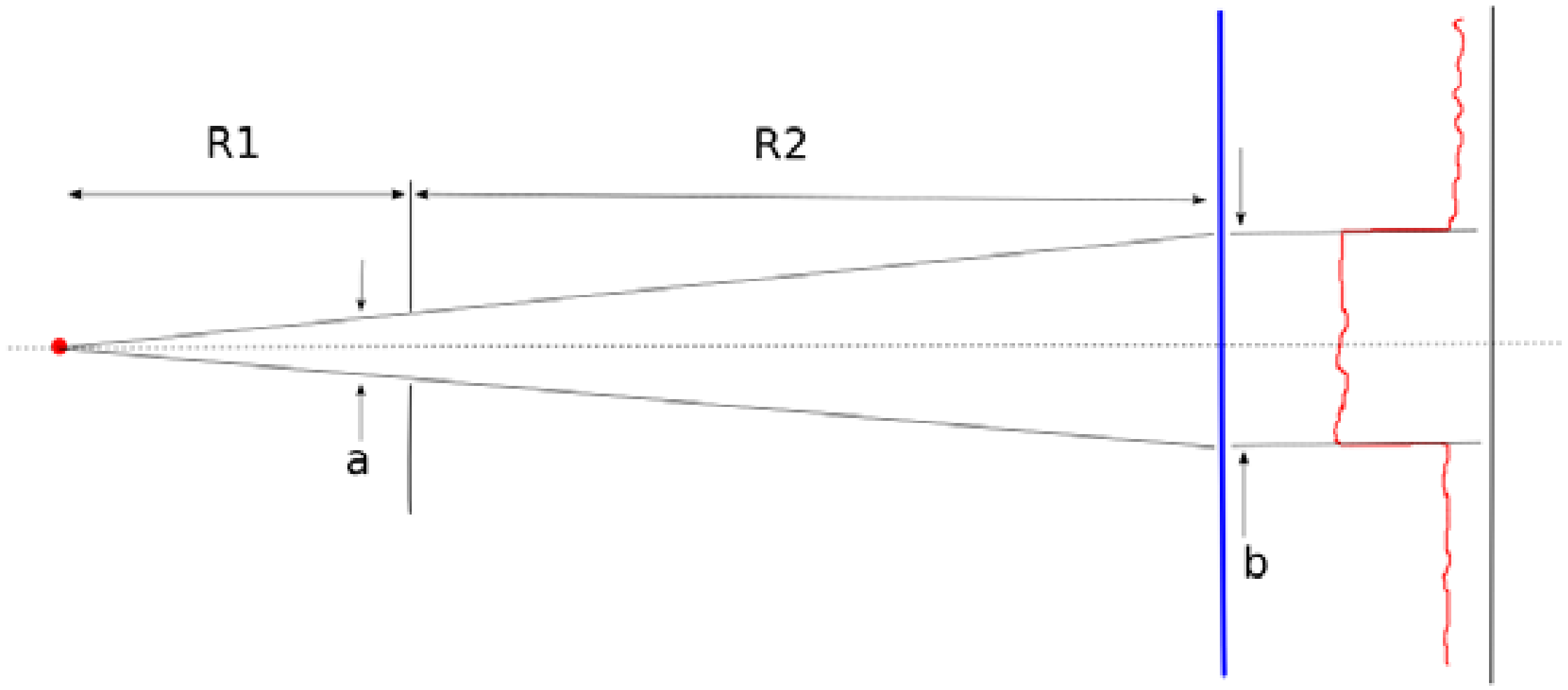

Figure 6.2-1. Point-size pinhole camera image.

When the source size is finite, but still small, the camera will still produce a pinhole image, but with a penumbra, as schematically shown in Figure 6.2-2.

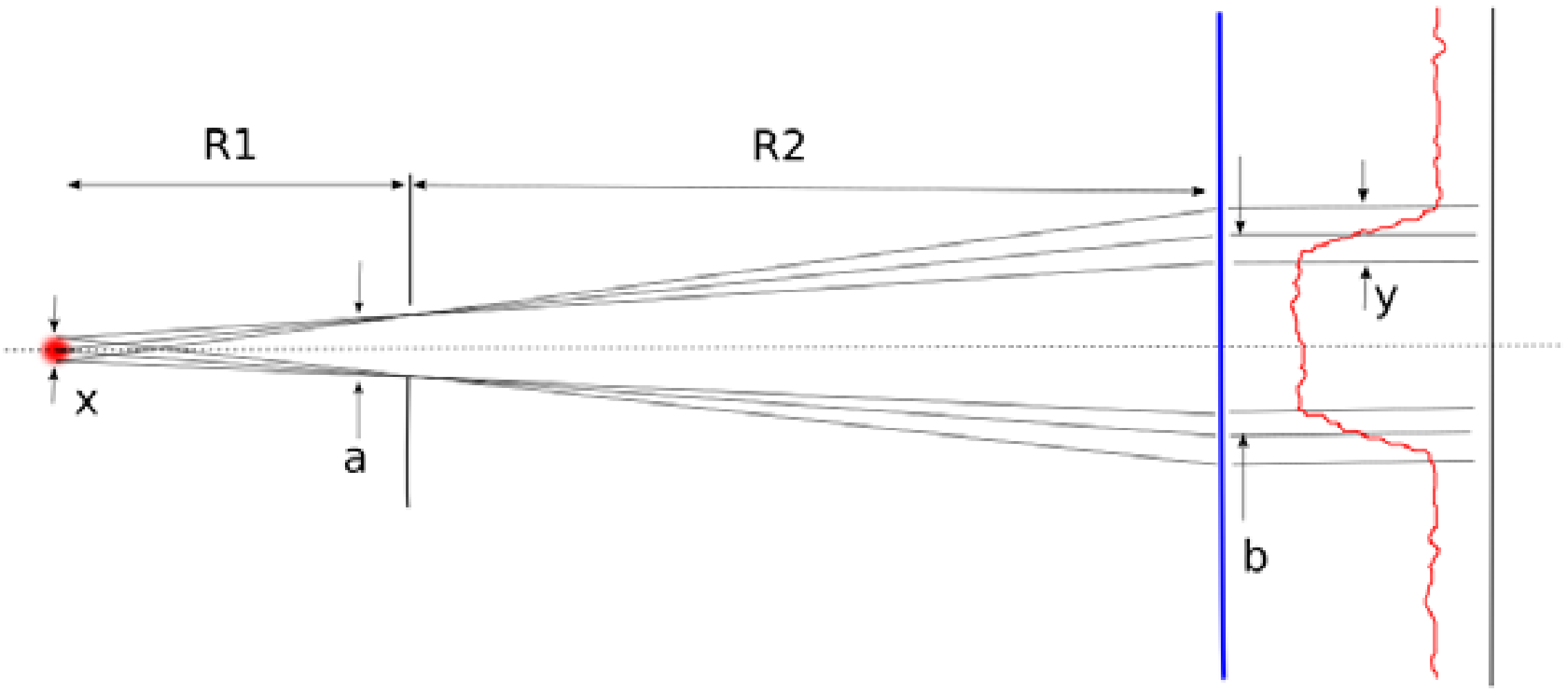

Figure 6.2-2. Finite-size pinhole camera penumbra image.

The magnification **M** of a pinhole camera, the pinhole image size **b**, and source penumbra **y** are related by simple geometrical relations:

$$M = \frac{R_1 + R_2}{R_1} \qquad (6.2\text{-}1)$$

$$b = M \times a \qquad (6.2\text{-}2)$$

$$y = M \times x \qquad (6.2\text{-}3)$$

**2x30-μm W pinhole *x*-pinch images:** The sizes of radiating sources can be estimated by analyzing the density profile of the penumbra image for a given camera geometry using relation (6.2-3). Figure 6.2-3 presents two, pinhole-camera images obtained from 2x30-μm W *x* pinches. The left picture shows a 1-mm-diameter pinhole image (shot ID 632) located 71-mm away from a source, and the right corresponds to a 220-μm-diameter pinhole (shot ID 634) located about 72-mm away from a source with magnification geometries of about 3.7 and 5.0, respectively. Both images were recorded with Gendex PSP plate enclosed in 17-μm-thick Al foil.

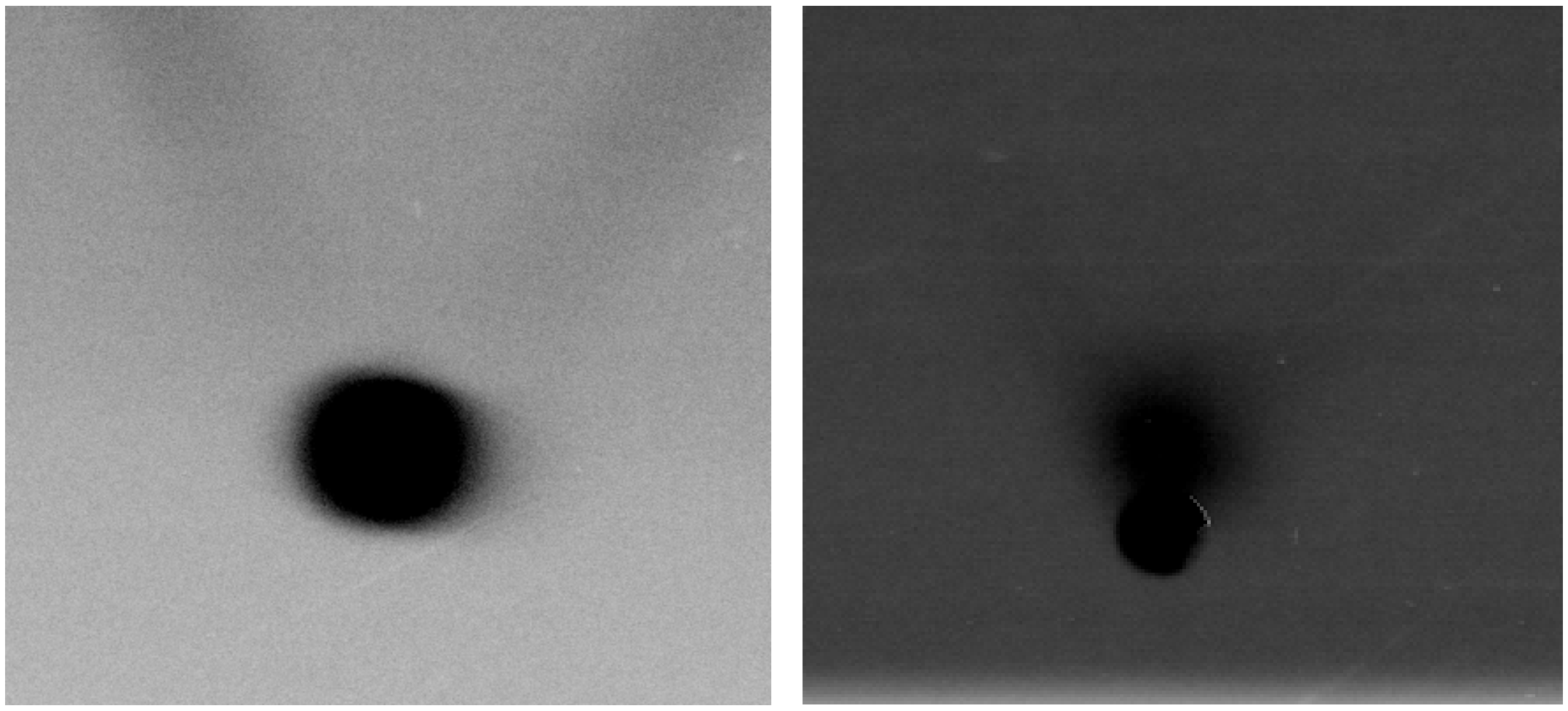

Figure 6.2-3. 2x30-μm W *x*-pinch images made with 1-mm (left) and 220-μm (right) pinhole cameras.

A clear pinhole image can be observed each cases which indicates that the radiation source size is much smaller than the pinhole diameter. Unfortunately, there were several problems with each shot, which makes it difficult to analyze the radiation source size. First, there were several XRD peak structures observed for the shot with 1-mm pinhole camera, and, second, the PSP plate was heavily saturated at the pinhole area for 220-µm pinhole image. The density profile of 1-mm and 220-µm pinhole images are shown in Figure 6.2-4. The radiation source size can be only roughly approximated to be about 30-µm or so, but no multiple source structures can be ruled out due to the shot complications mentioned above. Unfortunately, because of the limited supply of 30-µm-diameter W wire, we were not able to retake this data. This will be performed in future work outside this dissertation.

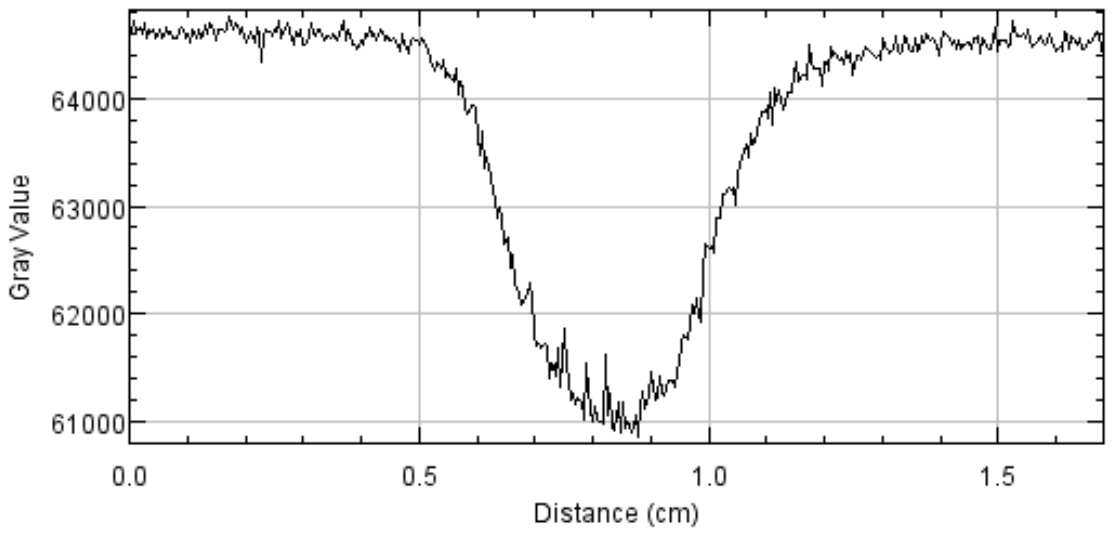

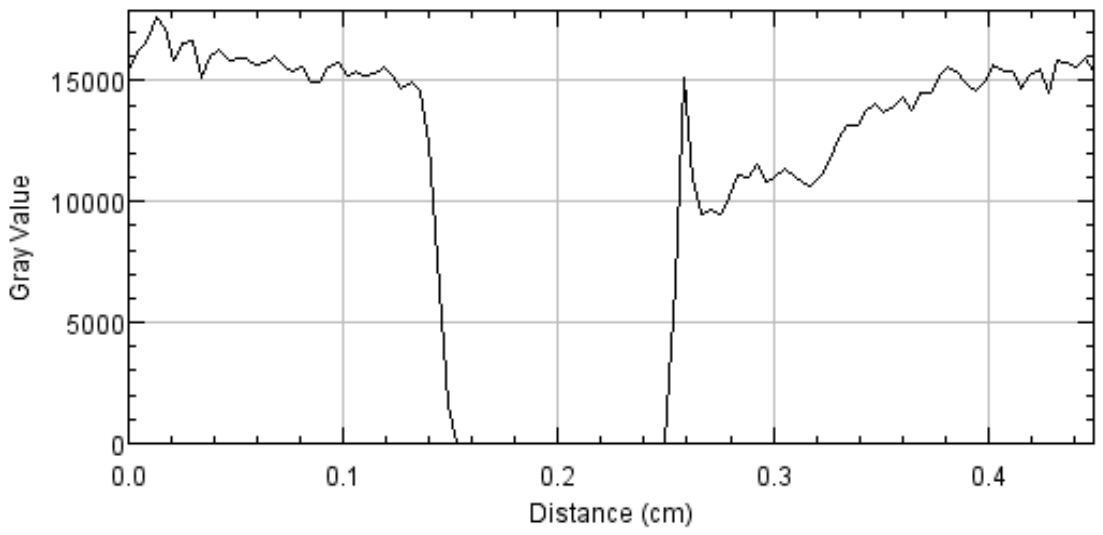

Figure 6.2-4. Density profile of 1-mm (left) and 220-µm (right) pinholes made with 2x30-µm W *x* pinches.

**2x30-µm Mo pinhole *x*-pinch images:** Figure 6.2-5 shows a 220-µm-diameter pinhole camera image obtained with 2x30-µm Mo *x* pinch for shot ID 657. The source-object distance was 65 mm, and magnification of pinhole camera for this shot was about 5.5. The image was recorded with Gendex PSP plate enclosed in 17-µm-thick Al foil. As can be seen, a very bright and high-contrast image of the pinhole is clearly observed (left image) with a density profile that is not saturated in the pinhole image region (right image).

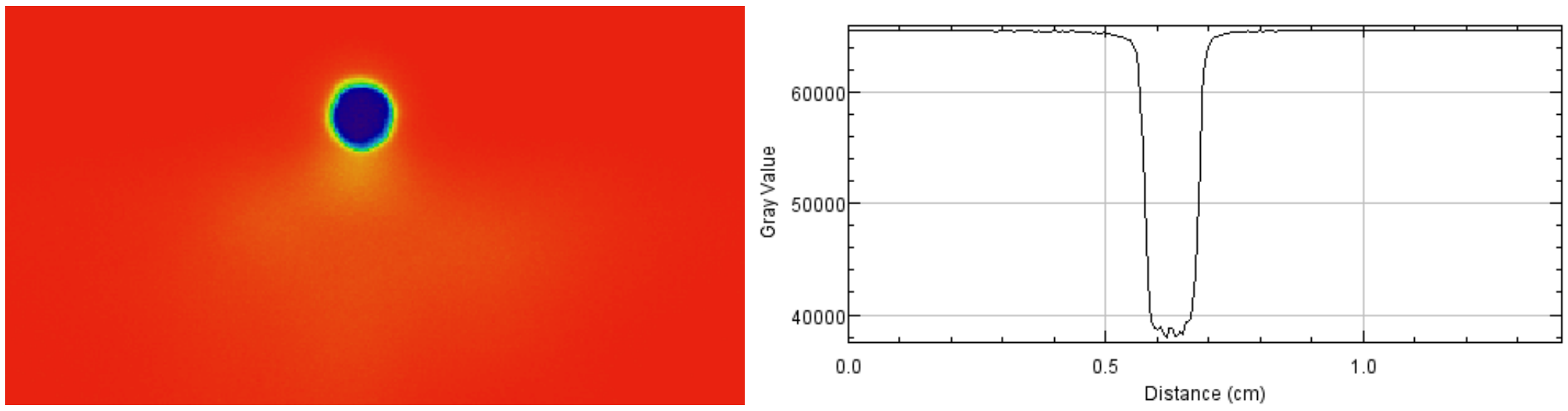

Figure 6.2-5. A 220-um-diameter pinhole image (left) and the density profile (right) from 2x30-µm Mo *x* pinch with 5.5X magnification.

The density profile of 220-µm pinhole image was analyzed with the earlier formula (6.2-3) to estimate the *x*-pinch radiation source size objects. The top plot presents the density profile of the pinhole image normalized to unity, and the bottom plot represents the derivative of the above density profile. The FWHM of the right-hand dip seen in the derivative is ~120 µm, which corresponds to about 22-µm radiation source size given the magnification of the camera.

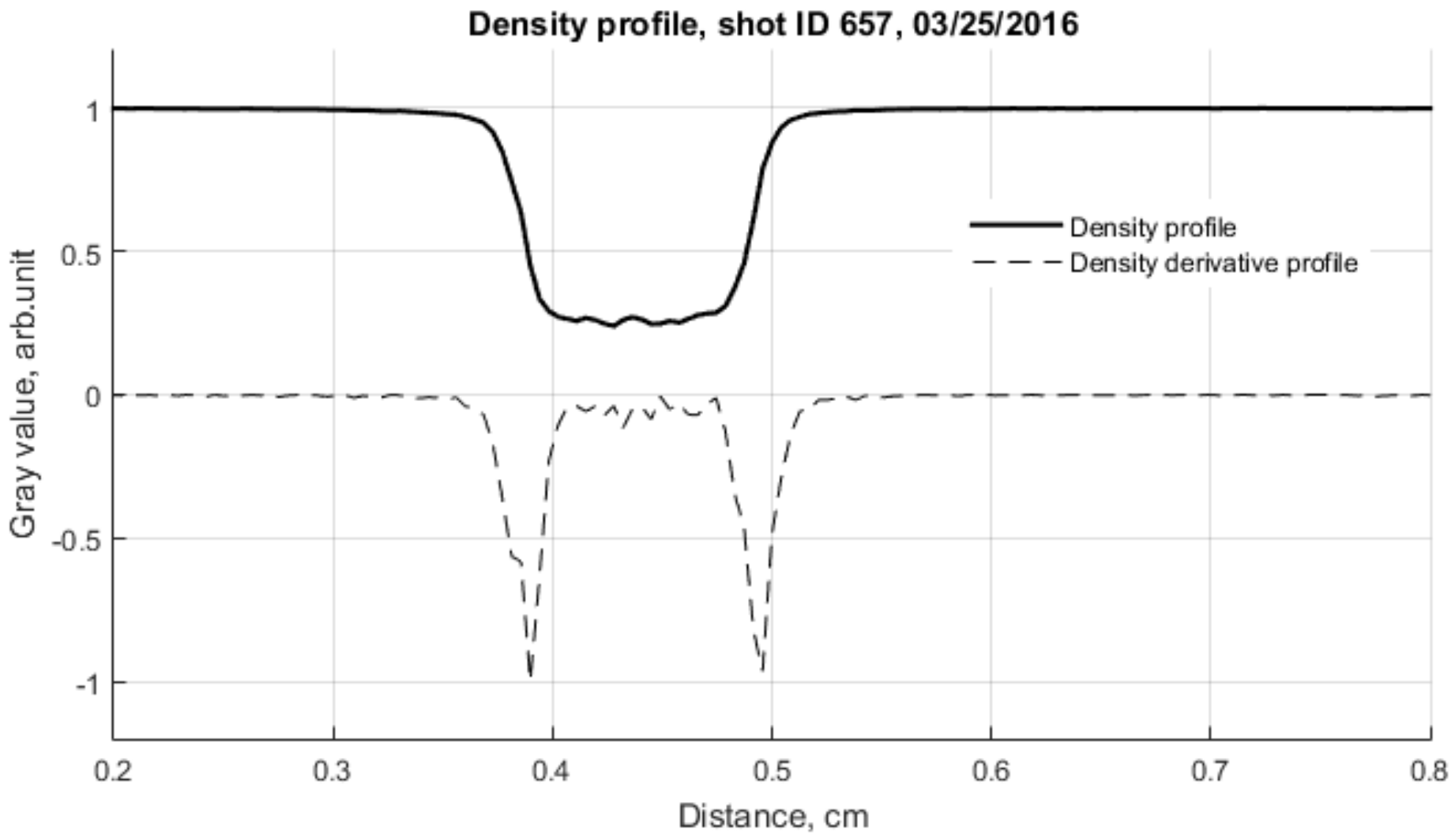

Figure 6.2-6. A 220-um-diameter pinhole-image density profile (top) and its derivative (bottom) from 2x30-µm Mo *x*-pinch with a 5.5 magnification.

Figure 6.2-7 shows exactly the same 220-µm-diameter pinhole camera image from 2x30-µm Mo *x* pinch, shot ID 659, **but with a magnification geometry of 10.3.**

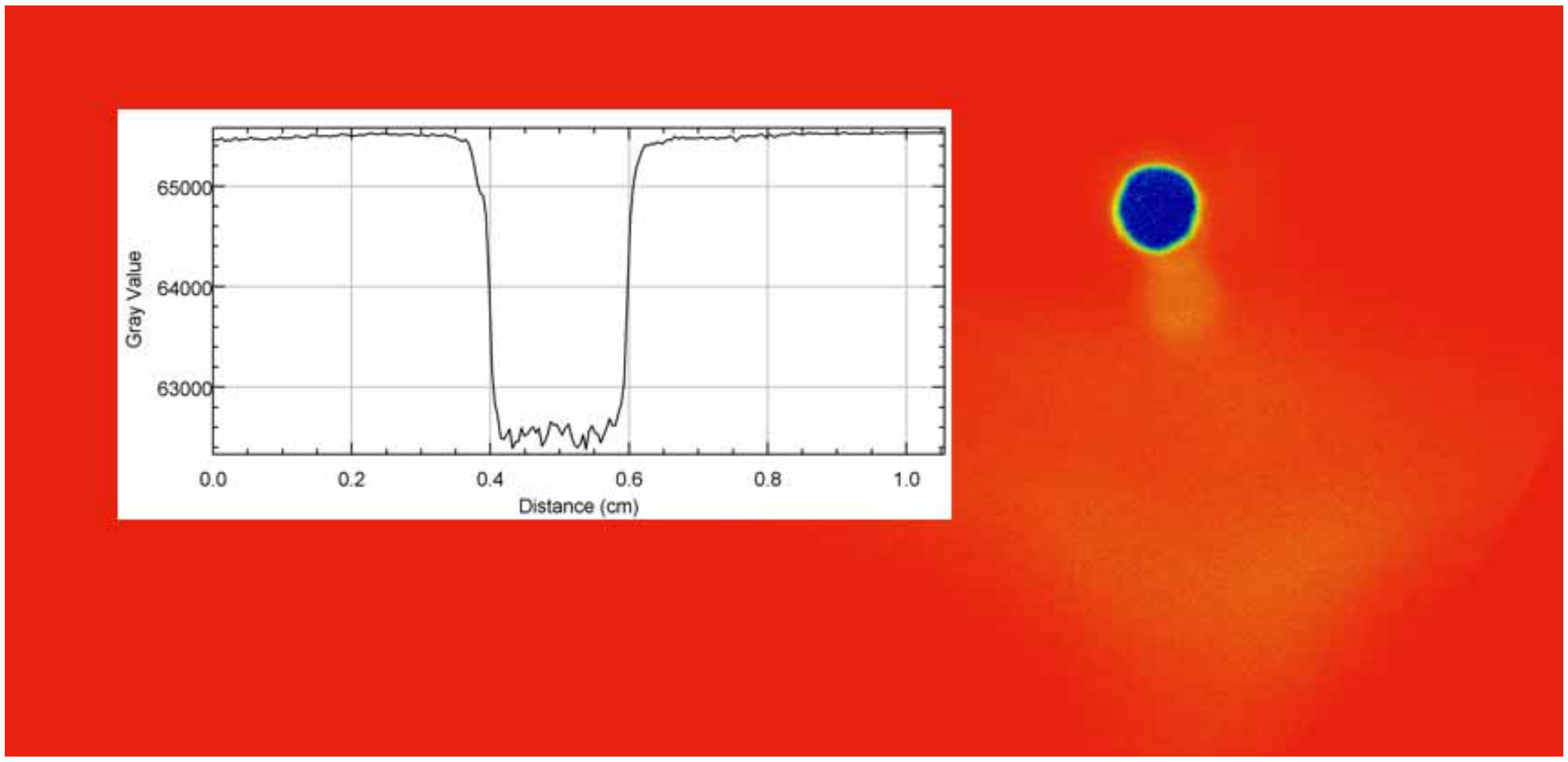

Figure 6.2-7. A 220-μm-diameter pinhole image and its density profile from 2x30-μm Mo *x* pinch with 10.3X magnification.

Again, by analyzing the density profile of the penumbra image, the radiation source size for this case can be estimated to be:

$$120\ \mu\text{m}/10.3 = 12\ \mu\text{m} \qquad (6.2\text{-}4)$$

The small values of radiation source size makes the *x* pinch a very attractive tool for high-resolution imaging of a small objects transparent to *x*-pinch radiation. As was indicated earlier (see the introduction for other *x*-pinch radiation machines and Table 5.7-1 for "hot spot" size predictions), the radiation source size is likely much smaller then evaluated here. Indeed, the penumbra size, we saw in our image experiment, was limited by the Gendex PSP plate spatial resolution (about 100 μm to 110 μm as was pointed in section 6.1), and higher quality image plates would be needed for a more precise *x*-pinch "hot spot" size calculation. In addition, the images obtained so far were time-integrated, and the overlapping effects from a moving "hot spot", or, even from multiple "hot-spot" objects, can blur the resulting image. Figure 6.2-8 presents a 220-μm-diameter pinhole image from 2x30-μm Mo *x* pinch (shot ID 656) when multiple spots can be observed. It is also interesting to see, that a bright neck structure and the *x*-pinch jet are formed at about 240-μm distance from the brightest "hot spot" object.

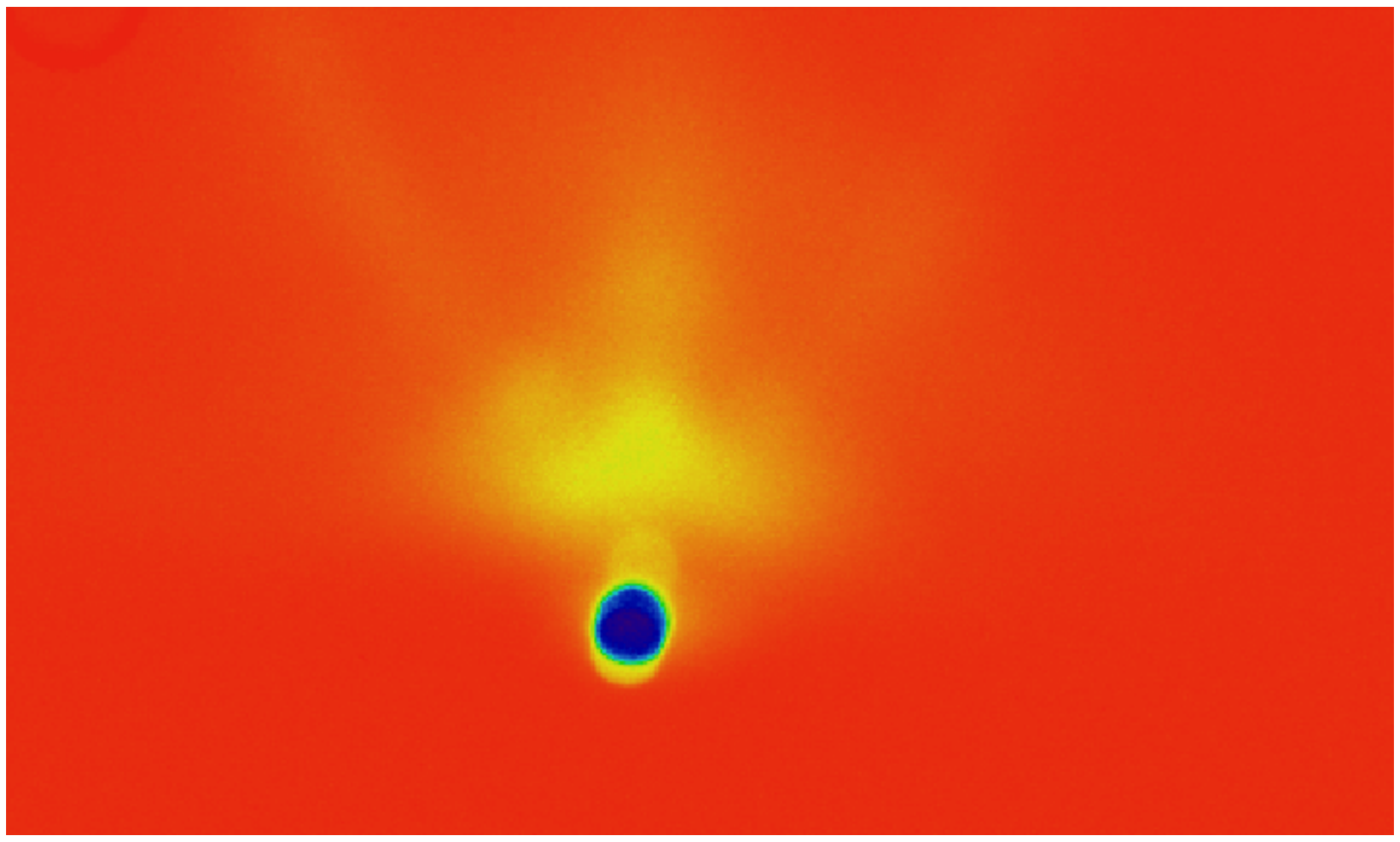

Figure 6.2-8. A time-integrated 220-μm-diameter pinhole image from 2x30-μm Mo *x* pinch.

## 6.3 Step-wedge spectrum deconvolution

A step-wedge-filter density profile can be described by the following convolution integral of the unknown source spectrum and the step-wedge response function as follow:

$$D_i = A \int_{E_1}^{E_2} S(E,t) F_i(E) dE dt \qquad (6.3\text{-}1)$$

where

$D_i$ is the measured density profile from a step-wedge filter;

A is the normalization constant;

S(E, t) is an unknown source spectrum;

$F_i$ is the response function of a step-wedge filter at particular step i;

$E_1$ and $E_2$ are the integration limits.

The image-plate response function is assumed to be constant, and so, can be included in a normalization constant A. The goal of this section is to evaluate the unknown time-integrated spectrum S(E) at the energy range of the step-wedge filter sensitivity.

Figure 6.3-1 presents a typical step-wedge image obtained from 2x30-μm W *x* pinch at shot ID 651. The Al step-wedge filter was placed at the distance of about 36 cm away from *x*-pinch radiation source, and the Gendex PSP image plate was placed just behind the filter.

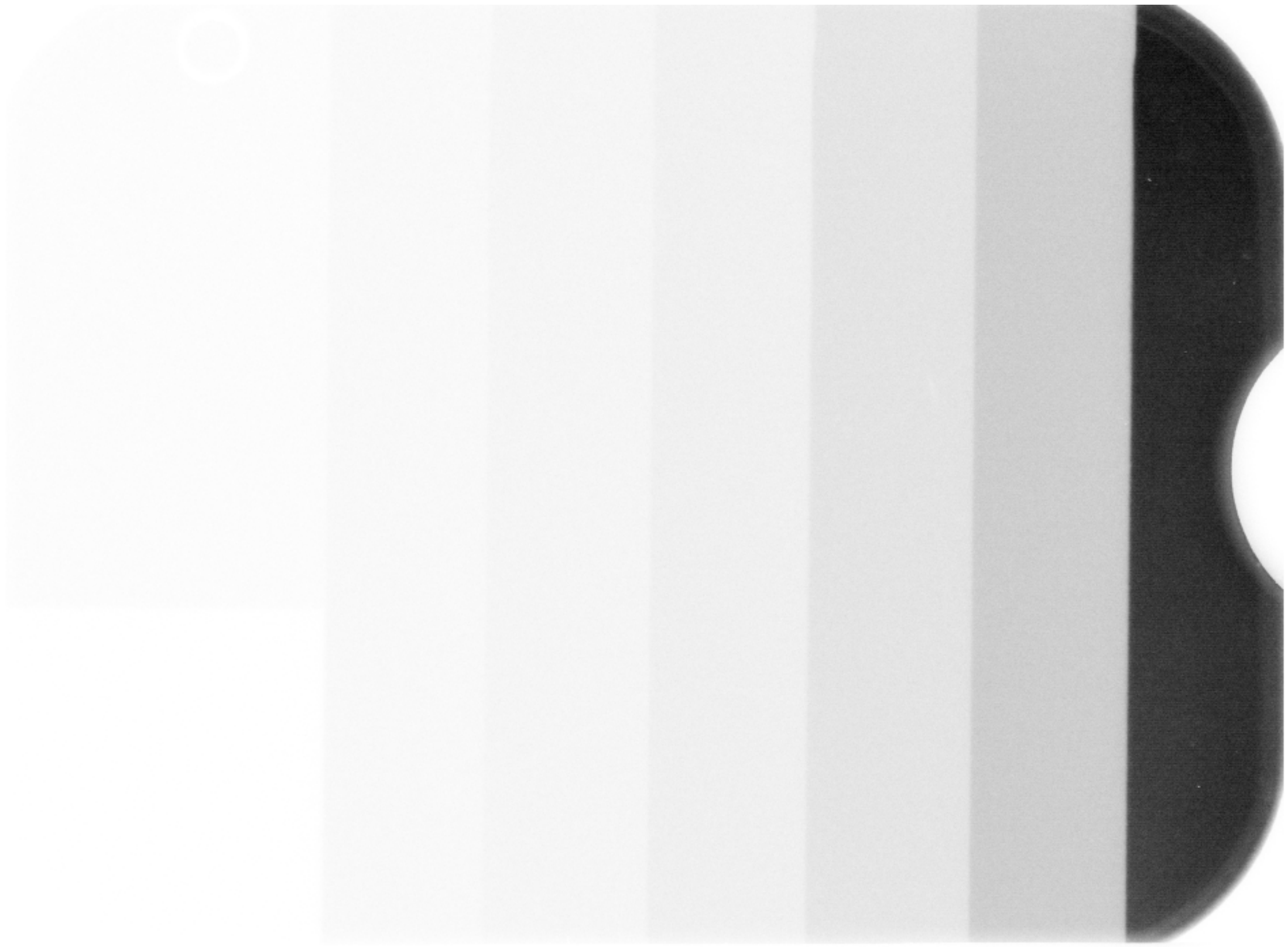

Figure 6.3-1. Al step-wedge image from 2x30-μm W *x* pinch.

The step-wedge filter response function for each of the filter steps was evaluated with XRD code [68] and presented in Figure 6.3-2.

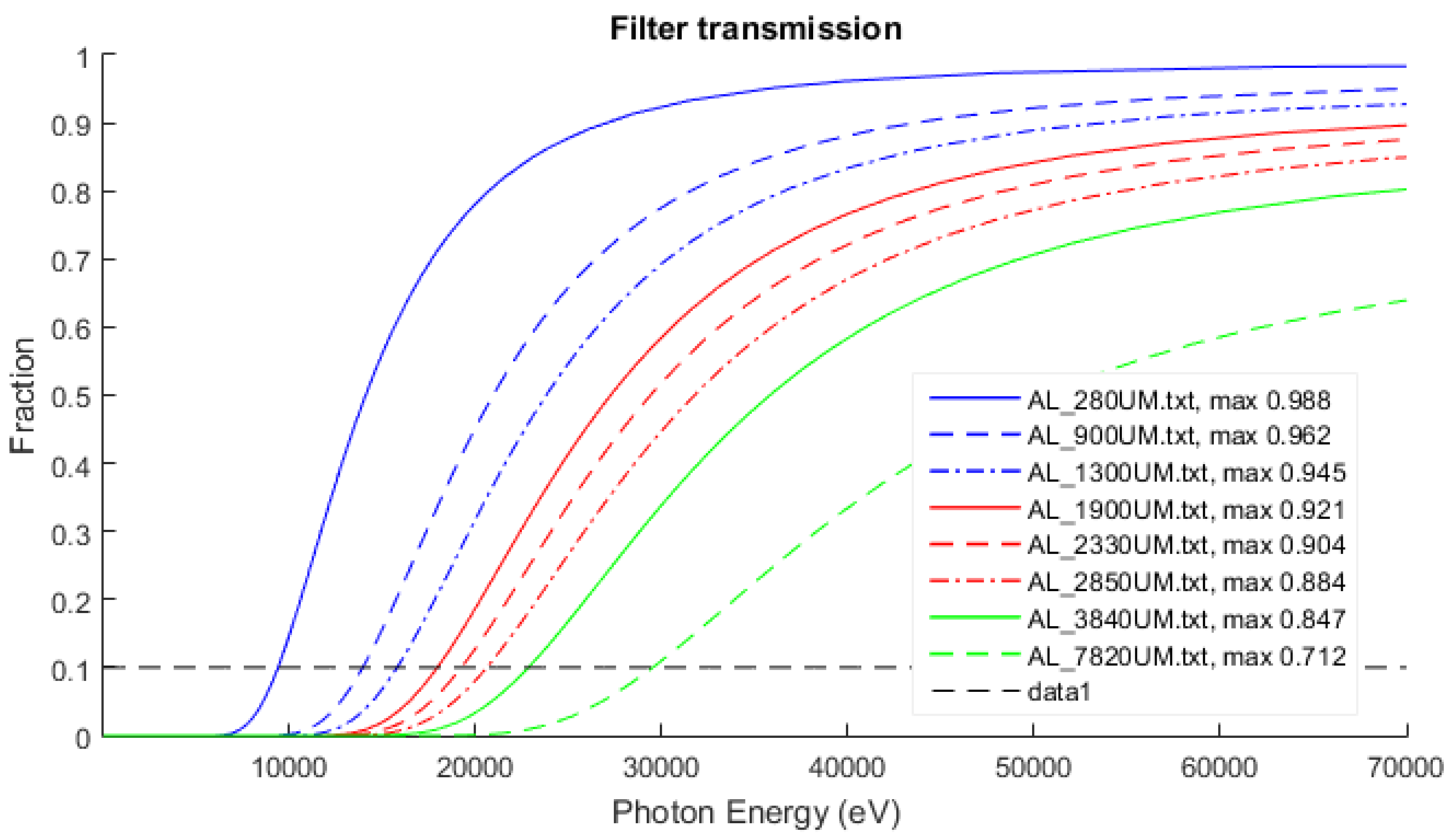

Figure 6.3-2. Al step-wedge response function.

The unknown spectrum in the energy range of the filter sensitivity was assumed to follow the exponential law with unknown temperature T:

$$S(E) = Be^{-\frac{h\nu}{kT}} \tag{6.3-2}$$

where

hν is the *x*-ray energy in eV;

k is Boltzmann's constant;

kT is the spectrum temperature in eV.

A series of simulations were run to evaluate the convolution integral (6.3-1) assuming spectrum (6.3-2) and to compare it with measured density profile for shot ID 651 presented above. Results of this analysis for some initial range of temperatures T are presented in Figure 6.3-3.

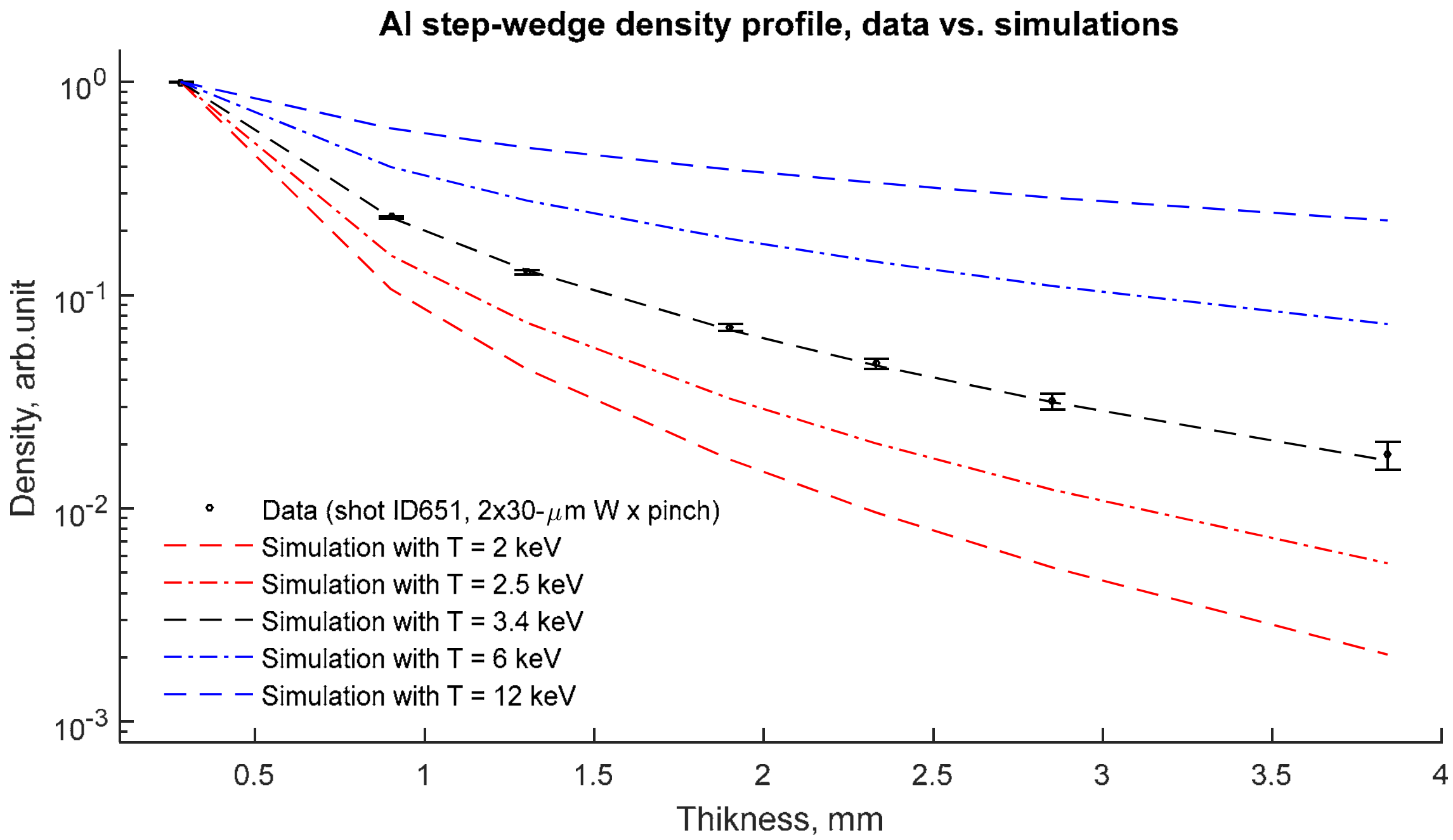

Figure 6.3-3. Density profile simulations for a 2x30-µm W *x* pinch.

As can be seen, the best fit to the experimental density profiles is observed when the exponential spectrum with a temperature of (3.4±0.1) keV is assumed in simulations. The error was evaluated by running the simulation with very fine temperature steps and by selecting all simulations which fit the experimental bars. This approximation is valid in the energy range of the step wedge filter sensitivity from about 9 keV up to 23 keV.

Figure 6.3-4 presents the density profile data for shot ID 653 performed with a 2x30-µm Mo *x* pinch and we compare these data with simulations as described above. As can be seen, the data can be fairly well explained when the exponential spectrum with temperature T of about (4.5±0.2) keV is assumed in simulation.

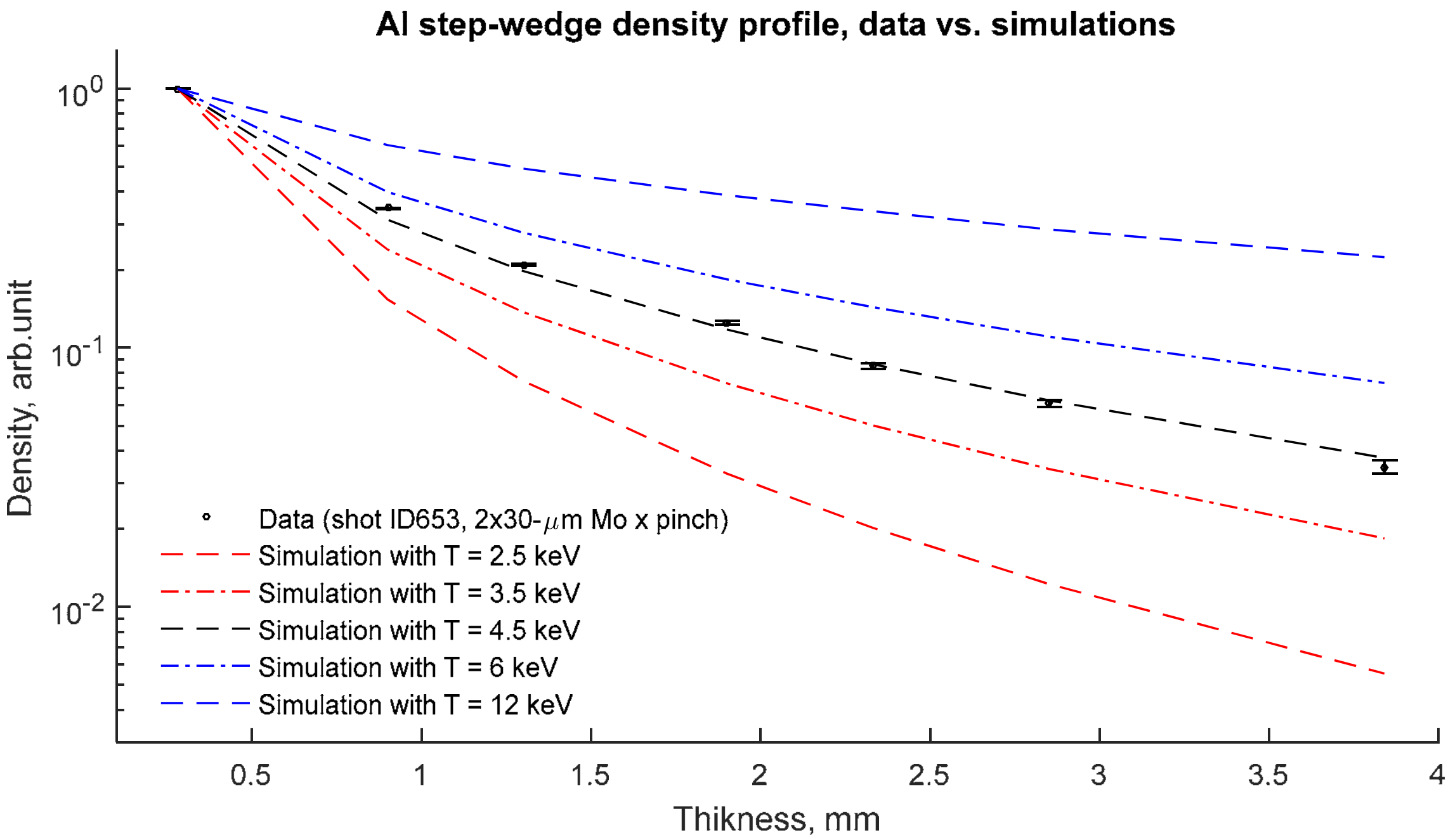

Figure 6.3-4. Density profile simulations for a 2x30-μm Mo *x* pinch.

To conclude, we developed a method to evaluate a time-integrated spectrum at the energy range from 9 keV to about 23 keV based on analysis of density profile function of Al step-wedge filter. For 2x30-μm W *x* pinches, the time-integrated x-ray spectrum can be approximated by an exponential shape having a temperature of about (3.4±0.1) keV, and for 2x30-μm Mo *x*-pinches the temperature is about (4.5±0.2) keV. We expect a higher-Z plasma to have a lower temperature due to radiative cooling, which is in qualitative agreement with a present measurements.

# 7 CONCLUSIONS AND REMARKS

The main objective of this work was to develop and test a new, compact and portable high-current pulse generator for *x*-pinch applications at the Idaho Accelerator Center. Below is a brief summary of what has been accomplished and what can be done in the future.

## 7.1 *X*-pinch driver development and simulations

The literature suggests the current rate-of-rise delivered to a low-inductance *x*-pinch load must be *at least* 1 kA/ns with a peak current amplitude of at least 100 kA. We singled out these parameters as the main design criteria that our new high-current pulse generator must satisfy to successfully drive an *x*-pinch load.

Three different designs for the development of an *x pinch*, plasma-radiation-source generator at the Idaho Accelerator Center were initially considered: 1) an *x pinch*, plasma-radiation-source generator that incorporates the existing ISIS Marx power supply and pulse-forming lines (PFLs); 2) an *x*-pinch driver based on low-inductance capacitors and switches; and 3) an *x*-pinch high-current pulse generator based on LTD technology. Each of these different designs was carefully evaluated and reported [50] - [54], and the last option was seen as the most promising one.

Our final design is based on two "slow", LTD bricks coupled to a low-inductance *x*-pinch load. A simple electrical model of 2-LTD-Brick driver was developed and SCREAMER simulations predicted that the LTD driver could deliver about 180-kA peak current into an *x*-pinch load with a 160-ns, 10-90%, rise time when it fully charged 100 kV. The corresponding maximum current rate of rise is about 1.4 kA/ns, which is above the required 1 kA/ns value.

## 7.2 Mechanical design and high-current diagnostics

Mechanically, we managed to place two LTD bricks inside one solid unit. Our design minimizes the total driver inductance and makes it compact and portable. The driver requires no oil or $SF_6$–gas systems to operate, and can be easily relocated to a new experimental site without being disassembled. The size of our driver is only 0.7x0.3x0.3 meters and it weighs about 90 kg.

We built electrostatically shielded, non-integrating Rogowski coil with a very good signal-to-noise ratio and placed it inside our driver to monitor the total current delivered to an *x*-pinch load. We developed a calibration method that allows us to calibrate our Rogowski coil with accuracy better than 1%.

In addition, a B-Dot current monitor was fabricated and was installed in the inner anode-cathode section, to directly monitor the current reaching the load. The B-Dot current monitor was cross-calibrated with the Rogowskii coil. This allowed us to determine if all driver energy reached the *x*-pinch load.

Many 2-LTD-Brick driver configurations (charging line, trigger line, etc,) were initially tested, and many technical problems related to driver failures were solved. In summary, after a lot of trial and error, a "good" working driver configuration was established that delivered a high-current pulse to low-inductance *x*-pinch loads. After a stable, working driver configuration was established, more than 200 *x*-pinch shots were performed without any driver failures. Once built and optimized, the driver was amazingly stable in operation and required little or no maintenance.

## 7.3 2-LTD-Brick driver electrical characteristics

To estimate driver electrical parameters a series of short-circuit tests were performed with 1.48-cm-long, 2.9-mm-diameter Ni-wire load installed in the anode-cathode gap and immersed into dielectric oil. The total driver inductance and resistance were evaluated by comparison the experimental data with LTSpice LCR model, and they were found to be 69 nH and 42 mΩ, respectively. After subtracting the wire inductance, the total internal inductance of our driver was estimated to be about 60 nH. This small driver inductance value allows for effective coupling of our driver to a low-inductance, *x*-pinch load with the required 1-kA/ns current rate-of-rise.

In addition, to estimate the 2-LTD-Brick driver performance at different charging voltages, a series of shot-circuit tests with 2.54-cm-long, 5.08-cm-diameter brass-cylinder load was performed and was reported [55] [56]. It was found that the driver could deliver about 150/180/210-kA peak-current when it was charged to 60/70/80-kV, respectively. The corresponding current rise times, 10-90%, were about 215/209/207 ns and max dI/dt was about 0.7/0.9/1.1 kA/ns.

## 7.4 Summary of the driver’s *x*-pinch radiation performance

The radiation performance of the constructed 2-LTD-Brick driver was evaluated for several different *x pinches* and reported [57] [58] [59] [60] [61]. A single *x*-ray diode (XRD) with different filters was used as the primary diagnostic tool to monitor *x*-pinch radiated *x*-ray power. A special method was developed to estimate the total *x*-pinch radiation power and energy.

It appears, that not all *x-pinch* materials and configurations are "good" choices for our 2-LTD-Brick driver at 80 kV. Shots with 80/127-µm Cu and 15-µm W wires reveal some potential problems and we addressed one method to select *x* pinches having a "good" radiation performance. The wire materials were either too thick (for 80/127-µm Cu *x* pinches), or too thin (for 15-µm W *x* pinches), and proper wire materials and their thickness were selected, based on driver scaling parameters.

Shots, performed with 2x20-µm W, 2x30-µm W and 2x30-µm Mo *x* pinches revealed "good" *x*-pinch radiation performance with our new 2-LTD-Brick driver: fast and bright XRD signals were always formed at a predictable time. The best *x*-pinch radiation performance was achieved with 30-µm-diameter Mo wires, when 2- to 6-ns wide XRD signals were reproducibly formed in the 190-ns to 250-ns time window after current start. The estimated *x*-pinch radiation power was at least 2-4.5 GW with a total *x*-ray energy delivered during the radiation burst of at least 8-15 J for photon energies between 120 eV and 285 eV and 400 eV and 2.4 keV. Shots with 2x20-µm W, 2x30-µm W *x* pinches also performed well, but not as reproducibly as the 30-µm Mo *x* pinches.

Shots, performed with 2x30-μm Cu wires, seem also to generate bright and fast *x*-pinch radiation signals. However, a multiple XRD peak structure was often observed immediately after the first *x*-pinch radiation peak. These multiple structures are not reproducible from shot-to-shot and, often, the second or later XRD peaks have a higher radiation power than the first peak.

## 7.5 Possible applications of a new 2-LTD-Brick *x*-pinch driver

The radiation sources, produced with our 2-LTD-Brick driver, are fast (a few ns), bright (a few GW), and quite small (a few μm). So, many applications can be considered, when all or some of these radiation parameters are needed. For example:

- A study of the dynamics of rapidly evolving plasmas. The time resolution of our *x*-pinch signal is about 2-3-ns (or less) and the time windows can be varied for different *x pinches*. We can study, for example, the dynamics of a single exploding wire. The wire can be placed in a separate generator or can be installed in a same *x*-pinch driver load.
- The *x*-pinch radiation source is bright and, potentially as small as a few μm. So, we can take high-resolution, phase-contrast images of a range of small biological samples - spiders, for example.
- And many more applications, where unique *x*-pinch radiation parameters can be utilized individually or simultaneously.

In addition to the above applications, but even more significant, *x* pinches are a very unique source of high-energy density plasma (HEDP), which can be relatively simply generated in laboratory environments. Previously, almost all *x*-pinch installations were massive, costly, and were sited at large universities and national laboratories. Our new *x*-pinch radiation source can provide new HEDP research opportunities at the Idaho Accelerator Center and many other places.

## 7.6 Final conclusion and remarks

Let us conclude, that we have been successfully developed and tested a new, compact and portable high-current pulse generator to drive a low-inductance *x*-pinch load. The short-circuit test data confirmed the desired 1-kA/ns driver parameter, and shots with *x*-pinch load reveal the potential for *x*-pinch applications. Our best *x pinches* are able to generate fast (a few ns) and bright (more than GW) radiation pulses from potentially small (a few μm) "hot spot" radiation objects.

Our pulsed-power design is elegant, works well, and is very reproducible. The design could be improved. A new generation of capacitors have been developed by General Atomics Electronic Systems that have a thinner profile and have a higher peak current. Building LTD bricks using these new capacitors would yield a system with even lower inductance and higher peak current. Such an *x*-pinch driver would deliver a dI/dt greater than 1 kA/ns, deliver a higher peak current, and do so in a shorter rise time.

The vacuum power flow leading to the *x*-pinch load has not been optimized. The height of the vacuum insulator, the local vacuum feed geometries, and the vacuum-feed gaps can all be optimized. These changes would lead to a lower inductance driver and further increases in *x*-pinch performance.

While our *x*-pinch experiments demonstrate excellent performance with Mo *x* pinches, the performance of tungsten and copper *x* pinches has been erratic. Significant improvements in controlling the consistency and uniformity of the initial conditions of the *x* pinch are needed.

It is clear that many more *x*-pinch experiments will be needed to fully characterize the radiation performance of our new 2-LTD-Brick *x*-pinch driver. Even though it is easy to operate and maintain, *x*-pinch shots are time consuming, and such a characterization will require a significant time investment in the future. In addition, many questions could be addressed in future *x*-pinch experiments, for example:

- What is the detailed radiation spectrum for different wire materials? What is the spectrum from the hot spot, from the mini-diode, and from the bulk of the *x*-pinch plasma?
- What are *x*-ray fluxes from line radiation for different *x* pinches?
- What are the best *x* pinches for point-projection radiography?
- How does one maximize the total *x*-pinch radiation yield?
- What are the current and dI/dt requirements as a function of *x*-pinch material?
- and more…

Finally, our new 2-LTD-Brick driver is very compact and portable. It has proven to be very reliable and has a shot rate limited only by the replacement of the *x*-pinch load. The excellent *x*-pinch performance suggests that many research projects using *x* pinches may be possible in the near future.

# 8 BIBLIOGRAPHY

[1] M. Zakharov, G. V. Ivanenkov, A. A. Kolomenskii, S. A. Pikuz, A. I. Samokhin, and J. Ullschmeid, *Sov. Tech. Phys. Lett.,* vol. 8, p. 456, 1982.

[2] S. V. Lebedev, F. N. Beg, S. N. Bland, J. P. Chittenden, A. E. Dangor, M. G. Haines, M. Zakaullah, S.A. Pikuz, T.A. Shelkovenko, and D.A. Hammer, "X-ray backlighting of wire array Z-pinch implosions using X pinch," *Rev. Sci. Instrum.,* vol. 72, no. 1, p. 671, 2001.

[3] Leonid E. Aranchuk, Jean Larour, and Alexandre S. Chuvatin, "Experimental Study of X-Pinch in a Submicrosecond Regime," *IEEE Transaction on Plasma Science,* vol. 33, no. 2, pp. 990-996, Apr. 2005.

[4] A. P. Artyomov, A. V. Fedyunin, S. A. Chaikovsky, V. I. Oreshkin, I. V. Lavrinovich, and N. A. Ratakhin, "Determining Thermodynamic Parameters of Aluminum X-Pinch Plasma," *Technical Physics Letters,* vol. 39, no. 1, p. 12, 2013.

[5] J. P. Chittenden, A. Ciardi, C. A. Jennings, S. V. Lebedev, D. A Hammer, S. A. Pikuz, and T. A. Shelkovenko, "Structural Evolution and Formation of High-Pressure Plasmas in X Pinches," *Phys. Rev. Lett.,* vol. 98, p. 025003, 2007.

[6] G. V. Ivanenkov, S. A. Pikuza, T. A. Shelkovenko, J. Greenly, D. B. Sinars, and D. A. Hammer, "Formation, Cascade Development, and Rupture of the X-Pinch Neck," *Journal of Experimental and Theoretical Physics,* vol. 91, no. 3, pp. 469-478, 2000.

[7] G. V. Ivanenkov, S. A. Pikuz, D. B. Sinars, V. Stepnievski, D. A. Hammer, and T. A. Shelkovenko, "Microexplosion of a Hot Point in an X-Pinch Constriction," *Plasma Physics Reports ,* vol. 26, no. 10, pp. 868-874, 2000.

[8] T. A. Shelkovenko, D. B. Sinars, S. A. Pikuz, and D. A. Hammer, "Radiographic and spectroscopic studies of X-pinch plasma implosion dynamics and x-ray burst emission characteristics," *Phys. Plasmas,* vol. 8, no. 4, p. 1305, Apr. 2001.

[9] S. A. Pikuz, D. B. Sinars, T. A. Shelkovenko, K. M. Chandler, D. A. Hammer, G. V. Ivanenkov, W. Stepniewski, and I. Yu. Skobelev, "High Energy Density Z-Pinch Plasma Conditions with Picosecond Time Resolution," *Phys. Rev. Let. ,* vol. 89, no. 3, p. 035003, 2002.

[10] D. B. Sinars, S. A. Pikuz, T. A. Shelkovenko, K. M. Chandler, and D. A. Hammer, "Temporal parameters of the X-pinch x-ray source," *Rev. Sci. Instrum.,* vol. 72, p. 2948, 2001.

[11] S. A. Pikuza, T. A. Shelkovenkoa, D. B. Sinarsa, and D. A. Hammer, "Temporal Characteristics of X-ray Emission from X-Pinches," *Plasma Physics Reports,* vol. 32, no. 12, p. 1020–1033, 2006.

[12] T. A. Shelkovenko, D. B. Sinars, S. A. Pikuz, K. M. Chandler, and D. A. Hammer, "Point-projection x-ray radiography using an X pinch as the radiation source," *Rev. Sci. Instrum.,* vol. 72, p. 667, 2001.

[13] Byung Moo Song, S. A. Pikuz, T. A. Shelkovenko, and D. A. Hammer, "Small size X-pinch radiation source for application to phase-contrast x-ray radiography of biological specimens," in *Nuclear Science Symposium Conference Record*, 2002.

[14] Leonid E. Aranchuk and Jean Larour, "Absolute Spectral Radiation Measurements from 200-ns 200-kA X-Pinch in 10-eV–10-keV Range With 1-ns Resolution," *IEEE Transaction on Plasma Science,* vol. 37, no. 4, p. 575, Apr. 2009.

[15] D.B. Sinarsa, S.A. Pikuza, T.A. Shelkovenko, K.M. Chandlera, D.A. Hammer, J.P. Apruzeseb, "Time-resolved spectroscopy of Al, Ti, and Mo X pinch radiation using an X-ray streak camera," *Journal of Quantitative Spectroscopy and Radiative Transfer,* vol. 78, p. 61, 2002.

[16] S. A. Pikuz, T. A. Shelkovenko, V. M. Romanova, D. A. Hammer, A. Ya. Faenov, V. A. Dyakin, and T. A. Pikuz, "High-luminosity monochromatic x-ray backlighting using an incoherent plasma source to study extremely dense plasmas (invited)," *Rev. Sci. Instrum.,* vol. 68, p. 740, 1997.

[17] T. A. Shelkovenko, S. A. Pikuz, A. R. Mingaleev, and D. A. Hammer, "Studies of plasma formation from exploding wires and multiwire arrays using x-ray backlighting," *Rev. Sci. Instrum.,* vol. 70, p. 667, 1999.

[18] S. A. Pikuz, T. A. Shelkovenko, A. R. Mingaleev, D. A. Hammer, and H. P. Neves, "Density measurements in exploding wire-initiated plasmas using tungsten wires," *Phys. Plasmas,* vol. 6, p. 4272, 1999.

[19] S. A. Pikuz, T. A. Shelkovenko, D. B. Sinars, J. B. Greenly, Y. S. Dimant, and D. A. Hammer, "Multiphase Foamlike Structure of Exploding Wire Cores," *Phys. Rev. Lett.,* vol. 83, p. 4313–4316, 1999.

[20] D. B. Sinars, Min Hu, K. M. Chandler, T. A. Shelkovenko, S. A. Pikuz, J. B. Greenly, D. A. Hammer, and B. R. Kusse, "Experiments measuring the initial

energy deposition, expansion rates and morphology of exploding wires with about 1 kA/wire," *Phys. Plasmas 8, 216 (2001).,* vol. 8, no. 1, p. 216, 2001.

[21] S.V.Lebedev, F.N.Beg, S.N.Bland, J.P.Chittenden, A.E.Dangor, M.G.Haines, S.A.Pikuz, T.A.Shelkovenko, "Effect of core-corona plasma structure on seeding of instabilities in wire array z-pinches," *Phys. Rev. Lett.,* vol. 85, no. 1, p. 98, 2000.

[22] J. D. Douglass, J. B. Greenly, D. A. Hammer, R. D. McBride, S. A. Pikuz, and T. A. Shelkovenko, "The Imaging of Z-Pinches Using X-Pinch Backlighting," in *DENSE Z-PINCHES: 6th International Conference on Dense Z-Pinches*, 25-28 July 2005.

[23] R. Zhang, T. Zhao, X. Zou, X. Zhu, and X. Wang, "X-pinch Applications In X-ray Radiography and Design of Compact Table-Top X-Pinch Device," in *Power Modulator and High Voltage Conference (IPMHVC)*, 2010 IEEE International.

[24] Xiaobing Zou, Xinxin Wang, Rui Liu, Tong Zhao, Naigong Zeng, Yongchao Zhao, and Yanqiang Du, "X-Pinch And Its Applications In X-ray Radiograph," in *AIP Conf. Proc.* , 2009.

[25] Byung Moo Song, T. A. Shelkovenko, S. A. Pikuz, M. A. Mitchell, K. M. Chandler, and D. A. Hammer, "X Pinch X-Ray Radiation above 8 keV for Application to High-Resolution Radiography of Biological Specimens," *IEEE Transaction on Nuclear Science, 51(5),,* vol. 51, no. 5, p. 1979, Oct. 2004.

[26] S.C.Bott, D.J.Ampleford, S.N.Bland, J.P.Chittenden, S.V.Lebedev, J.B.A.Palmer, "Use of X-pinches to diagnose behavior of low density CH foams in wire array z-pinches," *Rev. Sci. Instrum.,* vol. 75, no. 10, p. 3944, 2004.

[27] F.N. Beg, R.B. Stephens, H.-W. Xu, D. Haas, S. Eddinger, G. Tynan, E. Shipton, B. DeBono, and K. Wagshal, "Compact X-pinch based point x-ray source for phase contrast imaging of inertial confinement fusion capsules," *Appl. Phys. Let.,* vol. 89, p. 101502, 2006.

[28] N. A. Ratakhin and V. F. Fedushchak, "Capacitors for Pulsed Power Systems," in *Beams 2002: 14'h International Conference on High-Power Particle Beams*, 2002.

[29] A.H. Bushnell, B. Song, J. Ends, R. Miller, D. Johnsod, J. Maenchen, "Design optimization of linear transformer driver (LTD) stage cell capacitors," in *Power Modulator Symposium, High-Voltage Workshop. Conference Record of the Twenty-Sixth International*, 23-26 May 2004.

[30] *Fast Pulse Capacitors, General Atomics Electronic Systems, Inc., San Diego, CA.*

[31] J. R. Woodworth, K. Hahn, J. A. Alexander, G. J. Denison, J. J. Leckbee, S. Glover, P. E. Wakeland, J. R. Blickem, R. Starbird, M. J. Harden, H. D. Anderson, F. R. Gruner, D. Van DeValde, "Gas switch studies for Linear Transformer Drivers," in *Pulsed Power Conference, 16th IEEE International*, 2007.

[32] J. R. Woodworth, J. A. Alexander, F. R. Gruner, W. A. Stygar, M. J. Harden, J. R. Blickem, G. J. Dension, F. E. White, L. M. Lucero, H. D. Anderson, L. F. Bennett, S. F. Glover, D. Van DeValde, and M. G. Mazarakis, "Low-inductance gas switches for linear transformer drivers," *Phys. Rev. ST Accel. Beams,* vol. 12, p. 060401, 2009.

[33] Woodworth, J.R., Alexander, J.A., Stygar, W.A., Bennett, L.F., Anderson, H.D., Harden, M.J., Blickem, J.R., Gruner, F.R., White, R., "Low inductance switching

studies for Linear Transformer Drivers," in *Pulsed Power Conference, PPC '09. IEEE*, 2009.

[34] *High voltage switches, L-3 Applied Technology.*

[35] Chuan Wang, Xingdong Jiang, Jianzhong Wang, Sumin Wei, Naigong Zeng, Tianjue Zhang, "The redesign installation of light II-A pulsed power generator and its potential application," in *Proceedings of PAC09*, Vancouver, Canada., 2009.

[36] X. Zou, R. Liu, N. Zeng, M. Han, J. Yuan, X. Wang, and G. Zhangx, "A pulsed power generator for x-pinch experiments," *Laser and Particle Beams,* vol. 24, p. 504, 2006.

[37] H. Chuaqui, E. Wyndham, C. Friedli and M. Favre, "Llampüdkeñ: a high-current, low-impedance pulser employing an auxiliary exponential transmission line," *Laser and Particle Beams,* vol. 15, no. 2, pp. 241-248, 1997.

[38] J. Maenchen, H. T. Sheldon, G. D. Rondeau, J. B. Greenly, and D. A. Hammer, "Voltage and current measurements on high power self-magnetically insulated vacuum transmission lines," *Rev. Sci. Instrum.,* vol. 55, p. 1931, 1984.

[39] D.H. Kalantar and D.A. Hammer, "The x-pinch as a point source of x rays for backlighting," *Rev. Sci. Instrum.,* vol. 66, p. 779, 1995.

[40] R. M. Gilgenbach, M. R. Gomez, J. Zier, W. W. Tang, D. French, Y. Y. Lau, M. G. Mazarakis, M. E. Cuneo, M. D. Johnson, B. V. Oliver, T. A. Mehlhorn, A. A. Kim, and V. A. Sinebryukhov, "MAIZE: a 1 MA LTD-Driven Z-Pinch at The University of Michigan," in *AIP Conf. Proc*, Alexandria, Virginia (USA), 2009.

[41] T. A. Shelkovenko, S. A. Pikuz, J. D. Douglass, R. D. McBride, J. B. Greenly, and D. A. Hammer, "Multiwire X-Pinches at 1-MA Current on the COBRA Pulsed-Power Generator," *IEEE Transaction on Plasma Science,* vol. 34, no. 5, p. 2336, 2006.

[42] A.S. Safronova, V.L. Kantsyrev, A.A. Esaulov, N.D. Ouart, U.I. Safronova, I. Shrestha, and K.M. Williamson, "X-ray spectroscopy and imaging of stainless steel X-pinches with application to astrophysics," *Eur. Phys. J. Special Topics,* vol. 169, p. 155–158, 2009.

[43] H. Mitchell, J. M. Bayley, J. P. Chittenden, J. F. Worley, A. E. Dangor1, M. G. Haines, and P. Choi, "A high impedance mega-ampere generator for fiber z-pinch experiments," *Rev. Sci. Instrum.,* vol. 67, p. 1533, 1996.

[44] Jian Wu, Ai'ci Qiu, Gang Wu, Min Lv, Liangping Wang, Tianshi Lei, Ning Guo, Juanjuan Han, Xinjun Zhang, Hailiang Yang, Peitian Cong, and Mengtong Qiu, "X-Pinch Experiments on 1-MA "QiangGuang-1" Facility," *IEEE TRANSACTIONS ON PLASMA SCIENCE,* vol. 38, no. 4, p. 639, 2010.

[45] Yu. Kalinin, S. Anan'ev, Yu. Bakshaev, P. Blinov, V. Bryzgunov, A. Chernenko, S. Danko, E. Kazakov, A. Kingsep, V. Korolev, V. Mizhiritsky, S. Pikuz, T. Shelkovenko, V. Smirnov, S. Tkachenko, and A. Zelenin, "Study of the Multiwire X-Pinch as a Load for Mega-Ampere-Range Pulsed Power Generators," in *DENSE Z-PINCHES: Proceedings of the 7th International Conference on Dense Z-Pinches*, 12–21 August 2008.

[46] L. E. Aranchuk, A. S. Chuvatin, and J. Larour, "Compact submicrosecond, high current generator for wire explosion experiments," *Rev. Sci. Instrum.,* vol. 75, p. 69, 2004.

[47] S. C. Bott, D. M. Haas, R. E. Madden, U. Ueda, Y. Eshaq, G. Collins IV, K. Gunasekera, D. Mariscal, J. Peebles, and F. N. Beg, "250 kA compact linear transformer driver for wire array z-pinch loads," *Phys. Rev. ST Accel. Beams,* vol. 14, p. 050401, 2001.

[48] G. A. Mesyats, T. A. Shelkovenko, G. V. Ivanenkov, A. V. Agafonov, S. Yu. Savinov, S. A. Pikuz, I. N. Tilikin, S. I. Tkachenko, S. A. Chaikovskii, N. A. Ratakhin, V. F. Fedushchak, V. I. Oreshkin, A. V. Fedyunin, A. G. Russkikh, N. A. Labetskaya, ..., "X-Pinch Source of Subnanosecond Soft X-ray Pulses Based on Small-Sized Low-Inductance Current Generator," *Journal of Experimental and Theoretical Physics,* vol. 111, no. 3, pp. 363-370, 2010.

[49] "A compact submicrosecond, high current generator," *Rev. Sci. Instrum.,* p. 083504, 2009.

[50] R. V. Shapovalov, "Adaptation of the ISIS Induction-Cell Driver to a Low-Impedance X-Pinch Load," Internal report under DTRA1-11-1-0036 contract, Pocatello, ID, 2013.

[51] R. V. Shapovalov , W. Beezhold, V. I. Dimitrov (dec), "Adaptation of the ISIS Induction-Cell Driver to a Low-Impedance X-Pinch Drive," in *Proceedings of North America Particle Accelerator Conference*, Pasadena, CA, 2013.

[52] R. V. Shapovalov, "Design of a Compact, Portable Plasma-Radiation-Source Generator at the Idaho Accelerator Center," in *6th Canadian-American-Mexican Grad. Stud. Conf.*, Waterloo, Canada, Aug, 2013.

[53] R. V. Shapovalov, "A compact, portable 2-LTD-brick x-pinch driver at the Idaho Accelerator Center: design, fabrication and testing," Internal report under DTRA1-11-1-0036 contract, Pocatello, ID, 2014.

[54] R. V. Shapovalov, W. Beezhold, R. B. Spielman, "Design and Fabrication of Compact, Portable X-Pinch Driver based on 2 LTD bricks at the Idaho Accelerator Center," in *Bulletin of APS, 15th Ann. Meet. of Northwest Section of the APS*, Seattle, WA, 2014.

[55] R. V. Shapovalov, "A new high-current (200 kA, 200 ns) pulser for x-pinch applications: low-inductance load testing and results," in *Bulletin of APS, 16th Ann. Meet. of Northwest Section of the APS*, Pullman, WA, 2015.

[56] R. V. Shapovalov and R. B. Spielman, "Short-Circuit Test Data of a new 2-LTD-Bricks x-pinch Driver at the Idaho Accelerator Center," in *IEEE Pulsed Power Conference & Symposium on Fusion Engineering*, Austin, 2015.

[57] R. V. Shapovalov, "Radiation Performance of A New, 2-Ltd-Brick X-Pinch Driver," in *7th Canadian-American-Mexican Graduate Student Physics Conference*, Oaxaca, Oaxaca, Mexico, Sep. 2015.

[58] R. V. Shapovalov, R. B. Spielman, "A novel compact & portable x-pinch radiation source generator (200-kA, 200-ns) at the Idaho Accelerator Center: timelines, shots & discussions," in *Graduate Student Symposium*, Pocatello, ID, Apr. 2016.

[59] R. V. Shapovalov, R. B. Spielman, "Overview of a new compact and portable x-pinch radiation source generator (200-kA, 200-ns) for physics and industry applications," in *APS meeting*, Salt Lake, UT, Apr. 2016.

[60] R. V. Shapovalov, R. B. Spielman, "A Brief Report on Time-Integrated Spectrum Measurements From W X-Pinch with 1 kA/ns Current Rate in the Energy Range from 10 to 20 KeV," in *17th Annual Meeting of the APS Northwest Section*, Penticton, BC, Canada, May 2016.

[61] R. V. Shapovalov, R. B. Spielman, "Mo X-Pinch Performance from a New Compact and Portable 1-kA/ns 2-LTD-Brick Driver," in *43 IEEE International Conference on Plasma Science*, Banff, Alberta, Canada, June 2016.

[62] M. G. Mazarakis and R. B. Spielman, "A Compact, High-Voltage E-Beam Pulser," in *Proceeding of the 12th IEEE International Pulsed Power Conference*, Monterey, CA, 1999.

[63] M. G. Mazarakis, W. E. Fowler, F. W. Long, D. H. McDaniel, C. L. Olson, S. T. Rogowski, R. A. Sharpe, K. W. Struve, and A. A. Kim, "High current fast 100-ns LTD driver development in Sandia Laboratory," in *Proceeding of the 15th IEEE International Pulsed Power Conference*, Monterey, CA, 2005.

[64] "Naval Research Laboratory LTD Generator. User, Assembly and Maintenance Manual.," Ktech corp, Albuqurque, NM, 2011.

[65] Rick B. Spielman, Mark L. Kiefer, Kelley L. Shaw, Ken W. Struve and Mel M. Winde, "Screamer A pulsed Power Design Tool: User's Guide for Version 3.3.1," Sandia Corporation, 2014.

[66] Y. Gryzin and R. B. Spielman, "Screamer V4.0: a Powerful Circuit Analysis Code," in *Pulsed Power Conference*, Austin, 2015.

[67] "http://www.solidworks.com/," SOLIDWORKS CORP.. [Online].

[68] R. Spielman, *Notes on design of RG-223-based non-integrated Rogowski coil.*

[69] "Digital Phosphor Oscilloscope - TDS5034B, TDS5054B, TDS5104B Data Sheet," Tektronix, 2010.

[70] R. J. Adler, *Pulse Power Formulary,* Marana, AZ: North Start High Voltage, 2008.

[71] *LTspice IV Getting Started Guide, 2011 Linear Technology.*

[72] Day, R.H., Lee, P., Saloman, E.B., Nagel, D.J., "X RAY DIODES FOR LASER FUSION," Los Alamos Scientific Laboratory, Los Alamos, 1981.

[73] G. A. Chandler, C. Deeney, M. Cuneo, D. L. Fehl, J. S. McGurn, R. B. Spielman, J. A. Torres, J. L. McKenney, J. Mills and K. W. Struve, "Filtered x-ray diode diagnostics fielded on the Z accelerator for source power measurements," *Rev. Sci. Instrum. ,* vol. 70, no. 1, p. 561, 1999.

[74] R. B. Spielman, "X-Ray detector: an x-ray radiation detector design code.," Sandia National Laboratory, Albuquerque, New Mexico, 1990.

[75] T. A. Shelkovenko, S. A. Pikuz, D. B. Sinars, K. M. Chandler, and D. A. Hammer, "X Pinch Plasma Development as a Function of Wire Material and Current Pulse Parameters," *IEEE TRANSACTIONS ON PLASMA SCIENCE,* vol. 30, no. 2, p. 567, 2002.

[76] Zucchini F, Bland SN, Chauvin C, Combes P, Sol D, Loyen A, Roques B, Grunenwald J, "Characteristics of a molybdenum X-pinch X-ray source as a probe

source for X-ray diffraction studies.," *Rev Sci Instrum,* vol. 86, no. 3, p. 033507, 2015.

[77] S. Bland, *private communications.*

[78] "Gendex Software & Drivers," [Online]. Available: http://www.gendex.com/software-drivers.

[79] A. G. Farman and T. T. Farman, "A comparison of 18 different x-ray detectors currently used in dentistry," *Oral and Maxillofacial Radiology,* vol. 99, no. 4, pp. 485-489, 2005.

[80] K. Chouffani, *private communications.*

[81] "ImageJ An open platform for scientific analysis," [Online]. Available: http://imagej.net/Welcome.

[82] D. G. Pellinen, M. S. Di Capua, S. E. Sampayan, H. Gerbracht, and M. Wang, "Rogowski coil for measuring fast, high-level pulsed current," *Rev. Sci. Instrum.,* vol. 51, no. 11, p. 1535, 1980.

# APPENDIXES

# (BY REQUEST ONLY)